\documentclass[12pt, a4paper,twoside]{Thesis} % Paper size, default font size and one-sided paper
\usepackage{wrapfig}
\usepackage{lscape}
\usepackage{rotating}
\usepackage{graphicx}
\usepackage{caption}
\usepackage{amsmath}
\usepackage{cancel}
\usepackage{quotchap}
\usepackage{extarrows}
\usepackage{caption}
\allowdisplaybreaks
\def\bea{\begin{eqnarray}}
	\def\eea{\end{eqnarray}}
\def\be{\begin{equation}}
	\def\ee{\end{equation}}
\def\nn{\nonumber}

\graphicspath{{Pictures/}} % Specifies the directory where pictures are stored
\usepackage[square, numbers,sort&compress]{natbib} % Use the natbib reference package - read up on this to edit the reference style; if you want text (e.g. Smith et al., 2012) for the in-text references (instead of numbers), remove 'numbers' v

\hypersetup{urlcolor=black, colorlinks=true} % Colors hyperlinks in blue - change to black if annoyingv`	
\title{\ttitle} % Defines the thesis title - don't touch this

\begin{document}
\makeatletter
\renewcommand*{\NAT@nmfmt}[1]{\textsc{#1}}
\makeatother

% prints author names as small caps

\frontmatter % Use roman page numbering style (i, ii, iii, iv...) for the pre-content pages

\setstretch{1.6} % Line spacing of 1.6 (double line spacing)

% Define the page headers using the FancyHdr package and set up for one-sided printing
\fancyhead{} % Clears all page headers and footers
\rhead{\thepage} % Sets the right side header to show the page number
\lhead{} % Clears the left side page header

\pagestyle{fancy} % Finally, use the "fancy" page style to implement the FancyHdr headers

\newcommand{\HRule}{\rule{\linewidth}{0.5mm}} % New command to make the lines in the title page

% PDF meta-data
\hypersetup{pdftitle={\ttitle}}
\hypersetup{pdfsubject=\subjectname}
\hypersetup{pdfauthor=\authornames}
\hypersetup{pdfkeywords=\keywordnames}

%----------------------------------------------------------------------------------------
%	TITLE PAGE
%----------------------------------------------------------------------------------------

\begin{titlepage}
\begin{center}

\HRule \\[0.4cm] % Horizontal line
{\Large \bfseries \ttitle}\\[0.3cm] % Thesis title
\HRule \\[1.5cm] % Horizontal line
 
\large \textit{A thesis submitted in partial fulfillment of the requirements for the degree of	Doctor of Philosophy}\\[0.3cm] % University requirement text
\textit{by}\\[0.4cm]

{\large\bf PRITAM DAS}\\
Registration no.: TZ155506 of 2014

\vfill
\graphicspath{ {./Figures/} }
\begin{figure}[hb]
  \centering
  \includegraphics[width=0.25\linewidth]{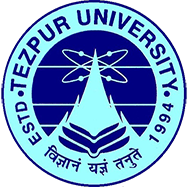}
\end{figure}

\vspace{0.8cm}
\DEPTNAME\\ % Research group name and department name
\textsc{ \UNIVNAME\\ India-784028}\\[1.5cm] % University name
\large \today\\[2cm] % Date

\end{center}

\end{titlepage}
%\afterpage{\null\newpage}
\newpage
\

\newpage

\section*{\Large List of publications: }
%		\skip
%\thispagestyle{empty}
%\rule{\textwidth}{0.4pt}
\begin{enumerate}
	\item
	{\bf Das P.}, Mukherjee A., Das M. K.; {\it Active and sterile neutrino phenomenology with $A_4$ based minimal extended seesaw}, {\bf Nuclear Physics B, Volume 941, April 2019, Pages 755-779}; arxiv:1805.092132
	\item{\bf Das P.}, Das M. K.,{\it  Phenomenology of $keV$ sterile neutrino within minimal extended seesaw}. arXiv: 1908.08417. {\bf IJMPA; Vol. 35, No. 22, 2050125 (2020)}
	\item {\bf Das P.}, Das M. K., Khan N. {\it Phenomenological study of neutrino mass, dark matter and baryogenesis within the framework of minimal extended seesaw}. arXiv: 1911.07243. {\bf JHEP; 2020, Article number: 18 (2020) }
	\item {\bf Das P.}, Das M. K., Khan N.; {\it A new feasible dark matter region in the singlet scalar scotogenic model.} {\bf  Nuclear Physics B;
		Volume 964, March 2021, 115307} ({\it 		arXiv:2001.04070})
	\item  Sarma L., {\bf Das P.}, Das M. K.; {\it Scalar dark matter and leptogenesis in the minimal scotogenic model.} {\bf Nuclear Physics B;
		Volume 963, February 2021, 115300
	} arXiv:2004.13762
	\item {\bf Das P.}, Das M. K., Khan N.; {\it Five-zero texture in neutrino-dark matter model within the framework of minimal extended seesaw.} {\bf arXiv:2010.13084} ({\it Under communication})
	\item {\bf Das P.}, Das M. K., Khan N.; {\it The FIMP-WIMP dark matter and Muon g-2 in the extended singlet scalar model} {\bf arXiv:2104.03271 } ({\it Under communication})
	\item {\bf Das P.}, Das M. K., Khan N.; {\it Extension of Hyperchargeless Higgs Triplet Model} {\bf arXiv:2107.01578} ({\it Under communication})
\end{enumerate}
\clearpage
%\vspace{.6cm}
%----------------------------------------------------------------------------------------
%	DECLARATION PAGE
%	Your institution may give you a different text to place here
%----------------------------------------------------------------------------------------

%\vspace{0.cm}

%	ABSTRACT PAGE
%----------------------------------------------------------------------------------------

\addtotoc{Abstract} % Add the "Abstract" page entry to the Contents
\abstract{\addtocontents{toc}{\vspace{1em}} % Add a gap in the Contents, for aesthetics
\setstretch{1.2}
{Discovery of neutrino oscillation \cite{Fukuda:1998ub, Ashie:2004mr,Abe:2016nxk,boger2000sudbury,Evans:2013pka,Abe:2011sj,Ahn:2012nd,Abe:2011fz,An:2012eh,Araki:2005qa, Eguchi:2002dm} imply the fact that neutrinos have mass and mixing, which can be regarded as one of the pioneer discoveries in the field of particle physics. In 2012, the Higgs mechanism was confirmed after the discovery of scalar boson at LHC, which completes the SM of particle physics. Even though this scalar boson was the same Higgs boson that the SM predicted, the experimental results still have room for the extended scalar sector. Despite of its many successes, the standard model (SM) of particle physics was not sufficient enough to address neutrino masses, along with other discrepancies like observed matter-antimatter asymmetry, presence of dark matter, {\it etc.}, therefore beyond the standard model (BSM) frameworks are inevitable to address current inconsistencies. 
	
	There are immense progresses in theory as well as experiments in the neutrino sector \cite{Araki:2005qa, Eguchi:2002dm,boger2000sudbury,Fukuda:1998ub, Ashie:2004mr,Abe:2011sj}. At the current date, the three active neutrino results have entered the era of precision measurements. Back in 2011, LSND (Liquid Scintillator Neutrino Detector) \cite{Aguilar:2001ty} individually reported the existence of a heavy generation of neutrinos, where an electron antineutrino was observed in a pure muon antineutrino beam. In the three neutrino paradigm, we have two mass squared differences with very small values ($\Delta m^2_{Sol}=7.4\times 10^{-5}$ eV$^2$ and $\Delta m_{Atm}^2=2.4\times 10^{-3}$ eV$^2$), on the other hand, what LSND incorporated was $\Delta m^2_{LSND}\sim \mathcal{O}(1)$ eV$^2$. Recently in 2018, MiniBooNE \cite{Aguilar-Arevalo:2018gpe} experiments also reported the same at $6\sigma$ CL. Simultaneously, the Gallium experiments GALLEX and SAGE and several observations of reactor antineutrino fluxes at a short baseline experiment also detected anomalies with eV scaled mass splitting \cite{Abdurashitov:2005tb,Abdurashitov:1999zd,giunti2012update, giunti2011statistical}. This extra generation(s) of a neutrino is termed as $sterile$ neutrino since it cannot oscillate into other generations, like the active neutrinos. However, it is still unclear about the exact mass scale or the exact number of generations of sterile neutrino. 
	
	Flavour symmetries play influential role in particle physics in BSM \cite{Babu:2009fd,Altarelli:2005yp}. We have used $A_4$ and $Z_n$ ($n, \text{an integer} \ge2$) discrete flavour symmetry to construct different neutrino mass models in our entire thesis. In realizing tiny neutrino masses, seesaw mechanisms are the convenient and efficient mechanism \cite{Minkowski:1977sc,Mohapatra:1979ia,King:2003jb}. We have considered a single generation of sterile neutrino along with three active neutrinos, therefore minimal extended seesaw (MES) is extensively used throughout this thesis. MES is the simplest extension of the canonical type-I seesaw, where along with the active neutrinos, sterile neutrino masses are realized for a wider range (eV to $keV$) \cite{Barry:2011wb,Zhang:2011vh}. Under this MES framework, we tried to incorporate the inconsistencies of the SM $like$, neutrino mass, baryogenesis, absolute neutrino mass and dark matter in this thesis.
	
	Baryogenesis is the process that can explain the observed asymmetry of matter-antimatter at the current date. There are several methods available in the literature by which baryogenesis can be obtained. Among them, leptogenesis is one of the popular processes to explain baryogenesis. The right-handed neutrino present in the seesaw mechanism can decay to a lepton doublet and a scalar (Higgs) doublet or an anti-lepton doublet and a CP conjugate scalar (Higgs) doublet, which may results in very small lepton asymmetry by violating the lepton number. A portion of this produced lepton asymmetry can convert to baryon asymmetry of the Universe {\it via} the $sphaleron$ process. This mechanism is known as thermal leptogenesis \cite{Davidson:2008bu} and this is valid for heavy RH neutrinos ($M_{R}\sim10^{9-14}$ GeV). On the other hand, if two RH neutrinos are of nearly degenerate mass, then {\it via} one-loop corrections might have a resonantly enhanced lepton asymmetry produced termed as resonant leptogenesis \cite{Pilaftsis:2003gt}. Resonant leptogenesis can be studied for light RH neutrinos also ($M_{Ri}\sim$ TeV). 
	
	Neutrino-less double beta decay ($0\nu\beta\beta$) study can solve the absolute neutrino mass problem. In this thesis, along with the active neutrino contribution, we also explore the influence of a single generation of the sterile neutrino with mass $m_S\sim\mathcal{O}$(eV-$keV$) on the effective electron neutrino mass associate with $0\nu\beta\beta$ \cite{Abada:2018qok}. Significant bounds on neutrino parameters as well as on active-sterile mixing parameters are obtained from the $0\nu\beta\beta$ study. 
	
	It is a well-known fact that the Universe consists of about 4$\%$ ordinary matter, 27$\%$ dark matter and the rest 	69$\%$ is mysterious unknown energy called dark energy, which is assumed to be the cause of the accelerated expansion of the Universe. The SM of particle physics fails to provide a viable dark matter candidate. Thus,
	new physics beyond the SM is essential to explain dark matter. There are various established options {\it viz.} extra $Z_n$-odd ($n\ge$2, is an integer) scalar, fermion, and combined of them with various multiplets, e.g., singlets, doublets, triplets, quadruplets, etc. have been studied to explain the dark matter phenomenology. In this thesis, we have also addressed a few viable dark matter candidates. Active neutrinos cannot behave as a dark matter candidate, but heavy sterile neutrino ($m_S\sim\mathcal{O}(keV)$) with small mixing with the active neutrinos can be a potential dark matter candidate. We also projected scalar singlet or doublet (real and complex) as potential dark matter candidates under suitable conditions. 
	
	Keeping these in view, a brief outline of our thesis chapters are presented as follows: 
	
	In {\bf Chapter 1}, starting with a brief discussion on neutrinos, we have addressed the SM and interactions there-in. We then discussed the beyond standard model (BSM) frameworks and subsequently discussed the BSM topics that were our prime motivation. We have also introduced discrete symmetries $A_4$ and $Z_n$ using which we have formulated mass models and worked out possible phenomenological aspects in subsequent chapters.
	
	In {\bf Chapter 2}, we have studied a model of neutrino mass within the framework of minimal extended seesaw (MES), which plays a vital role in active and sterile neutrino phenomenology in (3+1) scheme. In this model, $A_4$ flavour symmetry is augmented by an additional $Z_4\times Z_3$  symmetry to constraint the Yukawa Lagrangian of the model. We used a non-trivial Dirac mass matrix, with broken $\mu-\tau$  symmetry, as the origin of leptonic mixing. Interestingly, such mixing structure naturally leads to the non-zero reactor mixing angle $\theta_{13}$. Non-degenerate mass structure for right-handed neutrino $M_R$ is considered. We have also considered three different cases for sterile neutrino mass, $M_S$ to check the viability of this model, within the allowed $3\sigma$ bound in this MES framework.
	
	In {\bf Chapter 3}, to study neutrino and dark matter in a single framework, we have extended our previous model from chapter 2 with a few changes. In this model, $A_4$ flavour symmetry and the discrete $Z_4$ symmetry are used to stabilize the dark matter and construct desired mass matrices for neutrino mass.
	We use a non-trivial Dirac mass matrix with broken $\mu - \tau$ symmetry to generate the leptonic mixing. A non-degenerate mass structure for right-handed neutrinos is considered to verify the observed baryon asymmetry of the Universe via the mechanism of thermal Leptogenesis. The scalar sector is also studied in great detail for a multi-Higgs doublet scenario, considering the lightest $Z_4$-odd as a viable dark matter candidate. A significant impact on the region of DM parameter space and the fermionic sector are found in the presence of extra scalar particles.
	
	In {\bf Chapter 4}, we have explored the possibility of a single generation of $keV$ scale sterile neutrino ($m_S$) as a dark matter candidate within the minimal extended seesaw (MES) framework and its influence in neutrinoless double beta decay ($0\nu\beta\beta$) study. Three hierarchical right-handed neutrinos were considered to explain neutrino mass. We also address baryogenesis via the mechanism of thermal leptogenesis. A generic model based on $A_4\times Z_4\times Z_3$ flavour symmetry is constructed to explain both normal and inverted hierarchy mass pattern of neutrinos. Significant results on effective neutrino masses are observed in presence of sterile mass ($m_S$) and active-sterile mixing ($\theta_{S}$) in $0\nu\beta\beta$. To establish sterile neutrino as a dark matter within this model, we have checked the decay width and relic abundance of the sterile neutrino, which restricted sterile mass ($m_S$) within some definite bounds. Constrained regions on the CP-phases and Yukawa couplings are obtained from $0\nu\beta\beta$ and baryogenesis results.  Co-relations among these observable are also established and discussed within this model.
	
	In {\bf Chapter 5}, we have studied a model of neutrino and dark matter within the framework of a minimal extended seesaw (MES). Here, five-zero textures are imposed in the final $(4\times4)$ active-sterile mass matrix, which significantly reduces free parameter in the model. Three right-handed neutrinos were considered, two of them have degenerate masses which help us to achieve baryogenesis $via$ resonant leptogenesis. A singlet fermion (sterile neutrino) with mass $\sim\mathcal{O}$(eV) has also considered, and we put bounds on active-sterile mixing parameters via neutrino oscillation data. Resonant enhancement of lepton asymmetry is studied at the TeV scale, where we have discussed a few aspects of baryogenesis considering the flavour effects. The possibility of improvement in effective mass from $0\nu\beta\beta$ in the presence of a single generation of sterile neutrino flavour is also studied within the fermion sector. In the scalar sector, the imaginary component of the complex singlet scalar ($\chi$) is behaving as a potential dark matter candidate and simultaneously, the real part of the complex scalar is associated with the fermion sector for sterile mass generation. A broad region of dark matter mass have analyzed from various annihilation processes. The VEV of the complex scalar plays a pivotal role in achieving the observed relic density at the right ballpark.
	
	Finally, we have discussed the overall conclusions and summary of our work that we carried out in this thesis in {\bf Chapter 6}. In the same chapter, we have also proposed a future plan for our work in the neutrino and dark matter sector.}

\clearpage % Start a new page

%----------------------------------------------------------------------------------------
%	ACKNOWLEDGEMENTS
%----------------------------------------------------------------------------------------

\setstretch{1.3} % Reset the line-spacing to 1.3 for body text (if it has changed)

\acknowledgements{\addtocontents{toc}{\vspace{1em}} % Add a gap in the Contents, for aesthetics

I would gladly take this opportunity to acknowledge those few people without whose help and support this thesis would not have seen the light of the day. 
Firstly, I would extend my sincerest gratitude to my supervisor, {\it Prof. Mrinal Kumar Das}, who has been the constant pillar of support at every step of this PhD journey in the most real sense. His dedication to research enthusiasm and pursuit of physics has been an invaluable source of inspiration and encouragement. I feel fortunate enough to be supervised by him, who always ensured that I have not deviated from my goals. Apart from being a supervisor, he is a great human being. His approachable, caring and nourishing nature made me learn a lot from him.

My sincere thanks to my doctoral committee members, {\it Prof. Nilakshi Das}, and {\it Dr. Moon Moon Devi} for their insightful comments, questions and valuable suggestions helped me a lot to progress in my research and developed a scientific temperament.  I want to thank all the faculty members from the Department of Physics, Tezpur University, for their academic support and encouragement since my M.Sc. days.

I owe my deepest thanks and appreciation to my collaborator {\it Dr. Najimuddin Khan} from the Indian Association for the Cultivation of Science, Kolkata, for his contributory suggestions and tips that genuinely enriched my abilities to pursue further research. I shall remain ever grateful to {\it Najim da} to teach me so much about technical stuff. It is indeed a privilege for me to work with him, and I am ever grateful to him for his help.

I revere the patience and endless support extended with love from {\bf Maa-Deuta}. Words can't express how grateful I am to them for all of the sacrifices that they have made on my behalf. Their immense faith in me always keeps me moving, no matter how challenging the situation is. I want special thanks to my twin sister {\it Prity (Maina)}, for her constant support from the beginning of my journey. She always believed in me that I could do it. No matter how bad the situation was, she was still beside me. I solely owe them this thesis. I want to thank my elder sisters {\it Karabi Kalita Saharia} and {\it Gitanjali Kalita Barman} who are another source of constant support and influence. My heartfelt thanks to brother-in-laws {\it Ujjwal Jyoti Saharia} and {\it Uday Barman} for being always so supportive to me. I feel blessed to have a supportive family for the tremendous comfort and love. I am forever thankful to the Almighty for giving me strength and patience even in the hardest times.

It is with immense gratitude that I confess the support and help from my friends. {\it Dr. Bijoy Sankar Baruah, Aftab Ansari} and {\it Ankita Bhagawati}, who are the constants since my masters days at TU. I genuinely acknowledge them from the heart for being so supportive and tolerating me. I want to thank {\it Pratikshya baa} and {\it Happy baa}, for their encouraging words throughout the PhD period, and without their motivational words, I could not sum up this thesis in time. I cannot find words to express my recognition to my hostel mates {\it Pitambar Da, Manas Da, Kumar Da, Barkhang Da, Rewa Da, Homen Da, Chanda Da,} {\it GVS Bhagyaraj, Litul,  Dhruba, Satya, Ankur, Diganta} for always being there for me. I want to thank my lab mates {\it Ananya ba, Happy ba, Bichitra da, Nayana, Lavina and Jotin} for all the stimulating discussions about physics and bringing helping hands whenever needed and for all the fun times we had. In particular, it has been a pleasant experience to share a lab with {\it Ananya ba} and {\it Happy ba} during my initial days. I would also like to thank all my department colleagues and friends for their excellent company throughout the period. My sincere thanks to the department seniors {\it Iftak Hussain, Nabadweep Chamuah, Gautam Saikia, Pranjupriya Goswami, Dipankar Bora, Prarthana Borah, Mridusmita Buragohain} who were always there with their encouraging words and love. I express my warmest thanks to {\it Hiya Talukdar, Joydwip Sinha Roy, Bidyut Chutia, Bidisha Dowarah, Rajesh Kumar, Sonali Patnaik, Atashi Roy, Monal Alhuwalia, Dhanjyoti Barman, Subhraja Chakrabarty, Wahida Najnin} for endless support and encouragements. I owe my deepest gratitude to friends and seniors from IITG ({\it Dibyundu Da, Rishav Da, Debobrat, Madhurima, Prantik}) for the helping hands and fruitful discussions whensoever needed. Special thanks to {\it Manoranjan} and {\it Snehasish} from IITH and my labmates for carefully reading and suggesting promising improvements on this thesis.

I am thankful for {\it Patir Da} and {\it Narayan Da} for offering their unconditional helping hands in official works in the department without a second thought. I would also like to thank all other non-academic staffs of the department for their help at times. I must even acknowledge TU and DST-SERB for financial support for the last four years.

Finally, I would like to acknowledge the beautiful Tezpur University's entire fraternity for gifting me these most memorable years of my life, which I will cherish forever.

\vspace{.5cm}
\begin{flushright}
	\includegraphics[scale=0.2]{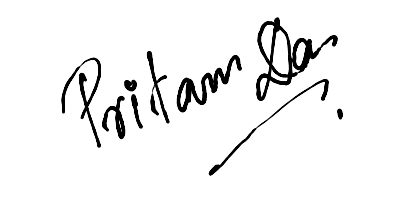}\\
\it \large	(Pritam Das)
\end{flushright}
}
\clearpage % Start a new page
%\newpage
%\pagebreak[1]
%----------------------------------------------------------------------------------------
%	LIST OF CONTENTS/FIGURES/TABLES PAGES
%----------------------------------------------------------------------------------------

\pagestyle{fancy} % The page style headers have been "empty" all this time, now use the "fancy" headers as defined before to bring them back

\lhead{\emph{Contents}} % Set the left side page header to "Contents"
%\vspace{-3cm}
\tableofcontents % Write out the Table of Contents
\lhead{\emph{List of Figures}} % Set the left side page header to "List of Figures"
\listoffigures % Write out the List of Figures

\lhead{\emph{List of Tables}} % Set the left side page header to "List of Tables"
\listoftables % Write out the List of Tables

%----------------------------------------------------------------------------------------
%	ABBREVIATIONS
%----------------------------------------------------------------------------------------

\clearpage % Start a new page
\setstretch{1.55} % Set the line spacing to 1.5, this makes the following tables easier to read
\lhead{\emph{Abbreviations}} % Set the left side page header to "Abbreviations"
\listofsymbols{ll}
 % Include a list of Abbreviations (a table of two columns)
{\vspace{-1.6cm}\\
\textbf{BAU} & \textbf{B}aryon  \textbf{A}symmetry of the \textbf{U}niverse\\
\textbf{BBN} & \textbf{B}ig \textbf{B}ang \textbf{N}ucleosynthesis\\
\textbf{BSM} & \textbf{B}eyond the  \textbf{S}tandard \textbf{M}odel \\
\textbf{CERN} & The European Organization for Nuclear Research\\
\textbf{CKM} & \textbf{C}abibbo \textbf{K}obayashi \textbf{M}askawa\\
\textbf{CPT} & \textbf{C}harge \textbf{P}arity \textbf{T}ime reversal\\
\textbf{CP} & \textbf{C}harge Conjugation \textbf{P}arity\\
\textbf{CMBR} & \textbf{C}osmic \textbf{M}icrowave \textbf{B}ackground \textbf{R}adiation\\
\textbf{CUORE} &  \textbf{C}ryogenic \textbf{U}nderground \textbf{O}bservatory for \textbf{R}are \textbf{E}vents\\
\textbf{DM} & \textbf{D}ark \textbf{M}atter \\
\textbf{DUNE} & 
\textbf{D}eep \textbf{U}nderground \textbf{N}eutrino \textbf{E}xperiment\\
\textbf{EWSB} & \textbf{E}lectro \textbf{W}eak \textbf{S}ymmetry \textbf{B}reaking\\
\textbf{eV} & \textbf{E}lectron \textbf{V}olt\\ 
\textbf{EXO} & \textbf{E}nriched \textbf{X}enon \textbf{O}bservatory\\
\textbf{GALLEX} & \textbf{G}allium \textbf{E}xperiment\\
\textbf{GERDA} & \textbf{Ger}manium \textbf{D}etector \textbf{A}rray\\
\textbf{GUT} & \textbf{G}rand \textbf{U}nified \textbf{T}heory\\
\textbf{IH} & \textbf{I}nverted  \textbf{H}ierarchy\\
\textbf{KATRIN} & \textbf{K}arlsruhe \textbf{T}ritium \textbf{N}neutrino \textbf{E}xperiment\\
\textbf{KamLAND-Zen} & \textbf{Kam}ioka \textbf{L}iquid Scintillator \textbf{A}nti- \textbf{N}eutrino \textbf{D}etector-\textbf{Xen}on\\
\textbf{$keV$} & \textbf{K}ilo \textbf{E}lectron \textbf{V}olt\\
\textbf{LHC} & \textbf{L}arge  \textbf{H}adron \textbf{C}ollider\\
\textbf{LH} & \textbf{L}eft \textbf{H}anded\\
\textbf{LNV} & \textbf{L}epton \textbf{N}umber \textbf{V}iolation\\
\textbf{LFV} & \textbf{L}epton \textbf{F}lavour \textbf{V}iolation\\
\textbf{LSND} & \textbf{L}iquid \textbf{S}cintillator \textbf{N}eutrino \textbf{D}itector\\
\textbf{MES} & \textbf{M}inimal \textbf{E}xtended \textbf{S}eesaw\\
\textbf{MiniBOONE} & \textbf{Mini} \textbf{Boo}ster \textbf{N}eutrino \textbf{E}xperiment\\
\textbf{MINOS} & \textbf{M}ain \textbf{I}njector \textbf{N}eutrino \textbf{O}scillation \textbf{S}earch\\
\textbf{NO$\nu$A} & \textbf{N}uMI \textbf{O}ff-axis $\nu_e$ \textbf{A}ppearance\\
\textbf{NH} & \textbf{N}ormal  \textbf{H}ierarchy\\
\textbf{NDBD/ $0\nu\beta\beta$} & \textbf{N}eutrinoless \textbf{D}ouble \textbf{B}eta \textbf{D}ecay \\
\textbf{NEMO}& \textbf{N}eutrino \textbf{E}ttore \textbf{M}ajorana \textbf{O}bservatory\\
\textbf{PMNS} & \textbf{P}ontecorvo \textbf{M}aki \textbf{N}akagawa \textbf{S}akata\\
\textbf{RENO} & \textbf{R}eactor \textbf{E}xperiment for \textbf{N}eutrino \textbf{O}scillation\\
\textbf{RH} & \textbf{R}ight \textbf{H}anded\\
\textbf{RL} & \textbf{R}esonant \textbf{L}eptogenesis\\
\textbf{SSB} & \textbf{S}pontaneous \textbf{S}ymmetry \textbf{B}reaking\\	
\textbf{SM} & \textbf{S}tandard \textbf{M}odel \\
\textbf{SDSS}& \textbf{S}loan \textbf{D}igital \textbf{S}ky \textbf{S}urvey\\
\textbf{SUSY} & \textbf{SU}per \textbf{SY}mmetry\\
\textbf{Super-K} & \textbf{Super} \textbf{K}amiokande\\
\textbf{SNO} & \textbf{S}udbery \textbf{N}eutrino \textbf{O}bservatory\\
\textbf{T2K} & \textbf{T}okai \textbf{to} \textbf{K}amioka\\
\textbf{TeV} & \textbf{T}era \textbf{E}lectron \textbf{V}olt\\
\textbf{VEV} & \textbf{V}acuum \textbf{E}xpectation \textbf{V}alue\\
\textbf{WMAP} & \textbf{W}ilkinson \textbf{M}icrowave \textbf{A}nisotropy \textbf{P}robe\\

}

%----------------------------------------------------------------------------------------
%	PHYSICAL CONSTANTS/OTHER DEFINITIONS
%----------------------------------------------------------------------------------------
%
%\clearpage % Start a new page
%
%\lhead{\emph{Physical Constants}} % Set the left side page header to "Physical Constants"
%
%\listofconstants{lrcl} % Include a list of Physical Constants (a four column table)
%{
%Speed of Light & $c$ & $=$ & $2.997\ 924\ 58\times10^{8}\ \mbox{ms}^{-\mbox{s}}$ (exact)\\
%% Constant Name & Symbol & = & Constant Value (with units) \\
%}

%----------------------------------------------------------------------------------------
%	SYMBOLS
%----------------------------------------------------------------------------------------
%----------------------------------------------------------------------------------------
%	DEDICATION
%----------------------------------------------------------------------------------------
%
\setstretch{1.3} % Return the line spacing back to 1.3
\pagestyle{empty} % Page style needs to be empty for this page
\dedicatory{\Huge \it Dedicated to {\Huge Maa-Deuta}\\
	(\Huge Joymati Das \& Bholanath Das)\\} % Dedication text
\addtocontents{toc}{\vspace{2em}} % Add a gap in the Contents, for aesthetics

%----------------------------------------------------------------------------------------
%	THESIS CONTENT - CHAPTERS
%----------------------------------------------------------------------------------------

\mainmatter % Begin numeric (1,2,3...) page numbering

\pagestyle{fancy} % Return the page headers back to the "fancy" style

% Include the chapters of the thesis as separate files from the Chapters folder
% Uncomment the lines as you write the chapters

% Chapter Template
\begin{savequote}[1\linewidth]
	\normalsize	``Home is behind, the world ahead,\\
	And there are many paths to tread\\
	Through shadows to the edge of night,\\
	Until the stars are all alight."
	\qauthor{{\large\it J.R.R. Tolkien}, {\rm  The Fellowship of the Ring} (1892--1973)}
\end{savequote}
\chapter{INTRODUCTION} % Main chapter title
%\vspace{-2.1cm}
%{\it ``It doesn't matter how beautiful your theory is, it doesn't matter how smart you are. If it doesn't agree with the experiment, it's wrong."--- {\Large Richard P. Feynman}}
%{\it ``Begin at the beginning," the King said gravely, ``and go on till you come to the end: then stop."---{ \Large Lewis Carroll}, {\rm Alice in Wonderland}}
\label{Chapter1} % Change X to a consecutive number; for referencing this chapter elsewhere, use \ref{ChapterX}

\lhead{Chapter 1. \emph{INTRODUCTION}} % Change X to a consecutive number; this is for the header on each page - perhaps a shortened title

%----------------------------------------------------------------------------------------
%	SECTION 1
%----------------------------------------------------------------------------------------
\section{A brief history of neutrinos}
The story of neutrinos had begun long before the scientists ever found any. In the year 1930, Wolfgang Pauli, the Austrian physicist proposed the existence of a neutral particle to conserve some physical quantities, while he was studying beta decay processes. He predicted a hypothetical particle which is neutral and might be carrying the missing energy and escaping detection. After that, the glorious journey of neutrino had begun. In 1934, during the study of radioactive decay, Enrico Fermi coined the name $neutrino$ for that hypothetical particle as it is neutral and the suffix $``-ino”$ indicating smallness in size. Pauli was worried about the new particle that interacts so weakly with matter and quoted {\it ``I have done a terrible thing, I have postulated a particle that cannot be detected”.} In fact, in 1956, neutrinos were discovered when Clyde Cowan and Frederick Reines detected antineutrinos emitted from a nuclear reactor and Pauli lost his bet.

Neutrinos come in three flavors: electron neutrino ($\nu_e$), muon neutrino ($\nu_{\mu}$) and tau neutrino ($\nu_{\tau}$); named after their respective fundamental charged leptons: electron, muon, and tau. Along with their tiny mass, what makes neutrinos intrigue scientists is that these tiny weakly interacting particles are the messenger from the universe's outer reach but very hard to detect and change flavour during their propagation. Electron neutrinos ($\nu_e$) produced by solar nuclear fusion reactions at our Sun change their flavour to muon neutrino ($\nu_{\mu}$) as they reach the earth’s atmosphere which is again converted into tau neutrino ($\nu_{\tau}$) within the atmosphere. This phenomenon of changing flavour is known by the term “neutrino oscillation”. 

In standard neutrino scenario three active neutrinos are involved with two mass square differences\footnote{order of $ 10^{-5}eV^{2} $ and  $ 10^{-3}eV^{2} $ for solar ($\Delta m^{2}_{21}$) and atmospheric ($\Delta m^{2}_{23}/\Delta m^{2}_{13}$) neutrino respectively.}, three mixing angles ($\theta_{ij};i,j=1,2,3$) and one Dirac CP phase ($\delta_{13}$). Several neutrino oscillation experiments like  SK\cite{Abe:2016nxk}, SNO\cite{boger2000sudbury} MINOS\cite{Evans:2013pka}, T2K\cite{Abe:2011sj}, RENO\cite{Ahn:2012nd}, DOUBLE CHOOZ\cite{Abe:2011fz}, DAYABAY\cite{An:2012eh}, $etc.$ have established the fact that neutrinos produced in a well-defined flavor eigenstate can be detected as a different flavor eigenstate while they propagate. This can be interpreted as, like all charged fermions, neutrinos have mass and mixing because their flavor eigenstates are different from mass eigenstates. The existence of neutrino mass was the first evidence for the new physics beyond the Standard Model (BSM). Some recent reviews on neutrino physics are put into references \cite{Mohapatra:2005wg,King:2003jb,King:2014nza}.
If neutrinos are Majorana particles, then there are two more CP-violating phases ($\alpha$ and $\beta$) come into the 3-flavour scenario. Majorana phases are not measured experimentally as they do not involve in the neutrino oscillation probability. The current status of global analysis of neutrino oscillation data \cite{Capozzi:2016rtj, Esteban:2020cvm} give us the allowed values for these parameters in $3\sigma$ confidence level, which are shown in Table \ref{ttab:d1}. Along with the Majorana phases, the absolute mass scale for the individual neutrino is still unknown as the oscillation experiments are only sensitive to the mass square differences, even though Planck data constrained the sum of the three neutrinos, $\Sigma m_{\nu}<0.12$ eV at $95{\%}$ confidence level \cite{Palanque-Delabrouille:2015pga}. The fact that the absolute neutrino mass scale for the individual neutrino is not known, as the oscillation probability depends on the mass square splittings but not the absolute neutrino mass. Moreover, neutrino oscillation experiments tell that the solar mass square splitting is always positive, which implies $m_2$ is always grater than $m_1$. However, we have not yet received the same confirmation regarding the atmospheric mass square splittings from the experiments. This fact allows us to have two possible mass hierarchy patterns for neutrinos; Normal Hierarchy (NH: $m_1\ll m_2<m_3$) as well as Inverted Hierarchy (IH: $m_3\ll m_1<m_2$). 

In the past few decades, there have been achievements in solar, reactor and accelerator experiments whose results are in perfect agreement with only three active neutrino scenario. Meanwhile, there are some anomalies which need explanation. The very first and most distinguished results towards new physics in the neutrino sector were from LSND results \cite{Athanassopoulos:1997pv,Aguilar:2001ty,Athanassopoulos:1996jb}, where electron anti-neutrino ($\overline{\nu_{e}}$) were observed in the form of muon anti-neutrino ($\overline{\nu_{\mu}}$) beam seemingly $\overline{\nu_{e}}$ was originally $\overline{\nu_{\mu}}$. Moreover, data from MiniBooNE \cite{Aguilar-Arevalo:2018gpe} results overlap with LSND results and indicate different generations of neutrino hypotheses. To make sure that these data are compatible with the current picture, one needs new mass eigenstates for neutrinos. These additional states must relate to right-handed neutrinos (RHN) for which bare mass term is allowed by all symmetries, $i.e.$ they should not be present in $SU(2)_{L}\times U(1)_{Y}$ interactions, hence are \textit{sterile}. Recently observed Gallium Anomaly observation\cite{Abdurashitov:2005tb,Abdurashitov:1999zd,giunti2012update, giunti2011statistical} is also well explained by sterile neutrino hypothesis. Although a few talks about the non-existence of extra neutrino, finally reactor anti-neutrino anomaly results \cite{Mention:2011rk, Kopp:2011qd} give a clear experimental proof that the presence of this fourth non-standard neutrino is mandatory. Moreover, cosmological observation \cite{Hamann:2010bk} (mainly CMB\footnote{cosmic microwave background} or SDSS\footnote{Sloan Digital Sky Survey}) also favor the existence of sterile neutrino. From cosmological consequences, it is said that the sterile neutrino has a potential effect on the entire Big-Bang Nucleosynthesis \cite{Izotov:2010ca}. LSND results predicted sterile neutrino with mass $\sim\mathcal{O}(1)eV$. To be more specific with the recent update with MiniBooNE experiment results \cite{Aguilar-Arevalo:2018gpe}, which combine the $\nu_e/\overline{\nu_{e}}$ appearance data with the LSND results to establish the presence of an extra flavour of neutrino up to 6.0$\sigma$ confidence level. However, these results from LSND/MiniBooNE are in tension with improved bounds on appearance/disappearance experiments results from IceCube/MINOS+ \cite{Dentler:2018sju}. Although, $\Delta{m_{41}^{2}}\sim1eV^2$ is consistent with global data from the $\nu_{e}$ disappearance channel which supports sterile neutrino oscillation at $3\sigma$ confidence level. However, exact mass scale or the exact number of sterile neutrino generations are not yet confirmed. Nevertheless, hints from different backgrounds point a finger towards the presence of at least a new generation of neutrinos.

Sterile neutrino is a neutral lepton which does not involve itself in weak interactions. However, they are induced by mixing with the active neutrinos, leading to an observable effect in the oscillation experiments. Furthermore, they could interact with gauge bosons which lead to some significant correction in non-oscillation processes, $e.g.$, in the neutrinoless double beta decay (NDBD) amplitude\cite{Goswami:2005ng,Goswami:2007kv}, beta decay spectra. Since RH neutrinos are SM gauge singlets\cite{deGouvea:2005er}, so it is possible that sterile neutrinos could fit in the canonical type-I seesaw as the RH neutrino if their masses lie in the eV regime. Some global fit studies have been carried out for sterile neutrinos at eV scale being mixed with the active neutrinos \cite{Kopp:2013vaa,Giunti:2013aea,Gariazzo:2015rra}. While doing this the Yukawa Coupling relating lepton doublets and right-handed neutrinos should be of the order $10^{-12}$ which implies a Dirac neutrino mass of sub-eV scale to observe the desired active-sterile mixing. These small Dirac Yukawa couplings are considered unnatural unless there is some underlying mechanism to follow. Thus, it would be captivating to choose a framework that gives low-scale sterile neutrino masses without tiny Yukawa coupling and simultaneously explaining active-sterile mixing.

\subsection{Theoretical advancement}
Theoretically, neutrino oscillation was first to come up in 1957, when Bruno Pontecorvo was inspired by kaon-antikaon oscillation. He suggested that $\nu_l-\bar{\nu}_l$ oscillation could occur if lepton number is violated. The idea of neutrino flavour mixing was conceptualized later by Maki, Nakagawa and Sakata \cite{Maki:1962mu} in their study of muon neutrino discovery. Later, in 1967 Pontecorvo proposed neutrino flavour oscillation and finally the modern formalism was developed in 1969 by Gribov and Pontecorvo \cite{Pontecorvo:1957qd,Pontecorvo:1957cp}.
In today's scenario, the flavour and mass eigenstates are related by the $3\times3$ unitary matrix, known as PMNS (named after Pontecorvo, Maki, Nakagawa and Sakata) matrix or the leptonic mixing matrix \cite{Giganti:2017fhf}. This matrix is parametrized by the mixing angle $\theta_{12}, \theta_{13}, \theta_{23}$ and CP-violating phases. 

Concerning the respective charge lepton generations, the three active neutrinos were named. Moreover, the number of active neutrinos constrained by the LEP measurement of invisible decay width of the $Z$ bosons is also suggesting three flavours of light neutrinos \cite{ALEPH:2005ab}:
\begin{equation}
	N_{\nu}=\frac{\Gamma_Z}{\Gamma(Z\rightarrow\nu\nu)_{SM}}=2.9840\pm0.0082,
\end{equation}
where, $\Gamma(Z\rightarrow\nu\bar{\nu})_{SM}$ gives the decay width of $Z$ boson into a single neutrino. This result leaves no room for extra generations of neutrino flavours. However, cosmological studies portrait a different picture for neutrino generations. For example, recent data from the Wilkinson Microwave Anisotropy Probe (WMAP) suggested that the total number of relativistic degree of freedom be $N_{eff}=3.84\pm0.40$ \cite{Ade:2015xua}. In the same footing, some experiments also suggested the presence of an extra generation of neutrinos. Hence, the concept of sterile neutrino was introduced in the scenario. Till now, the mass scale or the exact number of generations for sterile neutrinos are not known. However, in this thesis work, we have introduced a single generation of a sterile neutrino with different mass scale and study the consequences in current neutrino phenomenology. 

Even though the exact mass scale or numbers of sterile neutrino generations are still unknown, their presence may significantly contribute to the new physics. The presence of sterile neutrino is strongly motivated and highly influences the current reactor neutrino anomalies. Sterile neutrinos with different mass ranges play crucial role in astrophysics \cite{Petraki:2007gq}, cosmology \cite{Abazajian:2012ys,Abazajian:2017tcc}, collider physics \cite{Abada:2017jjx,Atre:2009rg,Deppisch:2015qwa}, $etc.$ Similar kind of studies were carried out in other context such as LRSM \cite{Borgohain:2019pya,Barry:2014ika}, extra dimensions \cite{Rodejohann:2014eka, Dev:2012bd}, in presence of exotic charged currents \cite{Ludl:2016ane} or in relation with $keV$ neutrino dark matter \cite{deVega:2011xh}. 
In order to accommodate sterile neutrino in the current SM mass pattern, various schemes were studied. In (2+2) scheme, two different neutrino mass states differ by $eV^2$, which is disfavored by current solar and atmospheric data \cite{Maltoni:2002ni}. Current status for mass square differences, corresponding to sterile neutrinos, dictates sterile neutrinos to be either heavier or lighter than the active ones. Thus, we are left with either (1+3) or (3+1) scheme. In the first case, three active neutrinos are in eV scale, and sterile neutrino is lighter than the active neutrinos. However, this scenario is ruled out by cosmology \cite{Hamann:2010bk,Giusarma:2011ex}. In the latter case, three active neutrinos are in sub-eV scale and sterile neutrino is in eV scale \cite{GomezCadenas:1995sj,Goswami:1995yq}. Numerous studies have been exercised taking this (3+1) framework with various prospects \cite{giunti2011statistical,Borah:2016xkc,Kopp:2011qd,Gariazzo:2015rra}. Meanwhile, a sterile neutrino with $keV$ scale also gained attention in active-sterile mixing. $keV$ sterile neutrinos significantly impact kinetic processes such as double $\beta$ decay, and they can also behave as a potential WIMP dark candidate with suitable active-sterile mixing. 

The analysis of neutrino phenomenology under the seesaw mechanism involves many more parameters than can be measured from the neutrino masses and mixing; hence, the situation becomes challenging to study. Texture zero in the final neutrino mass matrix implies some of the elements are much smaller than the other elements or zero; eventually, the number of free parameters are significantly reduced \cite{Kageyama:2002zw, Ghosh:2012pw, Borgohain:2019pya}. Flavour symmetries are widely used in neutrino phenomenology and play a crucial role in texture realization by reducing free parameters in the neutrino mass matrix \cite{Borah:2015vra,Babu:2009fd,Borah:2016xkc,Borgohain:2018lro, Kageyama:2002zw, Dziewit:2016qri}.
Flavon fields introduced under flavour symmetries to describe neutrino mixing phenomenology are customarily SM singlet.

\subsection{Experimental advancement}

The original intuition of neutrino oscillation proposed by Bruno Pontecorvo in 1957 \cite{Pontecorvo:1957cp}, the discoveries of solar and atmospheric oscillation were finally observed by  Super-Kamioka Neutrino Detector experiment (Super-Kamiokande) \cite{Fukuda:1998ub, Ashie:2004mr} and Sudbury Neutrino Observatory (SNO) \cite{boger2000sudbury} and shortly after them, it was confirmed by the Kamioka Liquid Scintillator Antineutrino Detector (KamLAND) \cite{Araki:2005qa, Eguchi:2002dm} in between the years 1998 and 2002. In 2015, these discoveries got recognized by the prestigious Nobel Prize in Physics to Takaaki Kajita from University of Tokyo, Japan and Arthur B. McDonald from Queen's University, Canada {\it “for the discovery of neutrino oscillations, which shows that neutrinos have mass”}. Since the milestone, there has been fast progress in the experimental sector. The non-zero value of the last unknown mixing angle, $\theta_{13}$ was first indicated by the long-baseline accelerator neutrino experiment Tokai-to-Kamioka (T2K) \cite{Abe:2011sj} and discovered by the reactor experiments Daya Bay \cite{An:2014bik}, reactor experiments for Neutrino Oscillations (RENO) \cite{Ahn:2012nd} and Double Chooz \cite{Abe:2011fz}, it was measured with incredible accuracy: $\theta_{13} \sim 8.5^0 \pm0.2^0$ \cite{An:2012eh}. 
	\begin{table}[t]
	\begin{center}
		%	\centering
		\begin{tabular*}{\columnwidth}{@{\extracolsep{\fill}}|l|ll|ll|@{}}
			\hline
			Parameters&bfp$\pm 1\sigma$&3$\sigma$ (NH)&bfp$\pm1\sigma$&3$\sigma$ (IH)\\
			\hline
			\hline
			$\Delta m^{2}_{21}[10^{-5}eV^2]$&$7.42^{+0.21}_{-0.20}$&6.82-8.04&$7.42^{+0.21}_{-0.20}$&6.82-8.04\\% \hline
			$\Delta m^{2}_{31}[10^{-3}eV^2]$&$2.517^{+0.026}_{-0.028}$&2.435-2.598&$-2.498^{+0.028}_{-0.028}$&-2.518-4.414\\% \hline
			$sin^{2}\theta_{12}/10^{-1}$&$3.04^{+0.012}_{-0.012}$&2.69-3.43&$3.04^{+0.013}_{-0.012}$&2.69-3.43\\% \hline
			$sin^{2}\theta_{13}/10^{-2}$&$2.219^{+0.062}_{-0.063}$&2.032-2.41&$2.238^{+0.063}_{-0.062}$&2.052-2.428\\ %\hline
			$sin^{2}\theta_{23}/10^{-1}$&$5.73^{+0.16}_{-0.20}$&4.15-6.16&$5.73^{+0.16}_{-0.19}$&4.19-6.17\\% \hline
			$\delta_{13}/\circ$&$197^{+27}_{-24}$&120-369&$282^{+26}_{-30}$&193-352\\ \hline
		\end{tabular*}
		\caption{Recent NuFIT 5.0 (2020) results for active neutrino oscillation parameters with best-fit and the latest global fit $3\sigma$ range for normal and inverted ordering mode \cite{Esteban:2020cvm}.}\label{ttab:d1}
	\end{center}
\end{table}
Today, almost all oscillation parameters have been measure with amazing accuracy. Most recent updated results by NuFIT 5.0 (2020) are shown in table \ref{ttab:d1}. Current and next-generation experiments are more focused on better accuracy with the current oscillation parameters and look up the remaining puzzles associated with neutrino. Like, currently the major concern with neutrino is whether it is
Dirac or Majorana? What is the absolute mass scale of a neutrino? What is the actual mass ordering? How much CP violation is there? What are the exact mass and mixing of sterile neutrinos?

A recent update from global fit analysis and atmospheric data from Super-Kamiokande favours normal hierarchy over inverted hierarchy pattern \cite{Abe:2016nxk}. However, new Long Base-Line (LBL) results from T2K \cite{Abe:2016tii} and NO$\nu$A \cite{NOvA:2018gge} favours IH pattern. Despite slightly disagreeing on some parameter regions, those results from T2K, No$\nu$A and reactor experiments are statistically in excellent agreements with each other. Apart from the active neutrino parameters, active-sterile mixing also gained some serious attention over the past few years. Since 2011, several new results have led to better experimental sensitivity in electron and muon neutrino disappearance channels. Reactor anti-neutrino experiments ruled out earlier results from reactor and gallium anomalies. However, a new best-fit region from the latest results shows a similar $\Delta m_{41}^2$ but smaller amplitude. Sterile neutrino with eV as well as $keV$ could be probed in future KATRIN experiment \cite{Osipowicz:2001sq,Mertens:2014nha}, and $keV$ sterile neutrino has a potential to affect electron energy spectrum in tritium $\beta$-decays \cite{Shrock:1980vy}. eV-scaled sterile neutrino has a characteristic kink on KATRIN experiments; hence, relevant results are expected in a few years. For $m_4\simeq2$ eV, the expected sensitivity corresponds to the active-sterile mixing element $|U_{e4}|^2\sim0.1 ~(90\% CL)$. Apart from KATRIN, future tritium and holmium bases experiments such as extended KATRIN \cite{Aker:2019uuj}, Project-8 \cite{Esfahani:2017dmu}, ECHo \cite{Gastaldo:2013wha}, and Holmes \cite{Alpert:2014lfa} are also on-going projects to test the sensitivity of sterile neutrinos. 
Moreover, sterile neutrinos with mass (0.4-50) $keV$ \cite{Boyarsky:2009ix} are considered as WIMP particles since they are relatively slow and much heavier compared to the active neutrinos.  Back in the 90s, Dodelson
and Widrow \cite{Dodelson:1993je} proposed $keV$ sterile neutrino as dark matter candidate produced via oscillation and collision from active neutrinos. Recently from various sources like, in the stacked spectrum of galaxy clusters \cite{Bulbul_2014}, individual spectra of nearby galaxy clusters \cite{Bulbul_2014,PhysRevLett.113.251301}, Andromeda Galaxy \cite{PhysRevLett.113.251301}, and in the Galactic Center region \cite{Boyarsky:2014ska,Riemer-Sorensen:2014yda} an unidentified line was reported. The position of the line is $E = 3.55$ $ keV$ with an uncertainty in position $\simeq\pm 0.05$ $keV$. If the line is interpreted as originating from a two-body decay of a DM particle, then the particle has its mass at about $m_S\simeq 7.1$ $keV$ and the lifetime $\tau_{DM} \simeq 10^{27.8\pm 0.3}$ sec \cite{PhysRevLett.113.251301}. 
\section{The Standard Model}
The curiosity of the human mind is a fantastic thing. The quest of finding the ultimate building blocks of matter and their interactions bring us to the field of $elementary$ $particle$ physics. At the current date, our knowledge of elementary particle and their interactions are having its fascinating period. There are four fundamental forces in Nature, $viz.,$ strong force, weak force, the electromagnetic force and gravitational force. The first three forces are gauge interactions, and the $Standard~Model$ (SM) of particle physics explain them all-together. They are explained via some internal symmetries, which we will be discussing in this thesis. On the other hand, the gravitational interaction originates from space-time symmetry and is explained by the general theory of relativity.

In the year 1961, Sheldon Glashow unified electromagnetic and weak interaction to electroweak interaction. Later, in 1967 Abdus Salam and Steven Weinberg  unified Higgs mechanism of $W^{\pm}$ and $Z$ boson into Glashow's electroweak interaction and giving it the modern form of the Standard Model. The SM of particle physics is considered one of the most successful theories of all time. It describes the nature and interactions of the fundamental particles with immense precision. The SM can be divided into two fundamental parts: the strong interaction and the electroweak interaction. The quarks experience the strong interaction, and gluons are the force carrier. This is a gauge theory, where the three coloured states of every flavour belong to the triplet representation of the group $SU(3)_c$. The electroweak interaction is a spontaneously broken gauge theory and it is two-fold interaction with a combination of electromagnetic and weak interaction. The electromagnetic interaction is associated with charged particles and mediated by the photon and it is a $U(1)_Q$ gauge group, where $Q$ represents the electric charge. The weak interactions are associated with the radioactive nuclear processes like nuclear beta decay and they are mediated via the weak bosons ($W^{\pm}$ and $Z$). They were combining these two interactions, the electroweak interaction described by an $SU(2)_L\times U(1)_Y$ interaction, which spontaneously broken down to $U(1)_Q$ at around 100 GeV. In any gauge theory, the SSB makes all gauge bosons massive except the photon, which remains massless. Finally, the SM gauge group be like $SU(3)_c\times SU(2)_L\times U(1)_Y$.

In this chapter, we will discuss the SM framework in a compact yet comprehensive manner and then look at the success of the theory and finally discuss the severe drawbacks, which are the primary concern of this thesis..
\subsection{Particle content and interaction}
The SM consists of six quarks ($u,d,c,s,t,b$) and three pair of leptons ($e,\mu,\tau$ and their respective neutrinos $\nu_e,\nu_{\mu},\nu_{\tau}$), and they are known as the $elementary~particles$. The SM also includes the mediator or the gauge bosons such as $\gamma$ (photon), $W^{\pm}, Z$, gluon and one neutral Higgs boson ($H$). The Higgs boson is responsible for giving mass to these SM particles. The particle content of the SM are arranged in table \ref{tsm1}.
\begin{table}[h]
	\centering
	\begin{tabular}{|c||c|c|c|c|c|}
		\hline
	Fields&$SU(3)_c$&$SU(2)_L$&$U(1)_Q$&$I_3$&$Y$\\
	\hline
	\hline
	$Q_{iL}=\begin{pmatrix}u_{iL}\\d_{iL}
	\end{pmatrix}$&3&2&1/3&$\begin{pmatrix}
	1/2\\-1/2
\end{pmatrix}$&-1\\
$u_{iR}$&3&1&4/3&0&4/3\\
$d_{iR}$&3&1&-2/3&0&-2/3\\
$L_{iL}=\begin{pmatrix}
	\nu_{iL}\\l_{iL}
\end{pmatrix}$&1&2&-1&$\begin{pmatrix}1/2\\-1/2
\end{pmatrix}$&-1\\
$l_{iR}$&1&1&-2&0&-2\\
\hline
$B_{\mu}$&1&1&0&0&0\\
$W_{\mu}^i$&1&3&0&$\pm1$&0\\
$G_{\mu}^a$&8&1&0&0&0\\
\hline
$\Phi=\begin{pmatrix}
	\phi^+\\\phi^0\\
\end{pmatrix}$&1&2&1/2&$\begin{pmatrix}
1/2\\-1/2
\end{pmatrix}$&1\\
\hline
	\end{tabular}
\caption{The SM particle content and charges under respective gauge groups. }\label{tsm1}
\end{table}
A quark doublet under $SU(2)$ is consists of a left-handed $up-type$ quark and a left handed $down-type$ quark. The doublet structure is, $Q_{iL}=\begin{pmatrix}u_{iL}\\d_{iL}
\end{pmatrix}$, where $i$ in the suffix corresponds for three generations of quark family. The $u$ stands for $up-type$ quark generations and it is  consists of $u,c,t$ quarks; and $d$ represents the three $down-type$ quark generations $d,s,b$ quarks. Similarly, a lepton doublet consists of a left-handed charged lepton and its corresponding left-handed neutrino. The doublet structure for lepton is given by $L_{iL}=\begin{pmatrix}
	\nu_{iL}\\l_{iL}
\end{pmatrix}$, here also $i$ stands for three lepton generations. Three charged leptons $e,\mu,\tau$ are represented by $l$ with corresponding left-handed neutrinos as $\nu_e, \nu_{\mu},\nu_{\tau}$. Right-handed quarks and charged leptons are always singlet under the same gauge group. The SM doesn't holds room for right-handed neutrinos.

The Higgs doublet under $SU(2)$ transform as a doublet and given by,$$\Phi=\begin{pmatrix}
	\phi^+\\\phi^0\\
\end{pmatrix},$$ where, the charged and the neutral components are complex scalars and they are represented as, $\phi^+=\frac{\phi_1+i\phi_2}{\sqrt{2}}$ and $ \phi^0=\frac{\phi_3+i\phi_4}{\sqrt{2}}$, respectively.

In the following, we will be discussing the SM Lagrangian symmetric under $SU(3)\otimes SU(2)_L\otimes U(1)_Y$ gauge group. This Lagrangian governs the masses and interactions of the SM particles and it is given by,
\begin{equation}
	\mathcal{L}_{SM}=\mathcal{L}_{gauge}+\mathcal{L}_{fermion}+\mathcal{L}_{Higgs}+\mathcal{L}_{Yukawa}.\label{tlag1}
\end{equation}
{\bf Gauge interaction:} The gauge Lagrangian regulates the dynamics of the gauge bosons and it can be written as,
\begin{equation}
	\mathcal{L}_{gauge}=-\frac{1}{4}G_{\mu\nu}^iG_i^{\mu\nu}-\frac{1}{4}W_{\mu\nu}^aW^{\mu\nu}_a-\frac{1}{4}B_{\mu\nu}B^{\mu\nu}.
\end{equation}
The field strength tensors are defined as,
\begin{eqnarray}
	G_{\mu\nu}^i&=&\partial_{\mu}G_{\nu}-\partial_{\nu}G_{\mu}-g_Sf^{ijk}G_{\mu}^jG_{\nu}^k,\\
	W_{\mu\nu}^a&=&\partial_{\mu} W_{\nu}^a-\partial_{\nu}W_{\mu}^a-g_W\epsilon^{abc}W_{\mu}^bW_{\nu}^c,\\
	B_{\mu\nu}&=&\partial_{\mu}B_{\nu}-\partial_{\nu}B_{\mu},
\end{eqnarray}
where $i,j,k=1,2,...,8$ and $a,b,c=1,2,3$. $f^{ijk}$ and $\epsilon^{abc}$ are the 
Levi-Civita symbol and $g_S, g_W$ are the gauge coupling constants under $SU(3)_c$ and $SU(2)_L$ respectively.

{\bf The fermionic interaction:} The kinetic energies of the quark and leptons and their interactions with the gauge bosons are embedded in the Lagrangian as,
\begin{equation}
	\mathcal{L}_{fermion}=i\bar{\Psi}_L\cancel{D}_L\Psi_L+i\bar{\Psi}_R\cancel{D}_R\Psi_R.
\end{equation}
The covariant derivative of the fermion-scalar gauge interaction is expressed as,
\begin{equation}
	D_{\mu}=\partial_{\mu}-ig^{\prime}B_{\mu}Y-ig_W W_{\mu^a}T^a-ig_SG^it^i,
\end{equation} 
and $D_L,D_R$ takes distinct structure depending upon respective gauge charges. The gauge coupling and generators under $SU(3)_c,SU(2)_L, U(1)_Y$ are $g_S,g_W,g^{\prime}$ and $t^i,T^a, Y$ respectively.
Under gauge groups, these fermion and gauge fields transform as,
\begin{eqnarray}
	\nn	U(1)_Y&:& \quad\quad\psi\rightarrow exp({i\lambda_Y(x)Y})\psi;\quad B_{\mu}\rightarrow B_{\mu}\frac{1}{g^{\prime}}\partial_{\mu}\lambda_Y(x),\\
	\nn SU(2)_L&:& \quad\quad\psi\rightarrow exp(i\lambda_L^a(x)T^a)\psi;\quad W_{\mu}^a\rightarrow W_{\mu}^a+\frac{1}{g_W}\partial_{\mu}\lambda_L^a(x)+\epsilon^{abc}W_{\mu}^b\lambda_L^c(x),\\
	SU(3)_c&:&\quad \quad\psi\rightarrow exp(i\lambda_c^i(x)t^i)\psi;\quad\quad G_{\mu}^i\rightarrow G_{\mu}^i+\frac{1}{g_S}\partial_{\mu}\lambda_c^i(x)+f^{ijk}G_{\mu}^j\lambda_c^k(x).
\end{eqnarray}
{\bf Higgs mechanism and SSB:} In the SM, the gauge bosons are massless due to the exact gauge symmetry, which restricts explicit mass term for them. However, Robert Brout, Francios Englert and Peter Higgs \cite{Higgs:1964pj,Higgs:1966ev,Englert:1964et} came up with the Higgs mechanism, that explain mass for all the SM particles (except photons and neutrinos) through the spontaneous symmetry breaking of the $SU(2)\otimes U(1)_Y$ gauge symmetry at the EW scale. The term spontaneous symmetry breaking (SSB) signifies that the Lagrangian is symmetric under the transformation of state vector but minimum energy state or the vacuum does not respect this symmetry. 

The Higgs Lagrangian contains both kinetic as well as the potential term and it is expressed as,
\begin{equation}
	\mathcal{L}_{Higgs}=(D_{\mu}\Phi)^{\dagger}(D^{\mu}\Phi)-V(\Phi),\label{lhigs}
\end{equation} 
where \begin{equation}
	\Phi=\begin{pmatrix}
		\phi^+\\\phi^0
	\end{pmatrix}=\begin{pmatrix}
		\frac{\phi_1+i\phi_2}{\sqrt{2}}\\ \frac{\phi_3+i\phi_4}{\sqrt{2}}
	\end{pmatrix} \label{thi1}
\end{equation} and $V(\Phi)$ is the Higgs potential, and it is given by,
\begin{eqnarray}
	V(\Phi)=m^2\Phi^{\dagger}\Phi+\lambda(\Phi^{\dagger}\Phi)^2.\label{tpot1}
\end{eqnarray}
In this potential given by \eqref{tpot1}, $\lambda$ is always a positive quantity, otherwise the potential goes to negative infinity and it goes unbounded from the below at very high field values. Since, the Lagrangian respect discrete symmetry like $Z_2~(\Phi\rightarrow-\Phi)$, the potential forbids any odd power terms of $\Phi$.
For $m^2>0$, the minima of the potential is found at $|\Phi|=\sqrt{\langle 0|\Phi^{\dagger}\Phi|0\rangle}=0$. On the other hand, for $m^2<0$, the minima occurred at $|\Phi|=\sqrt{\langle 0|\Phi^{\dagger}\Phi|0\rangle}=\sqrt{-\frac{m^2}{2\lambda}}=v/\sqrt{2}.$ 
\begin{figure}[ht]
	\centering
	\includegraphics[scale=0.6]{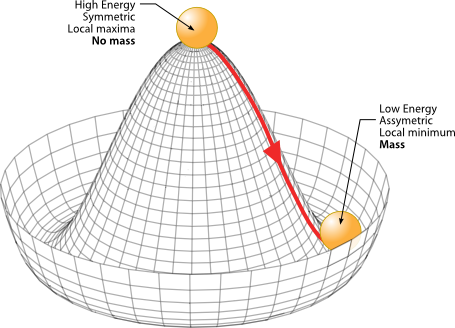}
	\caption{Schematic $maxican~hat$ representation of spontaneous symmetry breaking. The Higgs potential is symmetric at the top of hat, hence no mass. At the local minima, the symmetry is spontaneously broken and it gets mass \cite{img1}. }
\end{figure}

The fields $\phi_1,\phi_2$ and $\phi_4$ in equation \eqref{thi1} are non physical fields and by particular gauge choice they can be removed, known as unitary gauge choice. These nonphysical fields are the Goldstone bosons, and they are $`eaten$' up by the gauge bosons $W^{\pm}$, $Z$ and get massive. Hence, without any loss of generality to preserve the $U(1)_{em}$, we choose the total vacuum expectation value (VEV) in the direction of $\phi_3$. Now, the scalar field can be written as,
\begin{equation}
	\Phi=\begin{pmatrix}
		0\\ (h+v)/\sqrt{2},
	\end{pmatrix}
\end{equation}
where, $\phi_3=h+v$ and $h$ is the physical Higgs boson. Therefore the Higgs Lagrangian from equation \eqref{lhigs} takes the form as,
\begin{eqnarray}
	\nn	\mathcal{L}_{Higgs}=& \frac{1}{2}\partial_{\mu}h\partial^{\mu}+(h+v)^2\Big(\frac{g_2^2}{4}W_{\mu}^+ W^{\mu-}+\frac{g_1^2+g_2^2}{8}Z_{\mu}Z^{\mu}\Big)\\
	&-\frac{1}{4}\lambda(h+v)^2((h+v)^2-2v^2)+...\label{lhigs2}
\end{eqnarray}
Here, $W^{\pm}_{\mu}=(W_{\mu}^1\mp i W_{\mu}^2)/\sqrt{2}$  and it is straightforward to achieve the $W$ boson mass from equation \eqref{lhigs2} to be $M_W=g_2 v/2$. Photon and the $Z$ boson are the orthogonal combination of $W_{\mu}^3$ and $B_{\mu}$: $Z_W=c_WW_{\mu}^3-s_WB_{\mu}$ and $A_{\mu}=s_WW_{\mu}^3+c_WB_{\mu}$, with $c_W=\cos \theta_W$ and $s_W=\sin \theta_W$. The Weinberg angle or the weak mixing angle $\theta_W$ can be expressed as,
\begin{equation}
	\theta_W=\cos^{-1}\Big(\frac{g_2}{\sqrt{g_1^2+g_2^2}}\Big).
\end{equation}
The electric charge $e$, is the strength of the electromagnetic interaction can also be addressed using the gauge coupling constants as $e=\frac{g_1g_2}{\sqrt{g_1^2+g_2^2}}$. Finally, the mass terms obtained from equation \eqref{lhigs2} can be identified as,
\begin{eqnarray}
	\nn	M_h^2&=&2\lambda v^2,\\
	\nn M_W^2&=&\frac{1}{4}g_2^2 v^2,\\
	\nn M_Z^2&=& \frac{1}{4}(g_1^2+g_2^2)v^2.
\end{eqnarray}
The photon state $A_{\mu}$ does not have any coupling with the Higgs field, hence it remains massless.

{\bf Yukawa interaction:} After EWSB, quarks and the leptons get mass through the Yukawa terms. The most generic Yukawa Lagrangian connecting the Higgs doublet and the fermions is expressed as,
\begin{equation}
	\mathcal{L}_{Yukawa}=-\sum_{i=1;\\j=1}^{3}\Big[y_{ij}^u\bar{u}_{Ri}\tilde{\Phi^{\dagger}} Q_{Lj}+y_{ij}^d \bar{d}_{Ri}\Phi^{\dagger}Q_{Lj}+y_{ij}^l\bar{l}_{Ri}\Phi^{\dagger}L_{Lj}\Big]+h.c. \label{lyuk}
\end{equation}
Here, $\Phi$ is the Higgs doublet and $\tilde{\Phi}=i\sigma_2\Phi^*$ with $\sigma_2$ being the Pauli's second spin matrix, $\sigma_2=\begin{pmatrix}
	0&-i\\i&0
\end{pmatrix}$. $y_{ij}^u,y_{ij}^d$ and $y_{ij}^l$ are the elements of the $(3\times3)$ Yukawa matrix for respective fermions. $up-type$, $down-type$ quarks and $charged ~leptons$ get mass through equation \eqref{lyuk}, after the EWSB, and the mass matrix in the flavour basis are then given by,
\begin{equation}
	m_{ij}^f=\frac{1}{\sqrt{2}}y_{ij}^fv; ~~\quad\quad~\text{with,}~~f=u,d,l
\end{equation}  
Due to the absence of RH neutrinos, the flavour and mass bases are the same within the SM in the leptonic sector. However, the quark sector is not the same. To co-relate flavour and mass bases, we need to diagonalize the two complex matrices $m^u$ and $m^d$. This is done by multiplying the mass matrix on the left and right side by a unitary matrix. For the $up$ and $down$ type quarks, this can be expressed as,
\begin{equation}
	U^{-1}_Rm^uU_L=\begin{pmatrix}
		m_u&0&0\\0&m_c&0\\0&0&m_t
	\end{pmatrix};\quad D^{-1}_Rm^dD_L=\begin{pmatrix}
		m_d&0&0\\0&m_s&0\\0&0&m_b
	\end{pmatrix}.
\end{equation}
Hence, the flavour and mass basis are connect as,
\begin{eqnarray}
	\begin{pmatrix}
		u_1\\u_2\\u_3
	\end{pmatrix}_{L,R}=U_{L,R}\begin{pmatrix}
		u\\c\\t
	\end{pmatrix}_{L,R};\quad\quad\quad 	\begin{pmatrix}
		d_1\\d_2\\d_3
	\end{pmatrix}_{L,R}=D_{L,R}\begin{pmatrix}
		d\\s\\b
	\end{pmatrix}_{L,R}
\end{eqnarray}
These two matrices $U_L$ and $D_L$ are the matrix to rotate the left-handed $up-type$ and $down-type$ quarks. They take part in the charged-current weak interaction, which changes quark flavours. Therefore, the charged-current Lagrangian now can be written as,
\begin{eqnarray}
	-\frac{g_W}{\sqrt{2}}\begin{pmatrix}
		\bar{u}&\bar{c}&\bar{t}
	\end{pmatrix}_L\gamma^{\mu}W^+_{\mu}V_{CKM}\begin{pmatrix}
		d&s&b
	\end{pmatrix}^T+h.c.,
\end{eqnarray}
where, $V_{CKM}=U_L^{\dagger}D_L$ is the Cabibbo-Kobayashi-Maskawa (CKM) matrix. It is also convenient where all the rotations acted upon the $down-type$ quarks and the $up-type$ quarks are the mass eigenstates, $i.e.,$
\begin{equation}
	\begin{pmatrix}
		d^{\prime}&s^{\prime}&b^{\prime}
	\end{pmatrix}^T=V\begin{pmatrix}d&s&b\end{pmatrix}^T.
\end{equation} 
\subsection{Success of the SM}
The SM has been proved to be the most successful theory at describing and predicting various experiments. The fundamental prediction of the SM was the weak neutral current (WNC) and the presence of associated massive gauge bosons $viz.,$ $W^{\pm}$  and $Z$. After the successful discovery of WNC at 1973 and recently the discovery of Higgs boson at 2012 has completed the experimental verification of the all predicted particles of the SM. Among the numerous successes of the SM, we can count on the prediction of pull value, measurements of oblique parameters, Higgs signal strengths, $etc.$
\subsection{Shortcomings of the SM}
Despite being the most successful theory of the $20^\text{th}$ century in describing the elementary particles and their interactions, the SM suffers from some significant drawbacks in theoretical and experimental sectors. Gravitational interactions are not discussed under the SM. The SM fails to address neutrino mass, due to the absence of RH neutrinos, which is a severe drawback. Along with neutrino mass and mixing, there are several unanswered problems in the SM like flavour problem, mass hierarchy problem, matter dominance at our current Universe, presence of dark matter, $etc.$
In addition to these, CP problem, presence of extra generations of neutrinos, exact nature of neutrinos: Dirac of Majorana are not addressed in the SM framework. These shortfalls motivate us to go beyond the SM (BSM) framework. The primary objective of this thesis is focused on the BSM framework. Thus, in the following subsections, we will be discussing a few aspects of BSM frameworks that we study in this thesis.
\section{Beyond the SM (BSM)}
%With limited particle content and less flexibility with particle interactions, the SM failed to explain some crucial points.  
\subsection{Neutrino mass and mixing}
By nature, neutrinos are neutral particles and have tiny masses. Their interactions with matter are restricted to weak and gravitational interactions only; hence, they possess weak interaction. As the strength of gravitational interaction is far smaller than weak interaction, only weak interactions of neutrinos with matter are considered under study. The weak interaction of neutrinos with matter plays an important role in the field of particle physics as well as in the field of astrophysics and cosmology. Since they interact weakly, this makes the detection processes of neutrinos more challenging.

We know that neutrinos do changes flavour when they propagate via $neutrino~oscillation$. When a flavour of a neutrino is produced, it is a superposition of the mass eigenstates. In gauge theory, the physical states are achieved by diagonalizing the relevant mass matrices, which results from gauge symmetry breaking. In the neutrino sector, the mass and the flavour eigenstates are connected via the unitary matrix $U_{PMNS}$ or the PMNS matrix or the leptonic mixing matrix:
\begin{equation}
	\begin{pmatrix}
		\nu_e(x)\\\nu_{\mu}(x)\\\nu_{\tau}(x)
	\end{pmatrix}_L=U \begin{pmatrix}
	\nu_1(x)\\\nu_2(x)\\\nu_3(x)
\end{pmatrix}_L,
\end{equation}
where $\nu_{\alpha L}(x)$ and $\nu_{iL}(x)$ are the flavour and mass eigenstates respectively, with $\alpha=e,\mu,\tau$ and $i=1,2,3$ and $x$ signifies dependence of fields. In flavour mixing, neutrinos couple to the charged lepton of a specific flavour via weak charged current (CC) and the mass eigenstates are not involving separately but a coherent superposition of mass eigenstates that are associate with them:
\begin{equation}
	\mathcal{L}_{CC}=\frac{g_W}{2}W_{\mu}^-\sum_{\alpha=e,\mu,\tau}\bar{l}_{\alpha L}\gamma^{\mu}\nu_{\alpha L}+h.c.=\frac{g_w}{2}W_{\mu}^-\sum_{\alpha}\bar{l}_{\alpha L}\gamma^{\mu}\sum_{i=1}^{3}U_{\alpha i}\nu_{i L}+h.c..
\end{equation}

In the active neutrino scenario, the leptonic mixing matrix depends on three mixing angles $\theta_{13},\theta_{23}$ and $\theta_{12}$ and one CP-violating phase ($\delta$) for Dirac neutrinos and two Majorana phases $\alpha$ and $\beta$ for Majorana neutrino. 
If one considers the Dirac nature of neutrinos, there are six phases yet, the other five phases are not physical phases and can be absorbed by rephasing the lepton fields. However, the same is not possible in case of Majorna neutrinos, if the left-handed neutrino phases are rephased, their masses would come out to be complex; hence, two additional Majorana CP-phases $\alpha$ and $\beta$ are there. In case of Majorana neutrinos, phases do not involve in oscillation processes. Conventionally this Leptonic mass matrix for active neutrino is parameterized as \cite{Valle:2006vb},
\begin{eqnarray}
	\nn U_{PMNS}&=&\begin{pmatrix}
		1&0&0\\0&c_{23}&s_{23}\\0&-s_{23}&c_{23}
	\end{pmatrix}\begin{pmatrix}
	c_{13}&0&s_{13}e^{-i\delta}\\0&1&0\\-s_{13}e^{-i\delta}&0&c_{13}
\end{pmatrix}\begin{pmatrix}
c_{12}&s_{12}&0\\-s_{12}&c_{12}&0\\0&0&1
\end{pmatrix}P\\
	&&{\begin{pmatrix}
		c_{12}c_{13}&s_{12}c_{13}&s_{13}e^{-i\delta}\\
		-s_{12}c_{23}-c_{12}s_{23}s_{13}e^{i\delta}&c_{12}c_{23}-s_{12}s_{23}s_{13}e^{i\delta}&s_{23}c_{13}\\
		s_{12}s_{23}-c_{12}c_{23}s_{13}e^{i\delta}&-c_{12}s_{23}-s_{12}c_{23}s_{13}e^{i\delta}&c_{23}c_{13}\\
\end{pmatrix}}P.
\end{eqnarray}
The abbreviations used are $c_{ij}=\cos\theta_{ij}$ , $s_{ij}=\sin\theta_{ij}$ and $P$ would be a unit matrix \textbf{1} in the Dirac case but in Majorana case $P=\text{diag}(1,e^{i\alpha},e^{i(\beta+\delta)})$. As a result of new physics, the trivial $(3\times3)$ unitary leptonic mixing matrix, $U_{PMNS}$ may slightly deviate from its generic unitarity behaviour~\cite{Antusch:2006vwa,Akhmedov:2013hec}. In general, the mixing between active and sterile lead to non-unitarity in the $U_{PMNS}$ matrix. However, in our study, we have considered a minimal mixing between the active-sterile neutrinos, which does not bother the active neutrino scenario. Moreover, current neutrino oscillation data and electroweak precision measurements have contrived the $U_{PMNS}$ to be unitary at the $\mathcal{O}(10^{-2})$ level \cite{Xing:2019vks}. 

In a situation, where an additional flavour of neutrino $(sterile~neutrino)$ is added, we can use the $(4\times4)$ unitary PMNS matrix to diagonalize the neutrino mass matrix. This contains six mixing angles ($\theta_{12},\theta_{13},\theta_{23},\theta_{14},\theta_{24},\theta_{34}$), three Dirac CP phases ($\delta_{13},\delta_{14},\delta_{24}$) and three Majorana phases ($\alpha,\beta,\gamma$). The parametrization of the PMNS matrix can be represented as \cite{Gariazzo:2015rra},
\begin{equation}
	U_{PMNS}=R_{43}\tilde{R_{24}}\tilde{R_{14}}R_{23}\tilde{R_{13}}R_{12}P,
\end{equation}
where $R{ij},\tilde{R{ij}}$ are the rotational matrices with rotation in respective planes and $P$ is the diagonal Majorana phase matrix, which are given by,
\begin{eqnarray}
	\nn
	R_{34}=\begin{pmatrix}
		1&0&0&0\\
		0&1&0&0\\
		0&0&c_{34}&s_{34}\\
		0&0&-s_{34}&c_{34}\\
	\end{pmatrix},~\tilde{R_{24}}=\begin{pmatrix}
		1&0&0&0\\
		0&c_{24}&0&s_{24}e^{i\delta_{24}}\\
		0&0&1&0\\
		0&-s_{24}e^{i\delta_{24}}&0&c_{24}\\
	\end{pmatrix},\\P=\begin{pmatrix}
		1&0&0&0\\
		0&e^{i\alpha/2}&0&0\\
		0&0&e^{i(\beta/2-\delta_{13})}&0\\
		0&0&0&e^{-i(\gamma/2-\delta_{14})}\\
	\end{pmatrix}.
\end{eqnarray}

For active-sterile phenomenology, the most important elements in this unitary matrix are the fourth column elements. They are expressed as follows \cite{Hagstotz:2020ukm},
\begin{eqnarray}
	&|U_{e4}|^2=&\sin^2\theta_{14},\\
	&|U_{\mu4}|^2=&\cos^2\theta_{14}\sin^2\theta_{24},\\
	&|U_{\tau4}|^2=&\cos^2\theta_{14}\cos^2\theta_{24}\sin^2\theta_{34},\\
	&|U_{s4}|^2=&\cos^2\theta_{14}\cos^2\theta_{24}\cos^2\theta_{34}.
\end{eqnarray}
The active-sterile mixing angle $\theta_{i4}$ are essentially small, such that they do not influence the three neutrino mixing, which is allowed by current global fit limits. In principle, $|U_{s4}|^2$ be of the order of unity and other elements are expected to be very small. Hence, we can refer to the fourth neutrino mass eigenstate as the sterile neutrino itself, {\it i.e.,} $m_{4}\sim m_S$.

Therefore, the light neutrino mass matrix $M_{\nu}$ is diagonalized by the unitary PMNS matrix as,
\begin{equation}
	\label{eq:2}
	M_{\nu}=U_{PMNS}\ m_{\nu}^D\ U_{PMNS}^{T},
\end{equation}
where $m_{\nu}^D$ stands for diagonal mass eigenstates for the neutrino. In the case of active neutrinos it is $(3\times3)$ diagonal matrix whereas, for the active-sterile case, it possesses a $(4\times4)$ diagonal structure.

The sterile neutrino of mass of order eV, can be added to the standard 3-neutrino mass states in NH: $m_1\ll m_2<m_3\ll m_4$ as well as IH: $m_3\ll m_1<m_2\ll m_4$. The diagonal light neutrino mass matrix for NH emerges as, $m_{\nu}^{NH}=\text {diag}(0, \sqrt{\Delta m_{21}^{2}}, \sqrt{\Delta m_{21}^{2}+\Delta m_{31}^{2}},$ $\sqrt{\Delta m_{41}^{2}})$ and for IH as,  $m_{\nu}^{IH}=\text{diag}(\sqrt{\Delta m_{31}^{2}},\sqrt{\Delta m_{21}^{2}+\Delta m_{31}^{2}},0,\sqrt{\Delta m_{43}^{2}})$ . The lightest neutrino mass is zero in both the mass ordering as demanded by the MES framework. Here, $\Delta m_{41}^{2}(\Delta m_{43}^{2})$ is the active-sterile mass square difference for NH and IH respectively.

\subsubsection{Dirac and Majorana neutrino mass}
One of the most intriguing thing about neutrinos that makes them unique is the nature of their mass. The fact that neutrinos do not carry any electrical charges; hence, one can write two types of Lorentz invariant mass terms: Dirac and Majorana masses. When the particle and anti-particle are different, then one can write the Dirac mass term. In four-vector notation, the Dirac mass has the form $\bar{\psi}\psi$, and it connects the fields of opposite chirality, here $\psi$ is the four-component spinor. On the other hand, when the fermion $\psi$ is its own antiparticle, we can define the Majorana mass term as, $\psi^TC^{-1}\psi$ connecting the fields of same chirality, with $C$ being the charge conjugation matrix.

If the neutrinos are Dirac, the situation is similar to any fermion mass generation with electric charge conservation. However, if neutrinos are Majorana in nature, it could naturally explain why they are much lighter than the charged fermions. Lepton number is conserved if neutrinos are Dirac whereas lepton number is violated by two units in case of Majorana neutrinos, as they are their own anti-particle. When the Majorana mass term is addressed, the lepton number will be violated. This may have several significant consequences in neutrino phenomenology and cosmology, some of which we will try to address in this thesis. In a Lagrangian, where both Dirac and Majorana mass terms are addressed, a pair of Majorana fermions must be involved there, irrespective of the Majorana mass scale. As a conclusive discussion, since neutrinos are chargeless fermions, it is simple and more natural to address neutrinos as a `Majorana fermion' with lesser assumptions, rather than `Dirac fermions'. Throughout this thesis, we will address neutrino as a Majorana fermion, unless otherwise noted.
\subsection{Models of neutrino masses}
Unlike the other fermions in the SM, neutrinos are charge-less, massless, and they are colour singlet. SM also left no room for either any right-handed neutrinos or anti-neutrinos. Consequently, one can not write Dirac or Majorana mass term for neutrinos in the SM. Though the theoretical and experimental progresses so far, it is well established that neutrinos do oscillate and acquire $tiny$ masses. Thus we have to go beyond the SM to explain neutrino mass generation. There are various BSM mechanisms to explain neutrino mass, $viz.$, seesaw mechanism \cite{Minkowski:1977sc,Mohapatra:1979ia}, radiative seesaw mechanism \cite{Ma:2006km}, left-right symmetric model (LRSM) \cite{FileviezPerez:2008sr}, etc. 
\subsubsection{The seesaw mechanism}
The seesaw mechanism is the most appealing and convenient approach to introduce neutrino masses among the various neutrino mass models. An essential ingredient of the seesaw framework is the addition of very heavy Majorana masses. In a scenario with heavy neutrinos and other fermions, if neutrinos are also influenced by Higgs's mechanism, this will give rise to neutrino masses. These masses will get suppressed by the ratio of the other fermion masses and heavy Majorana masses. Hence, within the seesaw models, we can get tiny neutrino masses that we observe in Nature. 

Various types of seesaw mechanisms have been put in literature till date (for detail one may look at \cite{Minkowski:1977sc,Mohapatra:1979ia,Weinberg:1979sa,Babu:1993qv,Mukherjee:2015axj,Dev:2012bd,King:2003jb,Bernabeu:1987gr,Foot:1988aq,Mohapatra:1986bd}).
To address active and sterile neutrinos together under a single framework, the minimal extended seesaw (MES) \cite{Barry:2011wb, Zhang:2011vh} plays a key role. Here, the canonical type-I seesaw is extended with a singlet fermion and along with three active neutrinos, it can simultaneously explain sterile neutrino mass from eV to $keV$ range. Our whole thesis work is based on this MES framework, hence, we will briefly discuss other popular seesaw frameworks and give a detailed analysis for MES framework in the following subsections.

{\bf Type I seesaw :}
In type I seesaw mechanism, a minimum of two generations of heavy right-handed neutrinos are added to the SM particle content. The most general Lagrangian compatible with the SM gauge symmetry is given by,
\begin{equation}
	-\mathcal{L}\supset Y_{ij}^{\nu}\bar{L}_i\tilde{\Phi}\nu_{R_j}+\frac{1}{2}\bar{\nu}_{R_i}^cM_{ij}\nu_{R_j}+h.c.
\end{equation}
Here, $Y_{ij}^{\nu} $ is the Yukawa coupling associating left and right-handed fields through Higgs doublet $\Phi$. $M_{ij}$ is the Majorana mass matrix, which explicitly violates lepton number. The mass scale of the RH neutrinos are assumed to be very higher than the Dirac neutrino mass, $M>>Y^{\nu}\langle\phi^0\rangle$ within type-I seesaw and the mass for left-handed neutrinos are found as,
\begin{equation}
	\mathcal{M}_{\nu}\simeq-Y^{\nu}M^{-1}(Y^{\nu})^T\langle\phi^0\rangle.
\end{equation}

{\bf Type II seesaw :} There exist at least one scalar triplet with hypercharge 1 in the type II seesaw mechanism. Structure of the scalar triplet contains three complex scalars $\Delta^0,\Delta^+, \Delta^{++}$ and they can be cast as,
\begin{equation}
	\Delta=\begin{pmatrix}
		\Delta^+/\sqrt{2}&\Delta^{++}\\\Delta^0&-\Delta^{+}/\sqrt{2}
	\end{pmatrix}.
\end{equation}  
The Lagrangian compatible with the gauge symmetry is given as,
\begin{equation}
	-\mathcal{L}\supset\Big(Y_{ij}^{\Delta}L_i^TC\Delta L_j-\mu\tilde{\Phi}^T\Delta\tilde{\Phi}+h.c.\Big)+M_{\Delta}^2\text{Tr}(\Delta^{\dagger}\Delta),
\end{equation}
here, $Y^{\Delta}_{ij}$ is a Yukawa coupling, $\mu$ is a lepton-number violating coupling with mass dimension and $M_{\Delta}$ is the triplet mass. Like the type-I seesaw model, here also the mass of the triplet is assumed to be high, {\it i.e.,} $M_{\Delta}^2>>\mu\langle\phi^0\rangle$ and the neutrino mass is realized as,
\begin{equation}
	\mathcal{M}_{\nu}\simeq\frac{\mu\langle\phi^0\rangle^2}{M_{\Delta}^2}Y^{\Delta}.
\end{equation}

{\bf Type III seesaw :} The type III seesaw framework contains a fermion triplet with zero hypercharge, which can be represented as,
\begin{equation}
	\Sigma_i=\begin{pmatrix}
		\Sigma^0_i&\sqrt{2}\Sigma^+_i\\\sqrt{2}\Sigma^+_i&-\Sigma^0_i
	\end{pmatrix}.
\end{equation} 
The invariant Lagrangian is given by,
\begin{equation}
	-\mathcal{L}\supset Y_{ij}^{\Sigma}\bar{L}_i\tilde{\Phi}{\Sigma}_j+\frac{1}{2}M_{ij}\text{Tr}(\bar{\Sigma}_i{\Sigma}_j^c )+h.c.,
\end{equation}
here, $M_{ij}$ is the lepton-number violating Majorana mass term and  $Y_{ij}^{\Sigma}$ is the Yukawa coupling. Left-handed neutrino mass term can be found as,
\begin{equation}
	\mathcal{M}_{\nu}=-Y^{\Sigma}M_{\Sigma}^{-1}(Y^{\Sigma})^T\langle\phi^0\rangle^2.
\end{equation}

{\bf Inverse Seesaw (IS) :} The inverse seesaw contains TeV scaled RH neutrino, along with a singlet fermion $S$. Lagrangian in type III seesaw is given by,
\begin{equation}
	-\mathcal{L}\supset Y_{ij}^{\nu}L_i\tilde{\Phi}\nu_{R_j}+M^{ij}_R\bar{\nu}^C_{R_i}S_j-\frac{1}{2}\mu_R^{ij}\bar{\nu}_{Ri}^C\nu_{R_j}-\frac{1}{2}\mu_S^{ij}\bar{S}_i^CS_j+h.c.,
\end{equation}
where, $Y_{ij}^{\nu}$ is the neutrino Yukawa coupling, $M_R$ is lepton number conserving mass matrix, $\mu_R$ and $\mu_S$ are the lepton number violating symmetric mass matrices.
The neutral mass matrix in the basis ($\nu_{L}, \nu_{R}^c,S^c$) takes the form
\begin{equation}
	M_{\nu}^{IS}=
	\begin{pmatrix}
		0&M_D&0\\M_{D}^{T}&\mu_R&M_{R}^{T}\\0&M_{R}&\mu_S 
	\end{pmatrix}.
\end{equation}
In this case, $\mu\sim M_D<<M_S$ and this scenario has been termed as the inverse seesaw. Typically here $n(\nu_L)=n(\nu_R)=S=3$ however $n(\nu_R)=n(S)=2$ is also viable. The light neutrino mass can be evaluated in ISS as,
\begin{equation}
	\mathcal{M}_{\nu}\simeq M_D(M_R^T)^{-1}\mu_S M_R^{-1}M_D^T
\end{equation}

{\bf Minimal Extended Seesaw (MES) :} In MES scenario along with the SM particle, three extra right-handed neutrinos and one additional gauge singlet chiral field S is introduced. The Lagrangian of the neutrino mass terms for MES is given by: 
\begin{equation}
	-\mathcal{L}_{\mathcal{M}}= \overline{\nu_{L}}M_{D}\nu_{R}+\frac{1}{2}\overline{\nu^{c}_{R}}M_{R}\nu_{R}+\overline{S^c}M_{S}\nu_{R}+h.c. ,
\end{equation} 
The neutrino mass matrix will be a $7\times7$ matrix, which, in the basis ($\nu_{L},\nu_{R}^{c},S^c$), reads as
\begin{equation}
	M_{\nu}^{7\times7}=
	\begin{pmatrix}
		0&M_D&0\\M_{D}^{T}&M_{R}&M_{S}^{T}\\0&M_{S}&0 
	\end{pmatrix}\label{7b7}.
\end{equation}
Here $M_D$ and $M_R$ are $(3\times3)$ Dirac and Majorana mass matrices respectively whereas $M_S$ is a $(1\times3)$ matrix. As per the standard argument \cite{Schechter:1980gr}, the number of massless state is defined as $n(\nu_L)+n(S)-n(\nu_R)$, within our framework it is one ($3+1-3=1$), which is also verified by taking the determinant of the matrix \eqref{4b4} in the next paragraph of this section. The zeros at the corners of the $(7\times7)$ matrix of \eqref{7b7} have been enforced and motivated by some symmetry. This can be achieved with discrete flavor symmetry due to which it is clear that right handed neutrinos and $S$ carry different charges. Moreover, the MES structure could also be explained with the abelian symmetry. For example, one may introduce additional $U(1)^{\prime}$ under which along with the all SM particles we assumed and 3 RH neutrinos to be neutral. The RH singlet $S$ on the other hand carries a $U(1)^{\prime}$ charge $Y^{\prime}$ and we further introduce a SM singlet $\chi$ with hypercharge $-Y^{\prime}$. The matrix $M_S$ is generated by the gauge invariant coupling $S^c\chi\nu_R$ after $\chi$ acquires a VEV, while the Majorana mass for $S$($i.e., \overline{S^c}S$) and a coupling with the active neutrino $\nu_L$ are still forbidden by the $U(1)^{\prime}$ symmetry at the renormalizable level\cite{Barry:2011wb,Chun:1995js, Heeck:2012bz}. This explains the zeros in the $(7\times7)$ matrix. 
In the analogy of type-I seesaw, the mass spectrum of these mass matrices is considered as $M_{R}\gg M_{S}>M_{D}$, so that the heavy neutrinos decoupled at low scale. After diagonalizing, $(4\times4)$ neutrino mass matrix in the basis $(\nu_{L},S^c)$, is given by,
\begin{equation}
	M_{\nu}^{4\times4}=
	-\begin{pmatrix}
		M_{D}M_{R}^{-1}M_{D}^{T}&M_{D}M_{R}^{-1}M_{S}^{T}\\
		M_{S}(M_{R}^{-1})^{T}M_{D}^{T} & M_{S}M_{R}^{-1}M_{S}^{T}
	\end{pmatrix}\label{4b4}.
\end{equation}
Here in  $M_{\nu}^{4\times4}$ matrix \eqref{4b4}, there exists three eigenstates exists for three active neutrinos and one for the light sterile neutrino. Taking the determinant of equation \eqref{4b4}, we get,
% using $\text{det}\begin{pmatrix}A&B\\C&D \end{pmatrix}=\text{det}(A)\text{det}(D-CA^{-1}B)$ , we can get 
\begin{eqnarray}
	\nn		\text{det}(M^{4\times4}_{\nu})& =&
	\text{det}(M_{D}M_{R}^{-1}M_{D}^{T})\text{det}[ -M_{S}M_{R}^{-1}M_{S}^{T}\\\nn&&+ M_{S}(M_{R}^{-1})^{T}M_{D}^{T}( M_{D}M_{R}^{-1}M_{D}^{T})^{-1}(M_{D}M_{R}^{-1}M_{S}^{T})]\\
	\nn &=&\text{det}(M_{D}M_{R}^{-1}M_{D}^{T})\text{det}[M_{S}(M_{R}^{-1}-M_{R}^{-1})M_{S}^{T}]\\
	& =&0.
\end{eqnarray}
Here the zero determinant indicates that one of the eigenvalue is zero. Thus, the MES formalism demands one of the light neutrino mass be exactly vanished.
Proceeding for diagonalization, we face three choices of ordering of $M_S$ :
\begin{itemize}
	\item $M_{D}\sim M_{S}$: This indicates a maximal mixing between active and sterile neutrinos which is not compatible with the neutrino data.
	\item $M_{D}>M_{S}$: The light neutrino mass is obtained same as type-I seesaw $i.e.$, $m_{\nu}\simeq-M_{D}M_{R}^{-1}M_{D}^{T}$ and there would not be any sterile mass. Moreover, from the experimental bound on the active-sterile mass, the active neutrino masses would be in the eV scale which would contradict the standard Planck limit for the sum of the active neutrinos. 
	\item $M_{S}>M_{D}$: which would give the possible phenomenology for active-sterile mixing.
\end{itemize}
Now applying the seesaw mechanism to equation \eqref{4b4}, we get the active neutrino mass matrix as
\begin{equation}\label{tamass}
	m_{\nu}\simeq M_{D}M_{R}^{-1}M_{S}^T(M_{S}M_{R}^{-1}M_{S}^{T})^{-1}M_{S}(M_{R}^{-1})^{T}M_{D}^{T}-M_{D}M_{R}^{-1}M_{D}^{T}, 
\end{equation} and the sterile neutrino mass as
\begin{equation}\label{tsmass}
	m_{s}\simeq -M_{S}M_{R}^{-1}M_{S}^{T}.
\end{equation}
The first term of the active neutrino mass does not vanish, since $M_S$ is not a square matrix. Otherwise it would mutually cancel the two terms of the active neutrino mass term if $M_S$ were a square matrix. 
%In our study, we vary $m_s$ mass scale in between eV-$keV$ range, whereas $M_S$ scale is slightly greater than $M_D$ scale, which is near to EW scale.

The RH neutrino mass scale is near the GUT scale for MES, whereas for the inverse seesaw case, the RH mass scale is much smaller than the earlier one, hence the lepton number violating scale also.

\subsection{Neutrino oscillations}
Two massive neutral states with the same quantum number and slightly different mass can oscillating into each other. In matter-antimatter oscillation processes, there is an apparent violation of $C$ and $CP$, however, in neutrino oscillation case one flavour of a neutrino is oscillating to another flavour. They are associated with the mass eigenstates, resulting in a violation of the lepton number and the lepton flavour in the whole process. The oscillation processes will occur only when neutrinos do have mass and that mass difference is not too high. The neutrino oscillation process can be understood as follows: a definite flavour of a neutrino is generated in a charged-current interaction (a process involving an electron, will accompany with an electron neutrino, $\nu_e$ ). The physical state of this neutrino propagates with time and there is a probability that it may changes to either $\nu_{\mu}$ or $\mu_{\tau}$ flavour. When the two states have the same mass, there would not be any more oscillation and they will propagate in the same direction. If the initial and final states are of different mass but are in the same flavour states, then the flavour state will undergo evolution and not change to each other. Hence, there must be a mass difference between the states and the distinct difference between flavour with mass eigenstates in neutrino oscillation processes. Those flavour and mass eigenstates are connected through the $leptonic~ mixing~ matrix$, as discussed earlier in this chapter. In the following subsections, we will be discussing neutrino oscillation in a vacuum as well in matter.  
\subsubsection{Neutrino oscillation in vacuum}
An ideal oscillation experiment involves three steps. First, the production of neutrino flavour state from the charged state via charged-current processes. Next is the propagation step of the neutrino flavour and the last step is the detection process. Here, in this section, we will outline the essential steps involving neutrino oscillation in a vacuum in brief.

The flavour and mass eigenstates are related as,
\begin{equation}
	|\nu_{\alpha}\rangle=\sum_i U_{\alpha i}^*|\nu_i\rangle.
\end{equation}
The physical states will involve with time as,
\begin{equation}
	|\nu(t)\rangle=\sum_i U_{\alpha i}^*e^{-iE_it}|\nu_i\rangle=\sum_i U^*_{\alpha i}e^{-iE_it}\sum_{\beta}U_{\beta i}|\nu_{\beta}\rangle,
\end{equation}
where, $E_i=\sqrt{p_i^2+m_i^2}$ and for ultra-relativistic neutrinos ($p_i>>m_i$), it reduced to $E_i=p_i+\frac{m_i^2}{2p_i}$.  The probability of finding $\nu_{\beta}$ from the source $\nu_e$ after time $t$ is \cite{Sarkar:2008xir},
\begin{eqnarray}
	P_{\nu_{\alpha}\rightarrow\nu_{\beta}}=|\langle\nu_{\beta}(0)|\nu_{\alpha}(t)\rangle|^2=\sin^2(2\theta)\sin^2\Big(\frac{\Delta m^2 L}{4E}\Big),\label{osc1}
\end{eqnarray}
where, $E$ is the average energy of the neutrino, $\Delta m=|m_2^2-m_1^2|$ and $L$ is the length traversed by the beam before a $\nu_{\beta}$ detection was taking place. It is to be noted that, the oscillation probability depends upon the two parameters  $\sin^2(2\theta)$ and $\Delta m^2$. An important parameter to define here is the oscillation length. This is the distance, after which the neutrino flavour returns to its initial stage, $i.e.$, the phase become $2\pi$. This can be define as, $L_0=2.47\frac{E(\text{GeV})}{\Delta m^2(\text{eV}^2)}$ KM.

The three-generation oscillation also similar to the two flavour case and the oscillation probability of finding $\nu_{\beta}$ in a beam of $\nu_{\alpha}$ after a time $t$, can be expressed as,
\begin{equation}
	P_{\nu_{\alpha}\rightarrow\nu_{\beta}}=\sum_{ij}|U_{\alpha i}U^*_{\beta i}U^*_{\alpha j}U_{\beta j}|\cos \Big(\frac{2\pi L}{L_0^{ij}}\Big).
\end{equation}
In three generation case, there are two mass squared differences and three mixing angles. If one mass squared difference is zero, then we get the two-generation oscillation probability. Same is true if only one mixing angle is non-zero. 

If there is an extra generation of neutrino is present (sterile neutrino) along with the three generations of active neutrino, then also we can work out the oscillation probability. As discussed earlier, the ($1+3$) scheme of adding sterile neutrino in the active neutrino sector is ruled out, so the oscillation probability in the $(3+1)$ pattern can be worked out. If one considers short-baseline (SBL) oscillation for which $\Delta m^2 L/E<<\Delta m_{SBL}^2L/E\simeq1$ (here $\Delta m_{SBL}^2\equiv\Delta m_{41}^2\simeq\Delta m^2{42}\simeq\Delta m^2_{43}$ ), one can get the appearance probability as,
\begin{equation}
	P_{\nu_{\alpha}\rightarrow\nu_{\beta}}=\sin^2(2\theta^{SBL}_{\alpha\beta})\sin^2\Big(\frac{\Delta m^2_{41} L}{4E}\Big),~~~ \sin^2(2\theta^{SBL}_{\alpha\beta})\equiv4|U_{\alpha4U_{\beta4}}|^2 ~\text{with} ~(\alpha\ne\beta),
\end{equation}
and the disappearance probability as,
\begin{equation}
	P_{\nu_{\alpha}\rightarrow\nu_{\alpha}}=1-\sin^2(2\theta^{SBL}_{\alpha\alpha})\sin^2\Big(\frac{\Delta m^2_{41} L}{4E}\Big),~~~ \sin^2(2\theta^{SBL}_{\alpha\alpha})\equiv4|U_{\alpha4}|^2(1-U_{\alpha4}|^2).
\end{equation}
\subsubsection{Neutrino oscillation in matter}
When neutrinos propagate through matter, their mass could vary and many factors affecting them. For example, when $\nu_e$ and $\nu_{\mu}$ both are propagating through matter with varying density, $\nu_e$ will interact through matter via charged-current interaction and their mass will change; on the other hand there will not be any effect on $\nu_{\mu}$. Hence, oscillation through matter influences the mass squared differences, leading to the phenomenon of resonant oscillation, called the MSW effect, named after Mikheyev, Smirnov and Wolfenstein. 

While propagating through matter, the electron neutrino potential gets modified due to the charged-current interaction and this potential is proportional to the electron number density ($V_{\nu_e}=\sqrt{2}G_FN_e$). Hence, the effective mass of $\nu_e$ will be modified as,
\begin{equation}
	m_{nu_e}^2\rightarrow m_{\nu_e}^2+A=m_{\nu_e}^2+2\sqrt{2}G_FN_eE.
\end{equation}
Suppose two effective mass eigenvalues $m_1$ and $m_2$ are related to two mass eigenstates via a orthogonal mixing matrix $O$, with 
$$O=\begin{pmatrix}
	\cos\theta&\sin\theta\\-\sin\theta&\cos\theta
\end{pmatrix}$$
then the light neutrino mass matrix can be written as,
\begin{equation}
	M_{\nu}^2\simeq O^T(M_{\nu}^{\text{Diag}})^2O+\begin{pmatrix}
		A&0\\0&0
	\end{pmatrix}.
\end{equation}
Now, we can define a mass squared difference $\Delta m_{12}=|m_2^2-m_1^2|$ and $m_0=(m_1^2m_2^2+A)$ and the mass-squared matrix can be written as,
\begin{equation}
	M_{\nu}^2=\frac{m_0}{2}\begin{pmatrix}
		1&0\\0&1
	\end{pmatrix}+\frac{1}{2}\begin{pmatrix}
		A-\Delta m_{12}\cos2\theta&\Delta m_{12}\sin 2\theta\\\Delta m_{12}\sin 2\theta&-A+\Delta m_{12}\cos2\theta
	\end{pmatrix}
\end{equation}
The eigenvalue of this matrix are,
\begin{equation}
	m_{\nu_{1,2}}=\frac{m_0}{2}\pm\frac{1}{2}\sqrt{(\Delta m_{12}\cos2\theta-A)^2+\Delta m^2_{12}\sin^2 2\theta},
\end{equation}
and the effective mixing angle $\tilde{\theta}$ becomes,
\begin{equation}
	tan2\tilde{\theta}=\frac{\Delta m_{12}\sin2\theta}{\Delta m_{12}\cos2\theta-A}.
\end{equation}
Hence, the mixing angle and the effective physical mass of two neutrinos depend upon electron number density $N_e$. This effects have potential influence inside the sun, as the matter density is very high near the core. 

The three-generation and four-generation (three active+one sterile) neutrino oscillation are also essential to discuss and significantly influence the other oscillation parameters. In those cases, one needs to consider the non-adiabatic region of parameter spaces and all neutrinos' flavour to be included. However, those discussions need lots of approximations and complex calculations, yet the basic idea discussed here remains the same. Those discussions may deviate this thesis's original motivation; hence, we stop at this point and skip further discussion on this topic. 
\subsection{Baryogenesis}
Giving consideration to various established suggestions regarding the evolution of our Universe, it is confirmed that, at the very beginning, there were equal numbers of matter (protons and neutrons) and corresponding anti-matter (anti-protons and anti-neutrons). However, there is an asymmetry in the observed baryon number in the current scenario, and the scenario can be explained by the process, popularly known as baryon asymmetry of the Universe (BAU) baryogenesis. Baryogenesis could be the portal connecting particle physics and cosmology and opening up newer unexplored areas. 

Numerical definition for baryon asymmetry at the current date can be defined in two equivalent ways as \cite{Aghanim:2018eyx},
\begin{eqnarray}
	\eta_B\equiv\frac{n_B-n_{\bar{B}}}{n_{\gamma}}=(6.1\pm0.18)\times10^{-10};\\
	Y_{B}\equiv \frac{n_B-n_{\bar{B}}}{s}=(8.75\pm0.23)\times 10^{-11},
\end{eqnarray}
where, $n_B, n_{\bar{B}}, n_{\gamma}$ and $s$ are the densities of baryon, anti-baryon, photon and entropy respectively. As the Universe started with an equal amount of baryons and anti-baryons, these small primordial baryon asymmetries are not natural. There must have some underlying particle physics interactions that should have generated these small values before nucleosynthesis.

In 1967, Sakharov proposed three criteria involving elementary interactions to generate the currently observed baryon asymmetry from symmetric baryon universe. Those are considered as basic ingredients to explain baryogenesis \cite{Sakharov:1967dj},
\begin{itemize}
	\item Baryon number violation: This condition demands, baryon number must be violated to observe $Y_B\ne0$ starting from $Y_B=0$.
	\item $C$ and $CP$ violation: If either $C$ or $CP$ were conserved, then the rate of evolution involving baryons and anti-baryons would be same; hence, no asymmetry would have observed.
	\item Departure from thermal equilibrium: At thermal bath, there would not be any asymmetries in quantum numbers that are not conserved.  
\end{itemize}
The SM does have all these ingredients to explain this asymmetry, however, no SM mechanism have reported such large amount of baryon asymmetry yet. 
On the contrary, seesaw mechanisms are in pretty good agreement in accommodating baryogenesis, due to heavy RH neutrinos. Moreover, as seesaw demands lepton number violation, new CP-violating phases in the neutrino Yukawa interactions are eventually generated. It can be assumed that heavy singlet neutrinos decay out of equilibrium can produce this asymmetry. Thus, all three Sakharov conditions are satisfied naturally. Hence, baryogenesis becomes an integral part of the seesaw framework.

There are several models describing baryogenesis in different ways. Some of the possible mechanisms for baryogenesis are GUT baryogenesis \cite{Riotto:1998bt}, Electroweak baryogenesis \cite{Riotto:1999yt}, The Affleck-Dine mechanism \cite{Allahverdi:2012ju}, Leptogenesis \cite{Davidson:2008bu,Pilaftsis:2003gt}, $etc.$ Among these, the most popular mechanism to explain baryogenesis is $via$ lepton number violation or $leptogenesis$. In this thesis, we will be considering leptogenesis processes (thermal and resonant leptogenesis) to explain baryogenesis.
%In the following subsection, we will be discussing thermal and resonant leptogenesis mechanism in details. 
\subsubsection{Leptogenesis}
Fukugita and Yanagida first proposed the Leptogenesis mechanism  \cite{Fukugita:1986hr} back in 1986. They have proposed that during Majorana mass generation, the excess lepton number originated may transform into the baryon number excess through the baryon number violation of electroweak processes at high temperatures. Leptogenesis scenario appears naturally within the seesaw frameworks. The Yukawa coupling associated with the newly introduced heavy and hierarchical neutrinos (right-handed neutrinos in type-I seesaw) provides the necessary CP violation source. Leptogenesis becomes an essential study in seesaw frameworks, as it connects the neutrino sector and baryogenesis under a single frame. Leptogenesis is quite similar to the GUT baryogenesis \cite{Riotto:1998bt}, where the deviation from the equilibrium distribution of some heavy particle fuels up the necessary departure from thermal equilibrium. 

To execute leptogenesis, all three Sakharov conditions for baryogenesis must satisfy. The amount of CP violation needed in leptogenesis process can be achieved {\it via} tree-level and one-loop level diagrams. Slow Yukawa interactions above the electroweak scale associate with the decay of the right-handed neutrino(s) make it possible to departure from the thermal equilibrium. Hence, it is almost unavoidable when one involves seesaw frameworks to study the tininess of neutrino mass. There are several mechanisms to produce lepton asymmetry, {\it viz.} thermal leptogenesis, resonant leptogenesis, soft leptogenesis, Dirac leptogenesis, vanilla leptogenesis, {\it etc.} Among these choices of leptogenesis mechanism, in this thesis, we only study the first two from this list, {\it i.e.,} thermal and resonant leptogenesis.
\begin{figure}
	\centering
	\includegraphics[scale=0.65]{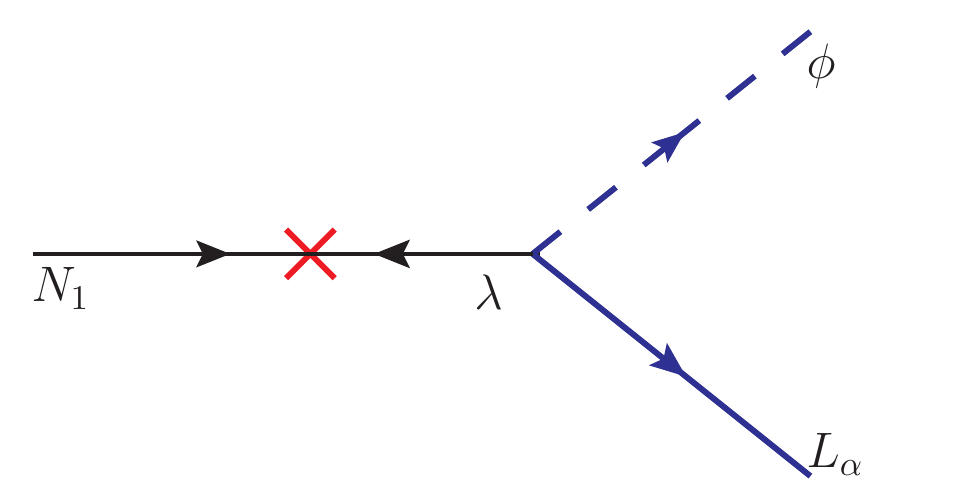}
	\includegraphics[scale=0.65]{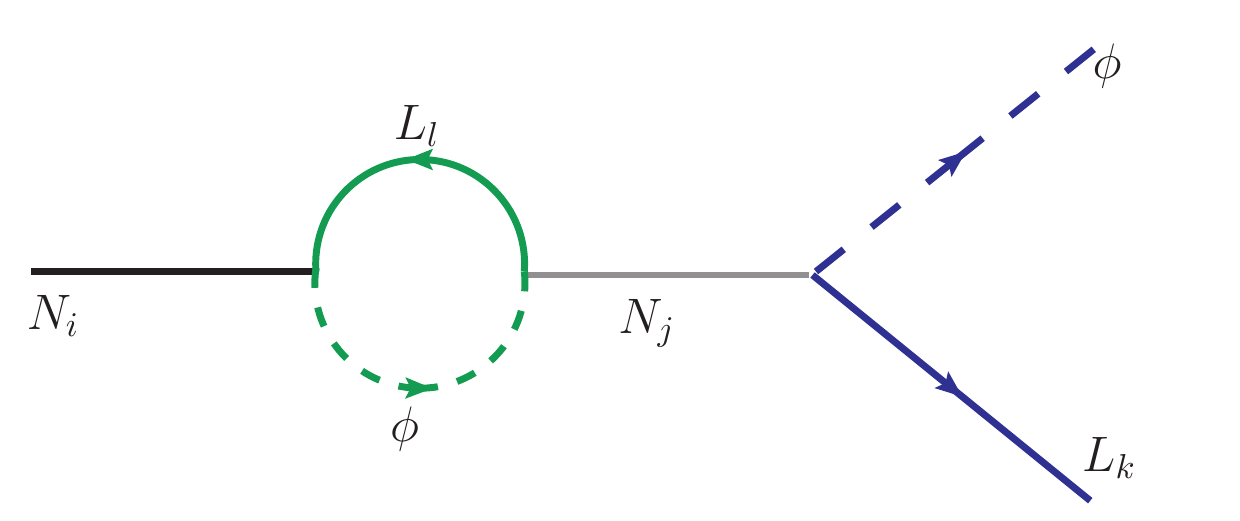}
	\caption{Tree level and one-loop level decay mode of leptogenesis involving right handed neutrino.}
\end{figure}
\subsubsection{Thermal leptogenesis}\label{thermlep}
In the context of seesaw frameworks, if the RH neutrinos are relatively heavy ($10^{14}~\text{GeV}>N_1\ge 10^{9}$ GeV) and they are hierarchical ($N_1<<N_{2,3}$), then thermal leptogenesis becomes feasible. 
When we consider a hierarchical mass pattern for RH neutrinos, the lightest one will decay to a Higgs and a lepton. This decay would produce sufficient lepton asymmetry to give rise to the observed baryon asymmetry of the Universe. Both baryon number ($B$) and lepton number ($L$) are conserved independently in the SM renormalizable Lagrangian. However, due to chiral anomaly, there are non-perturbative gauge field configurations \cite{Callan:1976je}, which produces the anomalous $B+L$ violation ($B-L$ is already conserved). These whole process of conversion of lepton asymmetry to baryon asymmetry via $B+L$ violation is popularly termed as ``sphalerons" \cite{Klinkhamer:1984di}.
To produce non-vanishing lepton asymmetry, the decay of $N_1$ must have lepton number violating process with different decay rates to a final state with particle and anti-particle. Asymmetry in lepton flavour $\alpha$ produced in the decay of $N_1$ is defined as,
	%%%%%%%%%%%%%%%%%%%%%%%%%%%%%%%%%%%%%%%%%%%%%%%%%
	%%%%%%%%%%%%%%%%%%%%%%%%%%%%%%%%%%%%%%%%%%%%%%%%%
	%%%%%%%%%%%%%%%%%%%%%%%%%%%%%%%%%%%%%%%%%%%%%%%%%
	%%%%%%%%%%%%%%%%%%%%%%%%%%%%%%%%%%%%%%%%%%%%%%%%%
	\begin{equation}
		\epsilon_{\alpha\alpha}=\frac{\Gamma(N_1 \rightarrow L_{\alpha}\phi_i)-\Gamma(N_1 \rightarrow \bar{L}_{\alpha} \bar{\phi}_i)}{\Gamma(N_1 \rightarrow L_{\alpha} \phi_i)+\Gamma(N_1 \rightarrow \bar{L}_{\alpha}\bar{\phi}_i)},
	\end{equation}
	%%%%%%%%%%%%%%%%%%%%%%%%%%%%%%%%%%%%%%%%%%%%%%%%%
	%%%%%%%%%%%%%%%%%%%%%%%%%%%%%%%%%%%%%%%%%%%%%%%%%
	%%%%%%%%%%%%%%%%%%%%%%%%%%%%%%%%%%%%%%%%%%%%%%%%%
	%%%%%%%%%%%%%%%%%%%%%%%%%%%%%%%%%%%%%%%%%%%%%%%%%
	where, $\overline{L}_{\alpha}$ is the antiparticle of $L_{\alpha}$ and $\phi_i$ is the lightest Higgs doublet present in our model.
\subsubsection{Resonant leptogenesis}\label{resolep}
 In resonant leptogenesis (RL), the leptonic asymmetry gets resonantly enhanced up to order of unity by Majorana neutrino self-energy effect \cite{Liu:1993tg}, when the mass splitting between the RH neutrinos are of the order of the decay rates ($\Delta M\sim \Gamma$). As a result in the thermal RL, the extra Majorana neutrino mass scale can be considered as low as the electroweak scale \cite{Pilaftsis:2003gt}, while satisfying agreement with the neutrino oscillation data. 
 However, thermal leptogenesis restricts the lower bound on the decaying RH neutrino ($M_{N_1}\ge10^9$ GeV). Such higher masses are not discoverable for recent/on-going experiments, thus study of low-scaled leptogenesis are preferable at the current situation. Resonant leptogenesis (RL) is popular in low scaled \cite{Pilaftsis:2003gt, Hambye:2001eu, Cirigliano:2006nu,Chun:2007vh,Kitabayashi:2007bs,Bambhaniya:2016rbb}, which allow nearly degenerate mass spectrum of RH neutrinos at the TeV scale.

The CP asymmetry contributions in resonant leptogenesis are arising from the self energy contribution. If two right handed neutrinos are quasi-degenerate or nearly degenerate ($N_1\simeq N_2$) then, considering $N_2$ as an intermediate state, the self energy contribution to the total CP asymmetry will read as\footnote{Here, the asymmetry is irrespective of any flavour and $``\lambda"$s are the respective Yukawa coupling associate with the RH neutrinos.}\cite{Davidson:2008bu},
\begin{eqnarray}
	\epsilon_{\nu_{R_1}}=-\frac{M_1\Gamma_2}{M_2^2}\frac{M_2^2\Delta M_{21}^2}{(\Delta M_{21}^2)^2+M_1^2\Gamma_2^2}\frac{\mathcal{I}m[(\lambda^{\dagger}\lambda)_{12}^2]}{(\lambda^{\dagger}\lambda)_{11}(\lambda^{\dagger}\lambda)_{22}}.
\end{eqnarray}
In the resonant case ($\Delta M_{21}\simeq\Gamma_2/2$),
\begin{equation}
	|\epsilon_{\nu_{R_1}}|\simeq\frac{1}{2}\frac{|\mathcal{I}m[(\lambda^{\dagger}\lambda)_{12}^2]|}{(\lambda^{\dagger}\lambda)_{11}(\lambda^{\dagger}\lambda)_{22}}.
\end{equation}
 Here, the asymmetry is governed by the Yukawa coupling only and neither mass nor mass splittings are associated with it.
%%%%%%%%%%%%%%%%%%%%%%%%%%%%%%%%%%%%%
\subsection{Neutrino-less double beta decay ($0\nu\beta\beta$)}\label{ndbd}
Absolute neutrino mass is yet another unknown to the physics community as oscillation experiments are sensitive to the mass squared difference ($\Delta m_{ij}^2$) and leptonic mixing angles ($\theta_{ij}$, with $i,j=1,2,3$) only. Apart from the oscillation studies, the kinematic study of reactions involving neutrino ($\nu$) and anti-neutrino ($\overline{\nu}$) can give us information about absolute mass. Considering Majorana nature of particles, Wendell Furry \cite{1939PhRv...56.1184F} studied a kinetic process similar to ``double-beta disintegration" without neutrino emission, popularly known as {\it neutrino-less double beta decay} $(0\nu\beta\beta)$ \cite{DellOro:2016tmg}. In simple word this can be expressed as, $$(A,Z)\rightarrow(A,Z+2)+2e^-.$$
From $0\nu\beta\beta$ integration, if the Majorana nature of the neutrino is verified, one can give conclusive remarks on absolute neutrino mass.
The  $(0\nu\beta\beta)$ process explicitly violets the lepton number by creating a pair of electron. Discovery of lepton number violation (LNV) process supported by existing theoretical picture and the $0\nu\beta\beta$ scenario allows leptons to take part in the process of matter-antimatter asymmetry of our Universe. Thus the observation of such a process is crucial for demonstrating baryogenesis idea \cite{Cline:2006ts} via lepton number violation. Many works on  $(0\nu\beta\beta)$ have been done considering the SM neutrinos \cite{Bilenky:2004wn,Bilenky:2012qi,Agostini:2018tnm}. Nevertheless, it is now clear that addition of a new fermion and studying its interactions within the SM particles can lead us to a broad range of new physics phenomenology \cite{Barry:2011wb,Abada:2018qok}.

We assumed that, heavy Majorana neutrinos mediate the observed $0\nu\beta\beta$ process at tree-level. Under the SM framework, the decay amplitude is proportional to \cite{Abada:2018qok, Gautam:2019pce}:
\begin{equation}
	\varSigma\ G_f^2\ U_{ei}^2\ \gamma_{\mu}P_R\frac{\cancel{p}+m_i}{p^2-m_i^2}\gamma_{\nu}P_L\simeq \varSigma\ G_f^2\ U_{ei}^2\ \frac{m_i}{p^2}\gamma_{\mu}P_R\gamma_{\nu},
\end{equation}
where, $G_F$ is the Fermi constant, $m_i$ the physical neutrino mass and $p$ is the neutrino virtual momentum such that $p^2= -(125 \text{MeV})^2$. The effective electron neutrino Majorana mass for the active neutrinos in the $0\nu\beta\beta$ process read as, 
\begin{equation}
	m^{3_{\nu}}_{eff}= m_1|U_{e1}|^2+m_2|U_{e2}|^2+m_3|U_{e3}|^2,
\end{equation}
The phase ``effective $electron$ neutrino" is used as only electrons were involved in the double decay process.
If the SM is extended by $n_S$ extra sterile fermions, the presence of those extra states will modify the decay amplitude which corrects the effective mass as \cite{Benes:2005hn},
\begin{equation}
	m_{eff} = \sum_{i=1}^{3+n_S}U_{ei}^2 \ p^2 \frac{m_i}{p^2-m_i^2},
\end{equation}
where, $U_{ei}$ is the $(3+n_S \times 3+n_S)$ matrix with extra active-sterile mixing elements. As we have considered only one sterile state, hence the effective electron neutrinos mass is modified as \cite{Barry:2011wb},
\begin{equation}
	m^{3+1}_{eff}= m^{3_{\nu}}_{eff}+m_4|\theta_{S}|^2,
\end{equation}
where, $|\theta_{S}|$ is the active-sterile mixing element and $m_4$ sterile mass.

Many experimental and theoretical progress were made so far and still counting in order to validate the decay process. Interestingly, till date no solid evidences from experiments confirmed $0\nu\beta\beta$ process. However, next-generation experiments \cite{Obara:2017ndb,Artusa:2014lgv,Hartnell:2012qd,Gomez-Cadenas:2013lta,Barabash:2011aa} are currently running in pursue of more accurate limit on the effective mass which may solve the absolute mass problem. Recent results from various experiments are shown in table \ref{teff}. These results give strong bounds on the effective mass $m_{eff}$. Kam-LAND ZEN Collaboration \cite{KamLAND-Zen:2016pfg} and GERDA \cite{Agostini:2018tnm} which uses Xenon-136 and Germanium-76 nuclei respectively gives the most constrained upper  bound upto 90\% CL with,
	$m_{eff}< 0.06-0.165 \ \text{eV}.$
\begin{table}[t]
	\centering
	\begin{tabular}{|c|c|c|c|}
		\hline
	Experiments (Isotope)&  $|m_{eff}|$ eV& Half-life (in years)& Ref.\\
		\hline
				\hline
		KamLAND-Zen(800 Kg)(Xe-136)&$0.025-0.08$&$1.9\times10^{25}$(90\%CL)&\cite{KamLAND-Zen:2016pfg}\\
		%	\hline
		KamLAND2-Zen(1000Kg)(Xe-136)&$<0.02$&$1.07\times10^{26}$ (90\%CL)&\cite{KamLAND-Zen:2016pfg}\\
		GERDA Phase II (Ge-76)& $0.09-0.29$&$4.0\times10^{25}$(90\%CL)&\cite{Agostini:2018tnm}\\
		CUORE (Te-130)& $0.051-0.133$&$1.5\times10^{25}$(90\%CL)&\cite{Artusa:2014lgv}\\
		SNO+ (Te-130)& $0.07-0.14$&$\sim10^{26-27}$&\cite{Hartnell:2012qd}\\
		SuperNEMO (Se-84)&$0.05-0.15$&$5.85\times10^{24}$(90\%CL)&\cite{Barabash:2011aa}\\
		AMoRE-II (M0-100)&$0.017-0.03$&$3\times10^{26}$(90\%CL)&\cite{Bhang:2012gn}\\
		EXO-200(4 Year)(Xe-136)& $0.075-0.2$&$1.8\times10^{25} $(90\%CL)&\cite{Tosi:2014zza}\\
		nEXO(5Yr+5Yr w/Ba Tagging)(Xe-136)& $0.005-0.011$&$\sim10^{28}$&\cite{Licciardi:2017oqg}\\
		\hline
	\end{tabular}
	\caption{Sensitivity of few past and future experiments with half-life in years. }\label{teff}	
\end{table}
\subsection{Dark matter}
%%%%%%%%%%%%%%%%%%%%%%%%%%%%%%%%%%%%%
The astrophysical observations like; gravitational lensing effects in bullet cluster, anomalies in the galactic rotation
curves, $etc.$, have confirmed the existence of dark matter in the Universe. Dark matter cannot be observed through its interactions with photons as they are charge-less. The satellite-based experiments~\cite{Freese:2008cz, Persic:1995ru,Bennett:2012zja} such as Planck and Wilkinson Microwave Anisotropy Probe
(WMAP) have measured the Cosmic Microwave Background
Radiation (CMBR) of the Universe with unprecedented accuracy. They have suggested that
the Universe consists of about 4$\%$ ordinary matter, 27$\%$ dark matter and the rest
69$\%$ is mysterious unknown energy called dark energy which is assumed to be the
cause of the accelerated expansion of the Universe. The SM of particle physics fails to provide a viable
dark matter candidate. Thus,
new physics beyond the SM is required to explain the observed presence of the dark matter.
The astrophysical and cosmological data so far can only tell us how much dark matter is
there in the Universe, $i.e.$, the total mass density.
% and it does not interact electromagnetically and strongly. 

Although we are convinced
that dark matter exists, still, there is no consensus about its composition as it does not interact electromagnetically or strongly. The possibilities incorporate the dense-baryonic matter and non-baryonic matter. The MACHO and the EROS \cite{Tisserand:2006zx, Alcock:1998fx} collaborations conclude that
dense-baryonic matter, $i.e.$, Massive Compact Halo Objects (MACHOs), black holes,  very faint stars, white dwarfs, non-luminous objects like planets
could add a few percent to the known mass discrepancy in the Galaxy halo observed in
galactic rotation curves. The non-baryonic dark matter components~\cite{Roszkowski:2017nbc} can be grouped into three categories based on their production mechanism and velocities, namely hot dark matter (HDM), warm dark matter
(WDM) and cold dark matter (CDM).
A simplified scheme in this regard is to introduce `weakly interacting massive particles' (WIMP) protected by a
discrete symmetry that ensures the stability of these particles. 
Various established options {\it viz.} extra $Z_n$-odd ($n\ge$2, is an integer) scalar, fermion, and combined of them with various multiplets, e.g., singlets, doublets, triplets, quadruplets, etc. have been studied to explain the dark matter phenomenology. One can see the recent review article on the dark matter~\cite{Hambye:2009pw, Bernal:2017kxu, Kahlhoefer:2017dnp, Tanabashi:2018oca} and the references therein. 
The DM particles may have different masses: massive
gravitons $\mathcal{O}(10^{-19})$ GeV~\cite{Dubovsky:2004ud}, axions $\mathcal{O}(10^{-5})$ GeV~\cite{Holman:1982tb}, sterile neutrino $\mathcal{O}(10^{-6})$ GeV~\cite{Dodelson:1993je}  and the point like WIMPs candidate having mass range $20$ GeV $-$ 340 TeV~\cite{Leane:2018kjk,Griest:1989wd, Hisano:2006nn}.
The upper bound on mass can be stretched up to 1 PeV for the composite dark matter candidate~\cite{Smirnov:2019ngs}. 

\subsubsection{$keV$ scale dark matter}
Since sterile neutrinos cannot thermalize easily, the most straightforward production mechanism is via mixing with the active neutrinos in the primordial plasma \cite{Dodelson:1993je}. 
Depending upon the production mechanism, one can discard the fact that the mixing of active neutrinos cannot generate sterile neutrinos to behave as a dark matter~\cite{Berlin:2016bdv}. Too large mixing between active-sterile correspond to a too large DM density; however, we can still consider the possibility by considering very small mixing angles\cite{Benso:2019jog, Adhikari:2016bei}.
The DM sterile neutrino production via mixing becomes most efficient at temperatures $T \sim 150-500$ MeV \cite{Dodelson:1993je, Asaka:2006nq, Adhikari:2016bei} resulting in the population of warm DM particles. Resonant production\footnote{Resonantly produced (RP) sterile neutrinos are typically much colder and the dispersion of their momentum distribution is also much smaller than thermal. Therefore, in some sense resonantly produced sterile neutrinos behave as a mixture of a cold and warm DM (CWDM) over some range of scales \cite{Adhikari:2016bei}} results into an efficient conversion of an excess of $\nu_e (\overline{\nu_e})$ into DM neutrinos $S$ \cite{Shi:1998km, Laine:2008pg}. One important thing to keep in mind here is that the overproduction of dark matter must be avoided to make them experimentally achievable with proper adjustment of the {\it critical temperature ($T_c$)}\footnote{Temperature at which dark matter production starts.}, we could avoid the overproduction of dark matter abundance. Above the critical temperature, the mixing parameter, $\sin^22\theta_{S}$ from equation \eqref{dm3} got heavily suppressed if sterile mass either vanishes or very high at that temperature \cite{Gelmini:2019clw, Benso:2019jog}. If one considers the mixing angle to be a dynamical quantity, it is impossible to obtain a relic of that quantity. 
Notwithstanding, in chapter \ref{Chapter4}, we have considered a small static active-sterile mixing angle ($\theta_S<10^{-6}$) such that they remain in the Universe as DM relic~\cite{Yeche:2017upn, Adhikari:2016bei, Baur:2017stq}. 
The important thing to note here is that sterile neutrino DM is practically always produced out of thermal equilibrium. Therefore, its primordial momentum distribution is in general, not given by a Fermi-Dirac distribution. Indeed, sterile neutrinos in equilibrium have the same number density as ordinary neutrinos, {\it i.e.}, 112  $cm^{-3} $.
With the sterile neutrino mass above $0.4~keV$ would lead to the energy density today $\rho_{sterile, eq}\simeq 45~ keV/cm^3$ , which significantly exceeds the critical density of the Universe $\rho_{crit} = 10.5 ~h^2 keV/cm^3$. Therefore, sterile neutrino DM cannot be a thermal relic (unless entropy dilution is exploited), and its primordial properties are in general different from such a particle.

\subsubsection{Detection of dark matter}
Recent results from various WMAP satellites and cosmological measurements, the relic density of the current Universe measures as $\Omega_{DM} h^2=0.1198\pm0.0012$ \cite{Aghanim:2018eyx}. %%%%%%%%%%%% 
Dark matter can be detected via direct as well as indirect detection experiments. As WIMP dark matter interacts with matters weakly, many experiments are focused on direct detection techniques.  
If WIMPs scatter from the atomic nucleus, then it deposits energy in the detector given by,
%%%%%%%%%%%%
\begin{equation}
	E_{deposit} = \frac{1}{2} M_{DM} v^2.
\end{equation}
%%%%%%%%%%%
The energy deposition can also be written as,
%%%%
%%%%%%%%%%%%
\begin{equation}
	E_{deposit} = \frac{\mu^2 v^2}{m_N}(1-\cos\theta).
\end{equation}
%%%%%%%%%%%%%%%
In the Earth frame, the mean velocity $v$ of the WIMPs relative to the target nucleus is about 220 km/s, $\mu$ is the reduced mass of the WIMP of mass $M_{DM}$ and the nucleus of mass $m_N$, and $\theta$ is the scattering angle.
As the dark matter is weakly interacting, it may rarely bump into the detector atom's nucleus and deposit energy which may create a signature at the detector. The amount of energy of a WIMP with mass $M_{DM}=100$ GeV would deposit in the detector is $E_{deposit}\simeq 27~ keV$ \cite{Aprile:2012nq}.
Presently non-observation of dark matter in direct detections set a limit on WIMP-nucleon scattering cross-section for a given dark matter mass from experiments XENON \cite{Aprile:2016swn}, LUX~\cite{Akerib:2019diq}.
%%%%%%%%%%%%%%%%%%%%

%%%%%%%%%%%%%%%%%%%%
Indirect detection of dark matter techniques are quite different. If the dark matter and its antiparticle are the same,
then they can annihilate to form known standard model particles such as photons ($\gamma$-ray), electrons ($e^-$), positrons ($e^+$) $etc.$
Various detectors were placed in the Earth's orbits, {\it e.g.}, Fermi Gamma-ray Space Telescope (FGST) \cite{Hooper:2010mq}, Alpha Magnetic Spectrometer (AMS) \cite{Aguilar:2016vqr}, PAMELA \cite{Adriani:2010rc} etc., observed the excess of gamma-ray and positron excess.

%From the particle physics point of view, the processes like $DM,~DM \rightarrow \gamma\gamma,~ e^+ e^-$ etc. have been used to explain such excess.They are model-dependent processes. The WIMP dark matter with different mass and coupling can explain these high energetic gamma-rays excess from the galactic centre and positron excess in the cosmic ray.

\section{Discrete flavour symmetries}
Symmetries play a salient role in particle physics. There are several types of symmetries in Nature, and they could be either continuous or discrete symmetries and global or local. Continuous and local symmetries are associated with all fundamental interactions like strong, weak, electromagnetic, and they are essential to understand those processes. Discrete symmetries such as $C, P$ and $T$ also plays a vital role in understanding particle nature and interaction. On the other hand, global symmetry is ubiquitous in Nature. 

When the quarks or lepton generations are mixed in the SM or BSM framework, their mass terms are generated from the flavour sector, $i.e.,$ Yukawa couplings. There would be numbers of free parameters that arise when these kinds of interactions take place. Hence, flavour symmetries are introduced to control these Yukawa couplings associated with flavour generations. Flavour symmetries proved their standpoint in past experimental results, such as the determination of quark mass and mixing and the discovery of oscillation parameters in the neutrino sector. Nevertheless, there are many unanswered questions related to flavour symmetries like the mass hierarchy problem, the difference in mixing pattern between the quark and lepton sectors. The experimental advancement in the quark sector is almost at the edge of completion; however, the experiments have entered the era of accuracy and precision in determining masses and mixing in the neutrino sector. Therefore, in the neutrino sector, it is essential to find {\bf The Model}, which can naturally fit itself with the experimental results. In the following, we will be discussing discrete flavour symmetries that we used to construct our models and those models based on which we have carried out this thesis work. 

\subsection{The $Z_n$ group} The group symmetry $Z_n$ is an Abelian discrete rotational symmetry group. It describes the symmetry of a plane figure that invariant after a rotation of $2\pi/n$ and always corresponds to the symmetry operations of physical objects. They are the finite subgroup of $SO(2)$ and can be generated by $e^{2\pi i/2}$. For example, $Z_2$ is a symmetry group of reflection from a fixed plane. It consists of two elements $+1$ and $-1$. Under $Z_2$ a field $x$ would transform as $x\rightarrow -x$. Similar way $Z_3$ and $Z_4$ also can be represented by triangular and square symmetry, respectively. Group elements of $Z_3$ and $Z_4$ are respectively $(1, \omega, \omega^2)$ and $(1,-1,i,-i)$. A pictorial representation of various $Z_n$ groups transformation associate with regular shapes are shown in figure \ref{gr1}. 
\begin{figure}
	\centering
	\includegraphics[scale=0.4]{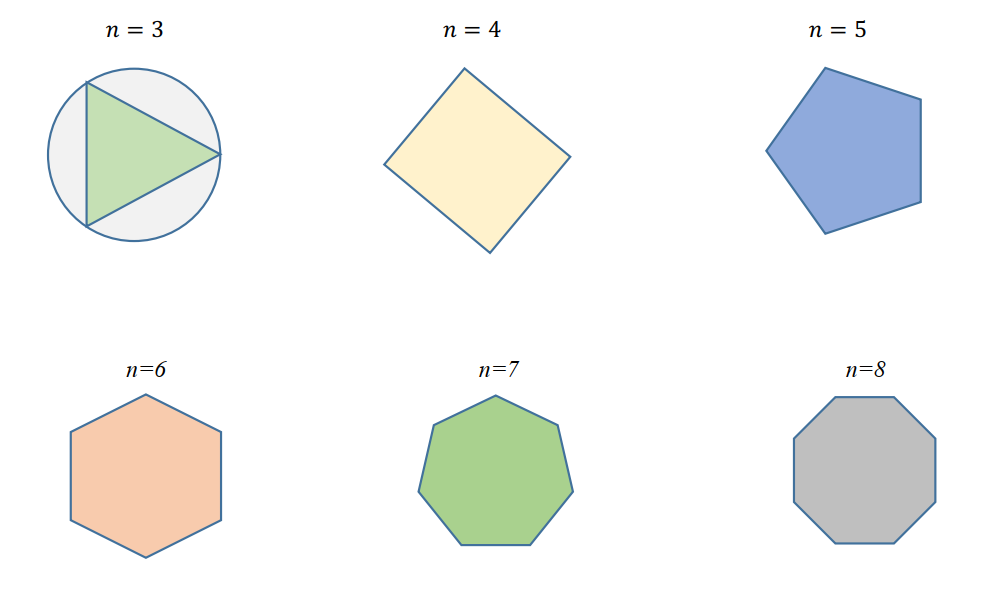}
	\caption{n-sided regular polygon associated with respective $Z_n$ groups.}\label{gr1}
\end{figure}

\subsection{The $A_4$ group:} $A_4$, the symmetry group of a tetrahedron (as shown in figure \ref{a4d}), is a discrete non-Abelian group of even permutations of four objects. It has $4!/2=12$ elements with four irreducible representations: three one-dimensional and one three-dimensional which are denoted by $\bf{1}, \bf{1'}, \bf{1''}$ and $\bf{3}$ respectively. $A_4$ can be generated by two basic permutations $S$ and $T$ given by $S=(4321)$ and $T=(2314)$ (For a generic (1234) permutation) such that,
$S^2=T^3=(ST)^3=1.3$.
\begin{figure}\centering
	\includegraphics[scale=0.69]{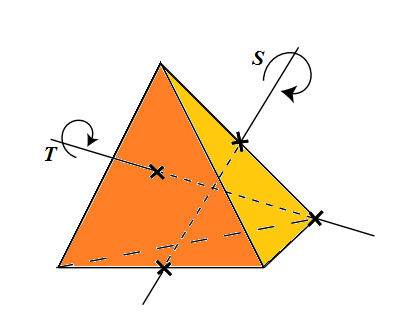}
	\caption{Pictorial representation of $A_4$ symmetry of tetrahedron.}\label{a4d}
\end{figure}

The irreducible representations for the $S$ and $T$ basis are different from each other. The 12 even permutations can be generated from $S$ and $T$ as follows,
\begin{eqnarray}
	\nn	I&=&(1234)\\
	\nn	T&=& (2314),~ST=(4132),~TS=(3241),~STS=(1423)\\
	\nn T^2&=& (314),~ST^2=(4213),~T^2S=(2431),~TST=(1342)\\
	S&=&(4321),~T^2ST=(3412),~TST^2=(2143)
\end{eqnarray}
Here, $I$ is the identity. 
We have already mentioned that among the four inequivalent representations: three of them are one dimensional and one is three dimensional and can be obtained as,
\begin{eqnarray}
	\nn 1&&\quad\quad S=1\quad\quad T=1\\
	1^{\prime}&&\quad\quad S=1\quad\quad T=e^{i2\pi/3}\equiv \omega\\
	\nn 1^{\prime\prime}&&\quad\quad S=1\quad\quad T=e^{i4\pi/3}\equiv \omega^2.
\end{eqnarray} 
Here, cube root of unity is defined as $\omega=exp(i\frac{2\pi}{3})$, such that $\omega^2=\omega^*,~1+\omega+\omega^2=0$. 

The three dimensional unitary representation in the basis where $S$ is diagonal, are represented as,
\begin{eqnarray}
	S=\begin{pmatrix}
		1&0&0\\0&-1&0\\0&0&-1
	\end{pmatrix}\quad\quad T=\begin{pmatrix}
		0&1&0\\0&0&1\\1&0&0
	\end{pmatrix}.
\end{eqnarray}

Throughout our analysis we have considered the $T$ diagonal basis as the charged lepton mass matrix is diagonal in our case. Their product rules are given as,
$$ \bf{1} \otimes \bf{1} = \bf{1}; \bf{1'}\otimes \bf{1'} = \bf{1''}; \bf{1'} \otimes \bf{1''} = \bf{1} ; \bf{1''} \otimes \bf{1''} = \bf{1'}$$
$$ \bf{3} \otimes \bf{3} = \bf{1} \otimes \bf{1'} \otimes \bf{1''} \otimes \bf{3}_a \otimes \bf{3}_s $$
where $a$ and $s$ in the subscript corresponds to anti-symmetric and symmetric parts respectively. Denoting two triplets as $(a_1, b_1, c_1)$ and $(a_2, b_2, c_2)$ respectively, their direct product can be decomposed into the direct sum mentioned above as,
\begin{equation}
	\begin{split}\label{a4r}
		& \bf{1} \backsim a_1a_2+b_1c_2+c_1b_2, \quad
		 \bf{1'} \backsim c_1c_2+a_1b_2+b_1a_2, \quad \bf{1''} \backsim b_1b_2+c_1a_2+a_1c_2,\\
		&\bf{3}_s \backsim (2a_1a_2-b_1c_2-c_1b_2, 2c_1c_2-a_1b_2-b_1a_2, 2b_1b_2-a_1c_2-c_1a_2),\\
		& \bf{3}_a \backsim (b_1c_2-c_1b_2, a_1b_2-b_1a_2, c_1a_2-a_1c_2).\\
	\end{split}
\end{equation}

\section{Scope of this thesis}
This thesis is organized as follows:

This chapter starts with a brief overview of the neutrinos in both theoretical and experimental sector. After that, we discuss the standard model of particle physics and the successes and the shortcomings for which we are motivated to go beyond the SM. In the BSM part, starting with the neutrino mass and mixing, we discuss trivial seesaw frameworks through which small neutrino masses can be generated. There we focus on the minimal extended seesaw (MES) framework, as all our works are based on this seesaw. Then we discuss neutrino oscillation for two, three and four flavour cases in a vacuum as well as in matter very briefly. After that, we discuss baryogenesis, neutrino-less double beta decay and dark matter, respectively.
Thereafter we dedicated a section to discrete symmetries based on which we have worked out neutrino model building. We have been using discrete $A_4$ flavour symmetry and $Z_n$ ($n\ge2$) extensively to construct our models.

In the second chapter, we study active-sterile phenomenology with a single generation of sterile neutrino ($m_S\sim \mathcal{O}$(eV)) along with the three active neutrinos. Three independent cases for sterile mass matrices are studied in both normal and inverted hierarchy mass ordering. Non zero reactor mixing angle was generated by breaking the trivial $\mu-\tau$ symmetry in the final neutrino mass matrix. 

Chapter three is an extension of the previous chapter. Here, we have divided the work into fermion and scalar sector. In the fermionic sector, along with the neutrino mass generation, we also study baryogenesis {\it via} thermal leptogenesis. In the scalar sector, we study an extended scalar sector in a multi Higgs doublet model. Among the three Higgs doublet, one of them does not acquire any vacuum expectation value (VEV); hence, it behaves as an inert Higgs, and the lightest component of this behaves as a viable dark matter candidate. 

Within MES, sterile mass can be stressed up to $keV$ scale. In the fourth chapter, we studied baryogenesis $via$ thermal leptogenesis, $0\nu\beta\beta$ considering sterile neutrino mass in $keV$ range. This $keV$ scaled sterile neutrino also plays the role of dark matter candidate in our study. Various correlations studies are carried out in both normal and inverted hierarchy mass ordering.  

The fifth chapter is dedicated to texture-zero neutrino dark matter model. Similar to chapter three, here also we study both fermion and scalar sector extensively. 
In the fermion sector, we study active-sterile mixing, baryogenesis $via$ resonant leptogenesis and $0\nu\beta\beta$ in the normal hierarchy mass pattern only. In the scalar sector, the complex scalar flavon that gives rise to sterile mass takes part in dark matter study. The imaginary component of that scalar flavon is behaving as a viable dark matter candidate and we can see the influence of VEV of that flavon itself in the dark matter parameter space. 

The final chapter is dedicated to the summary, conclusion and future scope of this thesis.  
%\input{Chapters/Chapter2} 
% Chapter Template
\begin{savequote}[1.02\linewidth]
	\normalsize	``We must not wait for things to come, believing that they are decided by irrescindable destiny.  If we
	want it, we must do something about it.''
	\qauthor{\large\it Erwin Schr{\"o}dinger (1887--1961)}
\end{savequote}
\chapter{Active and sterile neutrino phenomenology based on minimal extended seesaw} % Main chapter title

\label{Chapter2} % Change X to a consecutive number; for referencing this chapter elsewhere, use \ref{ChapterX}

\lhead{Chapter 2. \emph{Active and sterile neutrino phenomenology based on minimal extended seesaw}} % Change X to a consecutive number; this is for the header on each page - perhaps a shortened title

%----------------------------------------------------------------------------------------
%	SECTION 1
%----------------------------------------------------------------------------------------

In this chapter, we have studied a model of neutrino within the framework of minimal extended seesaw (MES), which
plays an important role in active and sterile neutrino phenomenology in (3+1) scheme. Here, the $A_4$ flavour symmetry is augmented by an additional $Z_4\times Z_3$  symmetry to constraint the Yukawa Lagrangian of the model. We use a non-trivial Dirac mass matrix, with broken $\mu-\tau$  symmetry, as the origin of leptonic mixing. Interestingly, such mixing structure naturally leads to the non-zero reactor mixing angle $\theta_{13}$. Non-degenerate mass structure for right-handed neutrino $M_R$ is considered. We have also considered three different cases for sterile neutrino mass, $M_S$ to check the viability of this model, within the allowed $3\sigma$ bound in this MES framework.\\ %This work is based on {\bf neutrino mass model I} . 

\section{Introduction}

There were a few works on MES based on $A_4$ which are available in literature \cite{Barry:2011wb,Zhang:2011vh} during the tenure of this work. Those studies were carried out before the discovery of non-zero reactor mixing angle $\theta_{13}$ \cite{Abe:2011fz}. Our model has considered different flavons to construct the non-trivial Dirac mass matrix ($M_D$), which is responsible for generating light neutrino mass. In this context, we have added a leading order correction to the Dirac mass matrix to accumulate non-zero reactor mixing angle($\theta_{13}$), instead of considering higher-order correcting term in the Lagrangian as mentioned in the ref. \cite{Altarelli:2005yp}.
We have introduced a new leading order correction matrix $M_P$ to the Dirac mass matrix produced from a similar kind of coupling term that accomplish the Dirac mass matrix ($M_D$). This matrix $M_P$ is added to $M_D$, such that there is a broken $\mu-\tau$ symmetry, which leads to the generation of the non-zero reactor mixing angle. The $M_D$ matrix constructed for NH does not work for IH, the explanation of which we have given in the model section. Thus, we have reconstructed $M_D$ by introducing a new flavon ($\varphi^{\prime}$) to the Lagrangian to study the case of the IH pattern.

In this chapter, we have extensively studied the consequences of taking the sterile mass pattern by altering the non-zero entry position in $M_S$. All these $M_S$ structures have been studied independently for both the mass ordering, and results are explained in section \ref{3sec4}. In the phenomenology part, we have constrained the model parameters in the light of current experimental data and shown a correlation between active and sterile mixing by considering three different $M_S$ structures.

This chapter is organized as follows. In section \ref{model1}, we have discussed the $A_4$ model and generation of the mass
matrices in the leptonic sector. We kept the section \ref{3sec4} and its subsections for the numerical procedure of NH and IH cases. In section \ref{3sec5}, we have carried out the numerical analysis. Finally, we have summarized this chapter in section \ref{con3}.

%In the diagonal charged lepton mass and Majorana light neutrino mass limit we have used the T-diagonal basis for the A4 product rule as given in the appendix. It would be S-diagonal basis if the lepton mass matrix is non-diagonal, which would lead to a different light neutrino mass matrix due to the difference in the product rule and the representations. 
\section{Model framework}\label{model1}
\subsection{Particle content and discrete charges}
Our present model is an extension of $A_4\times Z_4\times Z_3$ flavour symmetry. $A_4$ is quite popular in literature in explaining neutrino mass and mixing \cite{Altarelli:2009kr,Barry:2011wb,Chun:1995bb,Zhang:2011vh,Heeck:2012bz,Felipe:2013vwa,Babu:2009fd}. In this model, we have established active-sterile mixing pattern for the $(3+1)$ scheme. Along with the SM particle content, we have introduced a few extra fields to make the model workable. Three separate sterile mass matrices and their possible interactions were also studied under this model.
We have assigned left-handed (LH) lepton doublet $l$ to transform as $A_4$ triplet whereas right-handed (RH) charged leptons ($e^c,\mu^c,\tau^c$) transform as 1,$1^{\prime\prime}$ and $1^{\prime}$ respectively. The flavour symmetry is broken by the triplets $\zeta, \varphi $ and two singlet flavons $\xi$ and $\xi^{\prime}$. Besides the SM Higgs $H$, we have introduced two more Higgs ($H^{\prime},H^{\prime\prime}$)\cite{Felipe:2013vwa,Nath:2016mts} which remain invariant under $A_4$. We also have restricted non-desirable interactions while constructing the mass matrices. The particle content and the  $A_4 \times Z_4 \times Z_3$ charge assignment are shown in tables \ref{1tab1} and \ref{1tab2}. In the following subsections, we will be discussing normal and inverted mass hierarchy patterns separately.

\begin{table}[h!]\centering
	\begin{tabular}{|c|cccc|ccc|cccc|ccc|}
		\hline
		Fields & $l$ &$e_{R}$&$\mu_{R}$&$\tau_{R}$&$H$&$H^{\prime}$&$H^{\prime\prime}$&$\zeta$&$\varphi$&$\xi$&$\xi^{\prime}$&$\nu_{R1}$&$\nu_{R2}$&$\nu_{R3}$\\
		$\overline{\text{Charges}}$&&&&&&&&&&&&&&\\	
		\hline
		\hline
		SU(2)&2&1&1&1&2&2&2&1&1&1&1&1&1&1\\
%		\hline
		$A_4$&3&1&$1^{\prime\prime}$&$1^{\prime}$&1&1&1&3&3&1&$1^{\prime}$&1&$1^{\prime}$&1\\
%		\hline
		$Z_4$&1&-1&-1&-1&1&i&-1&-1&1&1&-1&1&-i&-1\\
%		\hline
		$Z_3$&1&1&1&1&1&$\omega$&1&1&1&1&$\omega^2$&1&$\omega^2$&1\\
		\hline
	\end{tabular}
	\caption{Particle content and their charge assignments under $SU(2),A_4$ and $Z_4\times Z_3$ groups.}\label{1tab1} 
\end{table}

%In this model, left handed lepton doublet $l$ are supposed to transform as $A_4$ triplet whereas right-handed charged leptons($e^{c},\mu^{c},\tau^{c}$) as $A_4$ singlets. Three $SU(2)$ doublets Higgs ($H_1,H_2,H_3$) are introduced in our model. To keep the charged lepton mass matrix diagonal we have introduced a flavon $\zeta$ which is a triplet under $A_4$ in this model. To construct a desired Dirac neutrino mass matrix, we have added two flavons $\varphi$ and $\eta $ to the SM scalar sector, which transform as $A_4$ triplet. On the other hand $\xi , \xi^{\prime}$ and $\chi$ singlets are introduced to obtain a diagonal Majorana neutrino mass matrix and the $M_S$ matrix.\\
\begin{table}[h!]\centering
	\begin{tabular}{|c|c|c|c|c|}
		\hline
		Fields & $S_1$&$S_2$&$S_3$&$\chi$\\
		$\overline{\text{Charges}}$&&&&\\
		\hline
		\hline
		$A_4$&$1^{\prime\prime}$&$1^{\prime}$&$1^{\prime\prime}$&$1^{\prime}$\\
	%	\hline
		$Z_4$&-i&1&i&i\\
	%	\hline
		$Z_3$&1&$\omega$&1&1\\
		\hline
	\end{tabular}
	\caption{Scalar singlet fields and their transformation properties under $A_4$ and $Z_4\times Z_3$ groups.}\label{1tab2}
\end{table}

\subsubsection{Normal hierarchical neutrino mass }

%We have considered the charged lepton mass matrix($m_l$) to be diagonal in our formalism,for that purpose we have assigned a triplet flavon $\zeta$ in our model. In order to construct the columns of $M_D$ we have introduced three triplet flavons $\varphi, \varphi^{\prime} and \varphi^{\prime\prime}$.On the other hand $\xi,\xi^{\prime} and \chi$ are introduced to form the diagonal right handed mass matrix and the $M_S$ matrix. 
The leading order invariant Yukawa Lagrangian for the lepton sector is given by,
\begin{equation}
	\mathcal{L} = \mathcal{L}_{\mathcal{M_\iota}}+\mathcal{L}_{\mathcal{M_D}}+\mathcal{L}_{\mathcal{M_R}}+\mathcal{L}_{\mathcal{M}_S}+h.c. .
\end{equation}
Where,
\begin{equation}
	\begin{split}
		\mathcal{L}_{\mathcal{M_\iota}} &= \frac{y_{e}}{\Lambda}(\overline{l} H \zeta)_{1}e_{R}+\frac{y_{\mu}}{\Lambda}(\overline{l} H \zeta)_{1^{\prime}}\mu_{R}+\frac{y_{\tau}}{\Lambda}(\overline{l} H \zeta)_{1^{\prime\prime}}\tau_{R}, \\
		\mathcal{L}_{\mathcal{M_D}}= & \frac{y_{1}}{\Lambda}(\overline{l}\tilde{H}\varphi)_{1}\nu_{R1}+\frac{y_{2}}{\Lambda}(\overline{l}\tilde{H^{\prime}}\varphi)_{1^{\prime\prime}}\nu_{R2}+\frac{y_{3}}{\Lambda}(\overline{l}\tilde{H^{\prime\prime}}\varphi)_{1}\nu_{R3},\\
		\mathcal{L}_{\mathcal{M_R}}= & \frac{1}{2}\lambda_{1}\xi\overline{\nu^{c}_{R1}}\nu_{R1}+\frac{1}{2}\lambda_{2}\xi^{\prime}\overline{\nu^{c}_{R2}}\nu_{R2}+\frac{1}{2}\lambda_{3}\xi\overline{\nu^{c}_{R3}}\nu_{R3}.\\
	\end{split}
\end{equation}
We have extended our study with three variety of $M_S$ structures, which is generated by the interaction of a singlet field $S_i$ and the right-handed neutrino $\nu_{Ri}$. The $A_4\times Z_4\times Z_3$ charge alignments for the scalar fields are given in table \ref{1tab2}. The effective mass terms for each of the above three cases are as follows, 
\begin{equation}
	\begin{split}
		\mathcal{L}_{\mathcal{M}_{S}^{1}}=& \frac{1}{2}\rho\chi\overline{S_1^{c}}\nu_{R1} ,\\
		\mathcal{L}_{\mathcal{M}_{S}^{2}}=& \frac{1}{2}\rho\chi\overline{S_2^{c}}\nu_{R2} ,\\
		\mathcal{L}_{\mathcal{M}_{S}^{3}}=& \frac{1}{2}\rho\chi\overline{S_3^{c}}\nu_{R3} .\\
	\end{split}
\end{equation}

In the Lagrangian, $\Lambda$ represents the cut-off scale of the theory, $y_{\alpha,i}$, $\lambda_{i}$ (for $\alpha=e,\mu,\tau$ and $i=1,2,3$) and $\rho$ representing the Yukawa couplings for respective interactions and all Higgs doublets are transformed as $\tilde{H} = i\tau_{2}H^*$ (with $\tau_{2}$ being the second Pauli's spin matrix)  to keep the Lagrangian gauge invariant. Following VEV alignments of the extra flavons are required to generate the desired light neutrino mas matrix\footnote{ A discussion on minimization of VEV alignment for the triplet fields($\zeta$ and $\varphi$) is added in the appendix section.}.
\begin{equation*}
	\begin{split}
		&\langle \zeta \rangle=(v,0,0),\\
		& \langle\varphi\rangle=(v,v,v),\\
		& \langle\xi\rangle=\langle\xi^{\prime}\rangle=v,\\
		& \langle\chi\rangle=u.
	\end{split}
\end{equation*}
Following the $A_4$ product rules and using the above-mentioned VEV alignment, one can obtain the charged lepton mass matrix as follows,
\begin{equation}
	M_{l} = \frac{\langle H\rangle v}{\Lambda}\text{diag}(y_{e},y_{\mu},y_{\tau}).
\end{equation}
The Dirac\footnote{ $M_D^{\prime}$ represents the uncorrected Dirac mass matrix which is unable to generate $\theta_{13}\neq 0$. The corrected $M_D$ is given by equation \eqref{tmd} } and Majorana neutrino mass matrices are given by,
\begin{equation}
	M^{\prime}_{D}=
	\begin{pmatrix}
		a&b&c\\
		a&b&c\\
		a&b&c\\
	\end{pmatrix},M_{R}=\begin{pmatrix}
		d&0&0\\
		0&e&0\\
		0&0&f\\
	\end{pmatrix};
	\label{temd}
\end{equation}
where, $a=\frac{\langle H\rangle v}{\Lambda}y_{1} , b=\frac{\langle H\rangle v}{\Lambda}y_{2} $ and $c=\frac{\langle H\rangle v}{\Lambda}y_{3}$. The elements of the $M_R$ are defined as $d=\lambda_{1}v, e=\lambda_{2}v$ and $f=\lambda_{3}v$. \\
Three different structures for $M_{S}$ reads as,   
\begin{equation}
	M_{S}^{1}= \begin{pmatrix}
		g&0&0\\
	\end{pmatrix},\;
	M_{S}^{2}= \begin{pmatrix}
		0&g&0\\
	\end{pmatrix}\
	\text{and} \
	M_{S}^{3}= \begin{pmatrix}
		0&0&g\\
	\end{pmatrix}.
\end{equation}
Considering only $M_{S}^{1}$ structure, the light neutrino mass matrix takes a symmetric form as,
\begin{equation}\label{tsymm}
	m_{\nu}= \begin{pmatrix}
		-\frac {b^2} {e}-\frac{c^2}{f} &-\frac {b^2} {e}-\frac{c^2}{f} &-\frac {b^2} {e}-\frac{c^2}{f} \\
		-\frac {b^2} {e}-\frac{c^2}{f} &-\frac {b^2} {e}-\frac{c^2}{f}&-\frac {b^2} {e}-\frac{c^2}{f} \\
		-\frac {b^2} {e}-\frac{c^2}{f}&-\frac {b^2} {e}-\frac{c^2}{f}&-\frac {b^2} {e}-\frac{c^2}{f}\\
	\end{pmatrix}.
\end{equation}
As we can see, this $m_{\nu}$\footnote{ We have used $M_D^{\prime}$ in lieu of $M_D$ and $M_S^1$ instead of $M_S$ in equation \eqref{tamass}} is a symmetric matrix (Democratic) generated by $M_D^{\prime},\ M_R$ and $M_S^1$ matrices. It can produce only one mixing angle and one mass square difference. This symmetry must be broken in order to generate two mass square differences and three mixing angles. In order to introduce $\mu-\tau$ asymmetry in the light neutrino mass matrix we introduce a new $SU(2)$ singlet flavon field ($\eta$), the coupling of which gives rise to a matrix \eqref{tpmatrix} which later on makes the matrix \eqref{tsymm} $\mu-\tau$ asymmetric after adding \eqref{tpmatrix} to the Dirac mass matrix($M_D^{\prime}$). This additional flavon and the new matrix \eqref{tpmatrix} have a crucial role in reproducing nonzero reactor mixing angles. The Lagrangian responsible for generating the matrix \eqref{tpmatrix} can be written as, 
\begin{equation}
	\mathcal{L}_{\mathcal{M_P}} = \frac{y_{1}}{\Lambda}(\overline{l}\tilde{H} \eta)_1 \nu_{R1}+\frac{y_{2}}{\Lambda}(\overline{l} \tilde{H^{\prime}}\eta)_{1^{\prime\prime}}\nu_{R2}+\frac{y_{3}}{\Lambda}(\overline{l}\tilde{H^{\prime\prime}} \eta)_{1^{\prime}}\nu_{R3}.
\end{equation}
The singlet flavon field ($\eta$) is supposed to take $A_4\times Z_4 \times Z_3$ charges as same as $\varphi$ (as shown in the table \ref{1tab1}). Now, considering VEV for the new flavon field as  $\langle \eta \rangle=(0,v,0)$, we get the matrix as, 
% We will not write like this The present structure of Dirac neutrino mass matrix does not produce non-vanishing reactor mixing angle. Therefore  
\begin{equation}\label{tpmatrix}
	M_{P}=
	\begin{pmatrix}
		0&0&p\\
		0&p&0\\
		p&0&0\\
	\end{pmatrix}.
\end{equation}
Hence $M_D$ from equation \eqref{temd} will take new structure as,
\begin{equation} \label{tmd}
	M_D=M^{\prime}_D+M_P=
	\begin{pmatrix}
		a&b&c+p\\
		a&b+p&c\\
		a+p&b&c\\
	\end{pmatrix}.
\end{equation}
\subsubsection{Inverted hierarchical neutrino mass} 
Earlier in the work \cite{Zhang:2011vh}, the author has explained the necessity of a new flavon to realize the IH within the MES framework. In this chapter, we also have modified the Lagrangian for the $M_D$ matrix by introducing a new triplet flavon $\varphi^{\prime}$ with VEV alignment as $\langle\varphi^{\prime}\rangle \sim (2v,-v,-v)$, which affects only the Dirac neutrino mass matrix and give desirable active-sterile mixing in IH. The invariant Yukawa Lagrangian for the $M_D$ matrix will be,
\begin{equation}
	\mathcal{L}_{\mathcal{M_D}}= \frac{y_{1}}{\Lambda}(\overline{l}\tilde{H_1}\varphi)_{1}\nu_{R1}+\frac{y_{2}}{\Lambda}(\overline{l}\tilde{H_2}\varphi^{\prime})_{1^{\prime\prime}}\nu_{R2}+\frac{y_{3}}{\Lambda}(\overline{l}\tilde{H_3}\varphi)_{1}\nu_{R3}.\\
\end{equation}
Hence the Dirac mass matrix will have the form, 
\begin{equation}
	M^{\prime}_{D}=
	\begin{pmatrix}
		a&-b&c\\
		a&-b&c\\
		a&2b&c\\
	\end{pmatrix},
\end{equation}
with, $a=\frac{\langle H\rangle v}{\Lambda}y_{1} , b=\frac{\langle H\rangle v}{\Lambda}y_{2} $ and $c=\frac{\langle H\rangle v}{\Lambda}y_{3}$. 

This Dirac mass matrix will also give rise to a symmetric $m_{\nu}$ like the NH case. Thus, the modified $M_D$ to break the symmetry will be given by,
\begin{equation}
	M_D=M^{\prime}_D+M_P=
	\begin{pmatrix}
		a&-b&c+p\\
		a&-b+p&c\\
		a+p&2b&c\\
	\end{pmatrix}.
\end{equation}
Other matrices like $M_{R},M_P,M_{S}^{1},M_{S}^{2},M_{S}^{3}$ will retain their same structure throughout the inverted mass ordering.

\section{Phenomenology of active-sterile neutrino}
\label{3sec4}
The active neutrino mass matrix is obtained using equation \eqref{tamass} and the sterile mass is given by equation \eqref{tsmass}. The complete matrix picture for NH and IH are presented in tables \ref{3tab:nh}-\ref{3tab:ih1} respectively. For numerical analysis, we have first fixed non-degenerate values for the right-handed neutrino mass parameters as $d=e=10^{13}$ GeV and $f=5\times10^{13}$ GeV so that they can exhibit successful leptogenesis without affecting the neutrino parameters, which we have studied in our next chapter. The mass matrix arises from equation~\eqref{eq:2} give rise to complex quantities due to Dirac and the Majorana phases. Since the leptonic CP phases are still unknown, we vary them within (0, 2$\pi$). The global fit $3\sigma$ values for other parameters like mixing angles, mass square differences are taken from \cite{Capozzi:2016rtj}. One interesting aspect of MES is that if we consider $M_{S}= 
\begin{pmatrix}g&0&0\\
\end{pmatrix},$ 
structure, then eventually the parameters from the first column of $M_D$ and $M_R$ matrices go away and do not appear in the light neutrino mass matrix given by \eqref{tamass}. The same argument justifies the disappearance of the model parameters in the other two cases also. Hence, the active neutrino mass matrix emerging from our model matrices is left with three parameters for each case. Comparing the model mass matrix with the one produced by light neutrino parameters given by equation \eqref{eq:2}, we numerically evaluate the model parameters satisfying the current bound for the neutrino parameters and establish correlation among various model and oscillation parameters within $3\sigma$ bound. Three assessment for each distinct structures of $M_D$ for both normal and inverted hierarchy cases are carried out in the following subsections. 
\subsection{Normal hierarchy}\label{ss1}
\begin{table}
	\begin{center}
		\begin{tabular}{|p{.5cm}|p{5cm}c|}
			\hline
			NH& ~~~~~~Structures & $-m_{\nu}$ \\
			\hline
			\hline
			I & 
			$\begin{aligned}
				&
				M_R=\begin{pmatrix}
					d&0&0\\
					0&e&0\\
					0&0&f\\
				\end{pmatrix}\\
				& M_{D}= \begin{pmatrix}
					a&b&c+p\\
					a&b+p&c\\
					a+p&b&c\\
				\end{pmatrix}\\
				& M_{S}^{1}= \begin{pmatrix}
					g&0&0\\
				\end{pmatrix}\\
			\end{aligned}$
			& $\begin{pmatrix}
				\frac {b^2} {e} + \frac {(c + p)^2} {f} &\frac {b (b + 
					p)} {e} + \frac {c (c + 
					p)} {f} &\frac {b^2} {e} + \frac {c (c + p)} {f} \\
				\frac {b (b + p)} {e} + \frac {c (c + p)} {f} &\frac {(b + 
					p)^2} {e} + \frac {c^2} {f} &\frac {b (b + 
					p)} {e} + \frac {c^2} {f} \\
				\frac {b^2} {e} + \frac {c (c + p)} {f} &\frac {b (b + 
					p)} {e} + \frac {c^2} {f} &\frac {b^2} {e} + \frac {c^2}
				{f} \\
			\end{pmatrix}$ \\
			\hline
			II& 
			$\begin{aligned}
				&
				M_R=\begin{pmatrix}
					d&0&0\\
					0&e&0\\
					0&0&f\\
				\end{pmatrix}\\
				& M_{D}= \begin{pmatrix}
					a&b&c+p\\
					a&b+p&c\\
					a+p&b&c\\
				\end{pmatrix}\\
				& M_{S}^{2}= \begin{pmatrix}
					0&g&0\\
				\end{pmatrix}\\
			\end{aligned}$
			& $\begin{pmatrix}
				\frac {a^2} {d} + \frac {(c + 
					p)^2} {f} &\frac {a^2} {d} + \frac {c (c + 
					p)} {f} &\frac {a (a + p)} {d} + \frac {c (c + p)} {f} \\
				\frac {a^2} {d} + \frac {c (c + 
					p)} {f} &\frac {a^2} {d} + \frac {c^2} {f} &\frac {a (a \
					+ p)} {d} + \frac {c^2} {f} \\
				\frac {a (a + p)} {d} + \frac {c (c + p)} {f} &\frac {a (a + 
					p)} {d} + \frac {c^2} {f} &\frac {(a + 
					p)^2} {d} + \frac {c^2} {f} \\
			\end{pmatrix}$ \\
			\hline
			III & 
			$\begin{aligned}
				&
				M_R=\begin{pmatrix}
					d&0&0\\
					0&e&0\\
					0&0&f\\
				\end{pmatrix}\\
				& M_{D}= \begin{pmatrix}
					a&b&c+p\\
					a&b+p&c\\
					a+p&b&c\\
				\end{pmatrix}\\
				& M_{S}^{3}= \begin{pmatrix}
					0&0&g\\
				\end{pmatrix}\\
			\end{aligned}$
			& $\begin{pmatrix}\frac {a^2} {d} + \frac {b^2} {e} &\frac {a^2} {d} + \frac {b (b + 
					p)} {e} &\frac {a (a + p)} {d} + \frac {b^2} {e} \\
				\frac {a^2} {d} + \frac {b (b + p)} {e} &\frac {a^2} {d} + \frac {(b +
					p)^2} {e} &\frac {a (a + p)} {d} + \frac {b (b + p)} {e} \\
				\frac {a (a + p)} {d} + \frac {b^2} {e} &\frac {a (a + 
					p)} {d} + \frac {b (b + p)} {e} &\frac {(a + 
					p)^2} {d} + \frac {b^2} {e} \\
			\end{pmatrix}$ \\
			\hline
		\end{tabular}
		\caption{The light neutrino mass matrices and the corresponding $M_D$ and $M_R$ matrices for three different structures of $M_S$ under NH pattern. }\label{3tab:nh}
	\end{center}
\end{table}
\begin{table}\centering
	\begin{tabular}{|c|c|c|c|}
		\hline
		Case&$M_S$&$m_s(eV)$&$R$\\
		\hline
		\hline
		I& $M_{S}^{1}= \begin{pmatrix}
			g&0&0\\
		\end{pmatrix}$
		&$m_s\simeq\frac{g^2}{10^4}$&$R^T\simeq{\begin{pmatrix}
				\frac{a}{g}& \frac{a}{g}& \frac{a+p}{g}\\
		\end{pmatrix}}^T$\\
		\hline
		II& $M_{S}^{2}= \begin{pmatrix}
			0&g&0\\
		\end{pmatrix}$&$m_s\simeq\frac{g^2}{10^4}$&$R^T\simeq{\begin{pmatrix}
				\frac{b}{g}& \frac{b+p}{g}& \frac{b}{g}\\
		\end{pmatrix}}^T$\\
		\hline
		III& $M_{S}^{3}= \begin{pmatrix}
			0&0&g\\
		\end{pmatrix}$&$m_s\simeq\frac{g^2}{5\times10^4}$&$R^T\simeq{\begin{pmatrix}
				\frac{c+p}{g}& \frac{c}{g}& \frac{c}{g}\\
		\end{pmatrix}}^T$\\
		\hline
	\end{tabular}
	\caption{Sterile neutrino mass and active-sterile mixing matrix for three different $M_S$ structures under NH pattern.}\label{3tab:msnh}
\end{table}

For the diagonal charged lepton mass, we have chosen a non-trivial VEV alignment, explained in previous section, resulting in a specific pattern Dirac mass; thus a broken $\mu-\tau$ symmetry along with non-zero reactor mixing angle is achieved. The complete picture for active neutrino mass matrices and the sterile sector for different cases are shown in tables \ref{3tab:nh} and \ref{3tab:msnh} respectively. For each $M_S$ structure, three variables are there in the light neutrino mass matrix. After solving them by comparing them with the light neutrino mass, we obtain some correlation plots which redefine our model parameters with more specific bounds. Correlations among various model parameters in NH are shown in fig. \ref{modelnh}. The fact one would notice in the active mass matrices from tables \ref{3tab:nh} and \ref{3tab:ih} is that in the limit, $p\rightarrow0$ and all the model parameters become equal to one, the matrix takes the form of a democratic mass matrix. Hence, the $p$ parameters bring out a phenomenological change and play an important role in our study. Various plots with the model parameter $p$ are shown below in fig. \ref{psnh} and \ref{psih}.

%The parameter $a$ is evaluated by choosing $10^{-5}<y_1<10^{-6}$, on the other hand $g$ comes into possession by Eq(19). $|V_{e4}|^2 vs.|V_{\tau 4}|^2$ plot is achieved and the allowed region is marked. 
As $m_s$ depends only on $M_R$ and $M_S$, so due to the non-degenerate value of $M_R$, the $m_s$ structure let us study the active-sterile mixing strength $R$. The active-sterile mixing matrix also has a specific form due to the particular $M_S$ structure.
%\pagebreak
\subsection{Inverted hierarchy}\label{ss2}
\begin{table}[h!]
	\begin{center}
		\begin{tabular}{|p{.4cm}|p{5.2cm}c|}
			\hline
			IH & ~~~~~~~~Structures & $-m_{\nu}$ \\
			\hline
			\hline
		I& 
			$\begin{aligned}
				&
				M_R=\begin{pmatrix}
					d&0&0\\
					0&e&0\\
					0&0&f\\
				\end{pmatrix}\\
				& M_{D}= \begin{pmatrix}
					a&-b&c+p\\
					a&-b+p&c\\
					a+p&2b&c\\
				\end{pmatrix}\\
				& M_{S}^{1}= \begin{pmatrix}
					g&0&0\\
				\end{pmatrix}\\
			\end{aligned}$
			& $\begin{pmatrix}
				\frac {b^2} {e} + \frac {(c + p)^2} {f} &  \frac {b(b - 
					p)} {e} + \frac {c (c + 
					p)} {f} & \frac {-2 b^2} {e} + \frac {c (c + p)} {f} \\
				\frac {b(b - p)} {e} + \frac {c (c + p)} {f} &\frac {(b - 
					p)^2} {e} + \frac {c^2} {f} &\frac {-2 b (b - 
					p)} {e} + \frac {c^2} {f} \\
				-\frac {2 b^2} {e} + \frac {c (c + p)} {f} & - \frac {2 b (b - 
					p)} {e} + \frac {c^2} {f} &\frac {4 b^2} {e} + \frac {c^2}
				{f} \\
			\end{pmatrix}$ \\
			\hline
			II & 
			$\begin{aligned}
				&
				M_R=\begin{pmatrix}
					d&0&0\\
					0&e&0\\
					0&0&f\\
				\end{pmatrix}\\
				& M_{D}= \begin{pmatrix}
					a&-b&c+p\\
					a&-b+p&c\\
					a+p&2b&c\\
				\end{pmatrix}\\
				& M_{S}^{2}= \begin{pmatrix}
					0&g&0\\
				\end{pmatrix}\\
			\end{aligned}$
			& $\begin{pmatrix}
				\frac {a^2} {d} + \frac {(c + 
					p)^2} {f} &\frac {a^2} {d} + \frac {c (c + 
					p)} {f} &\frac {a (a + p)} {d} + \frac {c (c + p)} {f} \\
				\frac {a^2} {d} + \frac {c (c + 
					p)} {f} &\frac {a^2} {d} + \frac {c^2} {f} &\frac {a (a \
					+ p)} {d} + \frac {c^2} {f} \\
				\frac {a (a + p)} {d} + \frac {c (c + p)} {f} &\frac {a (a + 
					p)} {d} + \frac {c^2} {f} &\frac {(a + 
					p)^2} {d} + \frac {c^2} {f} \\
			\end{pmatrix}$ \\
			\hline
			III & 
			$\begin{aligned}
				&
				M_R=\begin{pmatrix}
					d&0&0\\
					0&e&0\\
					0&0&f\\
				\end{pmatrix}\\
				& M_{D}= \begin{pmatrix}
					a&-b&c+p\\
					a&-b+p&c\\
					a+p&2b&c\\
				\end{pmatrix}\\
				& M_{S}^{3}= \begin{pmatrix}
					0&0&g\\
				\end{pmatrix}\\
			\end{aligned}$
			& $\begin{pmatrix}\frac {a^2} {d} + \frac {(b^2} {e} &\frac {a^2} {d} - \frac {b (-b + 
					p)} {e} &\frac {a (a + p)} {d} - \frac {2 b^2} {e} \\
				\frac {a^2} {d} - \frac {b (-b + 
					p)} {e} &\frac {a^2} {d} + \frac {(b - 
					p)^2} {e} &\frac {a (a + p)} {d} + \frac {2 b (b - p)} {e} \\
				\frac {a (a + p)} {d} + \frac {-2 b^2} {e} &\frac {a (a + 
					p)} {d} - \frac {2 b (b - p)} {e} &\frac {(a + 
					p)^2} {d} + \frac {4 b^2} {e} \\
			\end{pmatrix}$ \\
			\hline
		\end{tabular}
	\end{center}
	\caption{The light neutrino mass matrices and the corresponding $M_D$ and $M_R$ matrices for three different structures of $M_S$ under IH pattern.}\label{3tab:ih}
\end{table}
\begin{center}
	\begin{table}[h]\centering
		\begin{tabular}{|c|c|c|c|}
			\hline
			Case&$M_S$&$m_s(eV)$&$R$\\
			\hline
			\hline
			I& $M_{S}^{1}= \begin{pmatrix}
				g&0&0\\
			\end{pmatrix}$
			&$m_s\simeq\frac{g^2}{10^4}$&$R^T\simeq{\begin{pmatrix}
					\frac{a}{g}& \frac{a}{g}& \frac{a+p}{g}\\
			\end{pmatrix}}^T$\\
			\hline
			II& $M_{S}^{2}= \begin{pmatrix}
				0&g&0\\
			\end{pmatrix}$&$m_s\simeq\frac{g^2}{10^4}$&$R^T\simeq{\begin{pmatrix}
					\frac{-b}{g}& \frac{-b+p}{g}& \frac{2b}{g}\\
			\end{pmatrix}}^T$\\
			\hline
			III& $M_{S}^{3}= \begin{pmatrix}
				0&0&g\\
			\end{pmatrix}$&$m_s\simeq\frac{g^2}{5\times10^4}$&$R^T\simeq{\begin{pmatrix}
					\frac{c+p}{g}& \frac{c}{g}& \frac{c}{g}\\
			\end{pmatrix}}^T$\\
			\hline
		\end{tabular}
		\caption{Sterile neutrino mass and active-sterile mixing matrix for three different $M_S$ structures under IH pattern. }\label{3tab:ih1}
	\end{table}
\end{center}
To discuss the inverted mass ordering ($i.e., m_2>m_1>m_3$) of the neutrinos, referring to \cite{Zhang:2011vh}, we have introduced a new flavon as, $\langle\varphi^{\prime}\rangle=(2v,-v,-v)$ in the Yukawa Lagrangian for the Dirac mass term, so that this model can exhibit inverted hierarchy.
The numerical procedure for IH is analogous to the NH. We have also considered three distinguished cases for $M_S$, which is responsible for three separate $m_\nu$ matrices. A brief picture for the matrices has shown in table \ref{3tab:ih}.

In table \ref{3tab:ih1}, three different $M_S$ structures are shown, which lead to various $m_s$ and $R$ values. Unlike the normal ordering, a deviation from the common track is observed in $R$ matrix for the second case ($M_S^{2}=(0,g,0)$). This occurs due to the change in $M_D$ matrix structure for non-identical VEV alignment.

%

%\pagebreak
%\clearpage

%Here in equation \ref{eq:25}, six variables are there($b_1,b_2,b_3,c_1,c_2,c_3$). Solving these model parameters by comparing with the parameterized light neutrino mass matrix given by \ref{eq:2} within given $3\sigma$ bound (as given by Table 1), we have shown some plots between the model parameters with the Dirac phase and some among themselves in figure \ref{fig:7} .
%\begin{figure}

%\includegraphics[scale=.3]{gb1-b2.png}
%\includegraphics[scale=.3]{gb2-c3.png}
%\includegraphics[scale=.3]{gb2-delta.png}
%\includegraphics[scale=.3]{gc1-delta.png}

%\caption{Plots for the general case}\label{fig:7}
%\end{figure}

\section{Numerical analysis and results}
\label{3sec5} 
\begin{figure}[h]
	\includegraphics[scale=0.24]{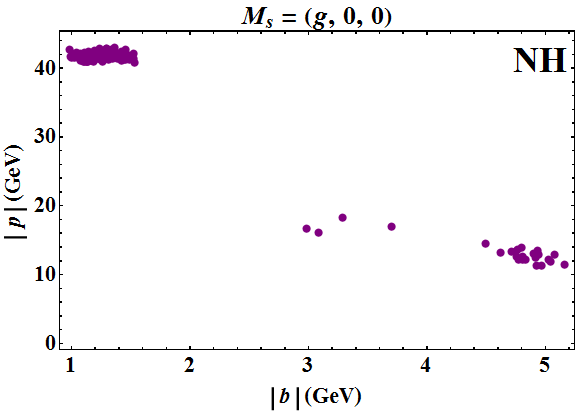}
	\includegraphics[scale=0.24]{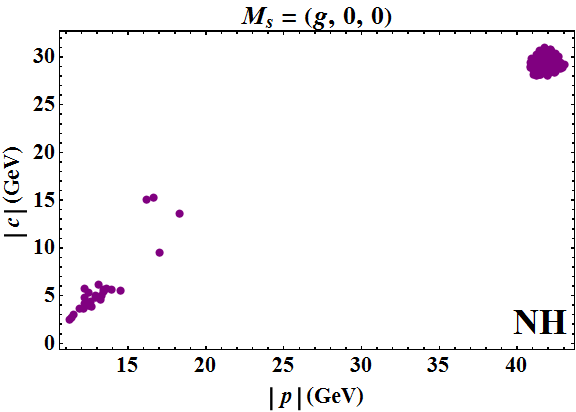}
	\includegraphics[scale=0.24]{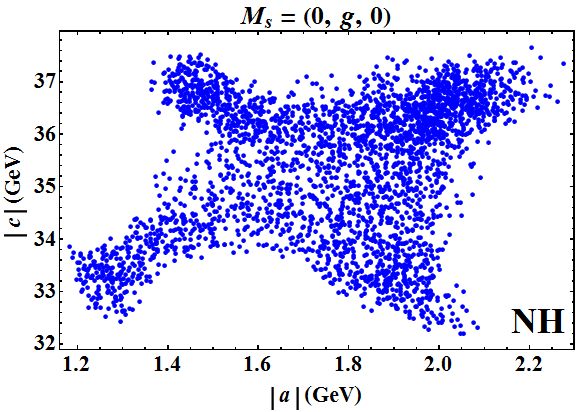}\\
	\includegraphics[scale=0.24]{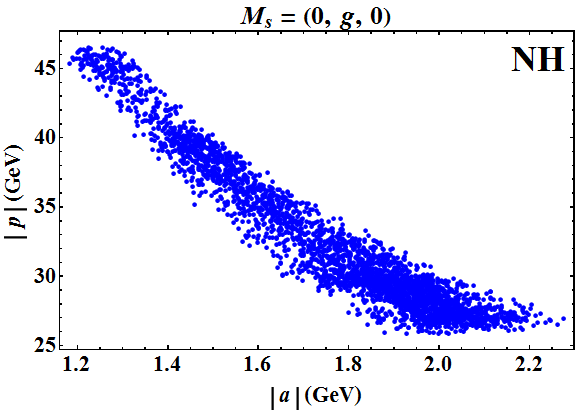}
	\includegraphics[scale=0.24]{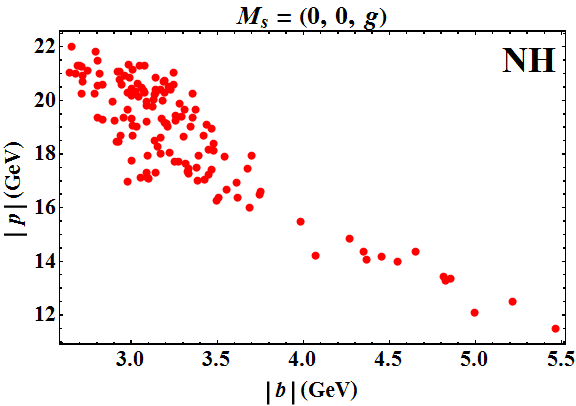}
	\includegraphics[scale=0.24]{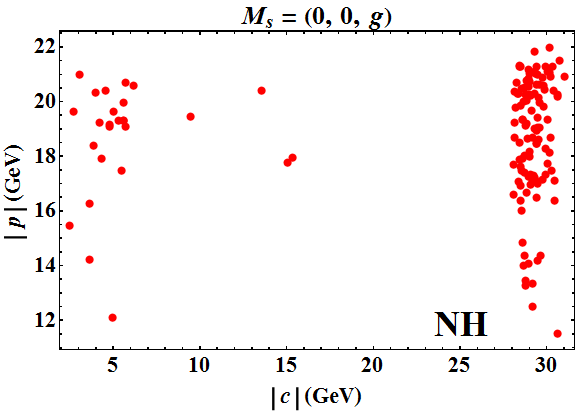}\\
	
	\caption{Variation of model parameters among themselves for the NH pattern.}\label{modelnh}
\end{figure}
%Here also we plot  $|V_{e4}|^2 vs.|V_{\mu 4}|^2$,but no mutual bound is obtained. $|V_{e4}|^2$ is in the $3\sigma$ bound but due to the larger $p$ value, $|V_{\mu 4}|^2$ value lies far away from the given bound.\\
\begin{figure}[h]
	\includegraphics[scale=0.24]{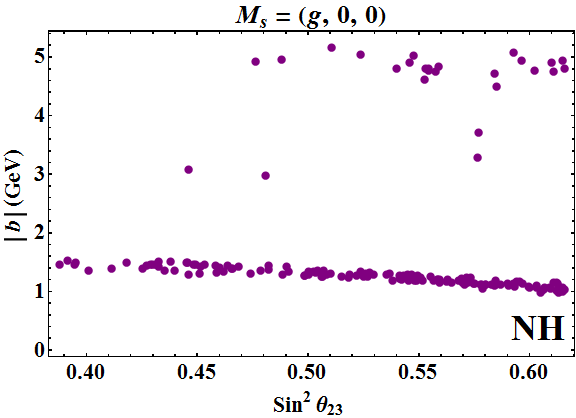}
	\includegraphics[scale=0.24]{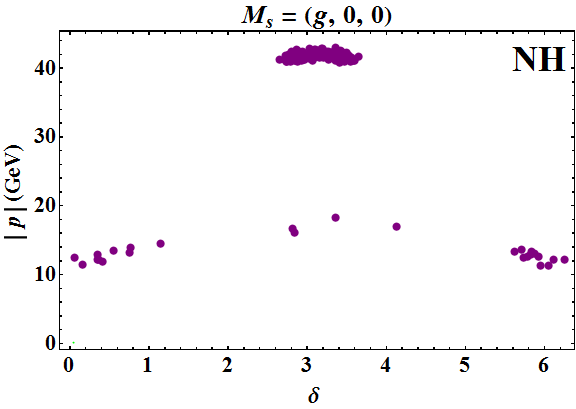}
	\includegraphics[scale=0.24]{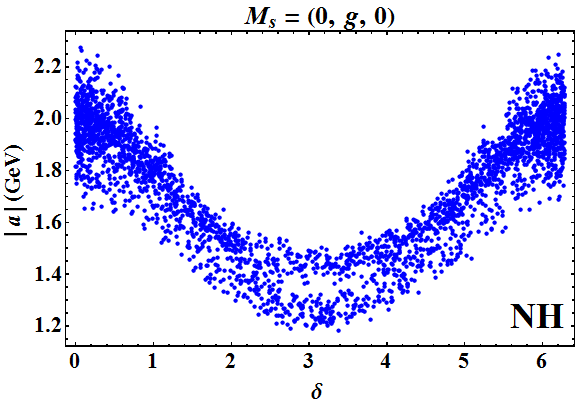}\\
	\includegraphics[scale=0.24]{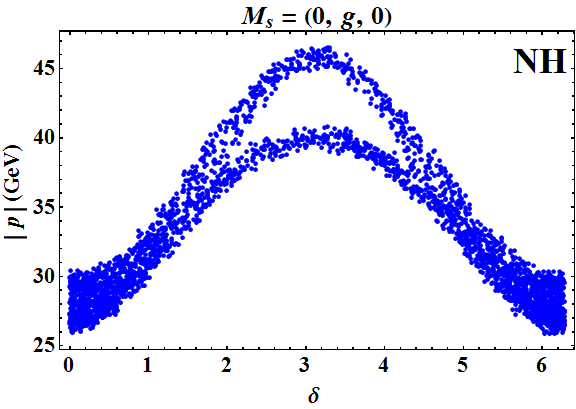}
	\includegraphics[scale=0.24]{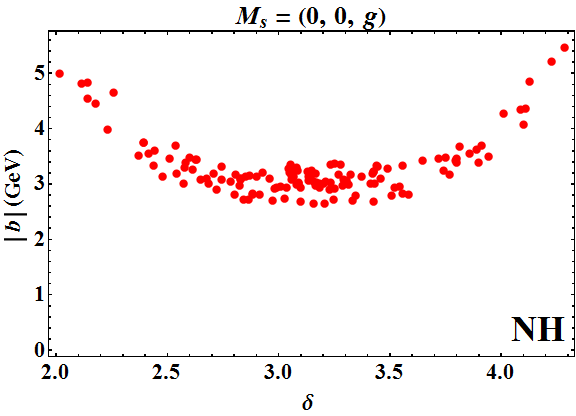}
	\includegraphics[scale=0.24]{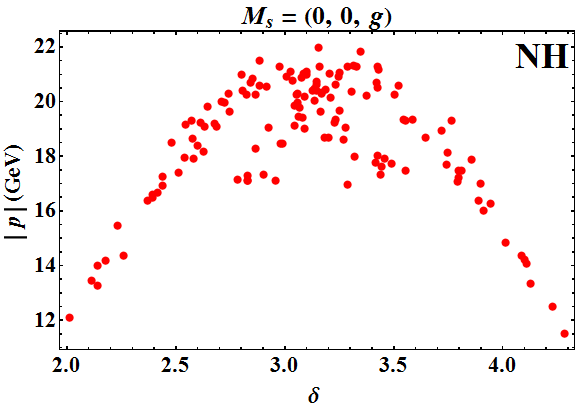}
	
	\caption{Correlation plots among various model parameters and light neutrino parameters (within $3\sigma$ bound) in NH. The Dirac CP phase shows a good correlation with the model parameters than the other light neutrino parameters.}\label{2nh}
\end{figure}
\begin{figure}[h]
	\includegraphics[scale=.24]{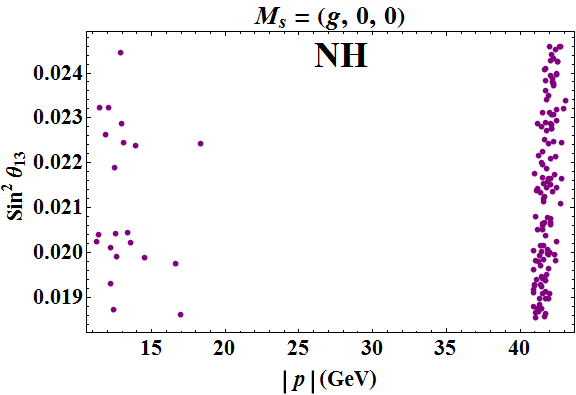}
	\includegraphics[scale=.24]{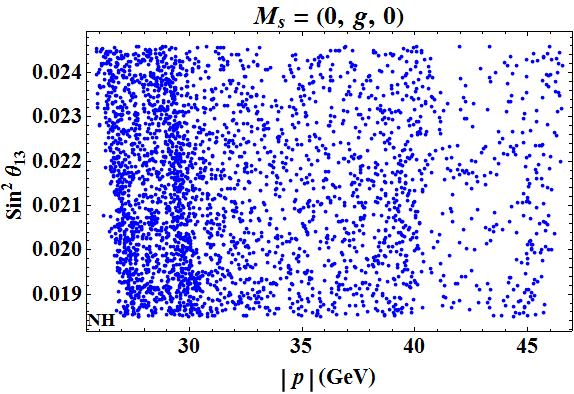}
	\includegraphics[scale=.24]{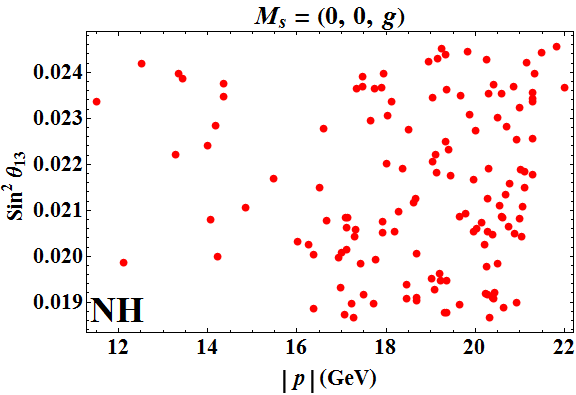}
	\caption{Variation of $\sin$ of reactor mixing angle with $p$, which is responsible for the generation of reactor mixing angle ($\theta_{13}$) for NH.}\label{psnh}
\end{figure}

\begin{figure}[h]
	\includegraphics[scale=.24]{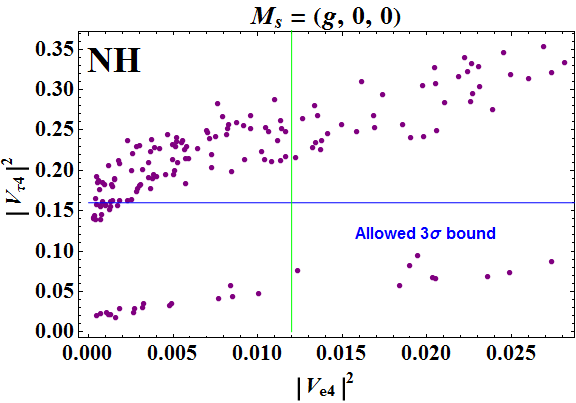}
	\includegraphics[scale=.24]{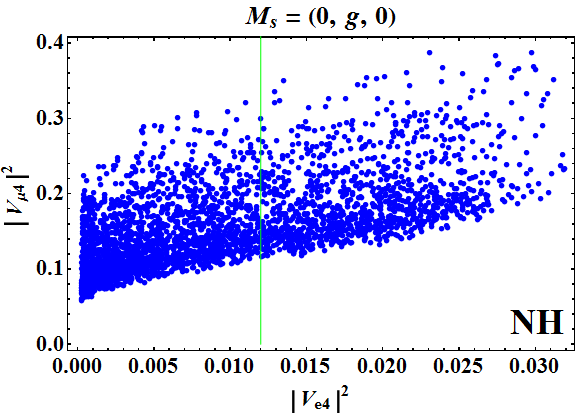}
	\includegraphics[scale=.24]{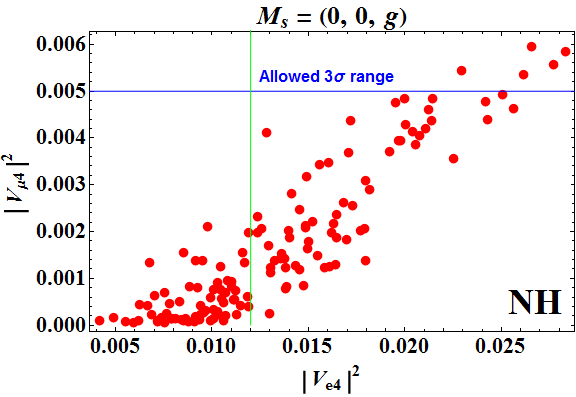}
	\caption{Allowed bound for the active-sterile mixing matrix elements in NH. The green line  is the lower bound for $|V_{e4}|^2$ and the blue line in the first plot gives the upper bound for $|V_{\tau4}|^2$ while in the third plot it gives the lower bound for $|V_{\mu4}|^2$}\label{4nh}
\end{figure}

\begin{figure}[h]
	\includegraphics[scale=0.24]{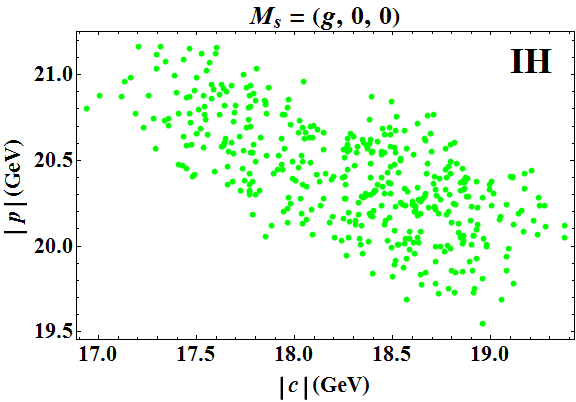}
	\includegraphics[scale=0.24]{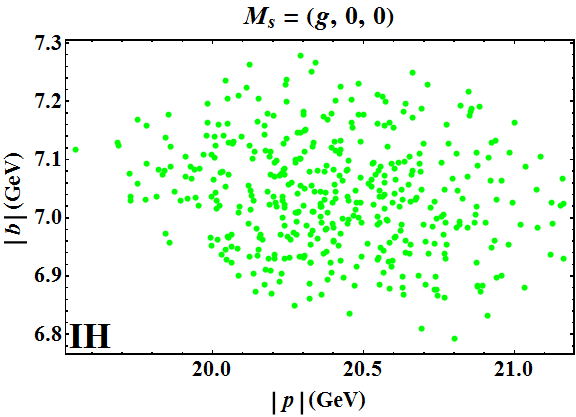}
	\includegraphics[scale=0.24]{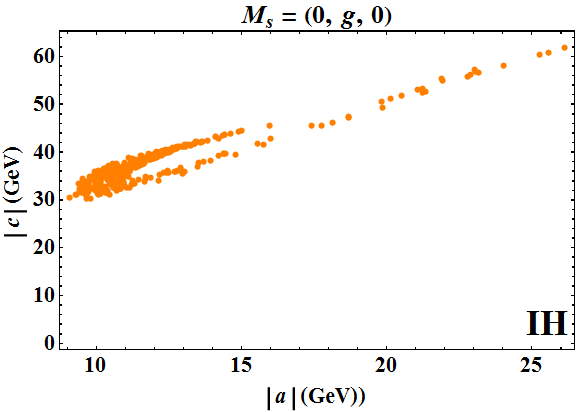}
	\includegraphics[scale=0.24]{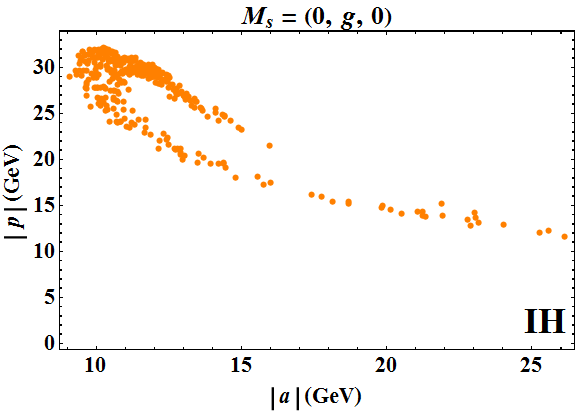}
	\includegraphics[scale=0.24]{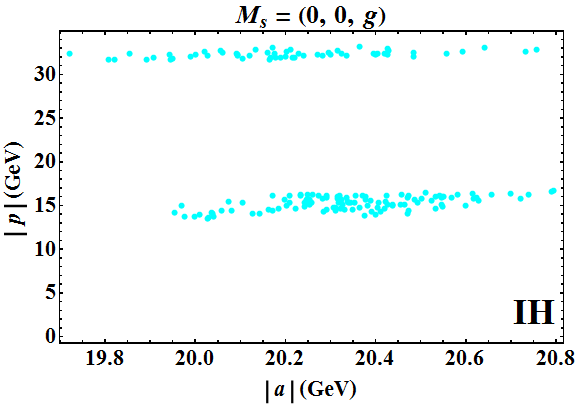}
	\includegraphics[scale=0.24]{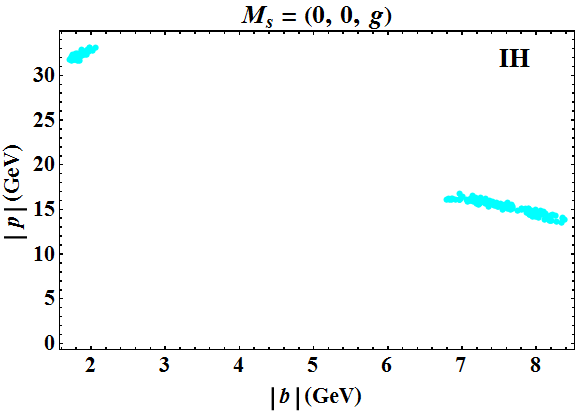}
	\caption{Constrained region of model parameters in case of IH pattern.}\label{modelih}
\end{figure}
\begin{figure}
	\includegraphics[scale=0.24]{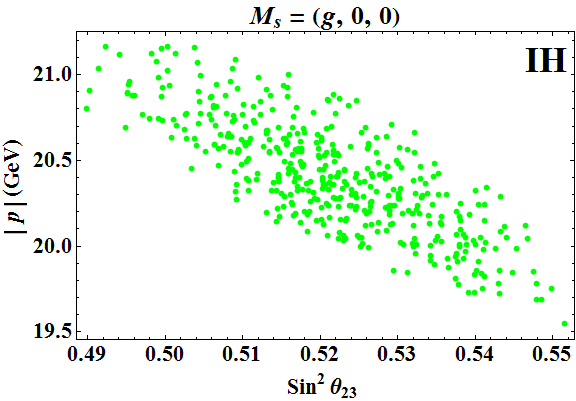}
	\includegraphics[scale=0.24]{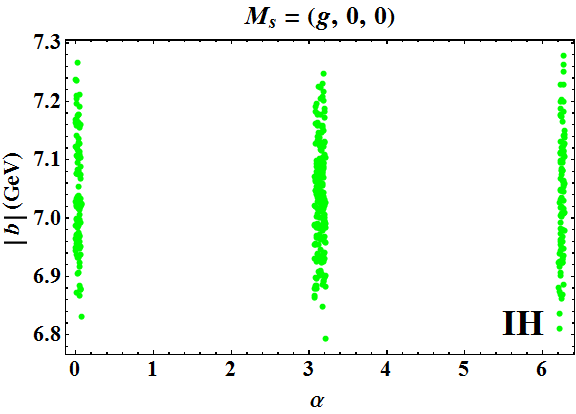}
	\includegraphics[scale=0.24]{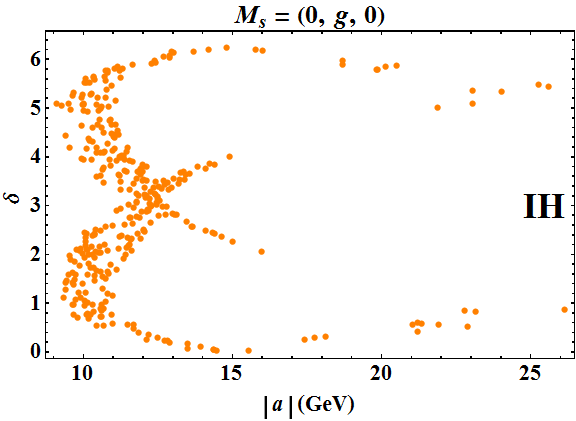}
	\includegraphics[scale=0.24]{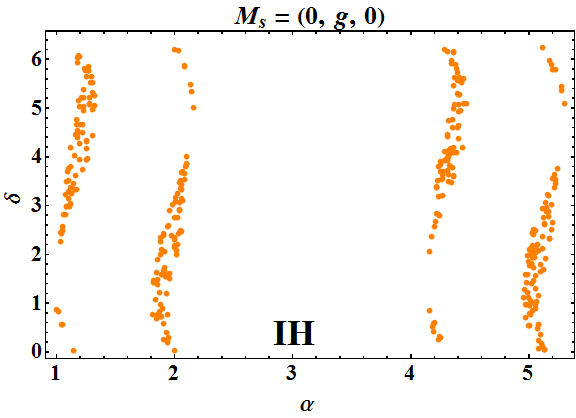}
	\includegraphics[scale=0.24]{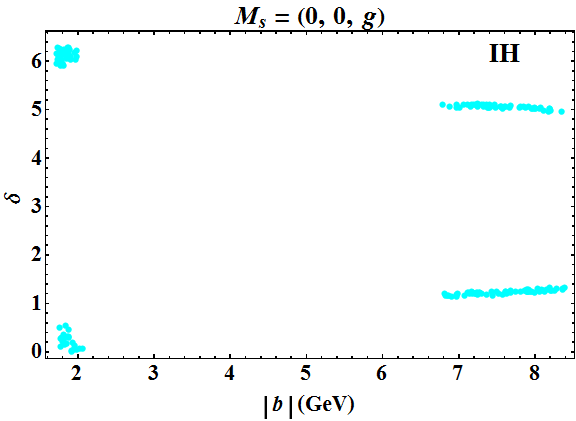}
	\includegraphics[scale=0.24]{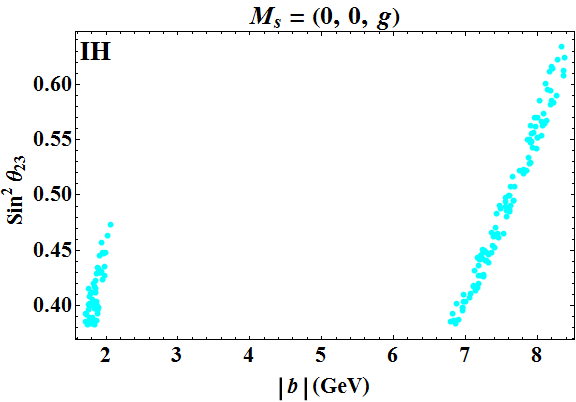}
	
	\caption{Correlation plots among various model parameters with light neutrino parameters in IH pattern.}\label{2ih}
\end{figure}

\begin{figure}[h]
	\includegraphics[scale=.240]{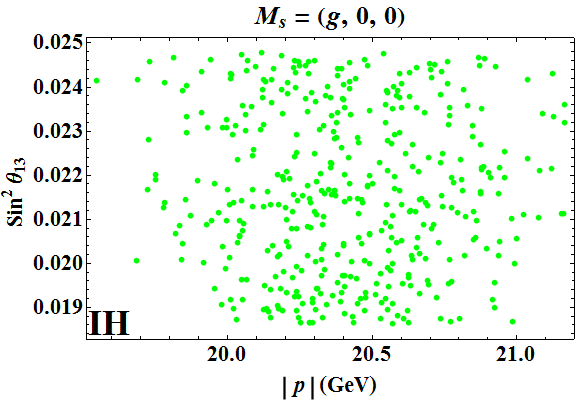}
	\includegraphics[scale=.24]{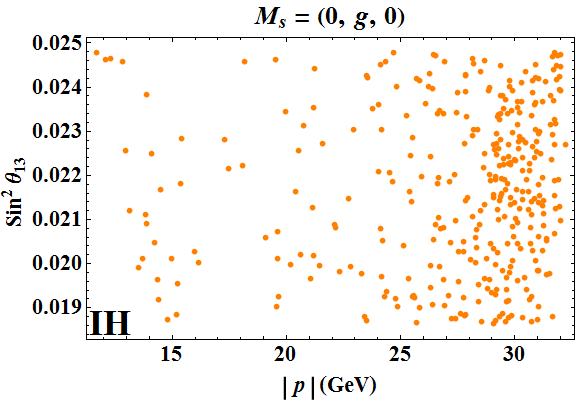}
	\includegraphics[scale=.24]{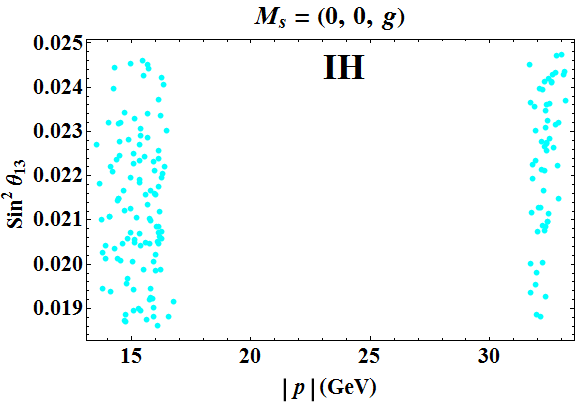}
	\caption{Variation of $p$ with the $\sin$ of reactor mixing angle for IH. The third structure of $M_S$ shows a constrained region for the model parameter.}\label{psih}
\end{figure}
\begin{figure}[h]\centering
	\includegraphics[scale=.3]{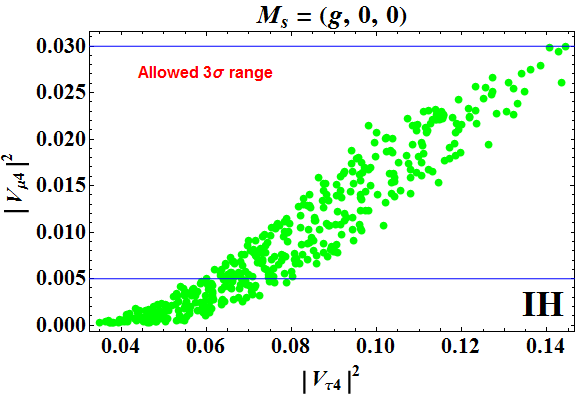}
	\includegraphics[scale=.3]{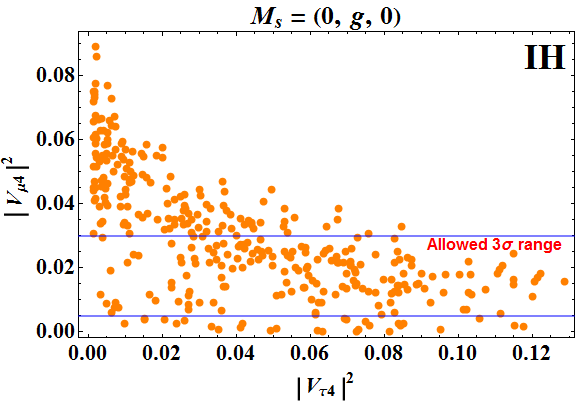}\\
	\includegraphics[scale=.3]{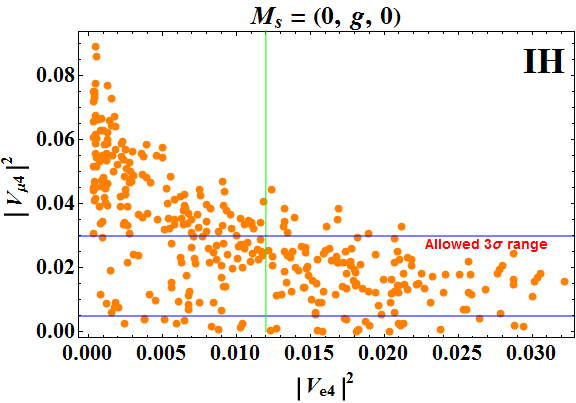}
	\includegraphics[scale=.3]{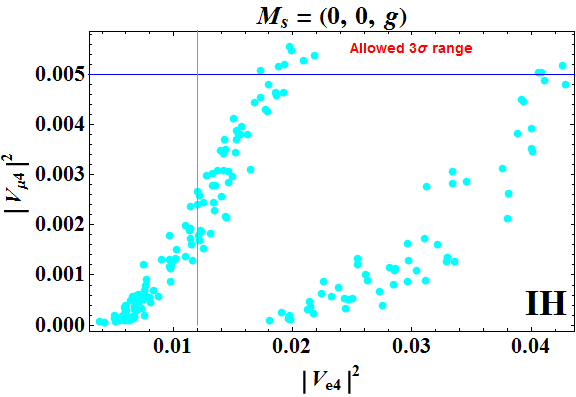}
	\caption{Allowed bound for active-sterile mixing matrix elements in IH. The blue solid line gives the upper and lower bound for $|V_{\mu4}|^2$ along the y-axix while solid green line gives the lower bound for $|V_{e4}|^2$ along the x-axis. }\label{4ih}
\end{figure}
\begin{figure}
	\includegraphics[scale=.24]{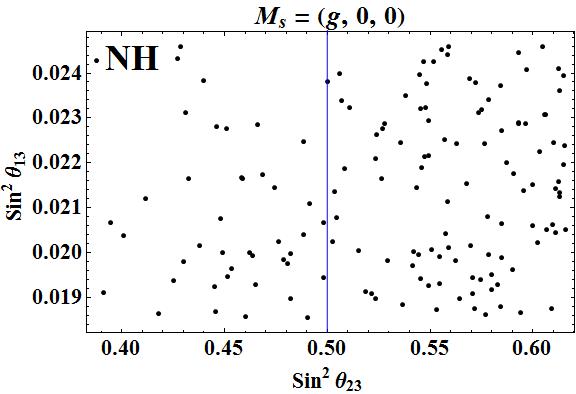}
	\includegraphics[scale=.24]{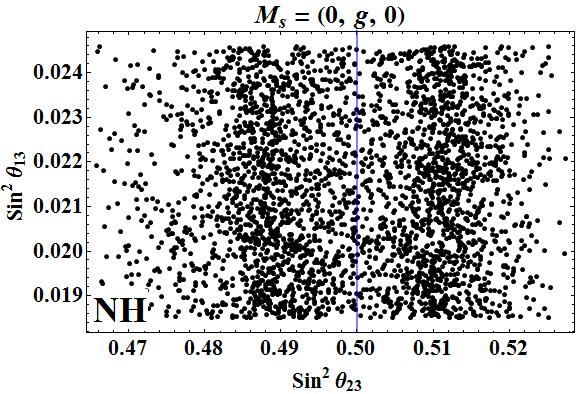}
	\includegraphics[scale=.24]{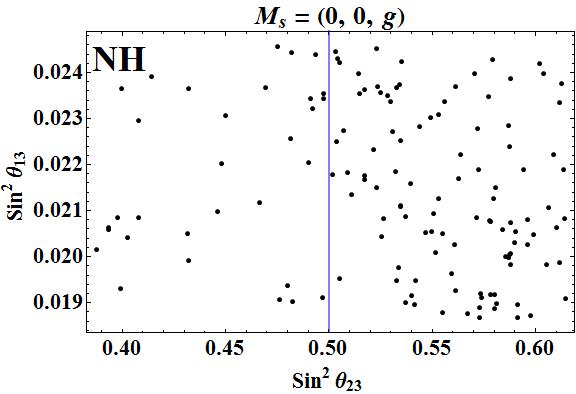}\\
	\includegraphics[scale=.24]{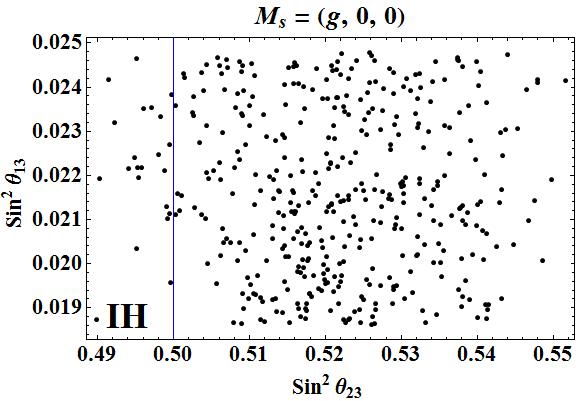}
	\includegraphics[scale=.24]{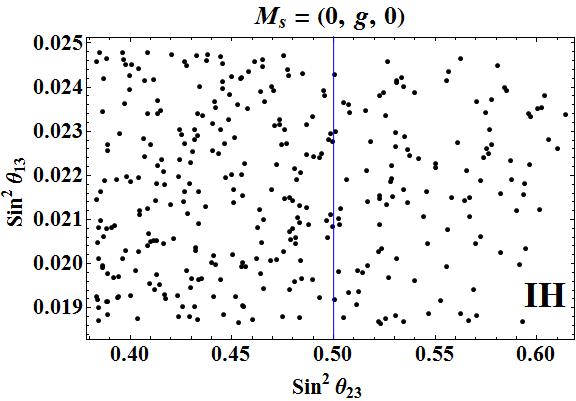}
	\includegraphics[scale=.24]{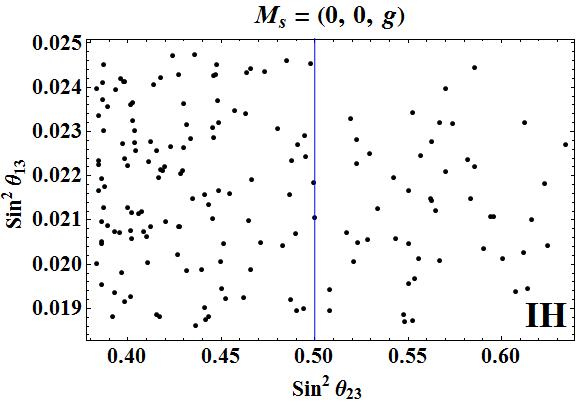}
	\caption{Variation of $\sin^2\theta_{13}$ vs. $\sin^2\theta_{23}$ in both the mass ordering for all the three $M_S$ structures. }\label{1323}
\end{figure}

Both normal and inverted cases are analyzed independently for three $M_S$ structures in this model. We have used similar numerical techniques for solving model parameters in both the cases (NH \& IH) and plotted them among themselves and the light neutrino parameters. The plots in fig. \ref{modelnh} , \ref{2nh}, \ref{modelih}, \ref{2ih} show constrained parameter space in the active neutrino sector in case of NH and IH for various $M_S$ structures. In most cases, the parameter space is narrow, which can be verified or falsified in future experiments. In the $m_{\nu}$ matrix, the $\mu-\tau$ symmetry is broken due to the additional term added to the Dirac mass matrix. The variation of $\sin^2\theta_{13}$ with $p$ plotted in fig. \ref{psnh} and \ref{psih} for NH \& IH respectively. Within NH, the first structure of $M_S$ shows a better-constrained region for the model parameter ($p$) than the other two structures. Whereas in the IH case, the third $M_S$ structure gives a relatively narrower region than that obtained for the other two structures. In fig. \ref{1323} we have plotted the $\sin$ squared of the reactor mixing angle against the atmospheric mixing angle. Within the NH mode, $M_S^1$ and $M_S^3$ structures favour the upper octant of $\sin^2\theta_{23}$ accommodating more numbers of data points, whereas, within IH mode, $M_S^2$ and $M_S^3$ structures are more favourable in the lower octant. The $M_S^1$ structure in IH mode is heavily constrained within the upper octant of the atmospheric mixing angle. On the other hand, the second structure of $M_S$ in the NH case shows deviation from the maximal atmospheric mixing, having dense regions in either octant thus not constrained for $\sin^2\theta_{23}$.

The active-sterile mixing phenomenology is also carried out under the same MES framework. The fourth column of the active-sterile mixing matrix is generated and solved with an acceptable choice of Yukawa coupling. Apart from generating non-zero $\theta_{13}$, the matrix element of $M_P$ has an important role in the active-sterile mixing. As we can see in table \ref{3tab:msnh} and \ref{3tab:ih1}, $p$ has an active participation in differentiating the elements of $R$ matrix. We have plotted the mixing matrix elements ($V_{e4},V_{\mu4},V_{\tau4}$) within themselves as shown in fig. \ref{4nh} and \ref{4ih}. The SK collaboration limits $|V_{\mu 4}|^2 < 0.04$ for $\Delta m_{41}^2>0.1eV^2$ at 90\% CL by considering $|V_{e4}|^2=0$ \cite{Abe:2016nxk}. For $\Delta m_{41}^2\sim 1eV^2$, the IceCube DeepCore collaboration suggested that $|V_{\mu 4}|^2 < 0.03$ with $|V_{\tau 4}|^2<0.15$ and $|V_{e4}|^2$ is around 0.012 at 90\% CL \cite{Aartsen:2017bap,Thakore:2018lgn}. In particular, for the various mass range of $\Delta m_{41}^2$ there are more fascinating results about active-sterile mixing, however, which we try to discuss in our subsequent chapters. These bounds are consistent with some of our model structures. In the NH case, the first and the third $M_S$ structure show an allowed $3\sigma$ range for the mixing parameters, but no such mutual allowed range is obtained for the second structure of $M_S$. The plots in fig. \ref{4ih} show the IH case for the mixing elements. The first structure of $M_S$ covers a wider range of allowed data points within the $3\sigma$ bound than the other two cases.

We have not shown any plots relating active-sterile mass squared difference with the active mass and active-sterile mixing elements because from table \ref{3tab:msnh} and \ref{3tab:ih1} one can see that the sterile mass is emerging as $g^2/10^4$, $g^2/10^4$ and $g^2/{5\times10^4}$ for respective $M_S^1$, $M_S^2$ and $M_S^3$ structures and the active sterile mixing matrix also contain $g$ exclusively. Since, $g$ is evaluated using active-sterile mass squared difference, therefore, a comparison of this mass squared difference with the active-sterile mixing matrix would be needless within this numerical approach. Since we are focusing on model building aspects, we are taking bounds for the light neutrino parameters from global fit data to verify/predict our model. Moreover, the plot between any two light neutrino observables would be simply a presentation between two global fit data sets that do not carry any significance in our study. This argument supports why we have not shown any plot between active-sterile mass squared differences/angles with the active neutrino masses/angles. Moreover, the active-sterile mixing matrix elements and mixing angles more or less represent the same phenomenology. However, we prefer to show correlations among the matrix elements over the mixing angles. The mixing elements' bounds are more auspicious than the angles (although the mixing elements depend on the mixing angles).\\
Authors in \cite{Barry:2011wb, Zhang:2011vh} have discussed active and sterile phenomenology by considering the same MES framework under $A_4$ flavour symmetry. The light neutrino mass matrix ($m_{\nu}$) is diagonalized using the tri-bimaximal mixing matrix and it is found to be $\mu-\tau$ symmetric matrix. On the other hand, as per current experimental demand \cite{An:2012eh}, in our current work, we have addressed non-zero reactor mixing angle by adding a correction term in the $M_D$, which breaks the trivial $\mu-\tau$ symmetry in $m_{\nu}$, that was considered zero in earlier studies. The additional correction term (added to $M_D$) substantially influences the reactor mixing angle discussed in the last paragraph. In addition to their previous work \cite{Barry:2011wb, Zhang:2011vh}, we have constructed our model with new flavons and extensively studied three non-identical $M_S$ structures separately in NH as well as in IH mode.

\section{Summary}\label{con3}
In this chapter, we have investigated the extension of low scale type-I seesaw $i.e.,$ the minimal extended seesaw, which restricts active neutrino masses to be within the sub-eV scale and generates an eV scale light sterile neutrino. $A_4$ based flavour model is extensively studied along with a discrete Abelian symmetry $Z_4$ and $Z_3$ to construct the desired Yukawa coupling matrices. Under this MES framework, the Dirac mass $M_D$ is a $(3\times3)$ complex matrix. The Majorana mass matrix $M_R$, which arises due to the coupling of right-handed neutrinos with the anti-neutrinos, is also a $(3\times 3)$ complex symmetric diagonal matrix with non-degenerate eigenvalues. A singlet $S_i$ (where $i=1,2,3$) is considered which couples with the right-handed neutrinos ($\nu_{Ri};i=1,2,3$) and produces a singled row  $(1\times 3)$ $ M_S$ matrix with one non-zero entry. In earlier studies like \cite{Zhang:2011vh,Nath:2016mts}, the $A_4$ flavor symmetry in MES was implemented with a little description. Our phenomenological study has addressed the non-zero reactor mixing angle with a detailed discussion on VEV alignment of the flavon fields discussed under the light of flavour symmetry within this MES framework. Three separate cases are carried out for both NH and IH for three $M_S$ structures. Within the active neutrino mass matrix, the common $\mu-\tau$ symmetry is broken along with $\theta_{13}\neq 0$ by adding a new matrix ($M_P$) to the Dirac mass matrix.

As a conclusive remark, the low scale MES mechanism is analyzed in this chapter. This model can also be used to study the connection between effective mass in neutrino-less double beta decay in a broader range of sterile neutrino mass from $eV$ to a few $keV$. Study of $keV$ scale sterile neutrino can be a portal to explain the origin of dark matter and related cosmological issues in this MES framework, which we will be discussing in our subsequent chapters.
% Chapter Template
\begin{savequote}[1\linewidth]
	\normalsize	``We do not need magic to change the world, we carry all the power we need inside ourselves already: we have the power to imagine better.''
	\qauthor{\large\it JK Rowling, (1965- Present)}
\end{savequote}
\chapter{Neutrino mass, dark matter and baryogenesis within the framework of minimal extended seesaw} % Main chapter title
%\vspace{-2.1cm}
%{\it ``It doesn't matter how beautiful your theory is, it doesn't matter how smart you are. If it doesn't agree with the experiment, it's wrong."--- {\Large Richard P. Feynman}}
%{\it ``All truths are easy to understandonce they are discovered; the point is to discover them."---{ \Large Galileo Galilei}}
\label{Chapter3} % Change X to a consecutive number; for referencing this chapter elsewhere, use \ref{ChapterX}

\lhead{Chapter 3. \emph{Neutrino mass, dark matter and baryogenesis within the framework of minimal extended seesaw}} % Change X to a consecutive number; this is for the header on each page - perhaps a shortened title

%----------------------------------------------------------------------------------------
%	SECTION 1
%----------------------------------------------------------------------------------------
\section*{}
\vspace{-1.7cm}
In this chapter, we have studied neutrino mass, dark matter and baryogenesis within the framework of a minimal extended seesaw.
A neutrino mass model has been constructed using an $A_4$ flavour symmetry along with the discrete $Z_4$ symmetry is used to stabilize the dark matter and establish desired mass matrices.
We use a non-trivial Dirac mass matrix with broken $\mu - \tau$ symmetry to generate the leptonic mixing. A non-degenerate mass structure for right-handed neutrinos is also considered to verify the observed baryon asymmetry of the Universe via the mechanism of thermal Leptogenesis. The scalar sector is also studied in great detail for a multi-Higgs doublet scenario, considering the lightest $Z_4$-odd field as a viable dark matter candidate. A significant impact on the region of DM parameter space, as well as in the fermionic sector are found in the presence of extra scalar particles.

\section{Introduction}

We have studied two important sectors in this chapter: the fermion and scalar sector. In the fermion sector, we have chosen the MES framework, where, along with the active neutrino mass generation, the validity of baryogenesis is also checked. Sterile neutrino phenomenology with the active neutrinos has already been discussed in previous chapter (Chapter \ref{Chapter2}). Within the scalar sector, we have considered two additional Higgs doublets along with the SM Higgs doublet, where one additional Higgs doublet acquires VEV. At the same time, another one does not acquire any VEV due to an additional $Z_4$ symmetry. The extension of scalar sector considering three Higgs doublets is quite popular in literature~\cite{Aranda:2012bv, Ivanov:2012ry, Chakrabarty:2015kmt}. The lightest neutral $Z_4$-odd Higgs doublet does not decay. Hence, it behaves as a potential candidate for dark matter. The other two scalar doublets are playing a crucial role in explaining the neutrino mass, mixing angles, and baryon asymmetry via leptogenesis. In the fermionic sector, they appear in the Dirac Lagrangian to give mass to the active neutrinos. In contrast, the scalar sector influence the DM phenomenology. All the theoretical and experimental constraints such as absolute stability, unitarity, EW precision, LHC Higgs signal strength, and dark matter density and direct detection are discussed here in details here.
% which was not being presented in the literature.

This chapter is organized as follows. The $A_4$ symmetry along with an additional $Z_4$ discrete symmetry to construct the model and generation of the mass matrices in the scalar and leptonic sector are addressed in \ref{model2} section. We have discussed active-sterile mass and mixing methodology and baryogenesis $via$ thermal leptogenesis followed by discussion regarding bounds on this model in section \ref{4s2} and its subsections. Numerical analysis of the dark matter and neutrino sector along with the BAU results are presented subsequently in section \ref{s4}. Finally, the summary of this chapter has put in section \ref{s5}.
%%%%%%%%%%%%%%%%%%%%%%%%%%%%%%%%%%%%%%%%%%%%%%%%%
%%%%%%%%%%%%%%%%%%%%%%%%%%%%%%%%%%%%%%%%%%%%%%%%%
%%%%%%%%%%%%%%%%%%%%%%%%%%%%%%%%%%%%%%%%%%%%%%%%%
%%%%%%%%%%%%%%%%%%%%%%%%%%%%%%%%%%%%%%%%%%%%%%%%%

%%%%%%%%%%%%%%%%%%%%%%%%%%%%%%%%%%%%%%%%%%%%%%%%%
%%%%%%%%%%%%%%%%%%%%%%%%%%%%%%%%%%%%%%%%%%%%%%%%%
%%%%%%%%%%%%%%%%%%%%%%%%%%%%%%%%%%%%%%%%%%%%%%%%%
%%%%%%%%%%%%%%%%%%%%%%%%%%%%%%%%%%%%%%%%%%%%%%%%%
\section{Model framework}\label{model2}
This model is an extension of our previous model with a slight change in fields and charges. Instead of three different sterile cases, in this model, we only have worked on a single case with $M_S=\begin{pmatrix}
	g&0&0
\end{pmatrix}$. In addition to active-sterile mixing and their phenomenological study, the scalar sector is also worked out in great details for a multi-Higgs model. 
\subsection{Particle content and discrete charges}
We display the particle content and their charges within our model in table ~\ref{2tab1}. The left-handed (LH) lepton doublet $l$ to transform as $A_4$ triplet, whereas right-handed (RH) charged leptons ($e^c,\mu^c,\tau^c$) transform as 1,$1^{\prime\prime}$ and $1^{\prime}$ respectively. The triplet flavons $\zeta, \varphi$ and two singlets $\xi$ and $\xi^{\prime}$  break the $A_4$ flavor symmetry by acquiring VEVs at large scale in the suitable directions\footnote{The chosen VEV alignments of the triplet flavons are obtained by minimizing the potential, can be found in the appendix \ref{appn1}.}. The Higgs doublets are assumed to be transformed as singlet under $A_4$. Additional $Z_4$ charges are assigned for the individual particles as per the interaction terms demands to restricts the non-desired terms.

The $SU(2)$ doublet Higgs ($\phi_3$) along with $\xi^{\prime} $ and $\nu_{R3}$ are odd under $Z_4$. However, the mass scale of the flavon $\xi^{\prime}$ and the RH particle $\nu_{R3}$ are too heavy in comparison to the SU(2) doublet, $\phi_3$. As $\phi_3$ is a $Z_4$ odd field,  it doesn't couple with any SM fields, and so the lightest neutral doublet does not decay to any SM particle directly. Indirect detection experiments constrain DM decay lifetime to be larger than $10^{27} -10^{28} ~s$ \cite{Cohen:2016uyg}, in the meanwhile, the dark matter can decay to the SM particles through $\xi^{\prime}$, $\nu_{R1}$ and $\nu_{R2}$ via 7,8-body decay processes, which is heavily suppressed by the propagator masses and decay width is almost zero. The lifetime, in this case, is much greater than $10^{60} ~s$. Hence, the lightest neutral particle of $\phi_3$ is stable, and it can serve as a viable cold-WIMP dark matter candidate. 
%%%%%%%%%%%%%%%%%%%%%%%%%%%%%%%%%%%%%%%%%%%%%%%%%
%%%%%%%%%%%%%%%%%%%%%%%%%%%%%%%%%%%%%%%%%%%%%%%%%
%%%%%%%%%%%%%%%%%%%%%%%%%%%%%%%%%%%%%%%%%%%%%%%%%
%%%%%%%%%%%%%%%%%%%%%%%%%%%%%%%%%%%%%%%%%%%%%%%%%
\begin{table}[h!]
	\begin{tabular}{|c|ccccc|cc|cccc|ccc|cc|}
		\hline
		Particles & $l$ &$e_{R}$&$\mu_{R}$&$\tau_{R}$&$\phi_1$&$\phi_2$&$\phi_3$&$\zeta$&$\varphi$&$\xi$&$\xi^{\prime}$&$\nu_{R1}$&$\nu_{R2}$&$\nu_{R3}$&$S$&$\chi$\\
		\hline
		\hline
		SU(2)&2&1&1&1&2&2&2&1&1&1&1&1&1&1&1&1\\
%		\hline
		$A_4$&3&1&$1^{\prime\prime}$&$1^{\prime}$&1&1&1&3&3&1&$1^{\prime}$&1&$1^{\prime}$&1&$1^{\prime\prime}$&$1^{\prime}$\\
%		\hline
		$Z_4$&1&1&1&1&1&i&-1&1&i&1&-1&1&-i&-1&i&-i\\
		\hline
	%	\hline		
	\end{tabular}
	\caption{Particle content and their charge assignments under SU(2), $A_4$ and $Z_4$ groups. The second block of the particle content ($l,e_R,\mu_R,\tau_R,\phi_1$) represents the left-handed lepton doublet, RH charged fermions and SM Higgs doublets respectively. $\phi_2$ and $\phi_3$ (inert) are the additional Higgs doublet. $\nu_{Ri}(i=1,2,3)$ and $S$ are RH neutrinos and chiral singlet. Rest of the particles ($\zeta,\varphi,\xi,\xi^{\prime},\chi$) are the additional flavons.}\label{2tab1} 
\end{table}
%%%%%%%%%%%%%%%%%%%%%%%%%%%%%%%%%%%%%%%%%%%%%%%%%
%%%%%%%%%%%%%%%%%%%%%%%%%%%%%%%%%%%%%%%%%%%%%%%%%
%%%%%%%%%%%%%%%%%%%%%%%%%%%%%%%%%%%%%%%%%%%%%%%%%
%%%%%%%%%%%%%%%%%%%%%%%%%%%%%%%%%%%%%%%%%%%%%%%%%
\subsubsection{Normal hierarchical neutrino mass }

The leading order invariant Yukawa Lagrangian for the lepton sector is given by,
%%%%%%%%%%%%%%%%%%%%%%%%%%%%%%%%%%%%%%%%%%%%%%%%%
%%%%%%%%%%%%%%%%%%%%%%%%%%%%%%%%%%%%%%%%%%%%%%%%%
%%%%%%%%%%%%%%%%%%%%%%%%%%%%%%%%%%%%%%%%%%%%%%%%%
%%%%%%%%%%%%%%%%%%%%%%%%%%%%%%%%%%%%%%%%%%%%%%%%%
\begin{equation}\label{2lag}
	\begin{split}
		\mathcal{L} =& \frac{y_{e}}{\Lambda}(\overline{l} \phi_1 \zeta)_{1}e_{R}+\frac{y_{\mu}}{\Lambda}(\overline{l} \phi_1 \zeta)_{1^{\prime}}\mu_{R}+\frac{y_{\tau}}{\Lambda}(\overline{l} \phi_1 \zeta)_{1^{\prime\prime}}\tau_{R} \\
		&+ \frac{y_{2}}{\Lambda}(\overline{l}\tilde{\phi_1}\zeta)_{1}\nu_{R1}+\frac{y_{2}}{\Lambda}(\overline{l}\tilde{\phi_1}\varphi)_{1^{\prime\prime}}\nu_{R2}+\frac{y_{3}}{\Lambda}(\overline{l}\tilde{\phi_2}\varphi)_{1}\nu_{R3}\\
		& +\frac{1}{2}\lambda_{1}\xi\overline{\nu^{c}_{R1}}\nu_{R1}+\frac{1}{2}\lambda_{2}\xi^{\prime}\overline{\nu^{c}_{R2}}\nu_{R2}+\frac{1}{2}\lambda_{3}\xi\overline{\nu^{c}_{R3}}\nu_{R3}\\
		&+ \frac{1}{2}\rho\chi\overline{S^{c}}\nu_{R1} .\\
	\end{split}
\end{equation}
%%%%%%%%%%%%%%%%%%%%%%%%%%%%%%%%%%%%%%%%%%%%%%%%%
%%%%%%%%%%%%%%%%%%%%%%%%%%%%%%%%%%%%%%%%%%%%%%%%%
%%%%%%%%%%%%%%%%%%%%%%%%%%%%%%%%%%%%%%%%%%%%%%%%%
%%%%%%%%%%%%%%%%%%%%%%%%%%%%%%%%%%%%%%%%%%%%%%%%%
$\Lambda$ in the Lagrangian, represents the cut-off scale of the theory, $y_{\alpha,i}$, $\lambda_{i}$ (for $\alpha=e,\mu,\tau$ and $i=1,2,3$) and $\rho$ representing the Yukawa couplings for respective interactions and all Higgs doublets are transformed as $\tilde{\phi_i} = i\tau_{2}\phi_i^*$ (with $\tau_{2}$ being the second Pauli's spin matrix)  to keep the Lagrangian gauge invariant.

The scalar flavons involved in the Lagrangian acquire VEV along  $ \langle \zeta \rangle=(v_m,0,0),~ \langle\varphi\rangle=(v_m,v_m,v_m), ~
\langle\xi\rangle=\langle\xi^{\prime}\rangle=v_m$ and $ \langle\chi\rangle=v_{\chi}$ by breaking the flavor symmetry, while $\langle \phi_i\rangle(i=1,2)$ get VEV ($v_i$) by breaking EWSB at electro-weak scale ($v_3=0$ due to additional $Z_4$ symmetry). The matrix structures are obtained from equation \eqref{2lag} using the VEV of these flavons (we followed the similar approach as the previous model). We achieve the light neutrino mass matrix as well as the sterile mass using equation \eqref{tamass} and \eqref{tsmass} respectively.

Following the Lagrangian from equation \eqref{lag}, the respective charged lepton, Dirac, Majorana and sterile mass matrix after acquiring the VEVs looks like,
%%%%%%%%%%%%%%%%%%%%%%%%%%%%%%%%%%%%%%%%%%%%%%%%%
%%%%%%%%%%%%%%%%%%%%%%%%%%%%%%%%%%%%%%%%%%%%%%%%%
%%%%%%%%%%%%%%%%%%%%%%%%%%%%%%%%%%%%%%%%%%%%%%%%%
%%%%%%%%%%%%%%%%%%%%%%%%%%%%%%%%%%%%%%%%%%%%%%%%%
\begin{equation}
	M_{l} = \frac{\langle \phi_1\rangle v_m}{\Lambda}\begin{pmatrix}
		y_e&0&0\\
		0&y_{\mu}&0\\
		0&0&y_{\tau}\\
	\end{pmatrix},\
	M^{\prime}_{D}=
	\begin{pmatrix}
		b&b&c\\
		0&b&c\\
		0&b&c\\
	\end{pmatrix},\ M_{R}=\begin{pmatrix}
		d&0&0\\
		0&e&0\\
		0&0&f\\
	\end{pmatrix}, \
	M_{S}= \begin{pmatrix}
		g&0&0\\
	\end{pmatrix}.
	\label{2emd}
\end{equation}
%%%%%%%%%%%%%%%%%%%%%%%%%%%%%%%%%%%%%%%%%%%%%%%%%
%%%%%%%%%%%%%%%%%%%%%%%%%%%%%%%%%%%%%%%%%%%%%%%%%
%%%%%%%%%%%%%%%%%%%%%%%%%%%%%%%%%%%%%%%%%%%%%%%%%
%%%%%%%%%%%%%%%%%%%%%%%%%%%%%%%%%%%%%%%%%%%%%%%%%
Here, $b=\frac{\langle \phi_2\rangle v_m}{\Lambda}y_{2} $ and $c=\frac{\langle \phi_3\rangle v_m}{\Lambda}y_{3}$. Other elements are defined as $d=\lambda_{1}v_m, e=\lambda_{2}v_m$, $f=\lambda_{3}v_m$ and $g=\rho v_{\chi}$. \\
Considering these structures, the light neutrino mass matrix generated from equation~\ref{tamass}, takes a symmetric form similar  to the previous model. Hence, we follow the similar approach like the previous case and add a perturbative matrix $M_P$ to the Dirac mass matrix.
%%%%%%%%%%%%%%%%%%%%%%%%%%%%%%%%%%%%%%%%%%%%%%%%%
%%%%%%%%%%%%%%%%%%%%%%%%%%%%%%%%%%%%%%%%%%%%%%%%%
%%%%%%%%%%%%%%%%%%%%%%%%%%%%%%%%%%%%%%%%%%%%%%%%%
%%%%%%%%%%%%%%%%%%%%%%%%%%%%%%%%%%%%%%%%%%%%%%%%%
New $SU(2)$ singlet flavon fields ($\zeta^{\prime}$ and $\varphi^{\prime}$) are considered which take $A_4\times Z_4$ charges same as $\zeta$ and $\varphi$ respectively. After breaking flavor symmetry, they acquire VEV along $\langle\zeta^{\prime}\rangle=(v_p,0,0)$ and $\langle\varphi^{\prime}\rangle=(0,v_p,0)$ directions. Scale of these VEV ($v_p$) in comparison to earlier flavon's VEV ($v_m$) are differ by an order of magnitude ($v_m>v_p$). The Lagrangian that generates the matrix \eqref{2pmatrix} can be written as,
%%%%%%%%%%%%%%%%%%%%%%%%%%%%%%%%%%%%%%%%%%%%%%%%%
%%%%%%%%%%%%%%%%%%%%%%%%%%%%%%%%%%%%%%%%%%%%%%%%%
%%%%%%%%%%%%%%%%%%%%%%%%%%%%%%%%%%%%%%%%%%%%%%%%%
%%%%%%%%%%%%%%%%%%%%%%%%%%%%%%%%%%%%%%%%%%%%%%%%%
\begin{equation}
	\mathcal{L}_{\mathcal{M_P}} =\frac{y_{1}}{\Lambda}(\overline{l}\tilde{\phi_1}\zeta^{\prime})_{1}\nu_{R1}+\frac{y_{1}}{\Lambda}(\overline{l}\tilde{\phi_1}\varphi^{\prime})_{1^{\prime\prime}}\nu_{R2}+\frac{y_{1}}{\Lambda}(\overline{l}\tilde{\phi_2}\varphi^{\prime})_{1}\nu_{R3}.
\end{equation}
%%%%%%%%%%%%%%%%%%%%%%%%%%%%%%%%%%%%%%%%%%%%%%%%%
%%%%%%%%%%%%%%%%%%%%%%%%%%%%%%%%%%%%%%%%%%%%%%%%%
%%%%%%%%%%%%%%%%%%%%%%%%%%%%%%%%%%%%%%%%%%%%%%%%%
%%%%%%%%%%%%%%%%%%%%%%%%%%%%%%%%%%%%%%%%%%%%%%%%%
Hence, the perturbed matrix looks like,
%%%%%%%%%%%%%%%%%%%%%%%%%%%%%%%%%%%%%%%%%%%%%%%%%
%%%%%%%%%%%%%%%%%%%%%%%%%%%%%%%%%%%%%%%%%%%%%%%%%
%%%%%%%%%%%%%%%%%%%%%%%%%%%%%%%%%%%%%%%%%%%%%%%%%
%%%%%%%%%%%%%%%%%%%%%%%%%%%%%%%%%%%%%%%%%%%%%%%%%
\begin{equation}\label{2pmatrix}
	M_{P}=
	\begin{pmatrix}
		0&0&p\\
		0&p&0\\
		p&0&0\\
	\end{pmatrix}.
\end{equation}
Therefore, $M_D$ from equation \eqref{2emd} will take new structure as,
\begin{equation} \label{2md}
	M_D=M^{\prime}_D+M_P=
	\begin{pmatrix}
		b&b&c+p\\
		0&b+p&c\\
		p&b&c\\
	\end{pmatrix}.
\end{equation}
%%%%%%%%%%%%%%%%%%%%%%%%%%%%%%%%%%%%%%%%%%%%%%%%%
%%%%%%%%%%%%%%%%%%%%%%%%%%%%%%%%%%%%%%%%%%%%%%%%%
%%%%%%%%%%%%%%%%%%%%%%%%%%%%%%%%%%%%%%%%%%%%%%%%%
%%%%%%%%%%%%%%%%%%%%%%%%%%%%%%%%%%%%%%%%%%%%%%%%%
\subsubsection{Inverted hierarchical neutrino mass }
We also modify the Lagrangian for the $M_D$ matrix by introducing a new triplet flavon $\varphi^{\prime\prime}$ with VEV alignment as $\langle\varphi^{\prime\prime}\rangle \sim (2v_m,-v_m,-v_m)$, which affects only the Dirac neutrino mass matrix and gives desirable active-sterile mixing in IH \cite{Zhang:2011vh, Das:2018qyt}. The invariant Yukawa Lagrangian for the $M_D$ matrix in IH mode will be,
%%%%%%%%%%%%%%%%%%%%%%%%%%%%%%%%%%%%%%%%%%%%%%%%%
%%%%%%%%%%%%%%%%%%%%%%%%%%%%%%%%%%%%%%%%%%%%%%%%%
%%%%%%%%%%%%%%%%%%%%%%%%%%%%%%%%%%%%%%%%%%%%%%%%%
%%%%%%%%%%%%%%%%%%%%%%%%%%%%%%%%%%%%%%%%%%%%%%%%%
\begin{equation}
	\mathcal{L}_{\mathcal{M_D}}= \frac{y_{2}}{\Lambda}(\overline{l}\tilde{\phi_1}\zeta)_{1}\nu_{R1}+\frac{y_{2}}{\Lambda}(\overline{l}\tilde{\phi_1}\varphi^{\prime\prime})_{1^{\prime\prime}}\nu_{R2}+\frac{y_{3}}{\Lambda}(\overline{l}\tilde{\phi_2}\varphi)_{1}\nu_{R3}.
\end{equation}
%%%%%%%%%%%%%%%%%%%%%%%%%%%%%%%%%%%%%%%%%%%%%%%%%
%%%%%%%%%%%%%%%%%%%%%%%%%%%%%%%%%%%%%%%%%%%%%%%%%
%%%%%%%%%%%%%%%%%%%%%%%%%%%%%%%%%%%%%%%%%%%%%%%%%
%%%%%%%%%%%%%%%%%%%%%%%%%%%%%%%%%%%%%%%%%%%%%%%%%
The Dirac mass matrix within IH mode (with perturbation matrix $M_P$) takes new structure as, 
%%%%%%%%%%%%%%%%%%%%%%%%%%%%%%%%%%%%%%%%%%%%%%%%%
%%%%%%%%%%%%%%%%%%%%%%%%%%%%%%%%%%%%%%%%%%%%%%%%%
%%%%%%%%%%%%%%%%%%%%%%%%%%%%%%%%%%%%%%%%%%%%%%%%%
%%%%%%%%%%%%%%%%%%%%%%%%%%%%%%%%%%%%%%%%%%%%%%%%%
\begin{equation}
	M_{D}=
	\begin{pmatrix}
		b&-b&c+p\\
		0&-b+p&c\\
		p&2b&c\\
	\end{pmatrix}.
\end{equation}

Complete matrix structures are shown in table \ref{2tab:nh}. In the following subsection, we have presented detailed theoretical as well as experimental bounds on this model.
\subsection{Extended scalars sector}
%\begin{equation}
%\mathcal{L}=\mathcal{L}_{Scalar}+\mathcal{L}_{fermion}.
%\end{equation}
The doublet Higgs scalars in this model are conventionally expressed as\cite{Keus:2013hya},
%%%%%%%%%%%%%%%%%%%%%%%%%%%%%%%%%%%%%%%%%%%%%%%%%
%%%%%%%%%%%%%%%%%%%%%%%%%%%%%%%%%%%%%%%%%%%%%%%%%
%%%%%%%%%%%%%%%%%%%%%%%%%%%%%%%%%%%%%%%%%%%%%%%%%
%%%%%%%%%%%%%%%%%%%%%%%%%%%%%%%%%%%%%%%%%%%%%%%%%
\begin{equation}
	\phi_1=\begin{pmatrix}
		H_1^+\\ \frac{(H_1+iA_1)}{\sqrt{2}}\\
	\end{pmatrix}\quad;\quad   \phi_2=\begin{pmatrix}
		H_2^+\\ \frac{(H_2+iA_2)}{\sqrt{2}}\\
	\end{pmatrix}\quad ; \quad   \phi_3=\begin{pmatrix}
		H_{3}^+\\ \frac{(H_{3}+iA_{3})}{\sqrt{2}}\\
	\end{pmatrix}.
\end{equation}
%%%%%%%%%%%%%%%%%%%%%%%%%%%%%%%%%%%%%%%%%%%%%%%%%
%%%%%%%%%%%%%%%%%%%%%%%%%%%%%%%%%%%%%%%%%%%%%%%%%
%%%%%%%%%%%%%%%%%%%%%%%%%%%%%%%%%%%%%%%%%%%%%%%%%
%%%%%%%%%%%%%%%%%%%%%%%%%%%%%%%%%%%%%%%%%%%%%%%%%
The kinetic part of the scalar is defined within SM paradigm as,
%%%%%%%%%%%%%%%%%%%%%%%%%%%%%%%%%%%%%%%%%%%%%%%%%
%%%%%%%%%%%%%%%%%%%%%%%%%%%%%%%%%%%%%%%%%%%%%%%%%
%%%%%%%%%%%%%%%%%%%%%%%%%%%%%%%%%%%%%%%%%%%%%%%%%
%%%%%%%%%%%%%%%%%%%%%%%%%%%%%%%%%%%%%%%%%%%%%%%%%
\begin{equation}
	\mathcal{L}_{KE}=\sum_{i=1}^3(D_{\mu}\phi_{i})^{\dagger}(D_{\mu}\phi_{i}),
\end{equation}
%%%%%%%%%%%%%%%%%%%%%%%%%%%%%%%%%%%%%%%%%%%%%%%%%
%%%%%%%%%%%%%%%%%%%%%%%%%%%%%%%%%%%%%%%%%%%%%%%%%
%%%%%%%%%%%%%%%%%%%%%%%%%%%%%%%%%%%%%%%%%%%%%%%%%
%%%%%%%%%%%%%%%%%%%%%%%%%%%%%%%%%%%%%%%%%%%%%%%%%
where, $D_{\mu}$ stands for the covariant derivative. The scalar potential of the Lagrangian is written in two separate parts. Among the three Higgs doublets, one of them does not acquire any VEV, so it behaves as inert while the other two are SM type Higgs doublet and acquire VEV by EWSB. The scalar potential of the Lagrangian is defined as\cite{Deshpande:1977rw},
%%%%%%%%%%%%%%%%%%%%%%%%%%%%%%%%%%%%%%%%%%%%%%%%%
%%%%%%%%%%%%%%%%%%%%%%%%%%%%%%%%%%%%%%%%%%%%%%%%%
%%%%%%%%%%%%%%%%%%%%%%%%%%%%%%%%%%%%%%%%%%%%%%%%%
%%%%%%%%%%%%%%%%%%%%%%%%%%%%%%%%%%%%%%%%%%%%%%%%%
\begin{equation}
	\mathcal{V}_{\phi_1,\phi_2,\phi_3}=\big(V_{\phi_1+\phi_2}+V_{\phi_3}\big),
\end{equation}
%%%%%%%%%%%%%%%%%%%%%%%%%%%%%%%%%%%%%%%%%%%%%%%%%
%%%%%%%%%%%%%%%%%%%%%%%%%%%%%%%%%%%%%%%%%%%%%%%%%
%%%%%%%%%%%%%%%%%%%%%%%%%%%%%%%%%%%%%%%%%%%%%%%%%
%%%%%%%%%%%%%%%%%%%%%%%%%%%%%%%%%%%%%%%%%%%%%%%%%
where,
%%%%%%%%%%%%%%%%%%%%%%%%%%%%%%%%%%%%%%%%%%%%%%%%%
%%%%%%%%%%%%%%%%%%%%%%%%%%%%%%%%%%%%%%%%%%%%%%%%%
%%%%%%%%%%%%%%%%%%%%%%%%%%%%%%%%%%%%%%%%%%%%%%%%%
%%%%%%%%%%%%%%%%%%%%%%%%%%%%%%%%%%%%%%%%%%%%%%%%%
\allowdisplaybreaks
\begin{equation}
	\begin{split}\label{12pot}
		V_{\phi_1+\phi_2}&=\mu^2_{11}\phi_1^{\dagger}\phi_1+\mu^2_{22}\phi_2^{\dagger}\phi_2-\frac{\mu^2_{12}}{2}(\phi_1^{\dagger}\phi_2+\phi_2^{\dagger}\phi_1)\\
		&+\kappa_1(\phi_1^{\dagger}\phi_1)^2+\kappa_2(\phi_2^{\dagger}\phi_2)^2+\kappa_3(\phi_1^{\dagger}\phi_1)(\phi_2^{\dagger}\phi_2)\\
		&+\kappa_4(\phi_2^{\dagger}\phi_1)(\phi_1^{\dagger}\phi_2)+\frac{\kappa_5}{2}((\phi_1^{\dagger}\phi_2)^2+(\phi_2^{\dagger}\phi_1)^2),
	\end{split}
\end{equation}
%%%%%%%%%%%%%%%%%%%%%%%%%%%%%%%%%%%%%%%%%%%%%%%%%
%%%%%%%%%%%%%%%%%%%%%%%%%%%%%%%%%%%%%%%%%%%%%%%%%
%%%%%%%%%%%%%%%%%%%%%%%%%%%%%%%%%%%%%%%%%%%%%%%%%
%%%%%%%%%%%%%%%%%%%%%%%%%%%%%%%%%%%%%%%%%%%%%%%%%
while potential for the inert Higgs is given as,
%%%%%%%%%%%%%%%%%%%%%%%%%%%%%%%%%%%%%%%%%%%%%%%%%
%%%%%%%%%%%%%%%%%%%%%%%%%%%%%%%%%%%%%%%%%%%%%%%%%
%%%%%%%%%%%%%%%%%%%%%%%%%%%%%%%%%%%%%%%%%%%%%%%%%
%%%%%%%%%%%%%%%%%%%%%%%%%%%%%%%%%%%%%%%%%%%%%%%%%
\allowdisplaybreaks
\begin{equation}
	\begin{split}
		V_{\phi_3}&=\mu^2_{33}\phi_3^{\dagger}\phi_3+\kappa_2^{DM}(\phi_3^{\dagger}\phi_3)^2+\kappa_3^{DM}((\phi_1^{\dagger}\phi_1)(\phi_3^{\dagger}\phi_3)+(\phi_2^{\dagger}\phi_2)(\phi_3^{\dagger}\phi_3))\nn\\
		&+\kappa_4^{DM}((\phi_1^{\dagger}\phi_3)^{\dagger}(\phi_1^{\dagger}\phi_3)+(\phi_2^{\dagger}\phi_3)^{\dagger}(\phi_2^{\dagger}\phi_3))+\kappa_5^{DM}((\phi_1^{\dagger}\phi_3)^2+(\phi_2^{\dagger}\phi_3)^2+h.c.).
	\end{split}
\end{equation}
%%%%%%%%%%%%%%%%%%%%%%%%%%%%%%%%%%%%%%%%%%%%%%%%%
%%%%%%%%%%%%%%%%%%%%%%%%%%%%%%%%%%%%%%%%%%%%%%%%%
%%%%%%%%%%%%%%%%%%%%%%%%%%%%%%%%%%%%%%%%%%%%%%%%%
%%%%%%%%%%%%%%%%%%%%%%%%%%%%%%%%%%%%%%%%%%%%%%%%%
In both the potentials, $\mu_{ij} (i=1,2)$, $\mu_{33}$ are the mass terms and $\kappa's$ are the scalar quartic couplings, responsible for mixing and masses of the physical scalar fields. The neutral CP-even fields of $\phi_1$ and $\phi_2$ get VEVs after electroweak symmetry breaking (EWSB), $i.e.$, $H_1=h_1+v_1$ and $H_2=h_2+v_2$.
The minimization conditions for the potential are,
%%%%%%%%%%%%%%%%%%%%%%%%%%%%%%%%%%%%%%%%%%%%%%%%%
%%%%%%%%%%%%%%%%%%%%%%%%%%%%%%%%%%%%%%%%%%%%%%%%%
%%%%%%%%%%%%%%%%%%%%%%%%%%%%%%%%%%%%%%%%%%%%%%%%%
%%%%%%%%%%%%%%%%%%%%%%%%%%%%%%%%%%%%%%%%%%%%%%%%%
\begin{eqnarray}
	\mu^2_{11}&=\mu_{12}^2\tan\beta-\frac{1}{2}v^2\big(2\kappa_1\cos^2\beta+\kappa_L\sin^2\beta\big),\\
	\mu^2_{22}&=\mu_{12}^2\cot\beta-\frac{1}{2}v^2\big(2\kappa_2\sin^2\beta+\kappa_L\cos^2\beta\big).\nn
\end{eqnarray}
%%%%%%%%%%%%%%%%%%%%%%%%%%%%%%%%%%%%%%%%%%%%%%%%%
%%%%%%%%%%%%%%%%%%%%%%%%%%%%%%%%%%%%%%%%%%%%%%%%%
%%%%%%%%%%%%%%%%%%%%%%%%%%%%%%%%%%%%%%%%%%%%%%%%%
%%%%%%%%%%%%%%%%%%%%%%%%%%%%%%%%%%%%%%%%%%%%%%%%%
Here, $\kappa_L=(\kappa_1+\kappa_2+\kappa_3)$, $v=\sqrt{v_1^2+v_2^2}$ and $\beta=\tan^{-1}\big(\frac{v_2}{v_1}\big)$. It is to be noted that there is no minimum along with the directions of the scalar fields in $\phi_3$ doublet due to an additional $Z_4$ symmetry.

A $(12\times12)$ mass matrix is obtained after EWSB, which is composed of four $(3\times3)$ sub-matrices with bases $(H_1^+,H_2^+,H_3^+)$ , $(H_1^-,H_2^-,H_3^-)$ , $(h_1,h_2,H_3)$ and $(A_1,A_2,A_3)$. The inert fields in these mass matrices remain decoupled as they do not get any VEV. The other fields give rise to five physical mass eigenstates $(H^{\pm},h,H,A)$ after rotation with the mass basis. Three other mass-less Goldstone bosons $(G^{\pm},G^0)$ are also generated, which are eaten up by the $W^{\pm}$ and $Z$ bosons to give mass them mass. The mass eigenstates for the physical scalars within 2HD (first two Higgs doublets) scalar sector are given by \cite{Branco:2011iw},
%%%%%%%%%%%%%%%%%%%%%%%%%%%%%%%%%%%%%%%%%%%%%%%%%
%%%%%%%%%%%%%%%%%%%%%%%%%%%%%%%%%%%%%%%%%%%%%%%%%
%%%%%%%%%%%%%%%%%%%%%%%%%%%%%%%%%%%%%%%%%%%%%%%%%
%%%%%%%%%%%%%%%%%%%%%%%%%%%%%%%%%%%%%%%%%%%%%%%%%
\allowdisplaybreaks
\begin{eqnarray}
	M_{A}^2&=& \frac{\mu_{12}^2}{2\cos\beta \sin\beta}-\kappa_5v^2,\\
	M_{H^{\pm}}^2&=&  M_A^2+\frac{1}{2}v^2(\kappa_5-\kappa_4),\\
	M_h^2&=&  \frac{1}{4}v^2\sec(\alpha+\beta)[(6\kappa_1+\kappa_L)\cos\alpha\cos\beta\nn\\
	&& +2(\kappa_1-\kappa_L)\cos\alpha\cos3\beta-\sin\alpha\sin\beta\, \{(6\kappa_2+\kappa_L)-(2\kappa_2+\kappa_L)\}],\\
	M_H^2&=&  \frac{1}{4}v^2\mathrm{cosec}(\alpha+\beta) \{2\cos\beta(2\kappa_1+\kappa_L+(2\kappa_1-\kappa_L)\cos2\beta\}\sin\beta\nn\\
	&&+\cos\alpha \, \{(6\kappa_2+\kappa_L)\sin\beta+(-2\kappa_2+\kappa_L)\sin3\beta\},\\
	\text{where,}\nn\\
	\nn	\mu_{12}^2&=& \frac{1}{2}v^2[\kappa_1+\kappa_2+\kappa_L+\mathrm{cosec}(2\alpha+2\beta)\{2(\kappa_1-\kappa_2)\sin2\alpha\\&&+(\kappa_1+\kappa_2-\kappa_L)\sin(2\alpha-2\beta)\}]\sin2\beta.\nn
\end{eqnarray}
%%%%%%%%%%%%%%%%%%%%%%%%%%%%%%%%%%%%%%%%%%%%%%%%%
%%%%%%%%%%%%%%%%%%%%%%%%%%%%%%%%%%%%%%%%%%%%%%%%%
%%%%%%%%%%%%%%%%%%%%%%%%%%%%%%%%%%%%%%%%%%%%%%%%%
%%%%%%%%%%%%%%%%%%%%%%%%%%%%%%%%%%%%%%%%%%%%%%%%%
Inert scalar sector remain decoupled from the other two scalar sector. After EWSB four physical mass eigenstates ($H_3,A_3,H_3^{\pm}$) can be written as,
%%%%%%%%%%%%%%%%%%%%%%%%%%%%%%%%%%%%%%%%%%%%%%%%%
%%%%%%%%%%%%%%%%%%%%%%%%%%%%%%%%%%%%%%%%%%%%%%%%%
%%%%%%%%%%%%%%%%%%%%%%%%%%%%%%%%%%%%%%%%%%%%%%%%%
%%%%%%%%%%%%%%%%%%%%%%%%%%%%%%%%%%%%%%%%%%%%%%%%%
\begin{equation}
	\begin{split}
		M_{H_3}^2=&\mu_{33}^2+\frac{1}{2}v^2\kappa_L^{DM},\\
		M_{A_3}^2=&\mu_{33}^2+\frac{1}{2}v^2\kappa_S^{DM},\\
		M_{H_3^{\pm}}^2=&\mu_{33}^2+\frac{1}{2}v^2\kappa_3^{DM},\\
	\end{split}
\end{equation}
%%%%%%%%%%%%%%%%%%%%%%%%%%%%%%%%%%%%%%%%%%%%%%%%%
%%%%%%%%%%%%%%%%%%%%%%%%%%%%%%%%%%%%%%%%%%%%%%%%%
%%%%%%%%%%%%%%%%%%%%%%%%%%%%%%%%%%%%%%%%%%%%%%%%%
%%%%%%%%%%%%%%%%%%%%%%%%%%%%%%%%%%%%%%%%%%%%%%%%%
where, $\kappa_{L,S}^{DM}=\kappa_3^{DM}+\kappa_4^{DM}\pm \kappa_5^{DM}$.
It is to be noted that the detailed study of the scalar potential and the interaction among the heavy scalar fields ($\varphi,\zeta,\xi,etc.$) are worked out in \cite{Das:2018qyt}, and the light scalar fields remains decoupled from these heavy scalar fields. 
These heavy scalar fields need to explain neutrino mass and mixing which we are going to discuss in the next subsection.
%%%%%%%%%%%%%%%%%%%%%%%%%%%%%%%%%%%%%%%%%%%%%%%%%
%%%%%%%%%%%%%%%%%%%%%%%%%%%%%%%%%%%%%%%%%%%%%%%%%
%%%%%%%%%%%%%%%%%%%%%%%%%%%%%%%%%%%%%%%%%%%%%%%%%
%%%%%%%%%%%%%%%%%%%%%%%%%%%%%%%%%%%%%%%%%%%%%%%%% 
%%%%%%%%%%%%%%%%%%%%%%%%%%%%%%%%%%%%%%%%%%%%%%%%%
%%%%%%%%%%%%%%%%%%%%%%%%%%%%%%%%%%%%%%%%%%%%%%%%%
%%%%%%%%%%%%%%%%%%%%%%%%%%%%%%%%%%%%%%%%%%%%%%%%%

%%%%%%%%%%%%%%%%%%%%%%%%%%%%%%%%%%%%%%%%%%%%%%%%%

\section{Methodology}\label{4s2}
\subsection{Neutrino mass and mixing angles}
Minimal extended seesaw is realized in this chapter to construct active and sterile masses. Recalling the model from \ref{model2} section, the mass matrices obtained for active neutrinos and active-sterile mixing elements are shown in table \ref{2tab:nh}. We have used similar numerical techniques like the previous chapter to solve these model parameters. It is to be noted that, we have used the light neutrino parameters satisfying the bounds from $\mu\rightarrow e\gamma$ \cite{Dohmen:1993mp}.
%%%%%%%%%%%%%%%%%%%%%%%%%%%%%%%%%%%%%%%%%%%%%%%%%
%%%%%%%%%%%%%%%%%%%%%%%%%%%%%%%%%%%%%%%%%%%%%%%%%
%%%%%%%%%%%%%%%%%%%%%%%%%%%%%%%%%%%%%%%%%%%%%%%%%
%%%%%%%%%%%%%%%%%%%%%%%%%%%%%%%%%%%%%%%%%%%%%%%%%
\begin{table}[ht] 
	%	\centering
	%	\begin{center}
	\resizebox{0.99\hsize}{!}{
		\begin{tabular}{|c|c|c|c|c|}
			\hline
			Ordering & Structures & $-m_{\nu}$&$m_S$& $R$ \\
			\hline
			\hline
			NH & 
			$\begin{aligned}
				&
				M_R=\begin{pmatrix}
					d&0&0\\
					0&e&0\\
					0&0&f\\
				\end{pmatrix}\\
				& M_{D}= \begin{pmatrix}
					b&b&c+p\\
					0&b+p&c\\
					p&b&c\\
				\end{pmatrix}\\
				& M_{S}= \begin{pmatrix}
					g&0&0\\
				\end{pmatrix}\\
			\end{aligned}$
			& $\begin{pmatrix}
				\frac {b^2} {e} + \frac {(c + p)^2} {f} &\frac {b (b + 
					p)} {e} + \frac {c (c + 
					p)} {f} &\frac {b^2} {e} + \frac {c (c + p)} {f} \\
				\frac {b (b + p)} {e} + \frac {c (c + p)} {f} &\frac {(b + 
					p)^2} {e} + \frac {c^2} {f} &\frac {b (b + 
					p)} {e} + \frac {c^2} {f} \\
				\frac {b^2} {e} + \frac {c (c + p)} {f} &\frac {b (b + 
					p)} {e} + \frac {c^2} {f} &\frac {b^2} {e} + \frac {c^2}
				{f} \\
			\end{pmatrix}$&$\simeq \frac{g^2}{10^4}$ & $\simeq{\begin{pmatrix}
					\frac{b}{g}\\ 0\\ \frac{p}{g}\\
			\end{pmatrix}}$\\
			\hline
			IH& 
			$\begin{aligned}
				&
				M_R=\begin{pmatrix}
					d&0&0\\
					0&e&0\\
					0&0&f\\
				\end{pmatrix}\\
				& M_{D}= \begin{pmatrix}
					b&-b&c+p\\
					0&-b+p&c\\
					p&2b&c\\
				\end{pmatrix}\\
				& M_{S}= \begin{pmatrix}
					g&0&0\\
				\end{pmatrix}\\
			\end{aligned}$
			& $\begin{pmatrix}
				\frac {b^2} {e} + \frac {(c + p)^2} {f} &  \frac {b(b - 
					p)} {e} + \frac {c (c + 
					p)} {f} & \frac {-2 b^2} {e} + \frac {c (c + p)} {f} \\
				\frac {b(b - p)} {e} + \frac {c (c + p)} {f} &\frac {(b - 
					p)^2} {e} + \frac {c^2} {f} &\frac {-2 b (b - 
					p)} {e} + \frac {c^2} {f} \\
				-\frac {2 b^2} {e} + \frac {c (c + p)} {f} & - \frac {2 b (b - 
					p)} {e} + \frac {c^2} {f} &\frac {4 b^2} {e} + \frac {c^2}
				{f} \\
			\end{pmatrix}$& $\simeq \frac{g^2}{10^4}$&$\simeq{\begin{pmatrix}
					\frac{b}{g}\\ 0\\ \frac{p}{g}\\
			\end{pmatrix}}$\\
			\hline
	\end{tabular}}
	%	\end{center}
	\caption{The light neutrino mass matrices ($m_{\nu}$), sterile mass ($m_S$) and active-sterile mixing patterns ($R$) with corresponding $M_D$, $M_R$ and $M_S$ matrices for NH and IH mass pattern.  }\label{2tab:nh}
\end{table}
%%%%%%%%%%%%%%%%%%%%%%%%%%%%%%%%%%%%%%%%%%%%%%%%%
%%%%%%%%%%%%%%%%%%%%%%%%%%%%%%%%%%%%%%%%%%%%%%%%%
%%%%%%%%%%%%%%%%%%%%%%%%%%%%%%%%%%%%%%%%%%%%%%%%%
%%%%%%%%%%%%%%%%%%%%%%%%%%%%%%%%%%%%%%%%%%%%%%%%%
\subsection{Baryogenesis $via$ thermal leptogenesis}\label{4bary}
We have considered a hierarchical mass pattern for RH neutrinos, among which the lightest will decay to a Higgs and lepton. This decay would produce sufficient lepton asymmetry to give rise to the observed baryon asymmetry of the Universe. Both baryon number ($B$) and lepton number ($L$) are conserved independently in the SM renormalizable Lagrangian. However, due to chiral anomaly, there are non-perturbative gauge field configurations \cite{Callan:1976je}, which produces the anomalous $B+L$ violation ($B-L$ is already conserved). These whole process of conversion of lepton asymmetry to baryon asymmetry $via$ $B+L$ violation is popularly termed as ``sphalerons" \cite{Klinkhamer:1984di}.
We have used the parametrization from \cite{Davidson:2008bu}, where, the working formula of baryon asymmetry produced is given by,
%%%%%%%%%%%%%%%%%%%%%%%%%%%%%%%%%%%%%%%%%%%%%%%%%
%%%%%%%%%%%%%%%%%%%%%%%%%%%%%%%%%%%%%%%%%%%%%%%%%
%%%%%%%%%%%%%%%%%%%%%%%%%%%%%%%%%%%%%%%%%%%%%%%%%
%%%%%%%%%%%%%%%%%%%%%%%%%%%%%%%%%%%%%%%%%%%%%%%%%
\begin{equation}\label{yb}
	Y_B=ck\frac{\epsilon_{11}}{g_{*}}.
\end{equation}
%%%%%%%%%%%%%%%%%%%%%%%%%%%%%%%%%%%%%%%%%%%%%%%%%
%%%%%%%%%%%%%%%%%%%%%%%%%%%%%%%%%%%%%%%%%%%%%%%%%
%%%%%%%%%%%%%%%%%%%%%%%%%%%%%%%%%%%%%%%%%%%%%%%%%
%%%%%%%%%%%%%%%%%%%%%%%%%%%%%%%%%%%%%%%%%%%%%%%%%
The quantities involved in this equation \ref{yb} can be explained as follows,
%%%%%%%%%%%%%%%%%%%%%%%%%%%%%%%%%%%%%%%%%%%%%%%%%
%%%%%%%%%%%%%%%%%%%%%%%%%%%%%%%%%%%%%%%%%%%%%%%%%
%%%%%%%%%%%%%%%%%%%%%%%%%%%%%%%%%%%%%%%%%%%%%%%%%
%%%%%%%%%%%%%%%%%%%%%%%%%%%%%%%%%%%%%%%%%%%%%%%%%
\begin{itemize}
	\item $c$ is the factor that measures the fraction of lepton asymmetry that being converted to baryon asymmetry. This value is approximately $12/37$.
	\item $k$ is the dilution factor produced due to wash out processes, which can be parametrized as,
	%%%%%%%%%%%%%%%%%%%%%%%%%%%%%%%%%%%%%%%%%%%%%%%%%
	%%%%%%%%%%%%%%%%%%%%%%%%%%%%%%%%%%%%%%%%%%%%%%%%%
	%%%%%%%%%%%%%%%%%%%%%%%%%%%%%%%%%%%%%%%%%%%%%%%%%
	%%%%%%%%%%%%%%%%%%%%%%%%%%%%%%%%%%%%%%%%%%%%%%%%%
	\begin{equation}\label{sk}
		\begin{split}
			k &\simeq \sqrt{0.1K}exp\Big[\frac{-4}{3(0.1K)^{0.25}}\Big], \quad \text{for} \quad K\geq10^6,\\
			&\simeq\frac{0.3}{K(lnK)^{0.6}}, \quad \text{for}\quad 10\leq K\leq 10^6,\\
			&\simeq\frac{1}{2\sqrt{K^2+9}}, \quad\text{for} \quad0\leq K \leq 10.\\
		\end{split}
	\end{equation}
	%%%%%%%%%%%%%%%%%%%%%%%%%%%%%%%%%%%%%%%%%%%%%%%%%
	%%%%%%%%%%%%%%%%%%%%%%%%%%%%%%%%%%%%%%%%%%%%%%%%%
	%%%%%%%%%%%%%%%%%%%%%%%%%%%%%%%%%%%%%%%%%%%%%%%%%
	%%%%%%%%%%%%%%%%%%%%%%%%%%%%%%%%%%%%%%%%%%%%%%%%%
	Here, $K$ is defined as,
	%%%%%%%%%%%%%%%%%%%%%%%%%%%%%%%%%%%%%%%%%%%%%%%%%
	%%%%%%%%%%%%%%%%%%%%%%%%%%%%%%%%%%%%%%%%%%%%%%%%%
	%%%%%%%%%%%%%%%%%%%%%%%%%%%%%%%%%%%%%%%%%%%%%%%%%
	%%%%%%%%%%%%%%%%%%%%%%%%%%%%%%%%%%%%%%%%%%%%%%%%%
	\begin{equation}
		K=\frac{\Gamma_1}{H(T=M_{\nu_{R1}})}=\frac{(\lambda^{\dagger}\lambda)_{11}M_{\nu_{R1}}}{8\pi}\frac{M_{Planck}}{1.66\sqrt{g_{*}}M^2_{\nu_{R1}}},
	\end{equation}
	%%%%%%%%%%%%%%%%%%%%%%%%%%%%%%%%%%%%%%%%%%%%%%%%%
	%%%%%%%%%%%%%%%%%%%%%%%%%%%%%%%%%%%%%%%%%%%%%%%%%
	%%%%%%%%%%%%%%%%%%%%%%%%%%%%%%%%%%%%%%%%%%%%%%%%%
	%%%%%%%%%%%%%%%%%%%%%%%%%%%%%%%%%%%%%%%%%%%%%%%%%
	here, $\Gamma_1$ is the decay width of $\nu_{R1}$, defined as,  $\Gamma_1=\frac{(\lambda^{\dagger}\lambda)_{11}M_{\nu_{R1}}}{8\pi} $ and the Hubble constant at $T=M_{\nu_{R1}}$ is defined as $H(T=M_{\nu_{R1}})=\frac{M_{Planck}}{1.66\sqrt{g_{*}}M^2_{\nu_{R1}}}$. 
	\item $g_{*}$ is the massless relativistic degree of freedom in the thermal bath and within SM, it is approximately $110$.
	\item $\epsilon_{11}$ is the lepton asymmetry produced by the decay of the lightest RH neutrino $\nu_{R1}$. This can be formulated as below.\\
	To produce non-vanishing lepton asymmetry, the decay of $\nu_{R1}$ must have lepton number violating process with different decay rates to a final state with particle and anti-particle. Asymmetry in lepton flavor $\alpha$ produced in the decay of $\nu_{R1}$ is defined as,
	%%%%%%%%%%%%%%%%%%%%%%%%%%%%%%%%%%%%%%%%%%%%%%%%%
	%%%%%%%%%%%%%%%%%%%%%%%%%%%%%%%%%%%%%%%%%%%%%%%%%
	%%%%%%%%%%%%%%%%%%%%%%%%%%%%%%%%%%%%%%%%%%%%%%%%%
	%%%%%%%%%%%%%%%%%%%%%%%%%%%%%%%%%%%%%%%%%%%%%%%%%
	\begin{equation}
		\epsilon_{\alpha\alpha}=\frac{\Gamma(\nu_{R1} \rightarrow l_{\alpha}\phi_i)-\Gamma(\nu_{R1} \rightarrow \overline{l}_{\alpha}\overline{\phi_i})}{\Gamma(\nu_{R1} \rightarrow l_{\alpha} \phi_i)+\Gamma(\nu_{R1} \rightarrow \overline{l}_{\alpha}\overline{\phi_i})},
	\end{equation}
	%%%%%%%%%%%%%%%%%%%%%%%%%%%%%%%%%%%%%%%%%%%%%%%%%
	%%%%%%%%%%%%%%%%%%%%%%%%%%%%%%%%%%%%%%%%%%%%%%%%%
	%%%%%%%%%%%%%%%%%%%%%%%%%%%%%%%%%%%%%%%%%%%%%%%%%
	%%%%%%%%%%%%%%%%%%%%%%%%%%%%%%%%%%%%%%%%%%%%%%%%%
	where, $\overline{l_{\alpha}}$ is the antiparticle of $l_{\alpha}$ and $\phi_i$ is the lightest Higgs doublet present in our model. Following the calculation for non-degenerate RH mass\footnote{For degenerate mass with mass spiting equal to decay width, one has to consider resonant leptogenesis.}, from the work of \cite{Davidson:2008bu}, we obtain the asymmetry term as,
	%%%%%%%%%%%%%%%%%%%%%%%%%%%%%%%%%%%%%%%%%%%%%%%%%
	%%%%%%%%%%%%%%%%%%%%%%%%%%%%%%%%%%%%%%%%%%%%%%%%%
	%%%%%%%%%%%%%%%%%%%%%%%%%%%%%%%%%%%%%%%%%%%%%%%%%
	%%%%%%%%%%%%%%%%%%%%%%%%%%%%%%%%%%%%%%%%%%%%%%%%%
	\begin{equation}
		\begin{split}
			\epsilon_{\alpha\alpha}= & \frac{1}{8\pi}\frac{1}{[\lambda^{\dagger}\lambda]_{11}}\sum_{j}^{2,3}\text{Im} {(\lambda_{\alpha 1}^{*})(\lambda^{\dagger}\lambda)_{1j}\lambda_{\alpha j}} g(x_{j})\\
			& + \frac{1}{8\pi}\frac{1}{[\lambda^{\dagger}\lambda]_{11}}\sum_{j}^{2,3}\text{Im} {(\lambda_{\alpha 1}^{*})(\lambda^{\dagger}\lambda)_{1j}\lambda_{\alpha j}} \frac{1}{1-x_j}.
		\end{split}\label{epsi}
	\end{equation}
	%%%%%%%%%%%%%%%%%%%%%%%%%%%%%%%%%%%%%%%%%%%%%%%%%
	%%%%%%%%%%%%%%%%%%%%%%%%%%%%%%%%%%%%%%%%%%%%%%%%%
	%%%%%%%%%%%%%%%%%%%%%%%%%%%%%%%%%%%%%%%%%%%%%%%%%
	%%%%%%%%%%%%%%%%%%%%%%%%%%%%%%%%%%%%%%%%%%%%%%%%% 
	Here, $x_j\equiv \frac{M_j^2}{M_1^2}$ and within the SM $g(x_j)$ is defined as,
	%%%%%%%%%%%%%%%%%%%%%%%%%%%%%%%%%%%%%%%%%%%%%%%%%
	%%%%%%%%%%%%%%%%%%%%%%%%%%%%%%%%%%%%%%%%%%%%%%%%%
	%%%%%%%%%%%%%%%%%%%%%%%%%%%%%%%%%%%%%%%%%%%%%%%%%
	%%%%%%%%%%%%%%%%%%%%%%%%%%%%%%%%%%%%%%%%%%%%%%%%%
	\begin{equation}
		g(x_j)=\sqrt{x_j}\Big(\frac{2-x_j-(1-x_j^2)\text{ln}(1+x_j/x_j)}{1-x_j}\Big).
	\end{equation}
	%%%%%%%%%%%%%%%%%%%%%%%%%%%%%%%%%%%%%%%%%%%%%%%%%
	%%%%%%%%%%%%%%%%%%%%%%%%%%%%%%%%%%%%%%%%%%%%%%%%%
	%%%%%%%%%%%%%%%%%%%%%%%%%%%%%%%%%%%%%%%%%%%%%%%%%
	%%%%%%%%%%%%%%%%%%%%%%%%%%%%%%%%%%%%%%%%%%%%%%%%%
	The second line from equation \eqref{epsi} violates the single lepton flavors, however, it conserves the total lepton number, thus it vanishes when we take the sum over $\alpha$,
	%%%%%%%%%%%%%%%%%%%%%%%%%%%%%%%%%%%%%%%%%%%%%%%%%
	%%%%%%%%%%%%%%%%%%%%%%%%%%%%%%%%%%%%%%%%%%%%%%%%%
	%%%%%%%%%%%%%%%%%%%%%%%%%%%%%%%%%%%%%%%%%%%%%%%%%
	%%%%%%%%%%%%%%%%%%%%%%%%%%%%%%%%%%%%%%%%%%%%%%%%%
	\begin{equation}\label{ep}
		\epsilon_{11}\equiv\sum_{\alpha}\epsilon_{\alpha\alpha}=\frac{1}{8\pi}\frac{1}{[\lambda^{\dagger}\lambda]_{11}}\sum_{j}^{2,3}\text{Im}{[(\lambda^{\dagger}\lambda)_{1j}]^2} g(x_{j}).
	\end{equation}
	%%%%%%%%%%%%%%%%%%%%%%%%%%%%%%%%%%%%%%%%%%%%%%%%%
	%%%%%%%%%%%%%%%%%%%%%%%%%%%%%%%%%%%%%%%%%%%%%%%%%
	%%%%%%%%%%%%%%%%%%%%%%%%%%%%%%%%%%%%%%%%%%%%%%%%%
	%%%%%%%%%%%%%%%%%%%%%%%%%%%%%%%%%%%%%%%%%%%%%%%%%
	The $\lambda$ used here is the Yukawa matrix generated from the Dirac mass matrix and the corresponding index in the suffix says the position of the matrix element.
\end{itemize} 
%%%%%%%%%%%%%%%%%%%%%%%%%%%%%%%%%%%%%%%%%%%%%%%%%
%%%%%%%%%%%%%%%%%%%%%%%%%%%%%%%%%%%%%%%%%%%%%%%%%
%%%%%%%%%%%%%%%%%%%%%%%%%%%%%%%%%%%%%%%%%%%%%%%%%
%%%%%%%%%%%%%%%%%%%%%%%%%%%%%%%%%%%%%%%%%%%%%%%%%
Now, baryon asymmetry of the Universe can be calculated from equation \eqref{yb} followed by the evaluation of lepton asymmetry using equation \eqref{ep}. The Yukawa matrix is constructed from the solved model parameters $b,c$ and $p$, which is analogous to the $(3\times3)$ Dirac mass matrix. Within our study the $K$ value lies within the range $10\leq K\leq 10^6$, hence, we have used the second parametrization of the dilution factor from equation \eqref{sk}.
%%%%%%%%%%%%%%%%%%%%%%%%%%%%%%%%%%%%%%%%%%%%%%%%%
%%%%%%%%%%%%%%%%%%%%%%%%%%%%%%%%%%%%%%%%%%%%%%%%%
\subsection{Bounds on this models}\label{4s3}
\subsubsection{Stability of the scalar potential}
The scalar potential should be bounded from the below in such a way that, even for large field values there is no other negative infinity that arises along any field space direction. The absolute stability condition for the potential [\ref{12pot}] are evaluated in terms of the quadratic coupling are as follows~\cite{Deshpande:1977rw},
%%%%%%%%%%%%%%%%%%%%%%%%%%%%%%%%%%%%%%%%%%%%%%%%%
%%%%%%%%%%%%%%%%%%%%%%%%%%%%%%%%%%%%%%%%%%%%%%%%%
%%%%%%%%%%%%%%%%%%%%%%%%%%%%%%%%%%%%%%%%%%%%%%%%%
%%%%%%%%%%%%%%%%%%%%%%%%%%%%%%%%%%%%%%%%%%%%%%%%%
\bea
\kappa_1>0, ~ \kappa_2>0,~ \kappa_2^{DM}>0, \kappa_{3,L,S}+2\sqrt{2\kappa_1\kappa_2}>0,&\nn\\
\kappa_{3,L,S}^{DM}+2\sqrt{\kappa_1\kappa_2^{DM}}>0,  \kappa_{3,L,S}^{DM}+2\sqrt{\kappa_2\kappa_1^{DM}}>0.\nn
\eea
%%%%%%%%%%%%%%%%%%%%%%%%%%%%%%%%%%%%%%%%%%%%%%%%%
%%%%%%%%%%%%%%%%%%%%%%%%%%%%%%%%%%%%%%%%%%%%%%%%%
%%%%%%%%%%%%%%%%%%%%%%%%%%%%%%%%%%%%%%%%%%%%%%%%%
%%%%%%%%%%%%%%%%%%%%%%%%%%%%%%%%%%%%%%%%%%%%%%%%%
\subsubsection{Unitarity bounds}
Unitarity bounds on the couplings are evaluated considering scalar-scalar, gauge boson-gauge boson, and
scalar-gauge boson scatterings \cite{Lee:1977eg}. In general, unitarity bounds are the couplings of the physical bases of the scalar potential. However, the couplings for the scalars are quite complicated, so we consider the couplings of the non-physical bases before EWSB. Then the S-matrix, which is expressed in terms of the non-physical fields, is transformed into an S-matrix for the physical
fields by making a unitary transformation \cite{Das:2014fea, Kanemura:1993hm}. The unitarity of the S-matrix demands the absolute eigenvalues of the scattering matrix should be less than $8\pi$ up to a particular scale. In our potential, bounds that come from the eigenvalues of the corresponding S-matrix are as follows,

%%%%%%%%%%%%%%%%%%%%%%%%%%%%%%%%%%%%%%%%%%%%%%%%%
%%%%%%%%%%%%%%%%%%%%%%%%%%%%%%%%%%%%%%%%%%%%%%%%%
%%%%%%%%%%%%%%%%%%%%%%%%%%%%%%%%%%%%%%%%%%%%%%%%%
%%%%%%%%%%%%%%%%%%%%%%%%%%%%%%%%%%%%%%%%%%%%%%%%%
\begin{minipage}[c]{0.50\textwidth}
	\begin{equation}
		\resizebox{0.98\hsize}{!}{ $
			\begin{split}
				& |\kappa_3\pm \kappa_4 |\leq8\pi,\\
				& |\kappa_3\pm \kappa_5 |\leq 8\pi,\\
				&|\kappa_3+2\kappa_4\pm 3\kappa_5|\leq 8\pi,\\
				&\Big|\kappa_1+\kappa_2\pm \sqrt{(\kappa_1-\kappa_2)^2+\kappa_4} \Big|\leq 8\pi,\\
				&\Big|3\kappa_1+3\kappa_2\pm \sqrt{9(\kappa_1-\kappa_2)^2+(2\kappa_3+\kappa_4)^2}\Big|\leq 8\pi,\\
				&\Big| \kappa_1+\kappa_2\pm \sqrt{(\kappa_1-\kappa_2)^2+\kappa_5} \Big|\leq 8\pi,\\
				&\Big| 3\kappa_2+3\kappa_2^{DM} \pm \sqrt{9(\kappa_2-\kappa_2^{DM})^2+(2\kappa_3^{DM}+\kappa_4^{DM})^2} \Big|\leq 8\pi,\\
				&\Big| \kappa_2+\kappa_2^{DM}\pm \sqrt{(\kappa_2-\kappa_2^{DM})^2+2\kappa_5^{DM}} \Big|\leq 8\pi,\nn
			\end{split}$}
	\end{equation}
\end{minipage}
\hspace{-0.4cm}
\begin{minipage}[c]{0.50\textwidth}
	\begin{equation*}
		\resizebox{0.98\hsize}{!}{ $
			\begin{split}
				&|\kappa_3^{DM}\pm \kappa_4^{DM}|\leq 8\pi,\\
				&|\kappa_3^{DM}\pm 2\kappa_5^{DM}|\leq 8\pi,\\
				&|\kappa_3^{DM}+2\kappa_4^{DM}\pm 6\kappa_5^{DM}|\leq 8\pi,\\
				&\Big|\kappa_1+\kappa_2^{DM}\pm \sqrt{(\kappa_1-\kappa_2^{DM})^2+\kappa_4^{DM}}\Big|\leq 8\pi,\\
				&\Big|3\kappa_1+3\kappa_2^{DM}\pm \sqrt{9(\kappa_1-\kappa_2^{DM})^2+(2\kappa_3^{DM}+\kappa_4^{DM})^2} \Big| \leq 8\pi,\\
				&\Big|\kappa_1+\kappa_2^{DM}\pm \sqrt{(\kappa_1-\kappa_2^{DM})^2+2\kappa_5^{DM}}\Big|\leq 8\pi,\\
				&\Big|\kappa_2+\kappa_2^{DM}\pm \sqrt{(\kappa_2-\kappa_2^{DM})^2+\kappa_4^{DM}}\Big|\leq 8\pi.\nn
			\end{split} $}
	\end{equation*}
\end{minipage}
%%%%%%%%%%%%%%%%%%%%%%%%%%%%%%%%%%%%%%%%%%%%%%%%%
%%%%%%%%%%%%%%%%%%%%%%%%%%%%%%%%%%%%%%%%%%%%%%%%%
%%%%%%%%%%%%%%%%%%%%%%%%%%%%%%%%%%%%%%%%%%%%%%%%%
%%%%%%%%%%%%%%%%%%%%%%%%%%%%%%%%%%%%%%%%%%%%%%%%%
\subsubsection{Bounds from electroweak precision experiments}
When we consider a new physics contribution above the EW scale, the effect of the virtual particles in loops does contribute to the electroweak precision bounds through vacuum polarization correction. Bounds from electroweak precision experiments are added in new physics contributions $via$ self-energy parameters $S,T,U$ \cite{Baak:2014ora}. The $S$ and $T$ parameters provide the new physics contributions to the neutral and the difference between neutral and charged weak currents, respectively. In contrast, the $U$ parameter is only sensitive to the mass and width of the W-boson; thus, in some cases, this parameter is neglected. In this model, inert scalars decouple from the other scalar fields. Contributions from first two doublet fields are \cite{Chakrabarty:2015kmt},
%%%%%%%%%%%%%%%%%%%%%%%%%%%%%%%%%%%%%%%%%%%%%%%%%
%%%%%%%%%%%%%%%%%%%%%%%%%%%%%%%%%%%%%%%%%%%%%%%%%
%%%%%%%%%%%%%%%%%%%%%%%%%%%%%%%%%%%%%%%%%%%%%%%%%
%%%%%%%%%%%%%%%%%%%%%%%%%%%%%%%%%%%%%%%%%%%%%%%%%
\bea
\Delta S_{2HD}&=&\frac{1}{\pi M_Z^2}\Big[\sin^2(\beta-\alpha)\mathcal{B}_{22}(M_Z^2,M_H^2,M_A^2)-\mathcal{B}_{22}(M_Z^2,M_{H^{\pm}}^2,M_{H^{\pm}}^2)\nn\\
&+&\cos^2(\beta-\alpha)\Big\{\mathcal{B}_{22}(M_Z^2,M_h^2,M_A^2)+\mathcal{B}_{22}(M_Z^2,M_Z^2M_H^2)-\mathcal{B}_{22}(M_Z^2,M_Z^2,M_h^2)\nn\nn\\
&-&M_Z^2\mathcal{B}_{0}(M_Z^2,M_Z^2,M_H^2)+M_Z^2\mathcal{B}_{0}(M_Z^2,M_Z^2,M_h^2)\Big\}\Big],\\
\Delta T_{2HD}&=&\frac{1}{16\pi M_W^2\sin^2_{\theta_W}}\Big[F(M^2_{H^{\pm}},M^2_A)+\sin^2(\beta-\alpha)\Big\{F(M^2_{H^{\pm}},M^2_H)-F(M^2_A,M^2_H) \}\nn\\
& +&\cos^2(\beta-\alpha)\Big\{F(M^2_{H^{\pm}},M^2_h)-F(M^2_A,M^2_h)+F(M^2_W,M^2_H)-F(M^2_W,M^2_h)\nn\\
&-&F(M^2_Z,M^2_H)+F(M^2_Z,M^2_h)+4M_Z^2\mathcal{\overline{B}}_{0}(M_Z^2,M_H^2,M_h^2)\nn\\
&-&M_Z^2\mathcal{\overline{B}}_{0}(M_W^2,M_H^2,M_h^2)\}\Big],\\
\Delta U_{2HD}&=&-S+\frac{1}{\pi M_Z^2}\Big[\mathcal{B}_{22}(M_W^2,M_A^2,M_{H^{\pm}}^2)-2\mathcal{B}_{22}(M_W^2,M^2_{H^{\pm}},M_{H^{\pm}}^2)\nn\\
&+&\sin^2(\beta-\alpha)\mathcal{B}_{22}(M_W^2,M_H^2,M_{H^{\pm}}^2)\nn\\
& +&\cos^2(\beta-\alpha)\Big\{\mathcal{B}_{22}(M_W^2,M_h^2,M_{H^{\pm}}^2)+\mathcal{B}_{22}(M_W^2,M_W^2,M_{H}^2)-\mathcal{B}_{22}(M_W^2,M_W^2,M_h^2)\nn\\
&-&M_W^2\mathcal{B}_{0}(M_W^2,M_W^2,M_{H}^2)+M_W^2\mathcal{B}_{0}(M_W^2,M_W^2,M_{H}^2)\Big\}\Big],
\eea
%%%%%%%%%%%%%%%%%%%%%%%%%%%%%%%%%%%%%%%%%%%%%%%%%
%%%%%%%%%%%%%%%%%%%%%%%%%%%%%%%%%%%%%%%%%%%%%%%%%
%%%%%%%%%%%%%%%%%%%%%%%%%%%%%%%%%%%%%%%%%%%%%%%%%
%%%%%%%%%%%%%%%%%%%%%%%%%%%%%%%%%%%%%%%%%%%%%%%%%
while the contributions from inert fields can be written as \cite{ Baak:2014ora},
%%%%%%%%%%%%%%%%%%%%%%%%%%%%%%%%%%%%%%%%%%%%%%%%%
%%%%%%%%%%%%%%%%%%%%%%%%%%%%%%%%%%%%%%%%%%%%%%%%%
%%%%%%%%%%%%%%%%%%%%%%%%%%%%%%%%%%%%%%%%%%%%%%%%%
%%%%%%%%%%%%%%%%%%%%%%%%%%%%%%%%%%%%%%%%%%%%%%%%%
\bea
\Delta S_{ID}&=&\frac{1}{2\pi}\Big[\frac{1}{6}\ln\frac{M_H^2}{M_{H^{\pm}}^2}-\frac{5}{36}+\frac{M_H^2M_A^2}{3(M_A^2-M_H^2)^2}+\frac{M_A^4(M_A^2-3M_H^2)}{6(M_A^2-M_H^2)^3}\ln \frac{M_A^2}{M_H^2} \Big],\\
\Delta T_{ID}&=&\frac{1}{32\pi^2\alpha v^2}\Big[F(M^2_{H^{\pm}},M^2_H)+F(M^2_{H^{\pm}},M_A^2)-F(M_A^2,M^2_H)\Big].
\eea
%%%%%%%%%%%%%%%%%%%%%%%%%%%%%%%%%%%%%%%%%%%%%%%%%
%%%%%%%%%%%%%%%%%%%%%%%%%%%%%%%%%%%%%%%%%%%%%%%%%
%%%%%%%%%%%%%%%%%%%%%%%%%%%%%%%%%%%%%%%%%%%%%%%%%
%%%%%%%%%%%%%%%%%%%%%%%%%%%%%%%%%%%%%%%%%%%%%%%%%
$\Delta U_{ID}$ is neglected in this case due to small mass differences $\Delta M_{H}=M_{A_3}-M_{H_3}$ and $\Delta M_{H^\pm}=M_{H_3^\pm}-M_{H_3}$ of the inert fields.
$F$ and $\mathcal{B}$'s are defined as,
%%%%%%%%%%%%%%%%%%%%%%%%%%%%%%%%%%%%%%%%%%%%%%%%%
%%%%%%%%%%%%%%%%%%%%%%%%%%%%%%%%%%%%%%%%%%%%%%%%%
%%%%%%%%%%%%%%%%%%%%%%%%%%%%%%%%%%%%%%%%%%%%%%%%%
%%%%%%%%%%%%%%%%%%%%%%%%%%%%%%%%%%%%%%%%%%%%%%%%%
\begin{eqnarray*}
	F(x,y)&=&\frac{x+y}{2}-\frac{xy}{x-y}\ln (\frac{x}{y}), \quad \text{for}\quad x\ne y \quad\text{otherwise}\quad 0.\\
	\mathcal{B}_{22}(q^2,m_1^2,m_2^2)&=&\frac{q^2}{24}[2\ln q^2+\ln(x_1x_2)+\{(x_1-x_2)^3-3(x_1^2-x_2^2)+3(x_1-x_2)\}\ln (x_1/x_2)\nn\\
	&&-\{2(x_1-x_2)^2-8(x_1+x_2)+\frac{10}{3}\}-\{(x_1-x_2)^2-x(x_1+x_2)+1\}f(x_1,x_2)\nn\\
	&&-6F(x_1,x_2)]  \xLongrightarrow{m_1=m_2}\frac{q^2}{24}[2\ln q^2+\ln x_1+(16x_1-\frac{10}{3})+(4x_1-1)G(x_1)],\\
	\mathcal{B}_0(q^2,m_1^2,m_2^2)&=& 1+\frac{1}{2}[\frac{x_1+x_2}{x_1-x_2}-(x_1-x_2)]\ln(x_1/x_2)+\frac{1}{2}f(x_1,x_2)\xLongrightarrow{m_1=m_2}2-2y\arctan\frac{1}{y},\\
	\mathcal{\overline{B}}_0(m_1^2,m_2^2,m_3^2)&=&\frac{m_1^2\ln m_1^2-m^2_3\ln m_3^2}{m_1^2-m_3^2}-\frac{m_1^2\ln m_1^2-m_2^2\ln m_2^2}{m_1^2-m_2^2},\\
	\text{with},&&x_i=m_i^2/q^2 \quad , \quad y=\sqrt{4x_1-1}\quad, \quad G(x_1)= -4y\arctan\frac{1}{y},\\ 
	&&f(x_1,x_2)=-2\sqrt{\Upsilon}[\arctan(\frac{x_1-x_2+1}{\sqrt{\Upsilon}})-\arctan(\frac{x_1-x_2-1}{\sqrt{\Upsilon}})]~\text{for}~\Upsilon>0,\\
	&&f(x_1,x_2)=\sqrt{-\Upsilon}\Big[\ln(\frac{x_1+x_2-1+\sqrt{-\Upsilon}}{x_1+x_2-1-\sqrt{-\Upsilon}})\Big]~\text{for}\quad \Upsilon<0,\\\text{and}\quad && f(x_1,x_2)=0 \quad \text{for}\quad  \Upsilon=0;\quad \text{where,} \quad \Upsilon=2(x_1+x_2)-x_1-x_2
	^2-1.
\end{eqnarray*}
%%%%%%%%%%%%%%%%%%%%%%%%%%%%%%%%%%%%%%%%%%%%%%%%%
%%%%%%%%%%%%%%%%%%%%%%%%%%%%%%%%%%%%%%%%%%%%%%%%%
%%%%%%%%%%%%%%%%%%%%%%%%%%%%%%%%%%%%%%%%%%%%%%%%%
%%%%%%%%%%%%%%%%%%%%%%%%%%%%%%%%%%%%%%%%%%%%%%%%%
One can add these contributions to the SM as,
%%%%%%%%%%%%%%%%%%%%%%%%%%%%%%%%%%%%%%%%%%%%%%%%%
%%%%%%%%%%%%%%%%%%%%%%%%%%%%%%%%%%%%%%%%%%%%%%%%%
%%%%%%%%%%%%%%%%%%%%%%%%%%%%%%%%%%%%%%%%%%%%%%%%%
%%%%%%%%%%%%%%%%%%%%%%%%%%%%%%%%%%%%%%%%%%%%%%%%%
\begin{equation}
	\begin{split}
		S=S_{SM}+\Delta S_{IDM}+\Delta S_{2HDM},\\
		T=T_{SM}+\Delta T_{IDM}+\Delta T_{2HDM},\\
		U=U_{SM}+\Delta U_{IDM}+\Delta U_{2HDM}.\\
	\end{split}
\end{equation}
%%%%%%%%%%%%%%%%%%%%%%%%%%%%%%%%%%%%%%%%%%%%%%%%%
%%%%%%%%%%%%%%%%%%%%%%%%%%%%%%%%%%%%%%%%%%%%%%%%%
%%%%%%%%%%%%%%%%%%%%%%%%%%%%%%%%%%%%%%%%%%%%%%%%%
%%%%%%%%%%%%%%%%%%%%%%%%%%%%%%%%%%%%%%%%%%%%%%%%%
We use the NNLO (Next to Next Leading Order) global electroweak fit results obtained by the Gfitter
group~\cite{Baak:2014ora}, $\Delta S_{IDM}+\Delta S_{2HDM}<0.05\pm0.11$, $T_{IDM}+\Delta T_{2HDM}<0.09\pm0.13$ and $\Delta U_{IDM}+\Delta U_{2HDM}<0.011\pm0.11$.
\subsubsection{LHC diphoton signal strength bounds}
At tree-level, the couplings of Higgs-like scalar $h$ to the fermions and gauge bosons in the presence of extra Higgs doublet ($\phi_2$) are modified due to the mixing. Loop induced decays will also have slight modification for the same reason. Hence, new contributions will be added to the signal strength. Using narrow width approximation, $\Gamma_h/M_h\rightarrow0$, the Higgs to diphoton strength is,
%%%%%%%%%%%%%%%%%%%%%%%%%%%%%%%%%%%%%%%%%%%%%%%%%
%%%%%%%%%%%%%%%%%%%%%%%%%%%%%%%%%%%%%%%%%%%%%%%%%
%%%%%%%%%%%%%%%%%%%%%%%%%%%%%%%%%%%%%%%%%%%%%%%%%
%%%%%%%%%%%%%%%%%%%%%%%%%%%%%%%%%%%%%%%%%%%%%%%%%
\begin{equation}
	\mu_{\gamma\gamma}=\frac{\sigma(gg\rightarrow h\rightarrow \gamma\gamma)_{BSM}}{\sigma(gg\rightarrow h\rightarrow \gamma\gamma)_{SM}}\approx\frac{\sigma(gg\rightarrow h)_{BSM}}{\sigma(gg\rightarrow h)_{SM}}\frac{Br(h\rightarrow\gamma\gamma)_{BSM}}{Br(h\rightarrow\gamma\gamma)_{SM}}.
\end{equation}
%%%%%%%%%%%%%%%%%%%%%%%%%%%%%%%%%%%%%%%%%%%%%%%%%
%%%%%%%%%%%%%%%%%%%%%%%%%%%%%%%%%%%%%%%%%%%%%%%%%
%%%%%%%%%%%%%%%%%%%%%%%%%%%%%%%%%%%%%%%%%%%%%%%%%
%%%%%%%%%%%%%%%%%%%%%%%%%%%%%%%%%%%%%%%%%%%%%%%%%
In presence of an extra inert Higgs doublet $\phi_3$, the signal strength does not change, however, due to mixing of $\phi_1$ and $\phi_2$, $h$ to flavon-flavon (or boson-boson) coupling become proportional to $\frac{\sin\alpha}{\cos\beta}(\text{or}\cos(\beta-\alpha))$. So, we may rewrite $\mu_{\gamma\gamma}$ as,
%%%%%%%%%%%%%%%%%%%%%%%%%%%%%%%%%%%%%%%%%%%%%%%%%
%%%%%%%%%%%%%%%%%%%%%%%%%%%%%%%%%%%%%%%%%%%%%%%%%
%%%%%%%%%%%%%%%%%%%%%%%%%%%%%%%%%%%%%%%%%%%%%%%%%
%%%%%%%%%%%%%%%%%%%%%%%%%%%%%%%%%%%%%%%%%%%%%%%%%
\begin{equation}
	\mu_{\gamma\gamma}=\frac{\sin^2\alpha}{\cos^2\beta}\frac{\Gamma(h\rightarrow\gamma\gamma)_{BSM}}{\Gamma(h\rightarrow\gamma\gamma)_{SM}}\frac{\Gamma^{total}_{h,SM}}{\Gamma^{total}_{h,BSM}}.
\end{equation}
%%%%%%%%%%%%%%%%%%%%%%%%%%%%%%%%%%%%%%%%%%%%%%%%%
%%%%%%%%%%%%%%%%%%%%%%%%%%%%%%%%%%%%%%%%%%%%%%%%%
%%%%%%%%%%%%%%%%%%%%%%%%%%%%%%%%%%%%%%%%%%%%%%%%%
%%%%%%%%%%%%%%%%%%%%%%%%%%%%%%%%%%%%%%%%%%%%%%%%%
Apart from the SM Higgs $h$, if the masses for the extra physical Higgses are greater than $M_h/2$, then $\frac{\Gamma^{total}_{h,SM}}{\Gamma^{total}_{h,BSM}}\approx\big(\frac{\sin^2\alpha}{\cos^2\beta}\big)^{-1}$. Hence, the modified signal strength will be written as,
%%%%%%%%%%%%%%%%%%%%%%%%%%%%%%%%%%%%%%%%%%%%%%%%%
%%%%%%%%%%%%%%%%%%%%%%%%%%%%%%%%%%%%%%%%%%%%%%%%%
%%%%%%%%%%%%%%%%%%%%%%%%%%%%%%%%%%%%%%%%%%%%%%%%%
%%%%%%%%%%%%%%%%%%%%%%%%%%%%%%%%%%%%%%%%%%%%%%%%%
\begin{equation}
	\mu_{\gamma\gamma}=\frac{\Gamma(h\rightarrow\gamma\gamma)_{BSM}}{\Gamma(h\rightarrow\gamma\gamma)_{SM}}.
\end{equation}
%%%%%%%%%%%%%%%%%%%%%%%%%%%%%%%%%%%%%%%%%%%%%%%%%
%%%%%%%%%%%%%%%%%%%%%%%%%%%%%%%%%%%%%%%%%%%%%%%%%
%%%%%%%%%%%%%%%%%%%%%%%%%%%%%%%%%%%%%%%%%%%%%%%%%
%%%%%%%%%%%%%%%%%%%%%%%%%%%%%%%%%%%%%%%%%%%%%%%%%
At one-loop level, the physical Higgs $H^{\pm}$ and $H_3^{\pm}$ add extra contribution to the decay width as,
%%%%%%%%%%%%%%%%%%%%%%%%%%%%%%%%%%%%%%%%%%%%%%%%%
%%%%%%%%%%%%%%%%%%%%%%%%%%%%%%%%%%%%%%%%%%%%%%%%%
%%%%%%%%%%%%%%%%%%%%%%%%%%%%%%%%%%%%%%%%%%%%%%%%%
%%%%%%%%%%%%%%%%%%%%%%%%%%%%%%%%%%%%%%%%%%%%%%%%%
\begin{equation}
	\Gamma(h\rightarrow \gamma\gamma)_{BSM}=\frac{\alpha^2M_h^3}{256\pi^3v^2}\Big|Q^2_{H^{\pm}}\frac{v\mu_{hH^+H^-}}{2M^2_{H^{\pm}}}F_0(\tau_{H^{\pm}})+Q^2_{H_3^{\pm}}\frac{v\mu_{hH_3^+H_3^-}}{2M^2_{H_3^{\pm}}}F_0(\tau_{H_3^{\pm}})+C\Big|,
\end{equation}
%%%%%%%%%%%%%%%%%%%%%%%%%%%%%%%%%%%%%%%%%%%%%%%%%
%%%%%%%%%%%%%%%%%%%%%%%%%%%%%%%%%%%%%%%%%%%%%%%%%
%%%%%%%%%%%%%%%%%%%%%%%%%%%%%%%%%%%%%%%%%%%%%%%%%
%%%%%%%%%%%%%%%%%%%%%%%%%%%%%%%%%%%%%%%%%%%%%%%%%
where, $C$ is the SM contribution, $ C=\sum_fN_f^cQ_f^2y_fF_{1/2}(\tau_f)+y_WF_1(\tau_W)$ and $\tau_x=\frac{M_h^2}{aM_X^2}$. $Q_i$ denote electric charge of corresponding particles and $N_f^c$ is the color factor. Higgs $h$ coupling to $f\overline{f}$ and $WW$ is denoted by $y_f=y_f^{\rm SM} \frac{\sin\alpha}{\cos\beta}$ and $y_W=y_w^{\rm SM}\cos(\beta-\alpha)$. $\mu_{hH^+H^-}$ and $\mu_{hH_3^+H_3^-}$ stands for corresponding couplings for the $hH_+H^-$ and $hH_3^+H_3^-$ respectively, which are defined below with the loop function $F_{(0,1/2,1)}(\tau)$~\cite{Djouadi:2005gj},
%%%%%%%%%%%%%%%%%%%%%%%%%%%%%%%%%%%%%%%%%%%%%%%%%
%%%%%%%%%%%%%%%%%%%%%%%%%%%%%%%%%%%%%%%%%%%%%%%%%
%%%%%%%%%%%%%%%%%%%%%%%%%%%%%%%%%%%%%%%%%%%%%%%%%
%%%%%%%%%%%%%%%%%%%%%%%%%%%%%%%%%%%%%%%%%%%%%%%%% 
\begin{equation*}
	\begin{split}
		\mu_{hH^+H^-}=&[2\kappa_4\sin\beta\cos\beta\cos\gamma+\cos\beta^2(\kappa_3v_1\cos\gamma+4\kappa_2v_2\sin\gamma)\nn\\
		&+\sin\beta^2(\kappa_4\sin\gamma+\kappa_1v_1\cos\gamma+\kappa_3v_2\sin\gamma)]\approx \kappa_3v_{SM},\\
		\mu_{hH_3^+H_3^-}=&\kappa_3^{DM}v_1[\sin\alpha\sin\beta-\cos\alpha\cos\beta],\\
		F_0(\tau)=&[\tau-f(\tau)]\tau^{-2},\\
		F_{1/2}(\tau)=&2[\tau+(\tau-1)f(\tau)]\tau^{-2},\\
		F_1(\tau)=&-[2\tau^2+3\tau+3(2\tau-1)f(\tau)]\tau^{-2},\\
		\text{with,}\quad f(\tau)=& \begin{cases} 
			(\sin^{-1}\sqrt{\tau})^2, \quad\quad\quad\quad\quad\quad \tau\leq 1,\\ 
			-\frac{1}{4}[\ln\frac{1+\sqrt{ 1-\tau^{-1}}}{1-\sqrt{1-\tau^{-1}}}-i\pi]^2\quad\quad \tau>1.
		\end{cases}\\
	\end{split}
\end{equation*}
%%%%%%%%%%%%%%%%%%%%%%%%%%%%%%%%%%%%%%%%%%%%%%%%%
%%%%%%%%%%%%%%%%%%%%%%%%%%%%%%%%%%%%%%%%%%%%%%%%%
%%%%%%%%%%%%%%%%%%%%%%%%%%%%%%%%%%%%%%%%%%%%%%%%%
%%%%%%%%%%%%%%%%%%%%%%%%%%%%%%%%%%%%%%%%%%%%%%%%%

%
%
%
%
%
%

%%%%%%%%%%%%%%%%%%%%%%%%%%%%%%%%%%%%%%%%%%%%%%%%%
%%%%%%%%%%%%%%%%%%%%%%%%%%%%%%%%%%%%%%%%%%%%%%%%%
\subsubsection{Bounds from dark matter } 
Various results from the WMAP satellite, combined with other cosmological measurements, we got the constrained dark matter relic density to be $\Omega_{DM} h^2=0.1198\pm0.0026$ \cite{Ade:2013zuv}. The dark matter sector of this model is behaving quite similar to the normal inert doublet model \cite{Honorez:2010re, Goudelis:2013uca, Arhrib:2013ela}.
However, in the presence of an extra SM type Higgs doublet, these pictures get slightly disturbed. Both the annihilation and co-annihilation are modified, and we find a larger region of allowed parameter spaces in this model than other inert Higgs doublet models. 
Within the inert Higgs doublet sector, the lightest neutral scalar ($H_3/A_3$) serves as an inert doublet dark matter candidate. In this model, we consider $H_3$ to be a dark matter candidate, and it is to be noted that the region of the dark matter parameter spaces will be slightly changed for the pseudoscalar as a dark matter candidate $A_3$. 
We have used {\tt FeynRules}~\cite{Alloul:2013bka} to construct our model and relic density calculations are carried out using {\tt MicrOMEGAs} \cite{Belanger:2018mqt}. We carry out detailed discussion about DM mass in the numerical analysis section.
%\subsection{Bounds from direct and indirect detection experiments}

At present, the most stringent limit on the spin-independent component of elastic
scattering cross section $\sigma^{SI}_{p}<4.1\times 10^{-47}~{\rm cm^2}$ for $M_{DM}\simeq 30$ GeV~\cite{Aprile:2018dbl}. 
In this analysis, we get the dark matter parameter space, which is allowed by direct detection data and other theoretical as well as experimental constraints. 
In this model, we took a minimal Higgs portal coupling, which is allowed by direct detection cross-section and invisible Higgs decay width (for $M_{DM}<M_h/2$) \cite{Duerr:2015aka}.
The dark matter indirect detection constraints such as Fermi-LAT data~\cite{FermiLAT:2011ab}, PAMELA~\cite{Adriani:2010rc} and AMS02~\cite{Aguilar:2016vqr} restrict arbitrary Higgs portal couplings~\cite{ Gaskins:2016cha}.
It is also possible to explain various observations in the indirect DM detection experiments like, combine results from LHC-14 Monojet + XENON1T + AMS02 antiproton flux~\cite{Arhrib:2013ela} from this model. Moreover, we have checked that the dark matter self-annihilation cross-sections for the allowed parameter space. The tiny Higgs portal coupling which is allowed by the direct detection data gives the self-annihilation cross-section, $<\sigma v>\, \lesssim \, \mathcal{O}(10^{-25})~{\rm cm^3s^{-1}}$. The relic density is adjusted by the other (co)annihilation processes and these allowed points do not exceed the indirect detection bounds. 
%It is also noted that we have checked that the dark matter self-annihilation cross-sections for the allowed parameter remain $<\sigma v>\, \lesssim \, \mathcal{O}(10^{-25})~{\rm cm^3s^{-1}}$ as the Higgs portal couplings are very small, allowed by the direct detection data. The relic density is adjusted by the other (co)annihilation processes. These allowed points do not exceeds the indirect detection bounds. 
However, we do not discuss these bounds here explicitly, as these estimations involve proper knowledge of the astrophysical backgrounds and an assumption of the DM halo profile which contains some arbitrariness.
%%%%%%%%%%%%%%%%%%%%%%%%%%%%%%%%%%%%%%%%%%%%%%%%%
%%%%%%%%%%%%%%%%%%%%%%%%%%%%%%%%%%%%%%%%%%%%%%%%%
\section{Numerical analysis and results}\label{s4}
\subsection{Dark matter}
%%%%%%%%%%%%%%%%%%%%%%%%%%%%%%%%%%%%%%%%%%%%%%%%%%%%%
%%%%%%%%%%%%%%%%%%%%%%%%%%%%%%%%%%%%%%%%%%%%%%%%%%%%%
%%%%%%%%%%%%%%%%%%%%%%%%%%%%%%%%%%%%%%%%%%%%%%%%%%%%%
%%%%%%%%%%%%%%%%%%%%%%%%%%%%%%%%%%%%%%%%%%%%%%%%%%%%% 
%%%%%%%%%%%%%%%%%%%%%%%%%%%%%%%%%%%%%%%%%%%%%%%%%%%%%
%%%%%%%%%%%%%%%%%%%%%%%%%%%%%%%%%%%%%%%%%%%%%%%%%%%%%
%%%%%%%%%%%%%%%%%%%%%%%%%%%%%%%%%%%%%%%%%%%%%%%%%%%%%
%%%%%%%%%%%%%%%%%%%%%%%%%%%%%%%%%%%%%%%%%%%%%%%%%%%%%
%%%%%%%%%%%%%%%%%%%%%%%%%%%%%%%%%%%%%%%%%%%%%%%%%%%%%
In this section, we discuss the numerical analysis and new bounds on the DM mass of the model.
As we know, the observed relic density through annihilation in this model mainly relies on the dark matter mass $M_{H_3}$, Higgs (both $h$ and $H$) portal coupling. These couplings are primarily depending upon the coupling $\kappa_L$ and the mixing angles $\alpha$ and $\beta$. The annihilation could also be affected by the mass of the more massive Higgs particle.
If we decrease the mass difference between the LSP and nLSPs (similar to the IDM), co-annihilation channels start to play a crucial role. The mass differences $\Delta M_{A_3}=M_{A_3}-M_{H_3}$ and $\Delta M_{H_3^\pm}=M_{H_3^\pm}-M_{H_3}$  are important here in calculating the relic abundance.
The other $Z_4$-even charged ($H_1^\pm$) and pseudoscalar ($A_1$) particles also come into this picture depending on their masses and the mixing angles $\alpha$ and $\beta$.

We performed scans in the four dimensional parameter space. Dark matter mass, $M_{H_3}$ is varied from 5 GeV to 1000 GeV and $\kappa_L$ from $-0.25$ to $0.25$ with a step size $0.001$,  $\Delta M_{A_3}$ from $0$ to $20$ GeV with a step size $0.2$ GeV. We also fixed $\Delta M_{H_3^\pm}$=100 GeV to avoid the collider constraints~\cite{ Djouadi:2011aa}.
The heavier Higgs masses are fixed at $M_H=400$ GeV and $M_{A,H^\pm}=430$ GeV. Two different regimes for a fixed value of the mixing angles  $\alpha$ and $\beta$ are obtained, and we define these as a low and high mass regime. These mass regimes for two different values of $\alpha$ and $\beta$ are shown in fig.~\ref{dmplot}.
Left plot stands for low mass regime whereas right one indicates high mass regime. The red points correspond to $\cos(\beta-\alpha)\sim 0.92$ and blue $\cos(\beta-\alpha)\sim 0.015$.
%%%%%%%%%%%%%%%%%%%%%%%%%%%%%%%%%%%%%%%%%%%%%%%%%
%%%%%%%%%%%%%%%%%%%%%%%%%%%%%%%%%%%%%%%%%%%%%%%%%
%%%%%%%%%%%%%%%%%%%%%%%%%%%%%%%%%%%%%%%%%%%%%%%%%
%%%%%%%%%%%%%%%%%%%%%%%%%%%%%%%%%%%%%%%%%%%%%%%%%
\begin{figure}[h!]
	\includegraphics[width=2.8in,height=2.8in, angle=0]{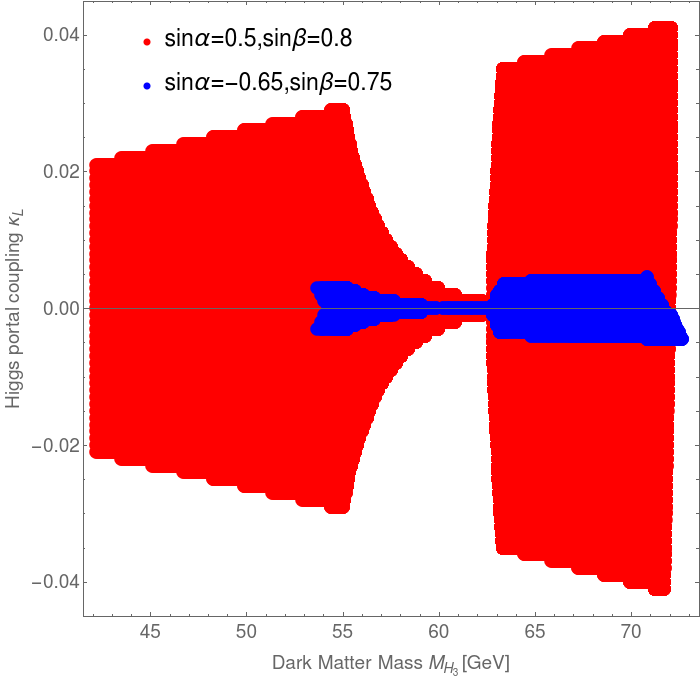}
	\includegraphics[width=2.8in,height=2.8in, angle=0]{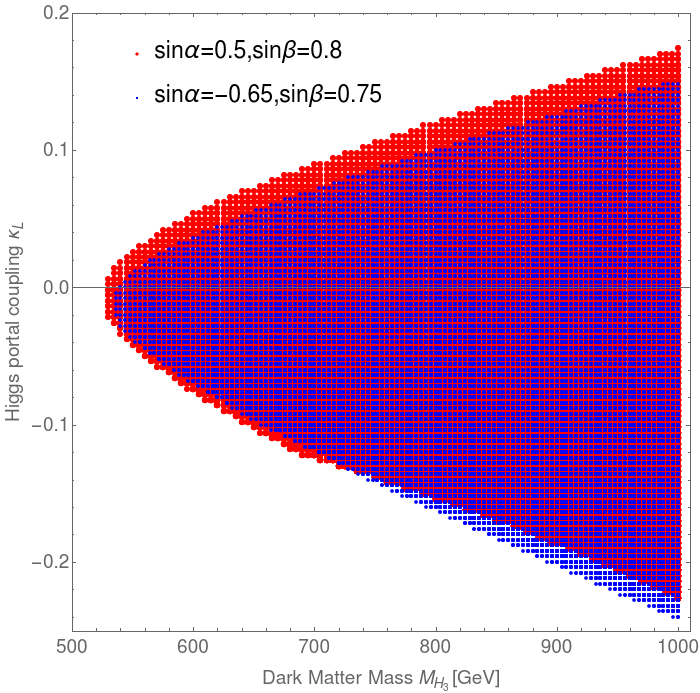}
	\caption{Left plot stands for low mass regime whereas right one indicates high mass regime. The red points correspond to $\cos(\beta-\alpha)\sim 0.92$ and blue $\cos(\beta-\alpha)\sim 0.015$. We keep the heavier Higgs masses fixed at $M_H=400$ GeV and $M_{A,H^\pm}=430$ GeV.}
	\label{dmplot}
\end{figure}
%%%%%%%%%%%%%%%%%%%%%%%%%%%%%%%%%%%%%%%%%%%%%%%%%
%%%%%%%%%%%%%%%%%%%%%%%%%%%%%%%%%%%%%%%%%%%%%%%%%
%%%%%%%%%%%%%%%%%%%%%%%%%%%%%%%%%%%%%%%%%%%%%%%%%
%%%%%%%%%%%%%%%%%%%%%%%%%%%%%%%%%%%%%%%%%%%%%%%%%
It is to be noted that, these points pass through all the experimental as well as theoretical constraints, that we have discussed in the previous section. For the low mass, a dominant part of the points ruled out by the Higgs/Z invisible decay width and direct detection constraints \cite{Aprile:2018dbl}. The STU parameter discards a part of the region, where $\Delta M_{A_3}$ is large, absolute stability of scalar potential and perturbativity, as $\Delta M_{A_3}$ is directly proportional to $\kappa_5^{DM}$. 

In the low mass regime $M_{H_3}<10$ GeV, DM dominantly annihilates into the SM fermions only, and this annihilation cross-section remains small due to the small coupling strength and mass. Hence, one would get overabundance as $\Omega_{DM} h^2 \propto \frac{1}{<\sigma v >} $. The abundance in this model is roughly
$1/\cos ^2(\beta-\alpha)$ times larger than the normal IDM abundance~\cite{Khan:2016sxm}.
In particular, within the mass regime $10-73$  GeV, correct values for $\Omega_{DM} h^2$ (within $3\sigma$) will be produced due to the contribution of the DM annihilation, co-annihilation or combined effect of these two processes. While the mass regime $10-42.2$ GeV is ruled out from the constraints of the decay width of SM gauge $W,Z$~\cite{Bertone:2004pz} and/or Higgs $h_1$ bosons ~\cite{Diaz:2015pyv}. The direct detection (DD) processes also played a strong role to ruled out this regime.
It is to be noted that in the normal IDM ~\cite{ Deshpande:1977rw, Arhrib:2013ela}, $M_{DM}<54$  GeV regime is ruled out from the similar constraints. In this model, the DM mass $42.2-54$ GeV up to $72.65$ GeV regime is still allowed due to the presence of the other $Z_4$-even scalar particles. We get the larger allowed region in the parameter spaces depending on the mixing angles $\alpha$ and $\beta$, and these regions are explained in fig.~\ref{dmplot}. A few examples of benchmark points presented in table.~\ref{BMPlow} and \ref{BMPhigh}. 
%%%%%%%%%%%%%%%%%%%%%%%%%%%%%%%%%%%%%%%%
%%%%%%%%%%%%%%%%%%%%%%%%%%%%%%%%%%%%%%%%
%%%%%%%%%%%%%%%%%%%%%%%%%%%%%%%%%%%%%%%%
%%%%%%%%%%%%%%%%%%%%%%%%%%%%%%%%%%%%%%%% 
\begin{table}
	\resizebox{0.98\hsize}{!}{
		\begin{tabular}{|c|c|c|c|c|c|c|c|}
			\hline
			BMP-low &$M_{DM}$ [GeV]&$\kappa_{L}$& $M_{A_3}$ [GeV]&$~\sin\alpha~$&$~\sin\beta~$&Relic density $~\Omega_{DM} h^2~$&DD cross-section [$\rm cm^2$]\\
			\hline
			\hline
			I & 42.20 & $-0.001$ & 53.4 &  0.50 & 0.80 & 0.1266 & $2.24\times 10^{-49}$\\
			II& 53.65 & $0.003$ &62.85&-0.65 & 0.75 & 0.1241  & $1.02\times 10^{-46}$\\
			\hline
			\hline
			III&$55.00$&$-0.029$&$73.80$&$0.50$&$0.80$&$0.1121$&$1.12\times 10^{-46}$\\
			IV&$55.00$&$0.001$&$63.65$&$-0.65$&$0.75$&$0.1161$&$1.08\times 10^{-47}$\\
			\hline
			\hline
			V&$65.00$&$-0.036$&$74.$&$0.50$&$0.80$&$0.1257$&$1.24\times 10^{-46}$\\
			VI&$65.00$&$-0.001$&$73.4$&$-0.65$&$0.75$&$0.1155$&$7.8\times 10^{-48}$\\
			\hline
			\hline
			VII& $72.05$ & $0.041$ &$91.85$ &$0.50$&$0.80$&$0.1152$&$1.31 \times 10^{-46}$\\
			VIII&$72.05$&$-0.001$&$93.85$&$-0.65$&$0.75$&$0.1140$&$6.38\times 10^{-48}$\\
			\hline
	\end{tabular}}
	\caption{Benchmark points for low dark matter mass regime. Each horizontal block contain two BMPs, corresponding to different set of mixing angles ($\alpha$ and $\beta$).}
	\label{BMPlow}
\end{table}
%%%%%%%%%%%%%%%%%%%%%%%%%%%%%%%%%%%%%%%%
%%%%%%%%%%%%%%%%%%%%%%%%%%%%%%%%%%%%%%%%
%%%%%%%%%%%%%%%%%%%%%%%%%%%%%%%%%%%%%%%%
%%%%%%%%%%%%%%%%%%%%%%%%%%%%%%%%%%%%%%%%

In the table~\ref{BMPlow}, a few benchmark points
have presented for low dark matter mass regime.
The first BMP-I corresponds to dark matter mass $42.2$ GeV with Higgs portal couplings $\kappa_L \cos(\beta+\alpha)= 0.12 \kappa_L$ ($\sin\alpha=0.5$ and $\sin\beta=0.8$) and $M_{A_3}=53.4$ GeV. As the $\Delta M_{A_3}$ ($=11.2$ GeV) is small, we get the relic density mainly dominated by the $Z$-mediated co-annihilation channels $H_3 A_3\rightarrow Z \rightarrow XX$, where $X={\rm SM~ fermions}$. One can find the corresponding diagrams (upper two) in fig.~\ref{Diag1}. Here, $M_{A_3}+M_{H_3}=95.6>M_Z$, hence this process is allowed by the $Z$-boson invisible decay width constraints. On the other hand, the Higgs portal coupling is also very small $\sim 10^{-3}$; hence, this point is also allowed by the invisible Higgs decay width and direct detection constraints. It is also to be noted that, in the normal inert doublet model, one may get the exact relic density for the DM mass below $54$ GeV, but one of these constraints will restrict this point. For our model, this may be considered as a new finding, as this small DM mass has not been discussed in the literature in details.

We changed the mixing angles $\sin\alpha=-0.65$ and $\sin\beta=0.75$, thus the Higgs portal coupling becomes $0.98 \kappa_L$. The dark matter masses $M_{H_3}=42.2$ GeV for this mixing angle get constrained by both the Higgs invisible decay width and direct detection cross-section. 
We get the allowed point for the next minimum dark matter mass $M_{H_3}=53.65$ GeV, for the Higgs coupling $0.98\kappa_L$. It could be understood as follows: as we change these angles, the Higgs portal coupling becomes too large, which violates the Higgs invisible decay width~\cite{Khachatryan:2016whc} and the direct detection cross-section bounds for the dark matter mass $42.2$ GeV. Hence, we increase the mass to get the allowed relic density. 
Here, $Z$-mediated  co-annihilation channels are contributing $78\%$ of the total processes and dominates the whole effective annihilation process. The annihilation processes $H_3 H_3 \rightarrow b\bar{b} (16\%)$ and $H_3 H_3 \rightarrow W^\pm W^{\mp *}(5\%)$ also played a significant role to achieve the exact relic density. All the diagrams in fig.~\ref{Diag1} ( upper and middle ) and fig. \ref{Diag2} (upper) are relevant here. It is to be noted that the processes  $H_3 H_3^\pm \rightarrow W^\pm \rightarrow XX$, where $X={\rm SM~ quarks/leptons}$ are not important in our case as we consider $\Delta M_{H_3^\pm}$=100 GeV.
\begin{figure}[h!]
	\begin{center}
		\includegraphics[width=5in,height=.8in, angle=0]{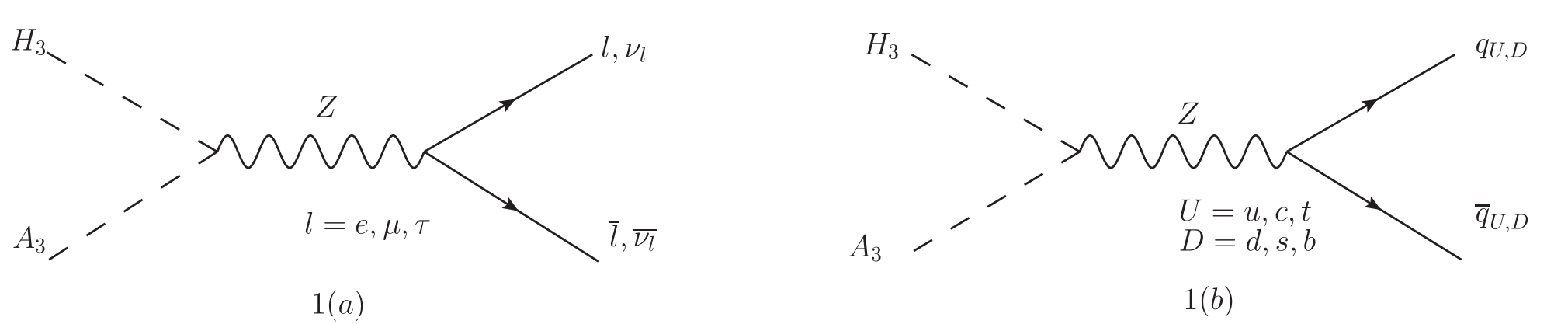}\\
		\includegraphics[width=5in,height=.8in, angle=0]{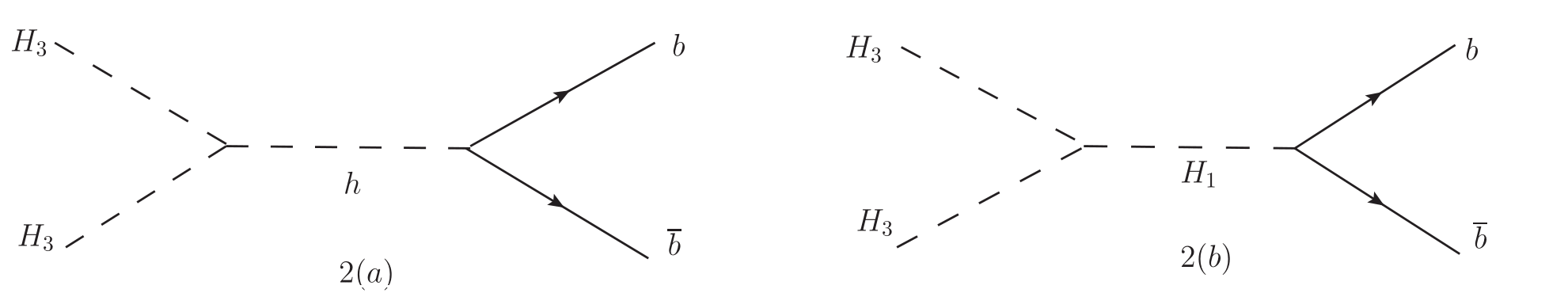}\\
		\includegraphics[width=5in,height=.8in, angle=0]{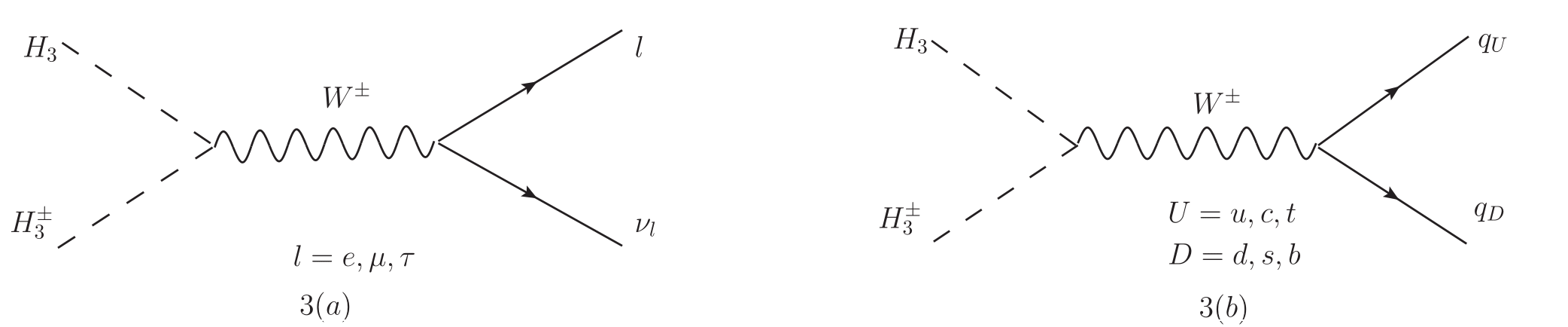}
	\end{center}
	\caption{The upper two diagrams stand for the co-annihilation channels $H_3 A_3 \rightarrow Z \rightarrow $ $ XX$, where $X={\rm SM~ quarks/leptons}$, the middles diagrams are $H_3 H_3^\pm \rightarrow W^\pm \rightarrow XX$. The dominant annihilation diagrams are $H_3 H_3 \rightarrow h/H \rightarrow {\rm b\bar{b}}$. These channels are effective and/or dominant for the dark matter mass regions $10-70$ GeV regime.}
	\label{Diag1}
\end{figure}
%%%%%%%%%%%%%%%%%%%%%%%%%%%%%%%%%%%%%%%%%%%%%%%%%%%%%
%%%%%%%%%%%%%%%%%%%%%%%%%%%%%%%%%%%%%%%%%%%%%%%%%%%%%
%%%%%%%%%%%%%%%%%%%%%%%%%%%%%%%%%%%%%%%%%%%%%%%%%%%%%
%%%%%%%%%%%%%%%%%%%%%%%%%%%%%%%%%%%%%%%%%%%%%%%%%%%%%
%%%%%%%%%%%%%%%%%%%%%%%%%%%%%%%%%%%%%%%%%%%%%%%%%%%%%
For BMP-III and IV, exact relic density is obtained with DM mass 55 GeV. The prime difference between these two BMPs arises due to the mixing angles, hence the Higgs portal coupling.
In the first case, $\cos(\alpha+\beta)=0.12$ and $\Delta M_{A_3}=18.80$ GeV with a quite large negative $\kappa_{L}=-0.029$. One can also get allowed relic density for similar positive $\kappa_{L}$ values and it can be understood from fig.~\ref{dmplot}. In this case, 
the exact relic density is obtained due to the Higgs mediated $H_3 H_3 \rightarrow b\bar{b}(88\%)$ annihilation channel. The other annihilation processes are $H_3 H_3 \rightarrow c\bar{c}, \tau^+\tau^-, WW^*(12\%)$. If we move towards the smaller values of $\kappa_{L}$, the co-annihilation channels become important to get the relic density. For the large values of  $\kappa_{L}$, the Higgs mass resonance region ($55-63$ GeV) of the dark matter are ruled out by one of the constraints as discuss earlier. For example, the dark matter mass $60$ GeV with $\kappa_{L}=\pm 0.01$, the relic density become $\Omega_{DM} h^2=0.01$.\\
For the BMP-IV, very small values of $\kappa_{L}$ are allowed as the total Higgs portal coupling is reduced by the mixing angles. We get exact relic density for 
$\Delta M_{A_3}=8.65$ GeV,
mainly through the co-annihilation channels $H_3 A_3 \rightarrow {\rm SM ~fermions} (92\%)$ which dominates over the DM annihilation channel $H_3 H_3 \rightarrow W^\pm W^{\mp *}$, (8\%). $|\kappa_{L}|>0.003$ region are ruled out by direct detection.
A similar analogy also works for $M_{DM}=65$ GeV.

For BMP-VII and VIII, $M_{DM}=72.05$ GeV, noticeable results are observed for two different mixing angles.
Since the DM mass is close to $W$-boson mass, which allowed to dominate via annihilation channel of $DM,DM\rightarrow W^\pm W^{\mp *}$ ($\sim 77\%$) for small Higgs portal coupling. On the other hand, for larger $\cos(\alpha+\beta)$, co-annihilation processes continues to dominates with $H_3 A_3 \rightarrow {\rm SM ~fermions}$, with a little contribution from dark matter annihilation process into $b\bar{b}$ ($\sim 3\%$) to give rise to the exact relic density. It is to be noted that the effective annihilation cross-section for the DM mass $72.65$ GeV, mixing angles $\sin\alpha=-0.65$ and $\sin\beta=0.75$ become large, hence, we got an under abundance. A negative value of $\kappa_{L}\sim -0.0045$ reduces the effective annihilation cross-section; hence, the exact relic density is observed. However, $\kappa_{L}< -0.0045$ region is ruled out by the direct detection cross-section. The region $72.65$ GeV to $536$ GeV is ruled out as the annihilation rates $H_3 H_3 \rightarrow W^\pm W^\mp, ZZ$ (see fig.~\ref{Diag2}) are very high, which reduces the relic abundance $\Omega_{DM} h^2 < 0.01$. The negative values of $\kappa_{L}$ may give the exact relic density. However, it will be ruled out the direct detection \cite{LopezHonorez:2010tb}. If we consider, minimal values of $\sin\beta$, one may get the allowed relic abundance, but these regions are again discarded by the constraints of the absolute stability and/or unitarity, perturbativity.\\
%%%%%%%%%%%%%%%%%%%%%%%%%%%%%%%%%%%%%%%%%%%%%%%%%%%%%
%%%%%%%%%%%%%%%%%%%%%%%%%%%%%%%%%%%%%%%%%%%%%%%%%%%%%
%%%%%%%%%%%%%%%%%%%%%%%%%%%%%%%%%%%%%%%%%%%%%%%%%%%%%
%%%%%%%%%%%%%%%%%%%%%%%%%%%%%%%%%%%%%%%%%%%%%%%%%%%%% 
\begin{figure}
	\begin{center}
		\includegraphics[width=5in,height=.8in, angle=0]{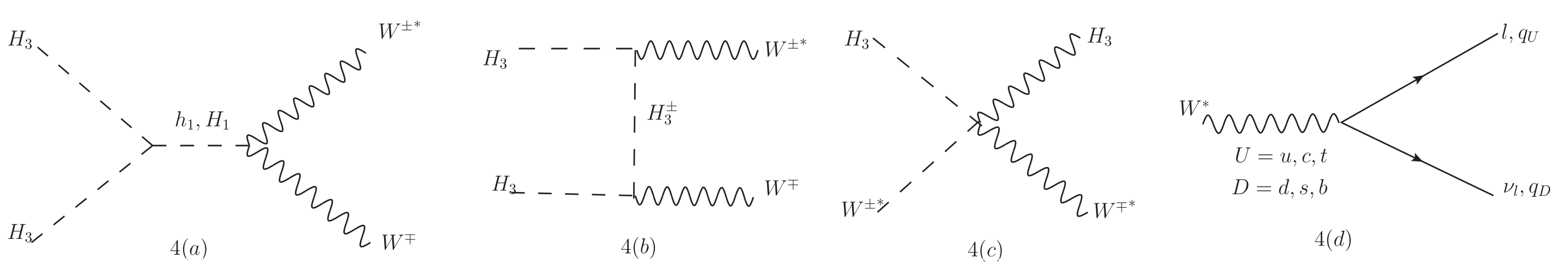}
		\includegraphics[width=5in,height=.8in, angle=0]{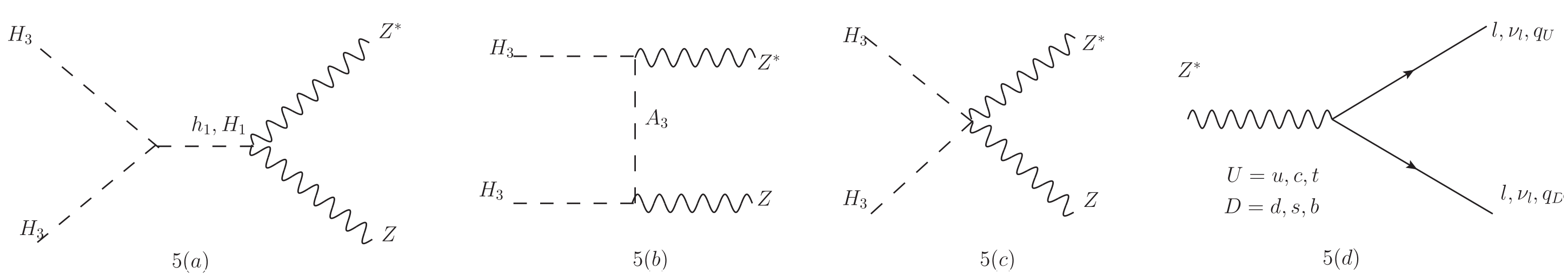}
	\end{center}
	\caption{These channels are effective and dominant for the dark matter mass regions $>70$ GeV regime. For the dark matter mass $M_{DM}<M_W / M_Z$, gauge bosons further decaying into SM fermions. The relic density is dominated by the $H_3 H_3 \rightarrow V V^*$, $ V^* \rightarrow {\rm SM~ quarks/leptons} $, where $V=W,Z$. For the high mass regime the dominant processes are $H_3 H_3 \rightarrow V V$.}
	\label{Diag2}
\end{figure}
%%%%%%%%%%%%%%%%%%%%%%%%%%
\begin{table}
	\resizebox{0.98\hsize}{!}{
		\begin{tabular}{|c|c|c|c|c|c|c|c|}
			\hline
			BMP-high &$M_{DM}$ [GeV]&$\kappa_{L}$& $M_{A_3}$ [GeV]&$~\sin\alpha~$&$~\sin\beta~$&Relic density $~\Omega_{DM} h^2~$&DD cross-section [$\rm cm^2$]\\
			\hline
			\hline
			I & 536 & $-0.02$ & 537 &  0.50 & 0.80 & 0.1127 & $5.783\times 10^{-49}$\\
			II & 530 & $-0.022$ & 530.5 &  -0.65 & 0.75 & 0.112 & $2.24\times 10^{-49}$\\
			\hline
			III & 760 & $0.044$ & 761 &  0.5 & 0.80 & 0.1195 & $1.39\times 10^{-48}$\\
			IV & 760 & $0.046$ & 765 &  -0.65 & 0.75 & 0.1122 & $1.24\times 10^{-46}$\\
			\hline
	\end{tabular}}
	\caption{Benchmark points for high dark matter mass regime. Each horizontal block contain two BMPs, corresponding to different set of mixing angles ($\alpha$ and $\beta$).}
	\label{BMPhigh}
\end{table}
Beyond 536 GeV, we obtain the points by satisfying the constrains discussed in previous subsections. For the high mass region, the DM annihilation channel almost equally contributes, and they get partially canceled out between various diagrams in the limit $M_{DM}\gg M_W$. Usually, $s- channel$ and $p- channel$ mediated processes like $DM,DM\rightarrow W^{\pm}W^{\mp}, ZZ$ get partially cancelled out by $u-channel$ and $t-channel$ processes to give rise to the correct relic bound.  One can find that the sum of the amplitude for these diagrams is proportional to $M_{H_3^{\pm}}^2-M^2_{DM}$~\cite{Khan:2015ipa}, hence, for high DM mass range such cancellation occurs for very small mass difference around 8 GeV. For a different set of $\alpha$ and $\beta$, DM mass vs. Higgs coupling for up to 1000 GeV mass for DM are shown in fig.~\ref{dmplot}. A slight shift in the allowed regions for two different sets of  $\alpha$ and $\beta$ values is observed. Few BMPs are also shown in table ~\ref{BMPhigh} for the high mass region. 

For the small Higgs coupling strength $0.12\kappa_{L}$, we get the satisfied relic density value with DM mass starting from 536 GeV, while for large $0.98\kappa_{L}$ the starting value for DM mass stands on 530 GeV. In both the processes, charge scalar decay to the $W$ boson ($H_3^+, H_3^-\rightarrow W^{\pm}W^{\mp}$) contributes around $16\%$ to $17\%$. However, DM annihilation processes like  $DM,DM\rightarrow W^{\pm}W^{\mp}$ ($\sim 13\%$ for BMP-I and $\sim15\%$ for BMP-II ) and $DM,DM\rightarrow ZZ$ ($\sim10\%$ for BMP-I and $\sim12\%$ for BMP-II) are contributing almost in equal amount, and we get desired range for relic density by partial cancellation among themselves. As we keep increasing the DM mass, the DM annihilation processes are contributing in modest fashion. For BMP-III and BMP-IV, we can find that annihilation processes like $DM,DM\rightarrow W^{\pm}W^{\mp}$($\sim16\%$ form BMP-III and $8\%$ for BMP-IV) and $DM,DM\rightarrow ZZ$ ($\sim9\%$ for BMP-III and $\sim11\%$ for BMP-IV) are contributing to achieving correct relic density. Interestingly, major co-annihilation channels are observed for $M_{DM}=760$ GeV, like $H_3^+H_3^-\rightarrow W^+W^-$. However, their contributions also get suppressed due to the partial cancellation among various diagrams. 
\subsection{Neutrino and baryogenesis}
In this chapter, along with the scalar sector, we also tried to shed light on the active neutrino sector in the presence of an extra Higgs doublet, which also takes part in the Lagrangian given by equation \eqref{lag} to give mass to the active neutrinos like the SM Higgs. According to our model, both the Higgs doublets acquire different VEVs. It is to be noted that we consider two different sets of $\sin\beta$. While there is very small distinction between these doublet VEVs for the two sets we have considered ($\sin\beta=0.8$ and $\sin\beta=0.75$), hence no significant difference can be found in neutrino sector for nearly identical $\tan\beta~(\sim1.33)$. As the involvement of the Higgs mass can be visualized through the Yukawa coupling of the Lagrangian, we tried to incorporate the result concerning the Yukawa couplings.
% A single flavor of sterile neutrino does exist in the scenario, however, a detailed work on active-sterile phenomenology has already been carried out in \cite{Das:2018qyt}.
%%%%%%%%%%%%%%%%%%%%%%%%%%%%%%%%%%%%%%%%%%%%%%%%%
%%%%%%%%%%%%%%%%%%%%%%%%%%%%%%%%%%%%%%%%%%%%%%%%%
%%%%%%%%%%%%%%%%%%%%%%%%%%%%%%%%%%%%%%%%%%%%%%%%%
%%%%%%%%%%%%%%%%%%%%%%%%%%%%%%%%%%%%%%%%%%%%%%%%%
\begin{figure}[h!]
	\includegraphics[width=2.8in,height=2.8in, angle=0]{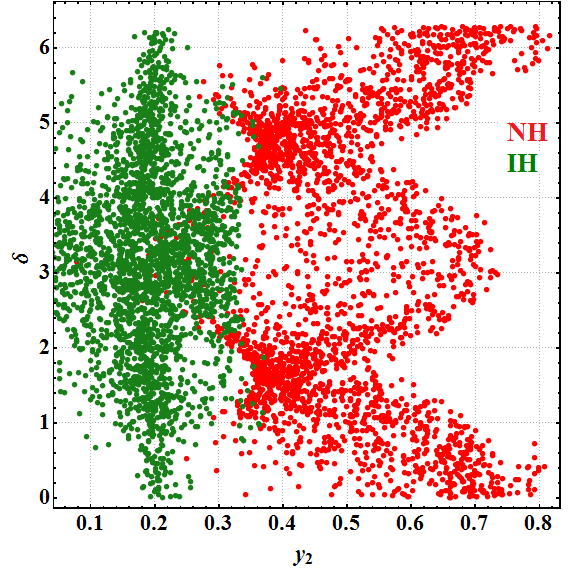}
	\includegraphics[width=2.8in,height=2.8in, angle=0]{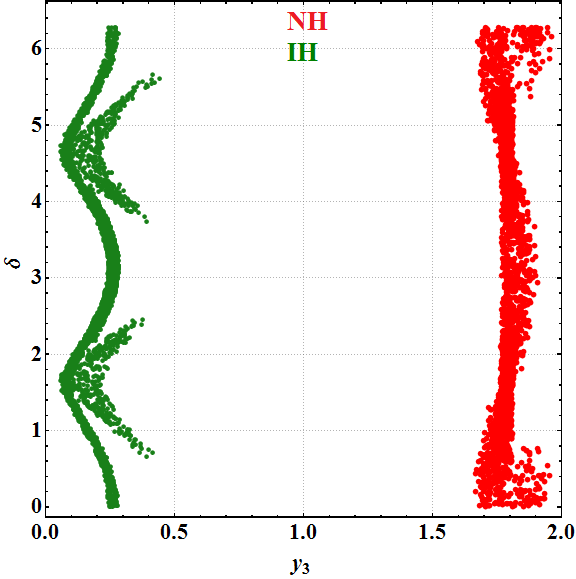}
	\caption{Constrained region in Dirac CP-phase can be seen with the Yukawa coupling. Red dots represents the normal hierarchy while green dots represents inverted hierarchy.} \label{ydel}
\end{figure}
\begin{figure}[h!]
	\includegraphics[width=2.8in,height=2.8in, angle=0]{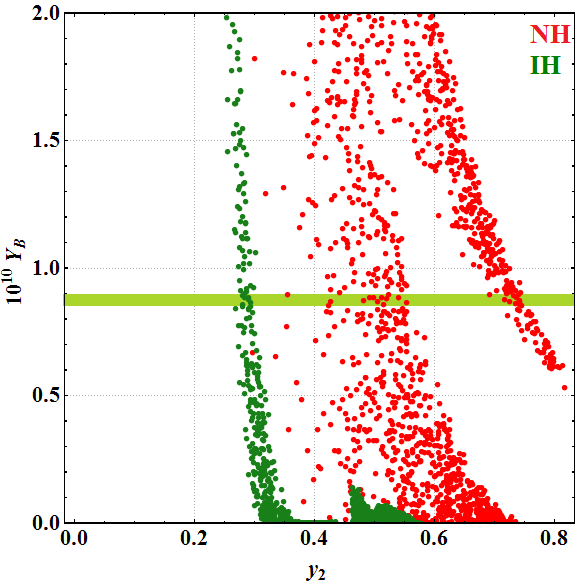}
	\includegraphics[width=2.8in,height=2.8in, angle=0]{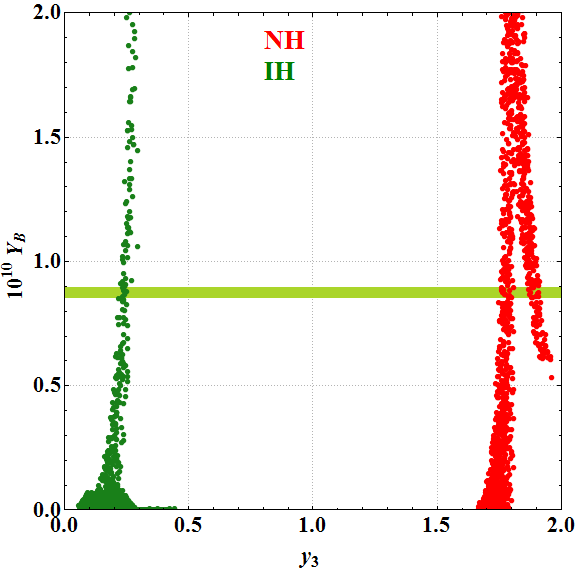}
	\caption{Variation of Yukawa coupling with the baryon asymmetry of the Universe. Red dots represents NH and black dots represents IH. The greenish band give the current BAU bound, which is $(8.75\pm0.23)\times10^{-11}$.}\label{ybau}
\end{figure}
%%%%%%%%%%%%%%%%%%%%%%%%%%%%%%%%%%%%%%%%%%%%%%%%%
%%%%%%%%%%%%%%%%%%%%%%%%%%%%%%%%%%%%%%%%%%%%%%%%%
%%%%%%%%%%%%%%%%%%%%%%%%%%%%%%%%%%%%%%%%%%%%%%%%%
%%%%%%%%%%%%%%%%%%%%%%%%%%%%%%%%%%%%%%%%%%%%%%%%%

Here, we present baryogenesis, including the neutrino mass and mixing angle constraints. To achieve the observed bound on the baryon asymmetry of the Universe, the Yukawa matrix must be non-zero and complex. 
After solving the model parameters using the latest global fit 3$\sigma$ values of the light neutrino parameters, we can construct the Yukawa matrix. 
Normal and inverted hierarchy are studied simultaneously in the model, and their results are discussed in this work. 
We found interesting bounds on Yukawa coupling with the Dirac CP-phase and the baryon asymmetry of the Universe. 
Even though all the light neutrino parameters along with the Higgs VEV depends upon the Yukawa couplings, the significant contributions can be observed in the reactor mixing angle ($\theta_{13}$) and all the mass squared differences.  

Here, $y_2$ corresponds to the SM Higgs while $y_3$ corresponds to the second Higgs doublet (see equation~\ref{lag}). 
Major constrains on experimental parameters like $\theta_{13}$ and $m_2$ are coming from $y_2$, which can be seen in the first plot of fig.~\ref{ydel} and \ref{ybau}.
Meanwhile, in the case of $y_3$, along with $\theta_{13}$ and $m_2$, drastic constraints are observed in $m_3$ as well. These dependencies reflect on the second plot of both the fig.~\ref{ydel} and ~\ref{ybau}. The Red dots represent the  NH, and green dots represent IH in both the plots.
In fig.~\ref{ydel}, variation of Yukawa couplings ($y_2,y_3$) are shown with the Dirac CP-phase.  We found that $y_2$ gets constrained between 0.3-0.7 in NH while for IH large numbers of points are accumulated in between 0.1 and 0.3. Similarly for $y_3$, NH mode is spread around 1.70-1.9 and for IH, its value lie within 0.1-0.5, due to the vanishing lightest neutrino mass ($m_3$). In current dim-5 situation, the Yukawa coupling of  $\mathcal{O}(10^{-2}-1)$~\cite{Abada:2007ux, Ibarra:2010xw} are in acceptable range, however, within our study, the large and small values of $y_3$ violates the experimental range of $\Delta m_{31}^2$. For example, large $y_3(\ge2.0)$, $\Delta m_{31}^2$ value exceed the current upper bound of $3\sigma$ value, whereas small $y_3(\le 0.2)$, $\Delta m_{31}^2$ value goes beneath the lower $3\sigma$ bound. 

As BAU value is highly sensitive to the experimental results, very narrow regions are observed with $y_2$ and $y_3$, which  are shown in fig.~\ref{ybau}. Results corresponding to the Yukawa couplings and BAU are shown in fig.~\ref{ybau}, which verify the successful execution of BAU within the MES framework for both the mass orderings. Large excluded regions in fig. \ref{ybau} are due to the bounds on light neutrino parameters imposed by the Yukawa matrix involved in the baryogenesis calculation. Similar analogy from fig. \ref{ydel}, regarding the bounds on $y_2$ and $y_3$ also works in fig. \ref{ybau}.
%%%%%%%%%%%%%%%%%%%%%%%%%%%%%%%%%%%%%%%%%%%%%%%%%
%%%%%%%%%%%%%%%%%%%%%%%%%%%%%%%%%%%%%%%%%%%%%%%%%
%%%%%%%%%%%%%%%%%%%%%%%%%%%%%%%%%%%%%%%%%%%%%%%%%
%%%%%%%%%%%%%%%%%%%%%%%%%%%%%%%%%%%%%%%%%%%%%%%%%
\section{Summary}\label{s5}
In this chapter, we have explored a $A_4$ based flavor model along with $Z_4$ discrete symmetry to establish tiny active neutrino mass. Along with this, the generation of non-zero reactor mixing angle ($\theta_{13}$), we simultaneously carried out multi Higgs doublet framework where one of the lightest odd particles behaves as DM candidate. This work is an extension of our previous work on active-sterile phenomenology, explained in chapter \ref{Chapter2}~\cite{Das:2018qyt}. Hence, we do not carry out the sterile neutrino phenomenology part in this chapter. Apart from neutrino phenomenology, the scalar sector is also discussed in great details. Three sets of SM like Higgs doublets are considered, where two of them acquire some VEV after EWSB and take part in the fermion sector, in particular, they involves in tiny neutrino mass generation. On the other hand, the third Higgs doublet does not acquire any VEV due to the additional $Z_4$ symmetry. 
As a result, the lightest odd particle becomes a viable candidate for dark matter in our model.

In this minimal extended version of the type-I seesaw, we successfully achieved the non-zero $\theta_{13}$ by adding a perturbation in the Dirac mass matrix for both the mass orderings.
The involvement of the high scaled VEVs of $A_4$ singlet flavons $\xi$ and $\xi^{\prime}$ ensure the $B-L$ breaking within our framework, which motivates us to study baryon asymmetry of the Universe within this framework. As the RH masses are considered in non-degenerate fashion, we have been able to produce desired lepton asymmetry (with anomalous violation of $B+L$ due to chiral anomaly), which eventually is converted to baryon asymmetry by the $sphaleron$ process.

The influence of the Higgs doublets ($\phi_1,\phi_2$) can be seen both in fermion as well as the scalar sector. In the fermion sector, the involvement of the Higgs doublets are related to the model parameters via the Yukawa couplings. Relations between the constrained Dirac CP-phase and satisfied baryogenesis results for two different Yukawa couplings are shown, which is related via two Higgs' VEV. On the other hand, in the case of the scalar sector, a large and new DM mass region in the parameter spaces is obtained due to the presence of other heavy particles. 
We are successfully able to stretch down the  DM mass at limit up to 42.2 GeV, satisfying all current bounds from various constraints.  
% Chapter Template
\begin{savequote}[1\linewidth]
	``There are two ways to slide easily through life; to believe everything or to doubt everything. Both ways save us from thinking.''
	\qauthor{\large\it 
		Alfred Korzybski (1879--1950)}
\end{savequote}
\chapter{Phenomenology of $keV$ sterile neutrino in minimal extended seesaw} % Main chapter title
%\vspace{-2.1cm}
%{\it ``It doesn't matter how beautiful your theory is, it doesn't matter how smart you are. If it doesn't agree with the experiment, it's wrong."--- {\Large Richard P. Feynman}}
%{\it ``There are two ways to slide easily through life; to believe everything or to doubt everything. Both ways save us from thinking."---{ \Large A. Korzybski}}
\label{Chapter4} % Change X to a consecutive number; for referencing this chapter elsewhere, use \ref{ChapterX}

\lhead{Chapter 4. \emph{Phenomenology of $keV$ sterile neutrino in minimal extended seesaw}} % Change X to a consecutive number; this is for the header on each page - perhaps a shortened title

%----------------------------------------------------------------------------------------
%	SECTION 1
%----------------------------------------------------------------------------------------
\section*{}
\vspace*{-2cm}
 In this chapter, we have explored the possibility of a $keV$ scale sterile neutrino ($m_S$) as a dark matter candidate within the minimal extended seesaw (MES) framework and it's influenced in neutrinoless double beta decay ($0\nu\beta\beta$) study. Three hierarchical right-handed neutrinos were considered to explain neutrino mass. We also address baryogenesis via the mechanism of thermal leptogenesis. A generic model based on $A_4\times Z_4\times Z_3$ flavour symmetry has constructed to explain both normal and inverted hierarchy mass pattern of neutrinos. Significant results on effective neutrino masses are observed in presence of sterile mass ($m_S$) and active-sterile mixing ($\theta_{S}$) in $0\nu\beta\beta$. To establish sterile neutrino as dark matter within this model, we checked decay width and relic abundance of the sterile neutrino, which restricted sterile mass ($m_S$) within some definite bounds. Constrained regions on the CP-phases and Yukawa couplings are obtained from $0\nu\beta\beta$ and baryogenesis results.  Co-relations among these observable are also established and discussed within this model.

\section{Introduction}

In this chapter, we are considering a sterile neutrino flavour with mass around $keV$ range in minimal extended seesaw (MES) \cite{Barry:2011wb,Zhang:2011vh,Rodejohann:2014eka, Das:2018qyt}, where an additional fermion singlet (sterile neutrino) is added along with three RH neutrinos. Sterile neutrino with eV as well as $keV$ could be probed in future KATRIN experiment \cite{Osipowicz:2001sq,Mertens:2014nha}. Moreover, $keV$ sterile neutrino has a potential to affect the electron energy spectrum in tritium $\beta$-decays \cite{Shrock:1980vy}. Typically, sterile neutrinos with mass (0.4-50) $keV$ \cite{Boyarsky:2009ix} are considered WIMP particles since they are relatively slow and much heavier than the active neutrinos. In fact for successfully observe $0\nu \beta\beta$ the upper bound for sterile neutrino mass should be 18.5 $keV$ \cite{Abada:2018qok, Benso:2019jog}.  %The lower bound on $m_S$ is kind of universal for sterile neutrino in order to establish itself as a DM candidate and
% Sterile neutrinos to be considered as DM candidate, they must satisfy the lower bound on the DM mass \cite{Tremaine:1979we} and the mixing angle between active and sterile neutrino must be very small such that it can decay only to an SM neutrino and a mono-energetic photon; otherwise, there would be an overabundance of DM. Taking all these points into consideration, 
Hence, choosing a mass range for the $keV$ regime sterile neutrino within (1-18.5) $keV$ and explore new possibilities to explain $0\nu\beta\beta$ within laboratory constraints along with DM signature is quite the right choice. Along with the sterile study, we have also verified baryogenesis produced $via$ the mechanism of thermal leptogenesis within our model. Finally, we have co-related all these observable under the same framework. 
%As seesaw demands lepton number violation, eventually new CP-violating phases in the neutrino Yukawa interactions is generated, and it is assumed that heavy singlet neutrinos decay out of equilibrium. Thus, all three Sakharov conditions are satisfied naturally in this scenario. One has to check whether this amount of lepton asymmetry would successfully able to give rise the appropriate amount of observed baryon asymmetry of the Universe. Non-hierarchical mass pattern for the RH neutrinos are considered satisfying current light neutrino bounds. Lightest of them is decaying to a lepton (anti-lepton) and a Higgs doublet producing lepton number violation, and then Standard Model sphaleron processes plays a crucial role in partially converting the lepton asymmetry into a baryon asymmetry of the Universe. 
\\
This chapter is organized as follows: the model has been addressed in section \ref{model3} then we have carried out the numerical procedure of the model in section \ref{5num}. Results of this study are discussed in the section \ref{result} and finally, we have summed up this chapter in section \ref{conc}. 
%\begin{table}[t]
%	\centering
%	\caption{Recent experiments results for active neutrinos parameters with best-fit and the latest global fit $3\sigma$ range \cite{Capozzi:2016rtj}.}\label{tab:d1}
%	\begin{tabular*}{\columnwidth}{@{\extracolsep{\fill}}llll@{}}
%		\hline

\section{Model framework}\label{model3}
This model has introduced a heavy chiral gauge fermion $S$, which behaves as a sterile neutrino. Unlike previous models, we have addressed sterile neutrino in $keV$ scale and study their consequences in this model. Discrete flavour symmetries like $A_4,Z_3$ and $Z_4$ are used to construct the model for normal and inverted hierarchy pattern.

\subsection{Particle content and discrete charges}
Apart from the type-I seesaw particle content, few extra flavours are added to construct the model.
Two triplets $\zeta, \varphi$, two singlets $\xi$ and $\xi^{\prime}$ are added to produce broken flavour symmetry. Besides the SM Higgs $H_1$, we also introduce an additional Higgs doublets ($H_2$) \cite{Felipe:2013vwa,Nath:2016mts}, to make the model work. Non-desirable interactions were restricted using extra $Z_4$ and $Z_3$ charges to the fields. To accommodate sterile neutrino into the framework, we add a chiral gauge singlet $S$, which interacts with the RH neutrino $\nu_{R1}$ via $A_4$ singlet ($1^{\prime}$) flavon $\chi$ to give rise to sterile mixing matrix. We used dimension-5 operators \cite{weinberg} for Dirac neutrino and charged lepton mass generation. One may notice, terms like $\frac{1}{\Lambda} \overline{S^c}S\varphi\varphi$ may ruin the current MES scenario by giving rise to unexpectedly higher mass term for the sterile neutrino \cite{Zhang:2011vh}. The $Z_3$ symmetry excludes those terms.
%\begin{table}[h]
%	\centering
%	\caption{Particle content and their charge assignments under SU(2), $A_4$ and $Z_4\times Z_3$ groups.}\label{tab1} 
%	\begin{tabular*}{\textwidth}{@{\extracolsep{\fill}}lrrrrrrrrrrrrrrrrl@{}}

As per the MES structure, the newly added singlet field $S$ does not interact with the active neutrinos, and they can be explained with the Abelian symmetries. For example, by introducing an additional $U(1)^{\prime}$ charge under which the SM particles and RH neutrinos are to be neutral. The singlet $S$, on the other hand, carries a $U(1)^{\prime}$ charge $Y^{\prime}$ and we further introduced a SM singlet $\chi$ with hypercharge $-Y^{\prime}$. Hence those coupling of $S$ with the active neutrinos are still forbidden by the $U(1)^{\prime}$ symmetry at the renormalizable level \cite{Barry:2011wb, Chun:1995js, Heeck:2012bz}. 

\subsubsection{Normal hierarchical neutrino mass}
The particle content with  $A_4 \times Z_4\times Z_3$ charge assignment under NH are shown in the table \ref{3tab1}.

\begin{table}[h]\centering
	\begin{tabular}{|c|cccccc|cccc|ccc|cc|}
%		\hline
		\hline
			Particles&  $l$ &  $e_{R}$&  $\mu_{R}$&  $\tau_{R}$&  $H_1$&  $H_2$&  $\zeta$&  $\varphi$&  $\xi$&  $\xi^{\prime}$&  $\nu_{R1}$&  $\nu_{R2}$&  $\nu_{R3}$&  $S$&  $\chi$\\
			\hline
			\hline
			SU(2)&2&1&1&1&2&2&1&1&1&1&1&1&1&1&1\\
			%	\hline
			$A_4$&3&1&$1^{\prime\prime}$&$1^{\prime}$&1&1&3&3&1&$1^{\prime}$&1&$1^{\prime}$&1&$1^{\prime\prime}$&$1^{\prime}$\\
			%	\hline
			$Z_4$&1&1&1&1&1&i&1&i&1&-1&1&-i&-1&i&-i\\
			%	\hline
			$Z_3$&1&1&1&1&1&1&1&1&1&1&1&1&1&$\omega^2$&$\omega$\\
			\hline
		%			\hline
		\end{tabular}
		\caption{Particle content and their charge assignments under SU(2), $A_4$ and $Z_4\times Z_3$ groups for NH mode.}\label{3tab1}
\end{table}

In lepton sector, the leading order invariant Yukawa Lagrangian is given by,
\begin{equation}\label{lag}
	\begin{split}
		\mathcal{L} =& \frac{y_{e}}{\Lambda}(\overline{l} H_1 \zeta)_{1}e_{R}+\frac{y_{\mu}}{\Lambda}(\overline{l} H_1 \zeta)_{1^{\prime}}\mu_{R}+\frac{y_{\tau}}{\Lambda}(\overline{l} H_1 \zeta)_{1^{\prime\prime}}\tau_{R} \\
		&+ \frac{y_{2}}{\Lambda}(\overline{l}\tilde{H_1}\zeta)_{1}\nu_{R1}+\frac{y_{2}}{\Lambda}(\overline{l}\tilde{H_1}\varphi)_{1^{\prime\prime}}\nu_{R2}+\frac{y_{3}}{\Lambda}(\overline{l}\tilde{H_2}\varphi)_{1}\nu_{R3}\\
		& +\frac{1}{2}\lambda_{1}\xi\overline{\nu^{c}_{R1}}\nu_{R1}+\frac{1}{2}\lambda_{2}\xi^{\prime}\overline{\nu^{c}_{R2}}\nu_{R2}+\frac{1}{2}\lambda_{3}\xi\overline{\nu^{c}_{R3}}\nu_{R3}\\
		&+ \frac{1}{2}\rho\chi\overline{S^{c}}\nu_{R1} .\\
	\end{split}
\end{equation}
In this Lagrangian, various Yukawa couplings are represented by $y_{\alpha,i}$, $\lambda_{i}$ (for $\alpha=e,\mu,\tau$ and $i=1,2,3$) and $\rho$ for respective interactions. Higgs doublets are transformed as $\tilde{H} = i\tau_{2}H^*$ ($\tau_{2}$ is the second Pauli's spin matrix) to keep the Lagrangian gauge invariant and $\Lambda$ is the cut-off scale of the theory, which is around the GUT scale. 
The scalar flavons involved in the Lagrangian acquire VEV along  $ \langle \zeta \rangle=(v,0,0), \langle\varphi\rangle=(v,v,v), 
\langle\xi\rangle=\langle\xi^{\prime}\rangle=v$ and $ \langle\chi\rangle=v_{\chi}$ by breaking the flavor symmetry, while $\langle H_i\rangle(i=1,2)$ get VEV ($v_i$) by breaking EWSB at electro-weak scale.
%VEV alignments of the extra flavons are required to generate the desired light neutrino mass matrix.
%\begin{equation*}
%\begin{split}
%&\langle \zeta \rangle=(v,0,0),\\
%& \langle\varphi\rangle=(v,v,v),\\
%& \langle\varphi^{\prime}\rangle=(0,v,-v),\\
%& \langle\xi\rangle=\langle\xi^{\prime}\rangle=v,\\
%& \langle\chi\rangle=u.
%\end{split}
%\end{equation*}
Following the $A_4$ product rules and using the above mentioned VEV alignment\footnote{The triplet VEV alignment of the scalars are the solution of the respective scalars at their minimal potential, discussed in appendix \ref{appn1}}, the Dirac neutrino mass\footnote{ $M_D^{\prime}$ is the unmodified Dirac neutrino mass matrix which is failed to generate $\theta_{13}\neq 0$. The modified $M_D$ is represented in equation \eqref{md} }, Majorana neutrino mass and the sterile mass matrices are given by,
\begin{equation}
	M^{\prime}_{D}=
	\begin{pmatrix}
		D_1&D_1&D_2\\
		0&D_1&D_2\\
		0&D_1&D_2\\
	\end{pmatrix},\ M_{R}=\begin{pmatrix}
		R_1&0&0\\
		0&R_2&0\\
		0&0&R_3\\
	\end{pmatrix}, \
	M_{S}= \begin{pmatrix}
		G&0&0\\
	\end{pmatrix}.
	\label{3emd}
\end{equation}
where, $D_1=\frac{\langle H_1\rangle v}{\Lambda}y_{2} $ and $D_2=\frac{\langle H_2\rangle v}{\Lambda}y_{3}$\footnote{We have assumed the VEV of the Higgs doublets to be identical for simplicity.}. Other elements are defined as $R_1=\lambda_{1}v, R_2=\lambda_{2}v$, $R_3=\lambda_{3}v$ and $G=\rho v_\chi$.  In order to achieve sterile mass in the $keV$ range, we have considered VEV for the $\chi$ flavon lie aroud TeV scale. A rough estimate of the mass scales of parameters are given as , $\Lambda\simeq 10^{14}$ GeV, $v\simeq 10^{13}$ GeV and $v_{\chi}\simeq 10$ TeV. \\
We have used similar approaches from our previous models from chapter \ref{Chapter2} and \ref{Chapter3} to break the trivial $\mu-\tau$ symmetry in the light neutrino mass matrix. We have introduced two new $SU(2)$ singlet flavon fields ($\zeta^{\prime}$ and $\varphi^{\prime}$)  which results the $M_P$ matrix \eqref{pmatrix} when they couple with the respective RH neutrinos. The active mass matrix gets modify by adding the matrix \eqref{pmatrix} to the Dirac neutrino mass matrix. The Lagrangian that generates the matrix \eqref{pmatrix} can be written as, 
\begin{equation}
	\mathcal{L}_{\mathcal{M_P}} =\frac{y_{1}}{\Lambda}(\overline{l}\tilde{H_1}\zeta^{\prime})_{1}\nu_{R1}+\frac{y_{1}}{\Lambda}(\overline{l}\tilde{H_1}\varphi^{\prime})_{1^{\prime\prime}}\nu_{R2}+\frac{y_{1}}{\Lambda}(\overline{l}\tilde{H_2}\varphi^{\prime})_{1}\nu_{R3}.
\end{equation}
New $SU(2)$ singlet flavon fields ($\zeta^{\prime}$ and $\varphi^{\prime}$) are considered and supposed to take $A_4\times Z_4\times Z_3$ charges as same as $\zeta$ and $\varphi$ respectively. After breaking flavor symmetry they acquire VEV along $\langle\zeta^{\prime}\rangle=(v_p,0,0)$ and $\langle\varphi^{\prime}\rangle=(0,v_p,0)$ directions, giving rise to the $M_P$ matrix as,
%The $SU(2)$ singlet flavon fields ($\eta,\eta^{\prime}$) are supposed to take $A_4\times Z_4$ charges as same as $\varphi$ and $\varphi^{\prime}$ respectively (as shown in the table \ref{tab1}). Now, considering VEV\footnote{This matrix $M_P$ is considered as a perturbation into the Dirac mass matrix system hence value of the VEV are considered a order less than the VEV that are involved in $M_D^{\prime}$} for the new flavon fields as  $\langle \eta \rangle=(0,v_p,0)$ and $\langle \eta^{\prime} \rangle=(0,0,v_p)$, we get the matrix as, 
% We will not write like this The present structure of Dirac neutrino mass matrix does not produce non-vanishing reactor mixing angle. Therefore  
\begin{equation}\label{pmatrix}
	M_{P}=
	\begin{pmatrix}
		0&0&P\\
		0&P&0\\
		P&0&0\\
	\end{pmatrix},
\end{equation}
with, $P=\frac{\langle H_{i}\rangle v}{\Lambda}y_{1}$ ($i=$1 or 2).
Scale of these VEV ($v_p$) in comparison to earlier flavon's VEV ($v$) are differ by an order of magnitude ($v>v_p$). Involvement of these new flavons are restricted in the leading order Lagrangian charged lepton mass matrix. Hence, the Dirac neutrino mass matrix, $M_D$ from equation \eqref{3emd} will take new structure as,
\begin{equation} \label{md}
	M_D=M^{\prime}_D+M_P=
	\begin{pmatrix}
		D_1&D_1&D_2+P\\
		0&D_1+P&D_2\\
		P&D_1&D_2\\
	\end{pmatrix}.
\end{equation}
\subsubsection{Inverted hierarchical neutrino mass} \label{cih}
The particle content with  $A_4 \times Z_4\times Z_3$ charge assignment under IH are shown in the table \ref{3tab2}.
\begin{table}[h]
	\begin{tabular}{|c|cccccc|ccccc|ccc|cc|}
%	\hline
	\hline
			Particles&  $l$ &  $e_{R}$&  $\mu_{R}$&  $\tau_{R}$&  $H_1$&  $H_2$&  $\zeta$&  $\varphi$& $\varphi^{\prime\prime} $&$\xi$&  $\xi^{\prime}$&  $\nu_{R1}$&  $\nu_{R2}$&  $\nu_{R3}$&  $S$&  $\chi$\\
			\hline
			\hline
			SU(2)&2&1&1&1&2&2&1&1&1&1&1&1&1&1&1&1\\
			%	\hline
			$A_4$&3&1&$1^{\prime\prime}$&$1^{\prime}$&1&1&3&3&1&$1^{\prime}$&1&1&$1^{\prime}$&1&$1^{\prime\prime}$&$1^{\prime}$\\
			%	\hline
			$Z_4$&1&1&1&1&1&i&1&1&i&1&-1&1&-i&-1&i&-i\\
			%	\hline
			$Z_3$&1&1&1&1&$\omega$&1&1&1&1&$\omega^2$&$\omega^2$&$\omega^2$&$\omega$&$\omega^2$&$\omega^2$&$\omega$\\
			\hline
			%		\hline
		\end{tabular}
		\caption{Particle content and their charge assignments under SU(2), $A_4$ and $Z_4\times Z_3$ groups for IH.}\label{3tab2}
\end{table}
Within MES, the situation is not that simple for IH mode \cite{Das:2018qyt, Zhang:2011vh}. A slight change in VEV arrangement is required in IH mode to give correct observed phenomenology \cite{Zhang:2011vh}. A new triplet flavon $\varphi^{\prime\prime}$ with VEV alignment along $\langle\varphi^{\prime\prime}\rangle \sim (2v,-v,-v)$ is introduced, which modifies the Dirac neutrino mass matrix. Particles and charges under symmetry groups ($SU(2)\times A_4\times Z_4 \times Z_2$ ) are shown in table \ref{3tab2}. The modified Yukawa Lagrangian for the $M_D$ is given by,
\begin{equation}
	\mathcal{L}_{\mathcal{M_D}}= \frac{y_{2}}{\Lambda}(\overline{l}\tilde{H_1}\zeta)_{1}\nu_{R1}+\frac{y_{2}}{\Lambda}(\overline{l}\tilde{H_1}\varphi^{\prime\prime})_{1^{\prime\prime}}\nu_{R2}+\frac{y_{3}}{\Lambda}(\overline{l}\tilde{H_1}\varphi)_{1}\nu_{R3}.
\end{equation}
Except the Dirac Lagrangian, other Lagrangian will retain the same form as per the equation \ref{lag}. The Dirac neutrino mass matrix takes new structure as, 
\begin{equation}
	M^{\prime}_{D}=
	\begin{pmatrix}
		D_1&-D_1&D_2\\
		0&-D_1&D_2\\
		0&2D_1&D_2\\
	\end{pmatrix}.
\end{equation}
Like the NH case, this Dirac neutrino mass matrix also gets modified by adding the $M_P$ matrix.  We have shown the complete matrix structure for both the mass ordering in the table \ref{5tab:nh}

\section{Phenomenology of $keV$ sterile neutrino}\label{5num}
%Following the minimal extended seesaw (MES) framework we set up active and sterile neutrino mass matrices. 

%\begin{table}[h]
%	\centering
\begin{table}[h]\tiny
	\centering
	\begin{tabular}{|p{3.7cm}cp{.4cm}p{0.6cm}|}
		\hline
	%	\hline
	~~~~~~~~	Structures&  $-m_{\nu}$&$m_s$ ($keV$)&$W$\\
		\hline
		\hline
		$\begin{aligned}& ~~~~~~~~~\text{Normal Heirarchy}\\&
			M_R=\begin{pmatrix}
				R_1&0&0\\
				0&R_2&0\\
				0&0&R_3\\
			\end{pmatrix}\\
			& M_{D}= \begin{pmatrix}
				D_1&D_1&D_2+P\\
				0&D_1+P&D_2\\
				P&D_1&D_2\\
			\end{pmatrix}\\
			& M_{S}= \begin{pmatrix}
				G&0&0\\
			\end{pmatrix}\\
		\end{aligned}$& $\begin{pmatrix}
			\frac {D_1^2} {R_2} + \frac {(D_2 + P)^2} {R_3} &\frac {D_1 (D_1 + 
				P)} {R_2} + \frac {D_2 (D_2 + 
				P)} {R_3} &\frac {D_1^2} {R_2} + \frac {D_2 (D_2 + P)} {R_3} \\
			\frac {D_1 (D_1 + P)} {R_2} + \frac {D_2 (D_2 + P)} {R_3} &\frac {(D_1 + 
				P)^2} {R_2} + \frac {D_2^2} {R_3} &\frac {D_1 (D_1 + 
				P)} {R_2} + \frac {D_2^2} {R_3} \\
			\frac {D_1^2} {R_2} + \frac {D_2 (D_2 + P)} {R_3} &\frac {D_1 (D_1 + 
				P)} {R_2} + \frac {D_2^2} {R_3} &\frac {D_1^2} {R_2} + \frac {D_2^2}
			{R_3} \\
		\end{pmatrix}$ &$\frac{G^2}{\lambda_{1}v}$&${\begin{pmatrix}
				\frac{D_1}{G}\\ 0\\ \frac{P}{G}\\
		\end{pmatrix}}$	\\
		\hline
		$\begin{aligned}
			&~~~~~~~~~\text{Inverted Hierarchy}\\&
			M_R=\begin{pmatrix}
				R_1&0&0\\
				0&R_2&0\\
				0&0&R_3\\
			\end{pmatrix}\\
			& M_{D}= \begin{pmatrix}
				D_1&-D_1&D_2+P\\
				0&-D_1+P&D_2\\
				P&2D_1&D_2\\
			\end{pmatrix}\\
			& M_{S}= \begin{pmatrix}
				G&0&0\\
			\end{pmatrix}\\
		\end{aligned}$
		& $ \begin{pmatrix}
			\frac {D_1^2} {R_2} + \frac {(D_2 + P)^2} {R_3} &  \frac {D_1(D_1 - 
				P)} {R_2} + \frac {D_2 (D_2 + 
				P)} {R_3} & \frac {-2 D_1^2} {R_2} + \frac {D_2 (D_2 + P)} {R_3} \\
			\frac {D_1(D_1 - P)} {R_2} + \frac {D_2 (D_2 + P)} {R_3} &\frac {(D_1 - 
				P)^2} {R_2} + \frac {D_2^2} {R_3} &\frac {-2 D_1 (D_1 - 
				P)} {R_2} + \frac {D_2^2} {R_3} \\
			-\frac {2 D_1^2} {R_2} + \frac {D_2 (D_2 + P)} {R_3} & - \frac {2 D_1 (D_1 - 
				P)} {R_2} + \frac {D_2^2} {R_3} &\frac {4 D_1^2} {R_2} + \frac {D_2^2}
			{R_3} \\
		\end{pmatrix}$	&$\frac{G^2}{\lambda_{1}v}$
		&${\begin{pmatrix}
				\frac{D_1}{G}\\ 0\\\frac{P}{G}\\
		\end{pmatrix}}$\\
		\hline
	\end{tabular}
	\caption{The active and sterile neutrino mass matrices and the corresponding Dirac ($M_D$), Majorana( $M_R$) and sterile( $M_S$) mass matrices for NH and IH mode. The active-sterile mixing matrices ($W$) and sterile mass for NH and IH mass pattern are also shown in respective columns. }\label{5tab:nh}
\end{table}
The active and active-sterile mass and mixing pattern under MES for both NH and IH patterns are illustrated in table \ref{5tab:nh}. A detailed discussion on the model and matrix structures are carried out in \ref{model3} section. 
We have assigned fixed non-degenerate values for the right-handed neutrino mass parameters as $R_1=\times10^{12}$ GeV, $R_2=10^{13}$ GeV and $R_3=5\times10^{13}$ GeV so that they can demonstrate favourable thermal leptogenesis without effecting the neutrino parameters. The mass matrix generated from equation~\eqref{eq:2} gives rise to complex parameters due to the Dirac and Majorana phases. As the leptonic CP phases are still unknown, we vary them within their allowed $3\sigma$ ranges (0, 2$\pi$).
We solved the model parameters of the active mass matrix using current global fit $3\sigma$ values for the light neutrino parameters, taken from \cite{Capozzi:2016rtj}.

This chapter mainly focuses on validating MES to study observable like neutrinoless double beta decay, dark matter, and baryogenesis in the presence of a $keV$ sterile neutrino ($m_S$) and finally, we will try to find correlation among those observable, which we have discussed in following sections. 

\subsection{$keV$ sterile neutrino as dark matter}
The most important criterion for a DM candidate is its stability, at least on the cosmological scale. The lightest sterile neutrino is not stable and may decay into SM particles. In the presence of sterile neutrinos, the leptonic weak neutral current is not diagonal in mass eigenstates \cite{PhysRevD.16.1444}, so the $S$ can decay at tree-level via $Z$-exchange, as $S\rightarrow \nu_i\overline{\nu_j}\nu_j$ , where $\nu_i,\nu_j$ are mass eigenstates. The $keV$ sterile neutrino decaying to the SM neutrinos (flavor eigenstates) via $S\rightarrow\nu_{\alpha}\nu_{\beta}\overline{\nu_{\beta}}$ gives the decay width as \cite{PhysRevD.16.1444,PhysRevD.25.766},
\begin{equation}
	\Gamma_{S \rightarrow 3_{\nu}}=\frac{G_F^2 m_{S}^5}{96\pi^3}\sin^2 \theta_{S}=\frac{1}{4.7\times 10^{10} sec}\Big(\frac{m_{S}}{50\ keV}\Big)^5\sin^2\theta_{S},
\end{equation}
where, $\theta_{S}$ and $m_{S}$ represents the active-sterile mixing angle and sterile mass respectively. This decay width must give a lifetime of the particle much longer than the age of the Universe. This put a bound on the mixing angle such that, 
\begin{equation}
	\theta_{S}< 1.1\times 10^{-7}\Big(\frac{50\ keV}{m_{S}}\Big)^5.
\end{equation}
\begin{figure}\centering
	\includegraphics[scale=0.45]{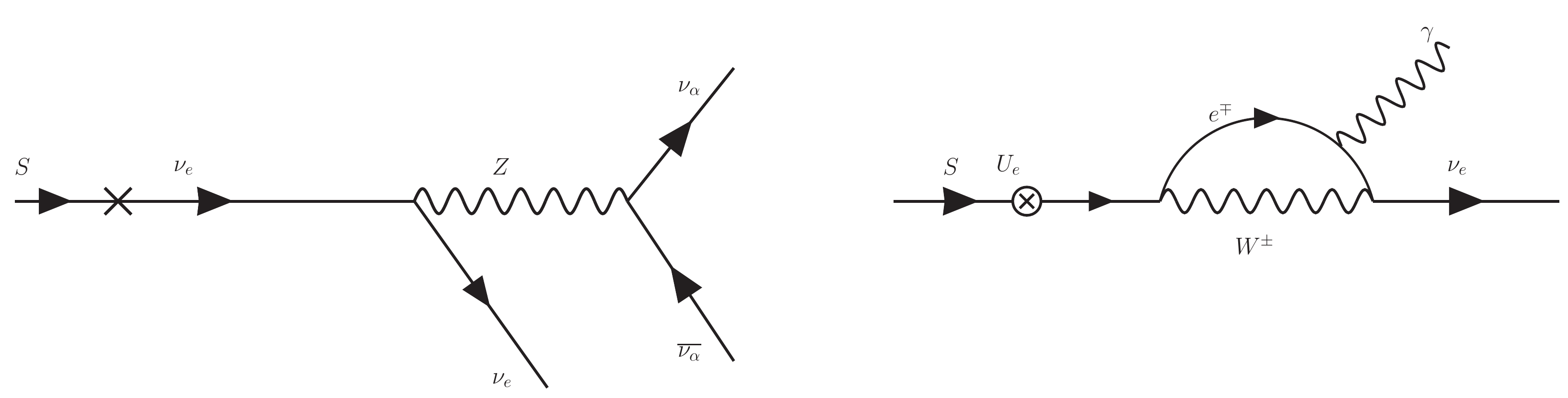}
	\caption{$S\rightarrow\nu_{\alpha}\nu_{\beta}\overline{\nu_{\beta}}\ $ (left) and $S\rightarrow \nu+\gamma$ (right) decay processed of the sterile neutrino \cite{Adhikari:2016bei}. Left figure gives dominant decay channel to three active neutrinos/anti-neutrinos and right figure shows loop mediated radiative decay channel that allows to look for the signal of sterile neutrino DM in the spectra of DM dominated objects.}\label{dm11}
\end{figure}
The mass squared difference emerging out of this bound is already much smaller than the current solar mass squared difference. To overcome this short come, either we use another sterile neutrino into the picture or consider one loop-mediated radiative decay process of $S\rightarrow \nu+\gamma$, as shown in fig. \ref{dm11}. This would put a stronger bound than the earlier $S \rightarrow 3_{\nu}$ decay process leading to a monochromatic X-ray line signal. However, as discussed in literature \cite{Adhikari:2016bei}, the decay rate is negligible on the cosmological scale because of the small mixing angle. The decay rate for the $S\rightarrow \nu+\gamma$ process is given as \cite{Abada:2014zra,Ng:2019gch}
\begin{equation}
	\Gamma_{S\rightarrow \nu\gamma}\simeq 1.32\times10^{-32}\Big(\frac{\sin^2 2\theta_{S}}{10^{-10}}\Big)\Big(\frac{m_{S}}{keV}\Big)^5\label{dm3}
\end{equation}
Relic abundance of the Universe can be worked out starting from the Boltzmann equation. We used results from \cite{Adhikari:2016bei,Abada:2014zra,Ng:2019gch} to check whether our model with $keV$ sterile neutrino can verify the observed relic abundance of DM. The working formula for relic abundance is given by,
\begin{equation}
	\Omega_{DM}h^2 \simeq 0.3 \Big(\frac{\sin^2 2\theta_{S\nu}}{10^{-10}}\Big)\Big(\frac{m_{S}}{100 keV}\Big)^2,
\end{equation}
with $\theta_{S\nu}$ is the sum of all the active-sterile mixing angles and $m_{S}$ represents the $keV$ ranged sterile neutrino mass. As seen from the above equations, decay rate and the relic abundance depend on the mixing and mass of the DM candidate. Hence, the same set of model parameters that are supposed to produce correct neutrino phenomenology can also evaluate the relic abundance and decay rate of the sterile neutrino. 
\section{Numerical analysis and results} \label{result}
\begin{figure}[h]\centering
	\includegraphics[scale=0.55]{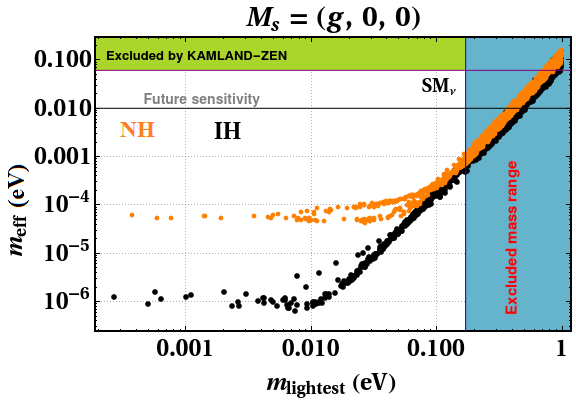}\\
	\includegraphics[scale=0.5]{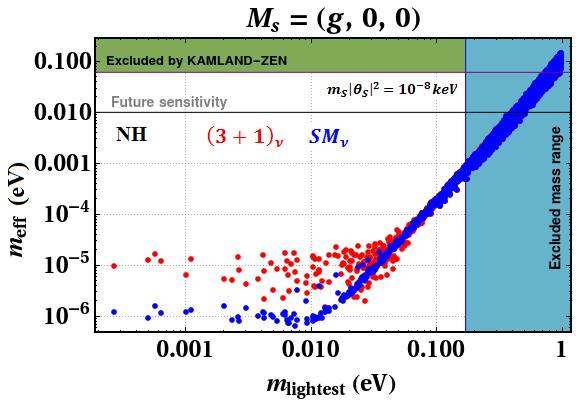}
	\includegraphics[scale=0.5]{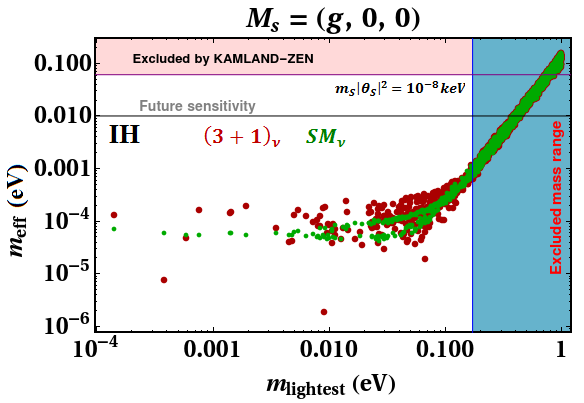}
	\caption{Variation of effective neutrino mass vs. the lightest neutrino mass. The upper plot represents the contribution from the active neutrinos only. The lower two plots represent NH and IH, respectively, for both active and active+sterile contributions. The horizontal grey line and vertical blue line represent the future $m_{eff}$ bound and the active masses' sum. In the presence of sterile neutrino, a much more comprehensive and significant impact is visible on the bottom plots. In both the mass orderings, we have fixed the mixing element $m_S|\theta_{S}|^2=10^{-8}~keV$.  }\label{eff}
\end{figure}
\begin{figure}\centering
	\includegraphics[scale=0.58]{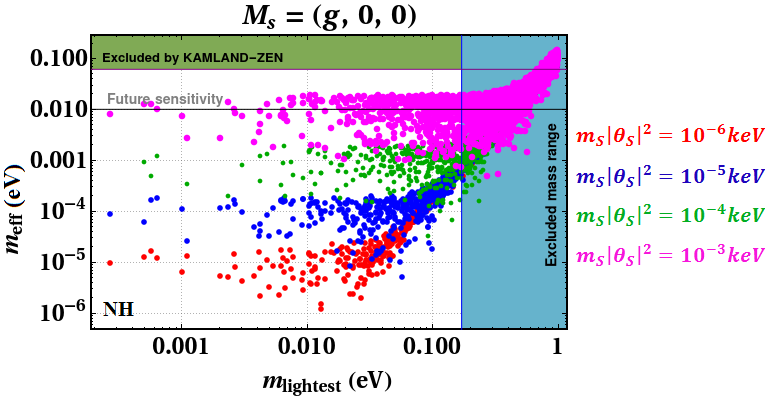}\\
	\includegraphics[scale=0.58]{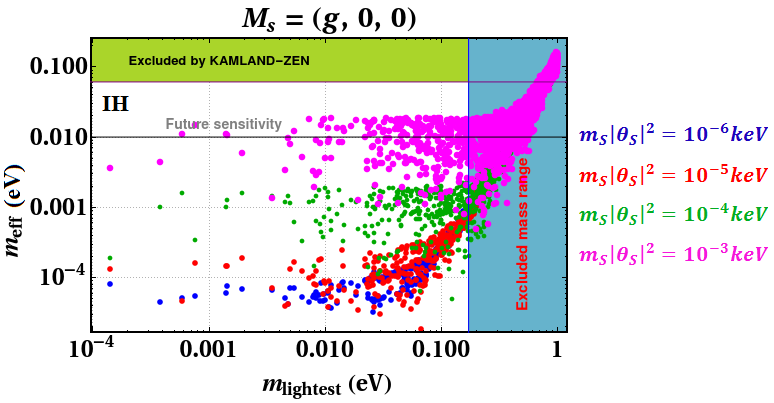}
	\caption{Variation of effective mass for different ranges of active-sterile mixing angle. The left plot represents the NH mode, while the right one represents the IH mode. In both the mass patterns, $m_S|\theta_{S}|^2>10^{-4}~keV$ fails to satisfy the future sensitivity bound of effective mass. This result gives upper bound on the active-sterile mixing element $``|\theta_{S}|^2"$ from the $0\nu\beta\beta$ study. }\label{u4}
\end{figure}
Under the hypothesis that future experiments will verify the existence of at least one heavy sterile neutrino in the $keV$ range, we worked out the possibility of its effect on $0\nu\beta\beta$ and verifying the fact that this sterile neutrino could behave as DM within the mass range of (1-18.5) $keV$. We have plotted effective neutrino mass ($m_{eff}$) against the lightest neutrino mass ($m_{lightest}$) in fig. \ref{eff}. The horizontal grey line gives the upper bound's future sensitivity on effective mass up to $10^{-2}$ eV, and the vertical blue line gives the upper bound on the sum of the active neutrino masses ($0.17$ eV). In the upper part of fig. \ref{eff}, NH (black) and IH (orange) contributions are coming only from the active neutrinos, whereas, in the lower two figures, NH and IH contributions are shown separately in the presence of $m_S$. In the sterile neutrino presence, one can observe a wider and improved data range in both the mass ordering. These extra contributions and improvements in effective mass are due to the sterile neutrino mass ($m_S$) and the active-sterile mixing ($\theta_{S}$). In fig. \ref{u4}, we completed the same analysis of $m_{eff}$ vs. $m_{lightest}$ for different orders of active-sterile mixing element. Exciting results are observed from both NH and IH mode. For $m_4|\theta_{S}|^2>10^{-4}$ $keV$, $0\nu\beta\beta$ fails the future experimental bound. From these results, we get the upper bound on the active-sterile mixing angles, and it is also obvious from the fact that the active-sterile mixing element must be tiny otherwise, there would be an overproduction of dark matter in our Universe \cite{Benso:2019jog}.

\begin{figure}\centering
	\includegraphics[scale=0.24]{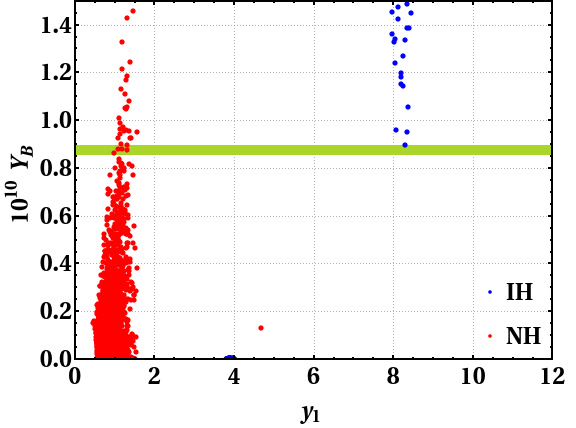}
	\includegraphics[scale=0.24]{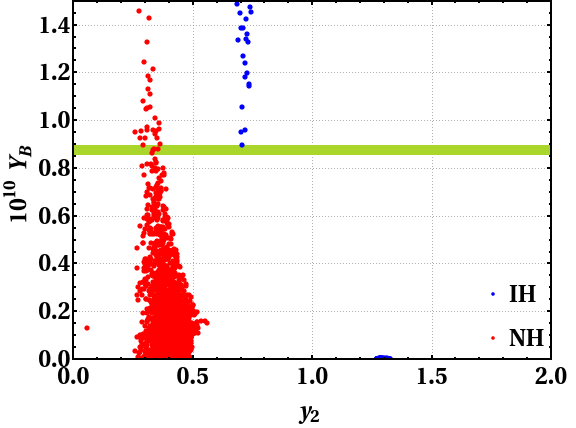}
	\includegraphics[scale=0.24]{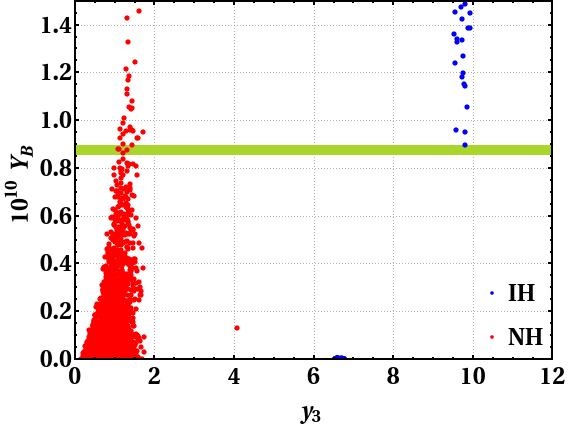}
	\caption{Variation of Yukawa coupling with BAU in both the mass ordering. Solid green band represents the current BAU value within 3$\sigma$ range, which is $(8.7\pm0.06)\times 10^{-11}$. Red and blue points represent NH and IH mass ordering, respectively. Stringent regions on the Yukawa couplings are due to recent bounds on the light neutrino parameters. }\label{yib}
\end{figure}
Variation of Yukawa couplings with the baryogenesis results are shown in fig. \ref{yib}. The red dots represent NH and the blue dots represent IH, respectively. The green bar gives the current allowed 3$\sigma$ value of BAU. As BAU value is highly sensitive to the experimental results, very narrow regions are observed in both the mass ordering satisfy baryogenesis in our model. NH shows more favourable results when we vary BAU with the Yukawa couplings. In current dimension-5 scenario, the Yukawa coupling of  $\mathcal{O}(10^{-2}-1)$~\cite{Abada:2007ux, Ibarra:2010xw, Das:2017nvm} are in acceptable range. Strong constrained regions in fig. \ref{yib} are due to the bounds on light neutrino parameters imposed by the Yukawa matrix involved in the baryogenesis calculation. These constrained regions of Yukawa couplings also put stringent bounds on the light neutrino parameters. For example, within NH, for large $y_3(\ge2.0)$, $\Delta m_{31}^2$ value exceed the current upper bound of $3\sigma$ value, whereas small $y_3(\le 0.2)$, $\Delta m_{31}^2$ value goes beneath the lower $3\sigma$ bound.
\begin{figure}\centering
%	\begin{minipage}[t]{0.37\textwidth}
		\includegraphics[scale=0.39]{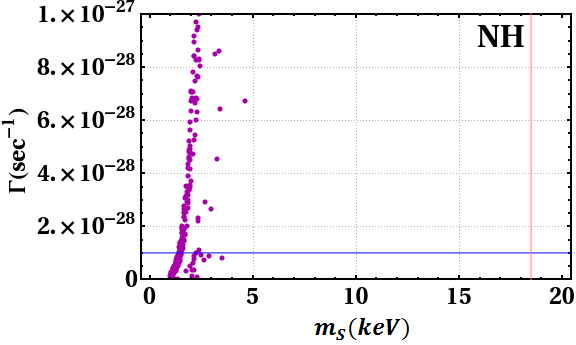}
		\includegraphics[scale=0.35]{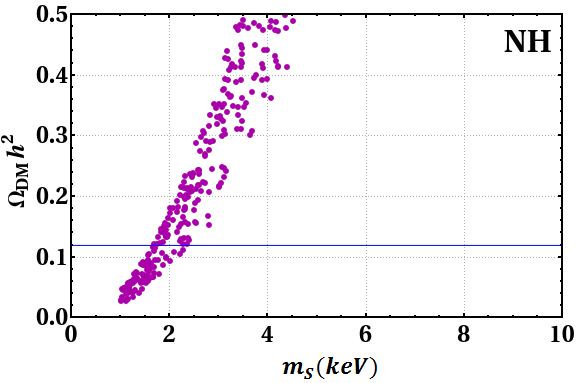}\\
		\includegraphics[scale=0.39]{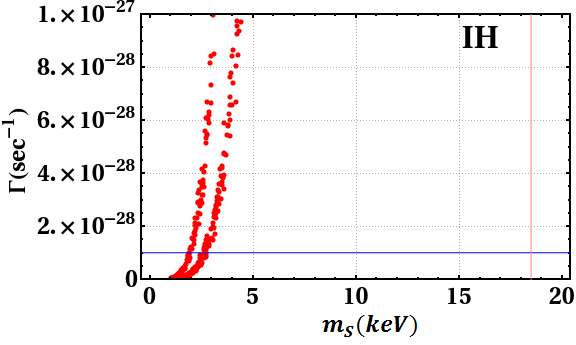}
		\includegraphics[scale=0.35]{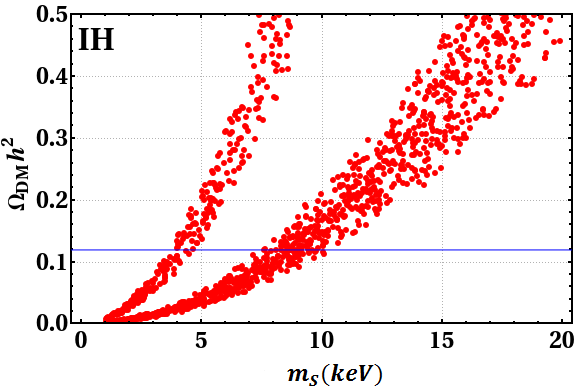}\\
%	\end{minipage}
%	\begin{minipage}[t]{0.42\textwidth}
%	\end{minipage}
	\caption{Variation of decay width ($\Gamma$) and the relic abundance ($\Omega_{DM}h^2$) of the Universe vs. the sterile neutrino mass ($m_S$). For decay width plots, the blue lines give the upper limit of the decay width, which is considered to be $\Gamma<10^{-28} sec$ and for relic abundance plots, the solid blue line gives the current best fit value for relic abundance of a particle to behave as a dark matter ($\Omega_{DM}h^2=0.119$). }\label{dm}
\end{figure}
Parallel to the $0\nu\beta\beta$ study, we have also examined the dark matter signature of the $keV$ sterile neutrino in fig. \ref{dm}. Decay width ($\Gamma$) and relic abundance of the sterile neutrino ($\Omega_{DM}h^2$) are plotted against the sterile mass ($m_{S}$) for both the mass ordering. Sterile neutrinos to behave as a DM, their lifetime must be greater than the age of the Universe so that their remnants remain in the Universe; hence, the decay width of the particle must be very less. In our study, we have considered the upper limit of decay width to be less than $10^{-28}\ (sec^{-1})$. The sterile neutrino mass is considered in a narrow region, $i.e.$, $(1-18.5)$ $keV$ to be a relic particle. Relic abundance obtained in both the mass ordering, satisfy the proper bound with different $m_S$ ranges. The allowed mass range for the sterile neutrino is very narrow $(1-3)$ $keV$ in the case of NH mode, while a broad mass spectrum satisfies the upper relic abundance bound in IH mode $(1-10)$ $keV$. 
Recent results suggested $m_S\simeq7.1~keV$ and lifetime $\tau_{DM}\simeq10^{27.8\pm0.3}$ second \cite{Boyarsky:2014ska}.
Even though the decay width and the relic abundance of the sterile neutrino are satisfied in both the mass ordering, NH results are consistent with a sterile mass within (1-3) $keV$. On the other hand, in IH mode, the relic abundance limit is within the sterile mass range from 4 $keV$ to 10 $keV$, while decay width is satisfied with a small mass up to 3 $keV$.

%The range for relic abduance lie within $0.112-0.129$, as the WIMP sololy cann't contribute to the whole DM so we consider 50\% of the DM contributions are coming from the WIMPs and we check relic abduance value  around 0.06.  
\begin{figure}\centering
	\includegraphics[scale=0.32]{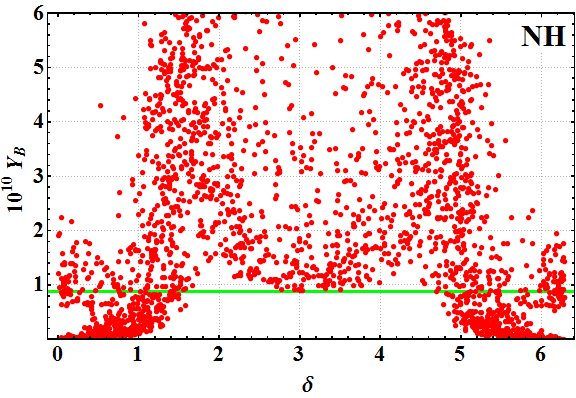}
	\includegraphics[scale=0.32]{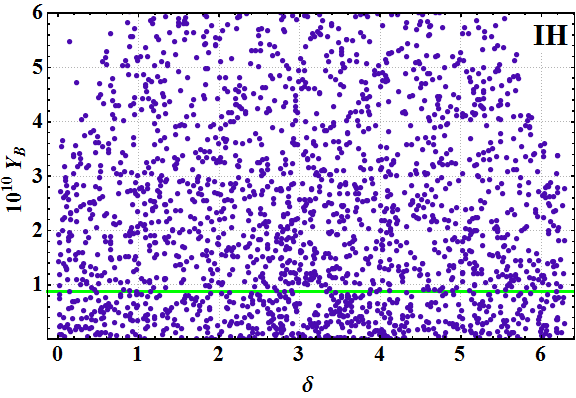}
	\caption{Variation of the Dirac delta phase with $Y_B$ in both the mass ordering. The solid green band represents the current BAU value, $Y_B=(8.7\pm0.23)\times 10^{-11}$. Both the mass orderings satisfy baryogenesis in our model and correlate with $\delta$. }\label{lep1}
\end{figure}
\begin{figure}\centering
	\includegraphics[scale=0.28]{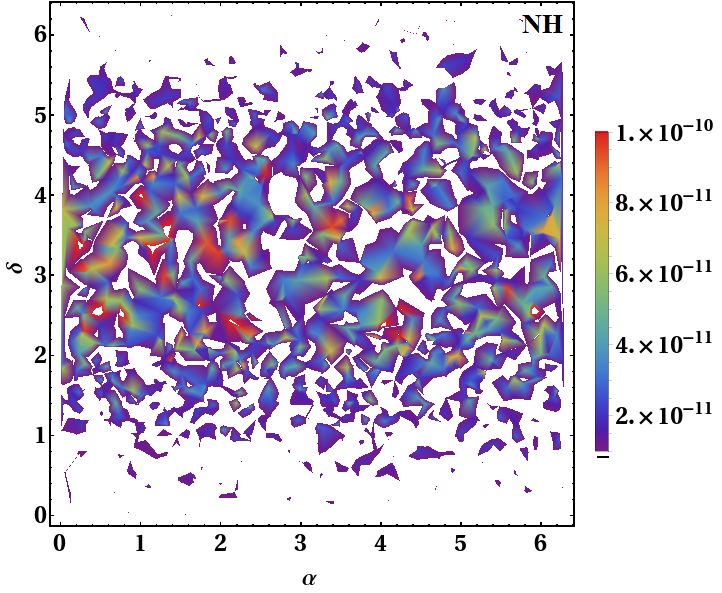}
	\includegraphics[scale=0.28]{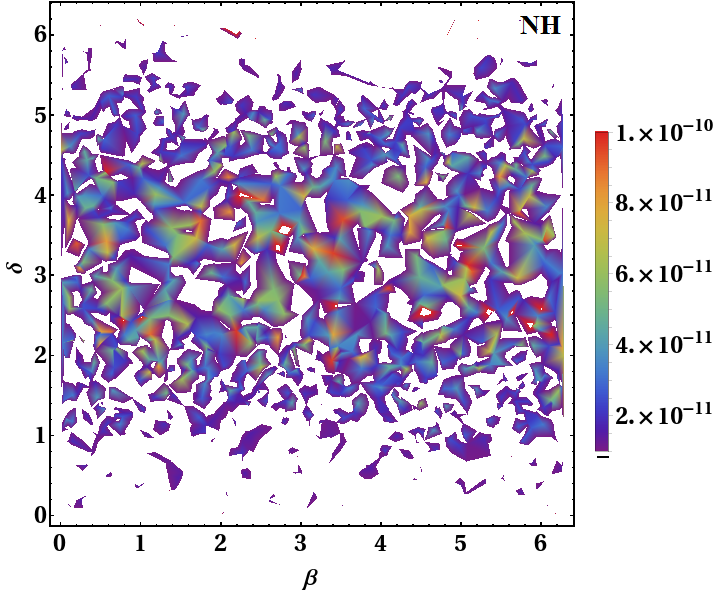}
	\caption{Projection of BAU value ($Y_B$) between Dirac CP-phase ($\delta$) along X-axis and Majorana phases ($\alpha$ and $\beta$ respectively) along Y-axis. Current BAU value range is around the red-orange colour band and this constrains the Dirac CP phase ($\delta$) in between the numerical values (2.0-4.0).}\label{delta}
\end{figure}
\begin{figure}\centering
	\includegraphics[scale=0.35]{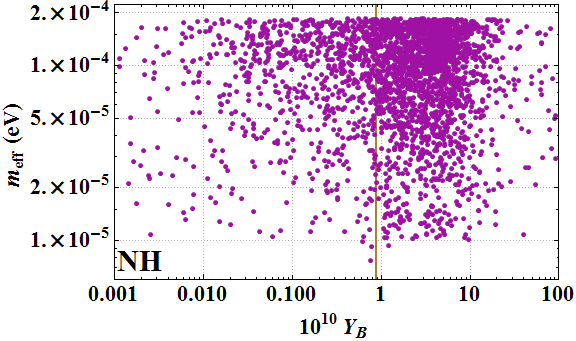}
	\includegraphics[scale=0.35]{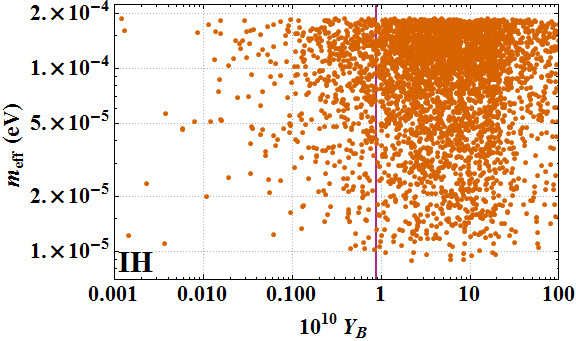}
	\caption{Correlation between effective neutrino mass ($m_{eff}$) with $Y_B$ in NH and IH, respectively. Solid vertical band represents the current BAU value, which is $(8.7\pm0.06)\times10^{-11}$. In both the cases, $m_{eff}$ lie well below the current upper bound and the solid vertical line indicates successful execution of baryogenesis and $0\nu\beta\beta$ in the model.}\label{ebau}
\end{figure}
\begin{figure}\centering
	\includegraphics[scale=0.29]{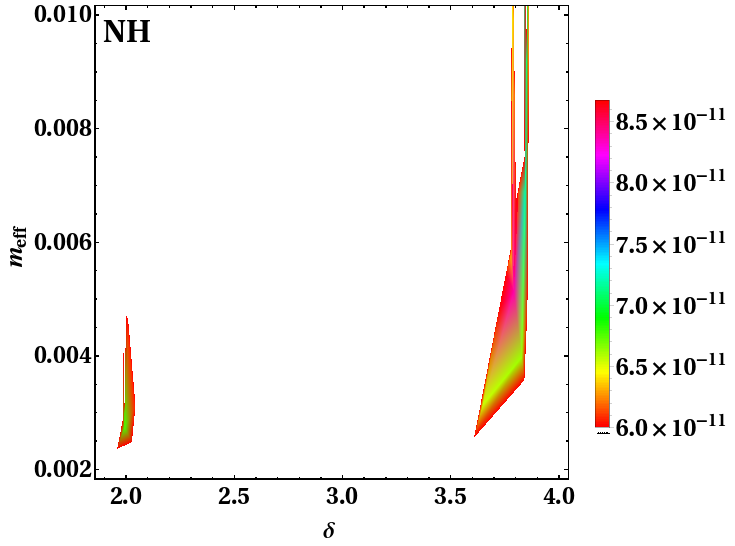}
	\includegraphics[scale=0.29]{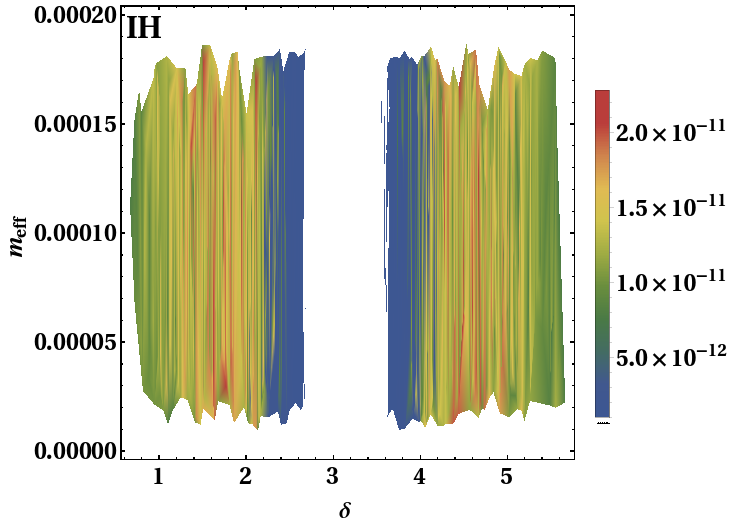}
	\caption{Projection of BAU value ($Y_B$) in a frame representing effective mass ($m_{eff}$) along Y-axis and the Dirac CP-phase ($\delta$) along X-axis. A precise constrained range for the Dirac CP-phase value around 3.5-4.0 is obtained for NH mode. At the same time, IH mode failed to reflect the exact $Y_B$ value. }\label{deb}
\end{figure}
\begin{figure}\centering
	\includegraphics[scale=0.28]{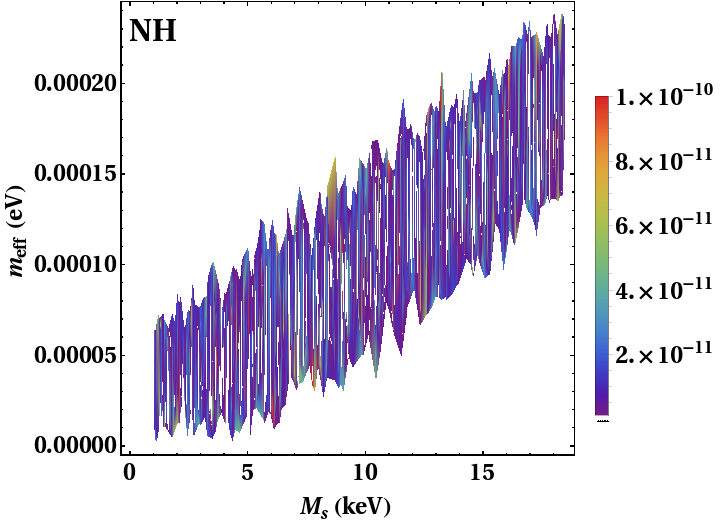}
	\includegraphics[scale=0.28]{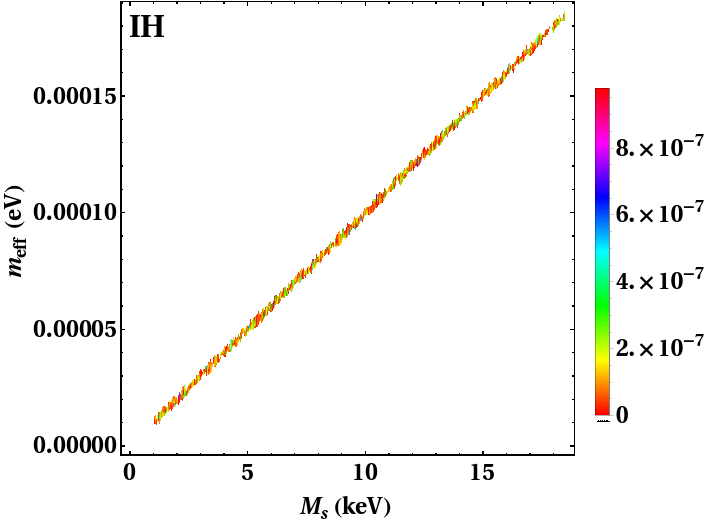}
	\caption{Projection of $Y_B$ in a frame representing effective mass ($m_{eff}$) along Y-axis and sterile mass ($m_S$ in $keV$) along X-axis.  }\label{seb}
\end{figure}

We also have checked the results for baryogenesis via thermal leptogenesis mechanism and showed a co-relation among other observable. In fig. \ref{lep1}, we varied the Dirac delta phase ($\delta$) with the baryon asymmetry of the Universe calculated in our model in both the mass orderings. Both these results show the validity of BAU within our model. Similar results can be seen from fig. \ref{delta}, where we project BAU in between Dirac and Majorana phases for NH mode only. Co-relation among the effective neutrino mass in the presence of $keV$ sterile neutrino with the BAU is also shown in fig. \ref{ebau}. Since the Dirac CP phase has influence in $0\nu\beta\beta$ as well as in BAU, we added a contour plot with $m_{eff}$ and $\delta$ along the axes and projected BAU value in the Z-plane in fig. \ref{deb}.  Constrained regions in the Dirac CP phase are observed in both the mass orderings. As we can see from the legends on the right-hand side of the figure, that IH pattern failed to project the currently observed value of BAU in a frame of $m_{eff}$ Dirac-CP phase $\delta$. We also present other contour plots in fig. \ref{seb}, where a measured BAU is projected in the frame between sterile neutrino mass ($m_S$) and effective electron neutrino mass. Since BAU results are very sensitive to the experiments, very narrow regions are observed. We can see that IH mode is almost ruled out in the presence of sterile mass and mixing while we get some region satisfying current BAU bound in NH mode. 
%\clearpage
\section{Summary} \label{conc}
In this chapter, we have studied the viability of $keV$ sterile neutrino to behave as a warm dark matter and giving an observable effect in $0\nu\beta\beta$ and baryogenesis $via$ the mechanism of thermal leptogenesis. We have constructed a mass model based on $A_4$ flavour model with discrete $Z_4\times Z_3$ charges. A singlet gauge fermion $S$ is considered, which couples with the right-handed neutrino, hence producing a single row  $(1\times 3)$  $ M_S$ matrix with one non-zero entry. The Dirac neutrino mass matrix is modified using a matrix, $ M_P$, which is generated via the same fashion as $M_D$ to make the active mass matrix $\mu-\tau$ asymmetric.
% By solving the model parameters, we have used their results to uphold further calculations.\\
%In our previous work \cite{Das:2018qyt}, we have considered the same framework with sterile neutrino mass lie within $eV$ range. The active phenomenology is isolated from the mass of the range of the sterile neutrino, so only active-sterile mixing bounds are considered in this work. The Dirac mass matrices in both the mass ordering are slightly modified to make the numerical analysis possible. Authors in \cite{Abada:2018qok} also carried out a similar kind of $0\nu\beta\beta$ discussion with one more sterile neutrino, while our study is limited to only one sterile state.  
 A Few interesting points based on the results are discussed as follows,
\begin{itemize}
	\item Presence of an extra heavy sterile flavour has a significant impact on effective neutrino mass. One can find a broader effective mass range in the active-sterile case than the active neutrino case.  Normal Hierarchy (NH) is more favourable than the inverted hierarchy (IH) mode for $0\nu\beta\beta$ in this MES framework.
	\item Consequential bound on active-sterile mixing angle is obtained for future sensitivity in effective mass from fig. \ref{u4}, which restricts the upper bound on the mixing element up to $10^{-4}$ for $|\theta_{S}|^2$.
	\item In fig. \ref{yib}, strongly constrained regions for the Yukawa couplings are obtained through baryogenesis calculation, which by the way, gives strict bounds on the choice model parameters.
	\item Dark matter analysis results from decay width and relic abundance restrict sterile neutrino mass within a few $keV$ to behave as dark matter. Among different bounds for the sterile neutrino's thermal relic mass, very few results are consistent with X-ray observations. Lyman-$\alpha$ forest of high resolution quasar spectra with hydrodynamical N-body simulations gives bounds ranging from $m_S \geq 1.8$ $keV$ to $m_S\geq 3.3$ $keV$ \cite{Seljak:2006qw,Viel:2013apy}. Regardless, these bounds may vary depending upon various uncertainties affecting the constraints \cite{Schultz:2014eia}. Within the MES framework, NH predicts sterile mass range from ($1-3$ $keV $) and IH results for relic abundance give mass up to 10 $ keV$ while the decay width constraints the mass within 3 $keV$.  Hence, from these results, we conclude that with current bounds on hand, sterile neutrino as a dark matter in minimal extended seesaw is still an unsettled aspect. A deeper discussion with new bounds on $keV$ sterile neutrino may resolve these issues, left for future studies.
	\item BAU is satisfied in this framework, and NH shows more efficient in producing the observed matter-antimatter density than the IH pattern. This model also successfully correlate $0\nu\beta\beta$ with the BAU result,
	which can be found in fig. \ref{ebau}. Projection of BAU on a plane between effective mass and Dirac CP phase, $\delta$ gives significant remark in our study. In fig. \ref{deb}, BAU results are constraining $\delta$ in both the mass ordering and NH results are more favourable with the current BAU value than the IH. Within NH, $\delta$ is tightly constrained in between ($2.0-4.0$) value.
	\item Projection of BAU with sterile mass and effective mass in the presence of sterile neutrino in fig. \ref{seb} gives an unsatisfactory remark while observing IH. Hence, IH fails to correlate them in a single frame. Although BAU value is very	small, NH manages to project the value and $keV$ sterile neutrino. 
	\item  In NH mode within this model, a constrained bound on the Dirac CP-phase is obtained from the baryogenesis study, which can be seen in the density plot of fig. \ref{delta} with Majorana phases in the X-axis. Majorana phases cover the whole $0-2\pi$ range, whereas the Dirac CP-phase is constrained between the value ($2.0-4.0$) satisfying observed BAU value.
\end{itemize}
%If the sterile neutrino is a main ingredient of the DM, it is potentially detectable in various X-ray observations [324, 566, 567]\\
In conclusion, the MES mechanism is analyzed in this chapter, considering a single flavour of a sterile neutrino in the $keV$ scale. Along with the active and sterile mass generation, this model can also be used to study the connection between effective mass in neutrinoless double beta decay ($0\nu\beta\beta$) in a wider range of sterile neutrino mass, simultaneously addressing the possibility of $keV$ scale sterile neutrino as dark matter particle. Although, results on $keV$ sterile neutrino as a dark matter candidate are still on the verge of uncertainty within the MES framework. We keep an optimistic hope to get better bounds from future experiments which may establish the same within MES. Results from baryogenesis via the mechanism of thermal leptogenesis are also checked and verified within this model. Finally, we have correlated all these observable under a single framework. Results in the NH mass pattern shows better consistency than the IH pattern. 
% Chapter Template
\begin{savequote}[1\linewidth]
\normalsize	``The career of a young theoretical physicist consists of treating the
	harmonic oscillator in ever-increasing levels of abstraction.''
	\qauthor{\large\it Sidney Coleman (1937--2007)}
\end{savequote}
\chapter{Five-zero texture in neutrino-dark matter model} % Main chapter title
%\vspace{-2.1cm}
%{\it ``It doesn't matter how beautiful your theory is, it doesn't matter how smart you are. If it doesn't agree with the experiment, it's wrong."--- {\Large Richard P. Feynman}}
%{\it ``The career of a young theoretical physicist consists of treating the	harmonic oscillator in ever-increasing levels of abstraction."---{ \Large Sidney Coleman}}
\label{Chapter5} % Change X to a consecutive number; for referencing this chapter elsewhere, use \ref{ChapterX}

\lhead{Chapter 5. \emph{Five-zero texture in neutrino-dark matter model}} % Change X to a consecutive number; this is for the header on each page - perhaps a shortened title

%----------------------------------------------------------------------------------------
%	SECTION 1
%----------------------------------------------------------------------------------------
\section*{}
\vspace*{-3.8cm}
In this chapter, we have studied a neutrino and dark matter model within the framework of minimal extended seesaw (MES). Five-zero textures are imposed in the final $(4\times4)$ active-sterile mass matrix, which significantly reduces the free parameter in the model. Three right-handed neutrinos were considered; two of them have degenerate masses, helping us to achieve baryogenesis $via$ resonant leptogenesis. A singlet fermion (sterile neutrino) with mass $\sim\mathcal{O}$(eV) is also considered, and we put bounds on active-sterile mixing parameters $via$ neutrino oscillation data. Resonant enhancement of lepton asymmetry has studied at the TeV scale, where we have discussed a few aspects of baryogenesis considering the flavour effects. The possibility of improvement in effective mass from $0\nu\beta\beta$ in the presence of a single generation of sterile neutrino flavour is also studied within the fermion sector. In the scalar sector, the imaginary component of the complex singlet scalar ($\chi$) is behaving as a potential dark matter candidate. Simultaneously, the complex scalar's real part is associated with the fermion sector for sterile mass generation. A broad region of dark matter mass is analyzed from various annihilation processes. The VEV of the complex scalar plays a pivotal role in achieving the observed relic density at the right ballpark.

\section{Introduction}

The origin of dark matter and neutrino mass in the current situation is a puzzle, and there are various BSM frameworks at tree level and loop level to address them under a single roof \cite{Hambye:2009pw, Bernal:2017kxu, Kahlhoefer:2017dnp, Tanabashi:2018oca, Magana:2012ph, Khan:2017xyh, Das:2014fea, Das:2019ntw}. We have considered a minimal model with a complex scalar singlet $\chi=(\chi^R+i\chi^I)$, which takes part in fermion and scalar sector. A $Z_4$ symmetry is chosen in such a fashion that the complex scalar singlet remains odd, which restricts its mixing with other SM particles. In the fermion sector, it couple to the additional singlet neutral fermion; sterile neutrino and RH neutrino, which help to generate masses of the sterile and active neutrinos and other experimental neutrino variables. On the other hand, in the scalar sector, the lightest component of $\chi$ (the imaginary part, $\chi^I$) serves as a viable weakly interacting massive particle (WIMP) dark matter candidate. Other scalar singlets and triplets in the model are decoupled from the dark matter analysis due to the large mass; nonetheless, they participate in neutrino mass generation and baryogenesis process.

There are ten elements in the (3+1) situation in the $(4\times4)$ symmetric mass matrix and 16 neutrino parameters (4 masses, six mixing angles and 6 phases). Typically, texture zero studies connect these parameters through zeros, and there we get definite bounds on them. Recent studies \cite{Borah:2017azf,Sarma:2018bgf,Zhang:2013mb,Borah:2016xkc} on this $(4\times4)$ mass have predicted various number of zeros with better accuracy on different aspects. In this chapter, we proposed the possibility with a maximum of five-zeros in the final mass matrix and studying active-sterile mixing phenomenology. Our main aim is to keep all these particles in the chosen model to explain neutrino parameters using the texture zero methods, dark matter, and the baryogenesis and neutrinoless double beta decay altogether. It was not done previously in the literature, which motivates us to carry out this detailed study.

This chapter is organized as follows: 
in section \ref{6sec2} we have discussed the theoretical background of the fermion as well as scalar sector. A detailed discussion of the model building aspect has been carried out in section \ref{model4}. In section \ref{6sec3}, we have discussed various constraints affecting the scalar sector. The numerical analysis has been carried out in section \ref{6sec4}, and finally we have summarized this chapter in section \ref{6sec5}.

%We organized this chapter as follows: In section \ref{sec2}, we discussed the complete theoretical framework and the model description in subsequent subsections. In the sub-sections of section \ref{sec3}, we discuss various constraints and numerical approaches for minimal extended seesaw and various parametrization, baryogenesis, neutrinoless double beta decay and dark matter respectively. In section \ref{sec4}, we discuss the numerical results of our work separately in each sub-sections, and finally, we concluded our work in section \ref{sec5}.
\section{Theoretical framework}\label{6sec2}
\subsection{Choice of five-zero texture}\label{s21}
Recent studies \cite{Borah:2016xkc,Borah:2017azf} shows a maximum of five-zeros in the $(4\times4)$ active-sterile mixing matrix were possible and beyond that, it fails to hold the latest bounds on the mixing parameters. In this chapter, we also focus on the maximum possible zero in the final $(4\times4)$ active-sterile mass matrix and study phenomenological consequences. In the five zeros texture, there are $^{10}C_5=252$ possibilities of zeros and 246 among them are ruled out due to the condition $(M_{\nu})_{i4}\ne0$ (with $i=e,\mu,\tau,4$).  Hence, we are left with six choices of mass matrices with zeros in the active sector. They are as follows,
\begin{eqnarray}\label{txz}
	\nn
	T_1=\begin{pmatrix}
		X&0&0&X\\
		0&0&0&X\\
		0&0&0&X\\
		X&X&X&X\\
	\end{pmatrix};~\quad
	T_2=\begin{pmatrix}
		0&0&0&X\\
		0&X&0&X\\
		0&0&0&X\\
		X&X&X&X\\
	\end{pmatrix}; ~\quad
	T_3=\begin{pmatrix}
		0&X&0&X\\
		X&0&0&X\\
		0&0&0&X\\
		X&X&X&X\\
	\end{pmatrix};\\
	T_4=\begin{pmatrix}
		0&0&X&X\\
		0&0&0&X\\
		X&0&0&X\\
		X&X&X&X\\
	\end{pmatrix};~\quad
	T_5=\begin{pmatrix}
		0&0&0&X\\
		0&0&X&X\\
		0&X&0&X\\
		X&X&X&X\\
	\end{pmatrix};~\quad
	T_6=\begin{pmatrix}
		0&0&0&X\\
		0&0&0&X\\
		0&0&X&X\\
		X&X&X&X\\
	\end{pmatrix}.
\end{eqnarray}
In this $(4\times4)$ active sterile mixing matrix, there are six mixing angles, six mass squared differences and four mass eigenvalues. For simplicity, we have fixed Dirac and Majorana phases. We use the latest $3\sigma$ global fit results for active neutrino parameters and equate zeros with the corresponding matrix obtained from the diagonalizing leptonic mixing matrix  $U_{PMNS}^{4\times4}$ matrix.  $m_{\nu}=U m^{\text{diag}}(m_1,m_2,m_3,m_4)U^T$. All the six structures were analyzed and it was found that only the $T_5$ structure able to satisfy the current bounds on active-sterile mixing parameters \cite{Abe:2016nxk,Ade:2015xua}. This result on five-zero texture agrees with past work \cite{Borah:2017azf}. Therefore, we have skipped the part of analyzing oscillation parameters in the active neutrino sector here and focused on digging on some other phenomenological aspects in this chapter.

\subsection{Model framework}\label{model4}
%\subsubsection{Model Description}
We have constructed this model considering one generation of eV scaled sterile neutrino, that couple with two RH neutrinos. This model is focused on implementing texture zero in the final $(4\times4)$ active-sterile mixing matrix and study their phenomenological aspects. In the scalar sector, the imaginary part of the complex scalar is behaving as a potential dark matter candidate. 

In this model set-up, we basically rely on $A_4$ flavour, and unwanted interactions were restricted using an extra $Z_3\times Z_4$ symmetry.
The particle content of this model are given in table~\ref{modelt}.
%%%%%%%%%%%%%%%%%%%%%%%%%%%%%%%%%
\begin{table}[h!]
	\begin{tabular}{|c|ccccc|cccccc|}
		\hline
		Fields$\rightarrow$ &&&Fermions&&&&Scalars&&&&\\
		\hline
		Charges$\downarrow$		& $L=(\nu_l \,\, l)^T $&$N_1$&$N_2$&$N_3$&$S$ &$\phi_1$&$\phi_2$&$\psi_1$&$\psi_2$&$\eta$&$\chi$\\
		\hline		
		$SU(2)$&2&1&1&1&1&2&2&1&1&1&1\\
		%	\hline
		$A_4$&3&1&$1$&$1$&$1^{\prime}$&1&1&3&3&1&$1^{\prime\prime}$\\
		$Z_3$&$\omega^2$&$\omega$&$\omega$&$\omega$&1&$\omega$&$\omega$&$\omega^2$&$\omega^2$&$\omega$&$\omega^2$\\
		$Z_4$&1&$-i$&$i$&$-i$&$-i$&$-i$&$i$&1&$1$&$i$&$-1$\\
		\hline
	%	\hline		
	\end{tabular}
	\caption{Particle content and their charge assignments under SU(2), $A_4$ and $Z_3\times Z_4$ groups.}\label{modelt}
\end{table}
%%%%%%%%%%%%%%%%%%%%%%%%%%%%%

We denotes the SM Higgs doublet as $\phi_1$ which transform as a singlet under $A_4$ symmetry and it is expressed as, $\phi_1=\frac{1}{\sqrt{2}}\begin{pmatrix}
	\phi_h^+,\,\phi_{h} + v_h+i \phi_h^0
\end{pmatrix}^T$, where $v_h$ is the vacuum expectation value (VEV) of the real part of the scalar doublet. 
Along with the SM Higgs boson, we have introduced a similar doublet $\phi_2=\frac{1}{\sqrt{2}}\begin{pmatrix}
	\phi_{2h}^+,\,\phi_{2h} + v_{2h}+i \phi_{2h}^0
\end{pmatrix}^T$ , which gets a tiny VEV ($v_{2h}\sim0.1$ eV) $via$ soft breaking mass term~\cite{Davidson:2009ha}. We have four more $SU(2)$ scalar singlet, $\psi_1$, $\psi_2$, $\eta$ and $\chi$. 
The $A_4$ triplet scalar flavons ($\psi_1,\psi_2$) get VEV\footnote{We follow the similar approach for evaluating triplet flavon VEVs by potential minimization as discussed in appendix \ref{appn1}} along $\langle\psi_1\rangle=(v,0,0)$, $\langle \psi_2\rangle=(0,v,0)$ and the singlets ($\eta,\chi$) achieve VEV along $\langle\eta\rangle=u$ and $\langle\chi\rangle=v_{\chi}$.
The VEV of these scalar fields breaks the $A_4$, including the $Z_3$ and $Z_4$ symmetries. This will help to understand the fermions (quarks, charged leptons and neutrino) mass matrix~\cite{Babu:2009fd, Altarelli:2005yp}.
Here, $\chi$ is a $SU(2)$ complex singlet and it is expressed as $\chi=(\chi^R+v_{\chi}+i\chi^I)/\sqrt{2}$. The scalar potential related to $ \phi_2,\psi_1,\psi_2$ and $\eta$ fields are completely decoupled from all the other scalar fields; however, these scalar play a crucial role to get the fermion mass spectrum.
The VEVs are considered as $v\sim 10^{12}$ GeV, $u\sim10^{10}$ GeV and $v_{\chi}\sim 1000$ GeV.

\subsubsection{Normal hierarchical neutrino mass}
The invariant Lagrangian can be written as,
\begin{eqnarray}\label{lag2}
	\mathcal{L}\supset\frac{y_l}{\Lambda}(\bar{L}\phi_1\psi_1)l_R+ \mathcal{L_{M_D}}+\mathcal{L_{M_R}}+\mathcal{L_{M_S}},
\end{eqnarray}
where,
\begin{equation}\label{lag1}
	\begin{split}
		\mathcal{L_{M_D}}=&\frac{Y_1}{\Lambda}(L\tilde{\phi_2}\psi_1)_{1}N_1+\frac{Y_2}{\Lambda}(L\tilde{\phi_1}\psi_1)_{1}N_2+\frac{Y_3}{\Lambda}(L\tilde{\phi_2}\psi_2)_{1}N_3,\\
		\mathcal{L_{M_R}}=&\lambda_1\eta\overline{N_1^c}N_3+\lambda_2\eta\overline{N_2^c}N_2+\lambda_3\eta\overline{N_3^c}N_1,\\
		\mathcal{L_{M_S}}=&q_1\chi SN_1+q_2\chi SN_3.
	\end{split}
\end{equation}
The first term in equation \eqref{lag2} will give rise to diagonal charged lepton mass matrix as $M_l=\frac{y_l}{\Lambda}\langle\phi_1\rangle\text{Diag}(v,v,v)$. Here, $l_R$ stands for $e_R,~\mu_R$ and $\tau_R$ and their charges under $(A_4\times Z_3\times Z_4)$ read as $(1\times \omega^2\times i)$ for each RH charged lepton.
The mass matrices generated will be of the form as follows,
\begin{itemize}
	\item {\bf Dirac mass matrix:} $M_D=\begin{pmatrix}
		a&0&0\\b&0&0\\0&0&c\\
	\end{pmatrix}$, with $a=\frac{Y_1\langle\phi_2\rangle v}{\Lambda}$, $b=\frac{Y_2\langle\phi_1\rangle v}{\Lambda}$, $c=\frac{Y_3\langle\phi_2\rangle v}{\Lambda}$.
	\item {\bf Majorana mass matrix:} $M_R=\begin{pmatrix}
		0&0&d\\0&e&0\\f&0&0\\
	\end{pmatrix}$, with $d\sim e\sim f=\lambda_iu$ ($i=1,2,3$ for respective positions).
	\item {\bf Sterile mass matrix:} $M_S=\begin{pmatrix}
		g&0&h\\
	\end{pmatrix}$ with $g=q_1v_{\chi}$ and $h=q_2v_{\chi}$. To get the sterile mass within eV scale with $v_{\chi}\simeq 1$ TeV and the Yukawa couplings $q_1,q_2$ takes value around $10^{-3}$.
	\item The final active-sterile mass matrix under minimal extended framework, takes the form,
	\begin{equation}
		M_{\nu}=\begin{pmatrix}
			M_DM_R^{-1}M_D^T&M_DM_R^{-1}M_S^T\\
			M_S{M_R^{-1}}^TM_D&M_SM_R^{-1}M_S^T\\
		\end{pmatrix}_{4\times4} =\begin{pmatrix}
			0&0&\frac{ac}{f}&\frac{ah}{f}\\
			0&0&\frac{bc}{f}&\frac{bh}{f}\\
			\frac{ac}{f}&\frac{bc}{f}&0&\frac{cg}{d}\\
			\frac{ah}{f}&\frac{bh}{f}&\frac{cg}{d}&\frac{(d+f)gh}{df}\\
		\end{pmatrix}_{4\times4}
	\end{equation}
	Due to the choice of small VEV to the additional Higgs doublet $\phi_2$ ($\sim$ eV), the (13) and (31) positions in the $M_{\nu}$ mass matrix are much smaller compared to the other elements (as $a,c\ll b$) and they lead to five-zero texture in the active-sterile mass matrix as,
	\begin{eqnarray}
		M_{\nu}\simeq\begin{pmatrix}
			0&0&0&\frac{ah}{f}\\
			0&0&\frac{bc}{f}&\frac{bh}{f}\\
			0&\frac{bc}{f}&0&\frac{cg}{d}\\
			\frac{ah}{f}&\frac{bh}{f}&\frac{cg}{d}&\frac{(d+f)gh}{df}\\
		\end{pmatrix}_{4\times4}
	\end{eqnarray}
\end{itemize}
It is to be noted that the dark matter candidate $\chi^I$ could not decay into $S$ and $N_1$ through the $\chi SN_1$ operators in equation \eqref{lag1}, as $M_{N_1}> M_{\chi^I}$. However, it could also decay into two active neutrinos through mixing. We have checked that the effective coupling strength for $\chi^I\nu_i\nu_j$ $(i,j=1,2,3)$ are very very less than $10^{-22}$, which implies that the dark matter decay lifetime is much greater than the lifetime of the Universe ($t_{Universe}\sim10^{-17}$ sec). Hence, in this model dark matter is stable, even we found the DM lifetime, $t_{DM}>10^{28}$ second \cite{Boyarsky:2014jta}.

\subsection{Extended scalar sector}
The potential for the complex scalar and Higgs boson can now be written as,
%%%%%%%%%%%%%%%%%%%%%%
%%%%%%%%%%%%%%%%%%%%%%
\begin{eqnarray}
	V &= &-\mu_{\phi_1}^2 \phi_1^{\dagger}\phi_1 + \lambda_1 (\phi_1^{\dagger}\phi_1)^2 + \lambda_2 (\phi_1^{\dagger}\phi_1) (\chi\chi^{\dagger}) \nn \\
	&& -\frac{1}{2} \mu_{\chi^R}^2 \chi \chi^{\dagger} - \mu_{\chi^I}^2 (\chi^2 + h.c.) + 
	\frac{\lambda_3}{4!} (\chi \chi^{\dagger})^2.
	\label{pot1}
\end{eqnarray}
%%%%%%%%%%%%%%%%%%%%%%
%%%%%%%%%%%%%%%%%%%%%%
The minimization condition can be written as,
%%%%%%%%%%%%%%%%%%%%%%%%
\begin{equation}
	\mu_{\phi_1}^2=  \lambda_1 v_h^2 + \frac{1}{2} \lambda_2 v_{\chi}^2,  ~~{\rm and }~~\mu_{\chi^R}^2=  - \lambda_2 v_h^2 - \frac{1}{12} \lambda_3 v_{\chi}^2
	\label{eq:min}
\end{equation}
%%%%%%%%%%%%%%%%%%%%%%%
The complex part of the singlet scalar $\chi$ do not mix with $\phi_h^0$ due to the charge assignment ($Z_3\times Z_4$ symmetry remains intact), hence neutral Goldstone boson takes the form $ G^0\approx \phi_h^0$ while the charged Goldstone bosons could be written as $G^\pm\approx \phi^\pm$. Corresponding gauge bosons eat these components. Only the CP-even, {\it i.e.}, the real component of these scalar gets mixed and the  mass matrix could be written as,
%%%%%%%%%%%%%%%%%%%%%%
\begin{equation}
	\begin{pmatrix}
		\phi_h & \chi^R\\
	\end{pmatrix}\begin{pmatrix}
		M_{\phi_h\phi_h}&M_{\phi_h \chi }\\M_{\chi\phi_h}&M_{\chi\chi}
	\end{pmatrix}\begin{pmatrix}
		\phi_h \\ \chi^R\\
	\end{pmatrix}\label{mm1}
\end{equation}
%%%%%%%%%%%%%%%%%%%%%%%% 

The mass terms are expressed as,
\begin{eqnarray}\label{mass1}
	&M_{\phi_h\phi_h} =&2 \lambda_1 v_h^2,\\
	&M_{\phi_h \chi }=&M_{\chi\phi_h} = \lambda_3 v_h v_{\chi},\\
	&M_{\chi\chi} =&- 2 \mu_{\chi^I}^2 + \frac{1}{12} \lambda_3 v_{\chi}^2.
\end{eqnarray}

The mass eigenstates are obtained by diagonalizing the mass matrix \eqref{mm1} with a rotation of $\phi_h-\chi^R$ basis,
\begin{equation}
	\begin{pmatrix}
		h\\H\\
	\end{pmatrix}=\begin{pmatrix}
		\cos\alpha&\sin\alpha\\-\sin\alpha&\cos\alpha\\
	\end{pmatrix}\begin{pmatrix}
		\phi_h \\ \chi^R\\
	\end{pmatrix}.
\end{equation} 
Here, the mixing angle $\alpha$ is defined as,
\begin{equation}
	\tan2\alpha=\frac{2 M_{\chi\phi_h}}{M_{\chi\chi}-M_{\phi_h\phi_h}}.
\end{equation}
The mass expression of the complex scalar is given by
\begin{equation} \label{mass2}
	M_{\chi^I}^2 = 2 \mu_{\chi^I}^2
\end{equation}

One can notice that the mass of the complex scalar directly depends on $\mu_{\chi^I}$, which also related to the mixing angle $\alpha$. Hence, we have to be very careful about the parameter space; else, it will directly affect the Higgs signal strength data \cite{Sirunyan:2018ouh}.
In this model, the CP-even scalar breaks the $Z_3,Z_4$ including $A_4$ symmetry. However, these symmetries for the singlet type pseudo scalar remains intact. Hence, this scalar remains stable and can serve as a viable dark matter candidate. One can see from equation \eqref{mass1}-\eqref{mass2} that the mass of the dark matter and the Higgs portal $\chi^I\chi^Ih(H)$ coupling strengths (the main annihilation channels) depend on the parameters $\mu_{\chi^I}$ and other quartic couplings especially $\lambda_3$ and VEVs, on the other hand, masses of the scalar fields ($h,H,\chi^I$) depends on $\mu_{\chi^I}$ mixing angle $\alpha$ and VEVs. We will discuss it in details in the dark matter section.  These parameters also affect the generation of fermionic mass and mixing angles.
Three generations of heavy right-handed (RH) neutrinos are also introduced, which gives respective masses to the active and sterile neutrinos\footnote{Within MES one active neutrino mass is always zero \cite{Zhang:2011vh,Barry:2011wb,Das:2018qyt}, in the NH mass ordering, $m_1$ will be zero. Hence two RH neutrinos will give mass to two active neutrinos, and one will give mass to a sterile mass.}. We have considered a nearly degenerate mass scale for two RH neutrinos ($N_1, N_3$)\footnote{Choice of the RH neutrino is arbitrary. We choose $N_3$ with $N_1$ instead of $N_2$ for the resonant production of CP asymmetry is due to the choice of sterile basis within the model. }, in such a way that they can exhibit resonantly enhanced leptogenesis in this work. 
\subsection{Yukawa coupling parametrization}
Without loss of generality, we have excluded the presence of sterile neutrino for baryogenesis study and use the conventional $(3\times3)$ matrix structures to parametrize the Yukawa matrix analogous to the CI (Casas-Ibarra) parametrization \cite{Casas:2001sr} as,
\begin{equation}
	Y=\frac{1}{v}U_{PMNS}^{3\times3}\sqrt{m^D_{\nu}}R^T\sqrt{M_R^{-1}},\label{ci}
\end{equation}
where $R$ is the rotational matrix with complex angle $z_i=x_i+iy_i$ ($x,y$ being real and free parameter with $x_i,y_i\in[0,2\pi]$ \cite{Ibarra:2003up}), and it can be parametrize as,
\begin{equation}
	R=\begin{pmatrix}
		1&0&0\\
		0&\cos z_1&\sin z_1\\
		0&-\sin z_1&\cos z_1\\
	\end{pmatrix}\begin{pmatrix}
		\cos z_2&0&\sin z_2\\
		0&1&0\\
		-\sin z_2&0&\cos z_2\\
	\end{pmatrix}\begin{pmatrix}
		\cos z_3&\sin z_3&0\\
		-\sin z_3&\cos z_3&0\\
		0&0&1\\
	\end{pmatrix}
\end{equation}
For our convenience, we have considered some random value (say $z_1=0^\circ-i90^\circ $) for the complex angles $z_i$ satisfying current value for the observed quantities. 
\subsection{Baryogenesis $via$ resonant leptogenesis}

We numerically solved the Boltzmann equations (BE) in the out-of-equilibrium scenario in the resonant leptogenesis study. With our choice of the nearly degenerate mass spectrum of TeV scaled RH neutrinos, the decay rates can produce sufficient asymmetry and at a temperature above the electroweak phase transition (EWPT), {\it i.e.}, for $T\ge T_c\sim147$ \cite{Dev:2014laa} GeV, $(B+L)$-violating interactions mediated by sphalerons are in thermal equilibrium, hence, the asymmetries in baryon number are generated. For $T<T_c$ , the produced baryon asymmetry gets diluted by photon interactions until the recombination epoch at temperature $T_0$. If there is no significant entropy release mechanism while the Universe is cooling down from $T_c$
to $T_0$, the baryon number in a co-moving volume, $\eta_B/s$, is constant during this epoch.

Before directly involving the BEs, let us first draw attention to a few quantities involved in the equations. The washout parameter, $K$, can be defined as the ratio of decay width to the Hubble parameter. Mathematically, $K=\Gamma_{N_i}/H(T=m_{N_i})$, with $\Gamma_{N_i}=\frac{(Y^{\dagger}Y)_{ii}}{8\pi}m_{N_i}$ and $H(T)=1.66 g_*\frac{T^2}{M_{Planck}}$. Here $M_{Planck}=1.12\times10^{19}$ GeV is the Planck mass, and $g_*$ is the relativistic degree of freedom with its value lies around 110. A small $K$ value corresponds to weak washout, and a large $K$ value corresponds to strong washout effects. In the case when flavours are taken under consideration \cite{Pilaftsis:2003gt},
\begin{equation}
	K_i=\frac{\Gamma (N_1\rightarrow l_i H)}{H(T=m_{N_1})}\simeq \frac{\Gamma (N_3\rightarrow l_i H)}{H(T=m_{N_3})} \quad\text{and}\quad K=\sum_{j=e,\mu,\tau}K_j.
\end{equation}
The relation among the number densities to the entropy density ($Y_x=\frac{N_x}{s}$) are expressed as follows \cite{Pilaftsis:2003gt},
\begin{equation}
	Y_B(T>T_c)=\frac{28}{79}Y_{B-L}(T>T_c)=-\frac{28}{51}Y_L(T>T_c).
\end{equation}
The CP asymmetry due to the heavy neutrino mixing in the flavoured case is given by, \cite{Dev:2014laa, Bambhaniya:2016rbb},
\begin{eqnarray}
	\epsilon_{il}&=&\epsilon_{il}^{mix}+\epsilon_{il}^{osc}.\\
	&=&\sum_{i\ne j}^{3}\frac{\mathcal{I}m\big[Y_{il}Y_{jl}^*(YY^{\dagger})_{ij}\big]+\frac{M_{i}}{M_{j}}\mathcal{I}m\big[Y_{il}Y_{jl}^*(YY^{\dagger})_{ji}\big]}{(YY^{\dagger})_{ii}(YY^{\dagger})_{jj}}\big(f_{ij}^{mix}+f_{ij}^{osc}\big).
\end{eqnarray}
The terms with superscripts inside the braces represents mixing and oscillation contributions respectively and they are expresses as,
\begin{eqnarray}
	f_{ij}^{mix}&=&\frac{(M_{i}^2-M_{j}^2)M_{i}\Gamma_j}{(M_{i}^2-M_{j}^2)^2+M_{i}^2\Gamma_j^2},\\
	f_{ij}^{osc}&=&\frac{(M_{i}^2-M_{j}^2)M_{i}\Gamma_j}{(M_{i}^2-M_{j}^2)^2+(M_{i}\Gamma_i+M_{j}\Gamma_j)^2\frac{|\text{Re}(YY^{\dagger})|}{(YY^{\dagger})_{ii}(YY^{\dagger})_{jj}}}
\end{eqnarray}
where, $\Gamma_i=\frac{M_{i}}{8\pi}(YY^{\dagger})_{ii}$ is the decay width at tree level. The typical time scale for the CP asymmetry variation is defined as,
\begin{equation}
	t=\frac{1}{2H}=\frac{z^2}{2H(M_1)}=\frac{Kz^2}{2\Gamma_{N_1}}\sim\frac{1}{\Delta M}.
\end{equation}
The CP asymmetry raises for $t\le\frac{1}{\Delta M}$ and exhibit its oscillation pattern only for $t\ge\frac{1}{\Delta M}$. They originate from the CP-violating decays of the two mixed states $N_1$ and $N_3$. The time dependence if the CP asymmetry is also neglected as in the strong washout regime, their contribution is minimal. 

The baryon to photon number density, ($\eta_B=\frac{N_B}{n_{\gamma}}$) in the RL scenario is also defined as,
\begin{equation}
	\eta_B\sim-\sum_{i}^{1,2,3} \frac{\epsilon_i}{200 K(z)}.
\end{equation}
One thing to note here, to achieve $\eta_B\sim6.1\times10^{-10}$ \cite{Davidson:2008bu}, the term $\frac{\epsilon_i}{K(z)}$ should be around $10^{-8}$. Hence, the resonant production of leptonic asymmetry $\mathcal{O}(1)$ is possible only when a strong wash-out region is there ($K\gg1$). This lead to a thermally dense plasma state, so the conditions required for kinetic equilibrium and decoherence of the heavy Majorana neutrinos in the BEs are comfortably satisfied.
%and $2\leftrightarrow2$\footnote{In the $2\leftrightarrow2$ processes, we neglected $N_iN_j\leftrightarrow LL$ and included collision terms like $LL\leftrightarrow \phi^{\dagger}\phi^{\dagger}$, $L\phi\leftrightarrow L^C\phi^{\dagger}$, where all external particles are massless. Moreover, $1\leftrightarrow3$, $2\leftrightarrow3$ processes are also neglected in this work, as they corresponds to higher order correction to the $1\leftrightarrow2$, $2\leftrightarrow2$ processes.} 

We have considered only $1\leftrightarrow2$ decays of the RH neutrinos and the $2\leftrightarrow2$ scatterings that describe $\Delta L=0,2$ transition processes \cite{Pilaftsis:2003gt}. 
%In particular, we neglect the contribution from $2\leftrightarrow2$ scattering and thermal effects, which become less significant for temperatures $T\le m_{N_1}$ relevant to leptogenesis.
Moreover, the decay rates of the two RH neutrinos are almost equal due to the nearly degenerate mass scheme ($\Gamma_{N_1}\sim\Gamma_{N_3}\sim\Gamma$). Thus, the CP asymmetries generated from both the decay processes gets resonantly enhanced. In the case of different decay rates, it would be wise to pick the only CP asymmetry contribution, which is resonantly enhanced. 
We have also included the flavour effect in our RL study, where the contribution from specific lepton flavours is accounted for. As leptogenesis is a dynamic process, the lepton asymmetry generated $via$ the decay and inverse decay of RH neutrinos is distributed among all three flavours. In some cases, the flavour dependent processes do wash out the lepton asymmetry $via$ inverse decay processes. Finally, the total lepton and the baryon asymmetry values are given by the sum of all three contributions. 
%To be specific, we include the gauge-mediated collision terms that describe processes such as $N_iL_j\leftrightarrow\phi^{\dagger}V_{\mu}$ and their crossing-symmetric reactions,where $V_{\mu}$ collectively denotes the $SU(2)_L$ and $U(1)_Y$ gauge bosons $W_{\mu}^a$ (with $a = 1, 2, 3$) and $B_{\mu}$ in the unbroken symmetric phase of the SM.
Hence, the Boltzmann equation for three flavor case can be written as \cite{Pilaftsis:2003gt},
\begin{eqnarray}
	&\frac{d\eta_{N_j}}{dz}=&\frac{z}{H(z=1)}\Big[\Big(1-\frac{\eta_{N_j}}{\eta_{N_j}^{eq}}\Big)\frac{1}{n_{\gamma}}\gamma_{L\phi}^{N_j}\Big],\\
	\nn&\frac{d\eta_{Li}}{dz}=&\frac{z}{H(z=1)\eta_{\gamma}}\Big[\sum_{i=1}^3\epsilon_i\Big(1-\frac{\eta_{N_j}}{\eta_{N_j}^{eq}}\Big)\epsilon_i\gamma_{L\phi}^{N_j}-\frac{2}{3}\eta_{L_i}\sum_{k=e,\mu,\tau}\Big(
	\gamma^{L_i\phi}_{L_k^c\phi^{\dagger}}+\gamma^{L_i\phi}_{L_k\phi}\Big)\\&&-\frac{2}{3}\sum_{k=e,\mu,\tau}\eta_{L_k}\Big(
	\gamma^{L_k\phi}_{L_k^c\phi^{\dagger}}-\gamma^{L_k\phi}_{L_i\phi})\Big],\label{be1}
\end{eqnarray}
where  $i=e,\mu,\tau$, represents three flavour case and $j=1,2,3$ represents RH neutrino generations. $\eta_{N_J}^{eq}$ is the equilibrium number density and $\gamma_{L\phi}^{N_j}$ is the $1\leftrightarrow2$ collision term. They are defined as follows.
\begin{eqnarray*}
	&\eta_{N_i}^{eq}=g_a\big(m_{N_i}T/2\pi\big)^{3/2} e^{z_i};~~(z_i=\frac{m_{N_i}}{T}),\\
	&\eta_{\gamma}=(2m_{N_i}^3/\pi^2z^3).
\end{eqnarray*}
The collision and scattering terms used in these equations are given in the appendix section \ref{appen2}. In the numerical section we present the evolution of RH neutrinos and the lepton number density with flavour effects.
%%%%%%%%%%%%%%%%%%%%%%%%%%%%%%%%%%%%%
 
\section{Constraints on this model}\label{6sec3}
There are various kinds of theoretical constraints in the model by which the parameter space of this model is constrained. In the following subsections, we will be discussing a few of them which we used in this model.

\subsection{Stability constraints}
The stability constraints demand that the potential be bounded from below, {\it i.e.}, it should not go negative infinity along any direction of the field space at large field values. For large field the quadratic terms of
the scalar potential in equation \eqref{pot1} are very small compared to the quartic terms,
therefore, the scalar potential can be written as,
\begin{equation}
	V(H, \chi^R)=-\frac{1}{4}\lambda_1H^4+\frac{\lambda_2}{4}H^2(\chi^R)^2+\frac{1}{24}\lambda_3 (\chi^R)^4.
\end{equation}
With further simplification, it will take the form,
\begin{equation}
	V(H,\chi^R)=\frac{1}{4}\Big[\sqrt{\lambda_1}H^2+\frac{\sqrt{\lambda_3}}{\sqrt{6}}(\chi^R)^2\Big]^2+\frac{1}{4}\Big[\lambda_2+\sqrt{\frac{2\lambda_1\lambda_3}{3}}\Big]H^2(\chi^R)^2.
\end{equation}
This scalar potential will be bounded from below if the following conditions are satisfied \cite{Deshpande:1977rw},
\begin{equation*}
	\lambda_1(\Lambda)>0;~~~\lambda_3(\Lambda)>0;~~~\text{and}~~\lambda_2(\Lambda)+\sqrt{\frac{2\lambda_1(\Lambda)\lambda_3(\Lambda)}{3}}>0.
\end{equation*}
Here, the coupling constants are evaluated at a scale $\Lambda$ using RG equations. 
\subsection{Perturbativity constraints}
This model, to remain perturbative at any given energy scale, one must impose upper bound on the coupling constants of the potential given by the equation \eqref{pot1} and they are as follows \cite{Lee:1977eg},
\begin{equation}
	|\lambda_1(\Lambda),~\lambda_2(\Lambda),~\lambda_3(\Lambda)|\le4\pi.
\end{equation}
\subsection{Unitarity constraints}
The scalar potential parameters of this model are severely constrained by the
unitarity of the scattering matrix (S-matrix), which consists of the quartic couplings of the scalar potential. For large field values, the scattering
matrix is obtained by using various scalar-scalar, gauge boson-gauge boson, and scalar-gauge
boson scatterings \cite{Lee:1977eg}. Following the unitarity condition, the S-matrix elements demand that the scattering matrix's eigenvalues should be less than $8\pi$ \cite{Das:2014fea, Kanemura:1993hm}. The unitary bounds in this model are,
\begin{equation}
	\lambda_1\le8\pi~~~\text{and}~~|12\lambda_1+\lambda_3\pm\sqrt{16\lambda_2^2+(\lambda_3-12\lambda_1)^2}|\le32\pi.
\end{equation}

\subsection{Bounds from Higgs signal strength data}

At tree-level, the couplings of Higgs-like scalar $h$ to the fermions and gauge bosons in the presence of extra Higgs doublet ($\phi_2$) and singlet scalar ($\chi$) are modified due to the mixing. Loop induced decays will also have slight modification for the same reason; hence, new contributions will be added to the signal strength \cite{Sirunyan:2018ouh}. As a particular case, we consider
the Higgs boson production cross-section via the gluon fusion mechanism and we use the narrow width
approximation  $\Gamma_h/M_h\rightarrow0$,
%%%%%%%%%%%%%%%%%%%%%%%%%%%%%%%%%%%%%%%%%%%%%%%%%
%%%%%%%%%%%%%%%%%%%%%%%%%%%%%%%%%%%%%%%%%%%%%%%%%
%%%%%%%%%%%%%%%%%%%%%%%%%%%%%%%%%%%%%%%%%%%%%%%%%
%%%%%%%%%%%%%%%%%%%%%%%%%%%%%%%%%%%%%%%%%%%%%%%%%
\begin{equation}
	\mu_{X}=\frac{\sigma(gg\rightarrow h\rightarrow X)_{BSM}}{\sigma(gg\rightarrow h\rightarrow X)_{SM}}\approx\frac{\sigma(gg\rightarrow h)_{BSM}}{\sigma(gg\rightarrow h)_{SM}}\frac{Br(h\rightarrow X)_{BSM}}{Br(h\rightarrow X)_{SM}},
\end{equation}
%%%%%%%%%%%%%%%%%%%%%%%%%%%%%%%%%%%%%%%%%%%%%%%%%
%%%%%%%%%%%%%%%%%%%%%%%%%%%%%%%%%%%%%%%%%%%%%%%%%
%%%%%%%%%%%%%%%%%%%%%%%%%%%%%%%%%%%%%%%%%%%%%%%%%
%%%%%%%%%%%%%%%%%%%%%%%%%%%%%%%%%%%%%%%%%%%%%%%%%
where, $X=b\bar{b}, \tau^+ \tau^-,\mu^+ \mu^-, W W^*, Z Z^*, \gamma\gamma$ is the standard model particle pairs. In presence of an extra Higgs doublet $\phi_2$, the signal strength does not change as it completely decoupled from the scalar sector, however, due to mixing of $\phi_1$ and $\chi$, $h$ to flavon-flavon ( or boson-boson) coupling become proportional to $\cos\alpha$. So, we may rewrite $\mu_{X}$ as,
%%%%%%%%%%%%%%%%%%%%%%%%%%%%%%%%%%%%%%%%%%%%%%%%%
%%%%%%%%%%%%%%%%%%%%%%%%%%%%%%%%%%%%%%%%%%%%%%%%%
%%%%%%%%%%%%%%%%%%%%%%%%%%%%%%%%%%%%%%%%%%%%%%%%%
%%%%%%%%%%%%%%%%%%%%%%%%%%%%%%%%%%%%%%%%%%%%%%%%%
\begin{equation}
	\mu_{X}=\cos^2\alpha \frac{\Gamma(h\rightarrow X)_{BSM}}{\Gamma(h\rightarrow X)_{SM}}\frac{\Gamma^{total}_{h,SM}}{\Gamma^{total}_{h,BSM}}.
\end{equation}
%%%%%%%%%%%%%%%%%%%%%%%%%%%%%%%%%%%%%%%%%%%%%%%%%
%%%%%%%%%%%%%%%%%%%%%%%%%%%%%%%%%%%%%%%%%%%%%%%%%
%%%%%%%%%%%%%%%%%%%%%%%%%%%%%%%%%%%%%%%%%%%%%%%%%
%%%%%%%%%%%%%%%%%%%%%%%%%%%%%%%%%%%%%%%%%%%%%%%%%
Apart from the SM Higgs $h$, if the masses for the extra physical Higgses are greater than $M_h/2$, $\frac{\Gamma^{total}_{h,SM}}{\Gamma^{total}_{h,BSM}}\approx\big(\cos^2\alpha \big)^{-1}$. Hence, the modified signal strength will be written as,
%%%%%%%%%%%%%%%%%%%%%%%%%%%%%%%%%%%%%%%%%%%%%%%%%
%%%%%%%%%%%%%%%%%%%%%%%%%%%%%%%%%%%%%%%%%%%%%%%%%
%%%%%%%%%%%%%%%%%%%%%%%%%%%%%%%%%%%%%%%%%%%%%%%%%
%%%%%%%%%%%%%%%%%%%%%%%%%%%%%%%%%%%%%%%%%%%%%%%%%
\begin{equation}
	\mu_{X}=\frac{\Gamma(h\rightarrow X)_{BSM}}{\Gamma(h\rightarrow X)_{SM}}.
\end{equation}
%%%%%%%%%%%%%%%%%%%%%%%%%%%%%%%%%%%%%%%%%%%%%%%%%
%%%%%%%%%%%%%%%%%%%%%%%%%%%%%%%%%%%%%%%%%%%%%%%%%
%%%%%%%%%%%%%%%%%%%%%%%%%%%%%%%%%%%%%%%%%%%%%%%%%
%%%%%%%%%%%%%%%%%%%%%%%%%%%%%%%%%%%%%%%%%%%%%%%%%
In this study, we notice that $\mu_{Z}$ is the most stringent, bounding $0.945\lesssim \cos\alpha\lesssim 1$ depending singlet scalar VEVs, although with little sensitivity. $\mu_{W}$ is less sensitive, yet bounding $0.92\lesssim \cos\alpha\lesssim 1$. In this work, we have kept fixed $\cos\alpha=0.95$ throughout the calculations.
\section{Numerical analysis and results}\label{6sec4}
\subsection{Neutrino mixing}
In this chapter, we have skipped active neutrinos analysis as there is vast literature available for this. Rather we focus on active-sterile mixing elements and sterile mass here. Our results show constancy with the previous results of \cite{Borah:2017azf}, and the structure $T_5$ being the only five-zero texture suitable to study the $(3+1)$ scenario. We have used the latest global fit 3$\sigma$ bounds on the active neutrino parameters and randomly solved for four unknowns. Four zeros of the structures from equation \eqref{txz} are equated to the $(4\times4)$ light neutrino mass matrix generated from the diagonalizing matrix given by the equation \eqref{4b4} and evaluate elements of the fourth column. We worked out the numerical analysis with such a choice of input parameters, which also satisfy the recent LFV data \cite{Parker_2018}. After solving for the active-sterile parameters, we have shown contour plots in fig. \ref{asf}. Interesting bounds on the mixing matrix and sterile mass are observed from the analysis. In the left figure, we have varied $\sin\theta_{24}$ and $\sin\theta_{14}$ along x-y plane and projected sterile mass $m_4^2$ on the z-plane. Similarly, in the plot next to this is also a projection of sterile mass in $\sin\theta_{34}-\sin\theta_{14}$ plane. In the left figure, $\theta_{14}$ and $\theta_{24}$ value lie around $2.29^{\circ}-11.4^{\circ}$ and $1.7^{\circ}-11.8^{\circ}$, respectively and they able to project $m_4^2$ value in between $0.1-3$ eV$^2$. On the other hand, the next figure gives bound on $\theta_{14}$ as $1.81^{\circ}-9.17^{\circ}$, with a wider range of $m_4^2=0.1-5$ eV$^2$ while satisfying $\theta_{34}$ value in between $0.57^{\circ}-9.74^{\circ}$.
\begin{figure}[h]
	\includegraphics[scale=.38]{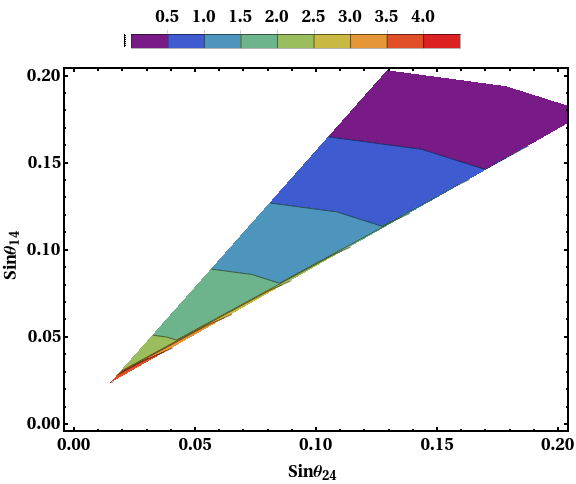}
	\includegraphics[scale=.37]{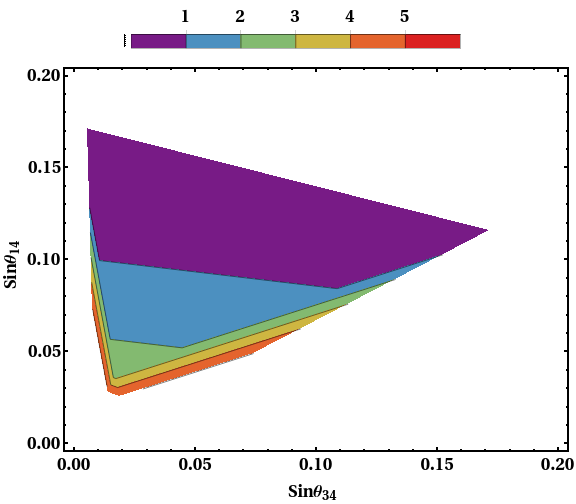}
	\caption{Projection of sterile mass with respect to active-sterile mixing angles. (Left) $\theta_{14}$ and $\theta_{24}$ able to project $m_S$ value around (0.1-3) eV$^2$ for $2^{\circ}-5^{\circ}$. (Right) Similar result for $\theta_{14}$ also satisfy for $m_S$ around (0.1-5) eV$^2$ and $\theta_{34}$ around $2^{\circ}-11^{\circ}$.}\label{asf}
\end{figure}
\begin{table}[h]\centering
	\begin{tabular}{|c|c|c|}
		\hline	Parameters&Experimental result \cite{An:2014bik}& Model result\\
		\hline$m_4^2$ (eV$^2$)&$0.7-2.5$&0.1-3\\
		$\theta_{14}$&$6^{\circ}-20^{\circ}$&$2.29^{\circ}-9.17^{\circ}$\\
		$\theta_{24}$&$3^{\circ}-11.5^{\circ}$&$1.7^{\circ}-11.8^{\circ}$\\
		$\theta_{34}$&$0^{\circ}-30^{\circ}$&$0.57^{\circ}-9.74^{\circ}$\\
		\hline
	\end{tabular}
	\caption{A comparison in between experimental results and results obtained from the model for the the active-sterile mixing parameters}\label{com1}
\end{table}
Combining these two results we can consider sterile mass around $0.1\le m_4^2(\text{eV}^2)\le3$ and $\theta_{14}$ value in between $2.29^{\circ}-9.17^{\circ}$. The current best fit value for $m_4^2$ is at 1.7 eV$^2$, and our results from the five-zero texture in the $(3+1)$ scenario are in good agreement with it. Other mixing parameters results are shown in table \ref{com1} for better understanding. These results are not verified completely; however, in future experiments, we may soon get solid bounds on these parameters.

\subsection{Baryogenesis $via$ resonant leptogenesis}
We have considered a scenarios with two nearly degenerate heavy Majorana neutrinos $N_{1,3}$ with masses at the TeV range. The mass of the third heavy Majorana neutrino $N_2$ will be taken to be of order $10^{12}$ GeV, so naturally, $N_2$ decouple from the low-energy sector of the theory. One flavour RL is not studied here, as it requires heavy decaying RH mass $\mathcal{O}(10^{12})$ GeV \cite{DeSimone:2007edo}. We analyse the Boltzmann equations by numerically solving the lepton number density values $N_L$ and the heavy-neutrino number densities $N_{N_i}$ as functions of the parameter $z=m_{N_1}/T$.
We set mass for $m_{N_1}\simeq m_{N_3}\simeq1$ TeV with $\frac{m_{N_3}}{m_{N_1}}-1=10^{-11}$ for our calculation. After solving for the RH and lepton number densities, the evolution patterns are projected in fig. \ref{nd}. We have also used leptogenesis equation solving tool {\rm ULYSSES} \cite{Granelli:2020pim} at some points to carry out few calculations.
\begin{figure}[h]
	\includegraphics[scale=.4]{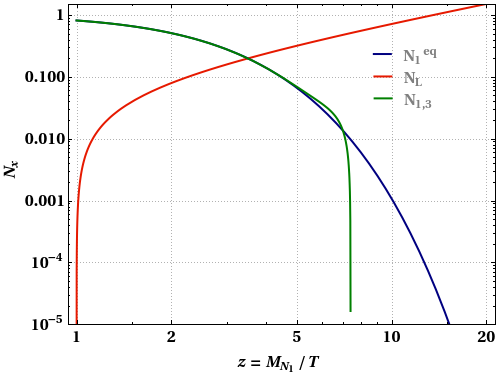}
	\includegraphics[scale=.45]{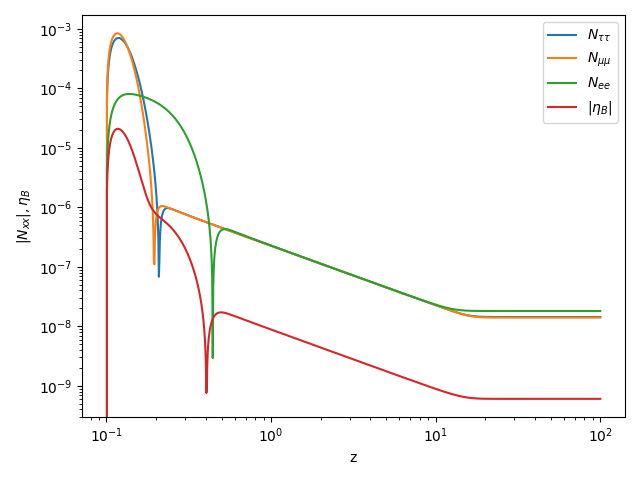}
	\caption{(Left) Evolution of total lepton number density, RH neutrino density and its equilibrium number density with $z=M_{N_1}/T$. (Right) Evolution of lepton flavor asymmetries generated for three flavours and the current baryon asymmetry of the Universe. For RL, we have considered nearly degenerate mass spectrum for the heavy RH neutrinos $M_{N_1}\sim$ 1 TeV and a mass splitting equivalent to the decay width $\Delta M\sim \Gamma$.}\label{nd}
\end{figure}
\begin{figure}
	\includegraphics[scale=.4]{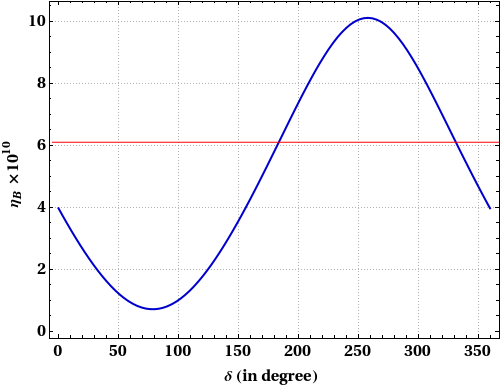}
	\includegraphics[scale=.4]{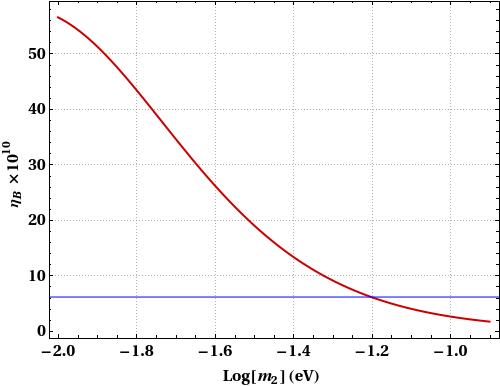}
	\caption{In the left figure, variation for Dirac CP phase ($\delta$) with the observed BAU value are shown. $\delta$ around $180^{\circ}$ and $330^{\circ}$ does satisfy the current BAU value. On the right figure, we have varied lightest neutrino mass with $\eta_B$ value. Lightest neutrino mass satisfies the best fit $\eta_B$ value with $6.1\times 10^{-10}$ at 0.063 eV.}\label{del}
\end{figure}

Results show an obvious pattern for our model with the particle number densities. In the left panel of fig. \ref{nd}, evolution of RH neutrino density ($N_{1,3}$) and lepton density along with the equilibrium number density of $N_1$ are projected against $z$. In the right panel, variation of lepton number density for three flavours and $\eta_B$ with $z$ are shown. The initial abundance for the leptonic and baryon number densities was assumed to be zero. Then by the decay of RH neutrinos, their concentration keeps increasing, at the same time, the $N_1$ number density keeps decreasing. As soon as the out-of-equilibrium is achieved, the decay process slows down, and the inverse decay excels, which never comes into thermal equilibrium. Thus the green curve deviates from the blue curve, and the lepton asymmetry is generated with increasing number density.

For $M_{N_1}\sim$ TeV, we have found that contribution from each lepton flavour is equally shared to give rise to the final lepton asymmetry. We have not shown the contributions from the off-diagonal terms ($N_{ij}$ with $i,j=e,\mu,\tau$ but $i\ne j$), as diagonal terms would dominate the whole situation here, due to the choice of our parameter space. Moreover,  the diagonal contributions $N_{ee}$ curve kink hits the lowest value near to $z\sim1$ is just a consequence of values of the input parameters. However, there is a saturation region beyond $z=1$ for all flavour contributions and coherence due to the RH neutrino mass ($N_{1,3}$). This result slightly contradict the RL$_{\tau}$ case, the significant contribution was coming from the $\tau$ lepton flavour, and other charge lepton contributions had less influence due to the larger washout rates \cite{Dev:2014laa}. Moreover, from previous studies, it is clear that within the strong washout regime, the final lepton or baryon asymmetry is independent of the initial concentration \cite{Dev:2014laa, Pilaftsis:2003gt}. Even if we start with considerable initial lepton asymmetry, the final asymmetry is achieved for $z\sim1$ within RL formalism by rapidly washing out the primordial asymmetry. %Hence, we consider the traditional way for number densities to the lepton as well as the RH neutrinos throughout our study.

We also varied other oscillation parameters associated with our model to the baryogenesis result. These model parameters are dictating the baryogenesis result through the CI parametrisation used in equation \eqref{ci} via the $U_{PMNS}$ matrix. In fig. \ref{del}, a variation of Dirac CP-phase with the observed BAU value is shown in the left panel. Delta ($\delta$) value around $180^{\circ}$ and $330^{\circ}$ are consistent with the current observed value, which is $\eta_B=(6.1\pm0.18)\times 10^{-10}$. In the right panel, we checked the $next$ lightest neutrino mass\footnote{Within MES, the lightest neutrino mass is zero naturally, so the second lightest neutrino eigenvalue ($m_2$) is considered as lightest with a definite value.} bound with the baryogenesis result. For an increase in $m_2$, there is a gradual decrease in $\eta_B$ value. With our parameter space choice, the $next$ lightest neutrino mass is coming out as 0.063 eV with the baryogenesis bound.   
\subsection{Neutrinoless double beta decay ($0\nu\beta\beta$)}
In this section, we study the numerical consequences on $0\nu\beta\beta$ using the bounds obtained from the previous section results. We use global fit light neutrino parameters for active neutrinos from table \ref{ttab:d1}, sterile parameters from texture zero bounds and CP phase from baryogenesis result. Variation of effective mass with the lightest neutrino mass for active neutrinos and active+sterile neutrino contributions are shown in fig. \ref{ndplot1}. The red region above the horizontal yellow line represents the upper bound on effective mass given by KAMLAND-ZEN, and the dashed grey line gives the future sensitivity on the upper bound. The Blue region on the right side of the horizontal blue line gives the upper bound on the sum of all three active neutrinos ($=0.12$ eV). A much wider region (green) satisfying the effective mass is achieved in active+sterile neutrino contribution, whereas a thin region (purple) is observed for active neutrino contribution only. Hence, a strong and impressive contribution from the sterile sector is observed in the five-zero texture structure. Even though there is a wider range covered in the presence of sterile neutrino, yet it goes beyond the current upper bound by KAMLAND-ZEN. From previous studies \cite{Abada:2018qok,Das:2019kmn}, it was clear that with a small mixing angle between active and sterile flavour can resolve this issue. 
\begin{figure}[h]\centering
	\includegraphics[scale=0.5]{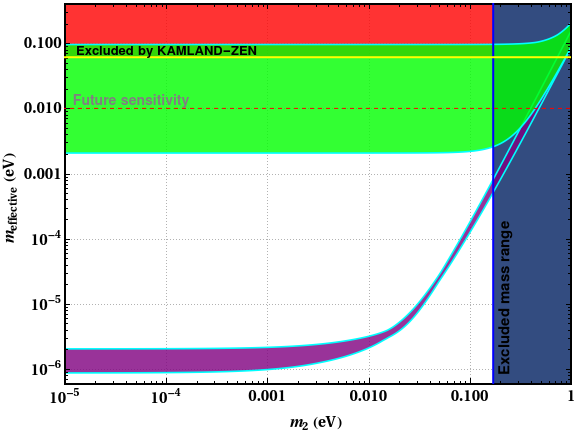}
	\caption{Results of effective mass {\it vs.} the lightest neutrino mass. Green shaded region represents the effective mass contribution from active+sterile case and the purple region gives only active neutrino contribution.}\label{ndplot1}
\end{figure}
\subsection{Dark matter}
This section will discuss various region of DM parameter space, satisfying the current relic density. The relic density in this model mainly come through the annihilation channels ($\chi^I\chi^I\rightarrow SM ~particle$), the intact $Z_{3,4}$ symmetry on $\chi^I$ and the mass gap ($M_{\chi^I}-M_H, M_{\chi^I}-M_{N_1}$) do not allow the co-annihilation diagrams. Also, the decay of the $\chi^I$ is greater than the lifetime of the Universe ($t_{\chi^I}\gg t_{Universe}$). The annihilation diagrams are shown in fig. \ref{ann1}.
\begin{figure}
	\includegraphics[scale=.52]{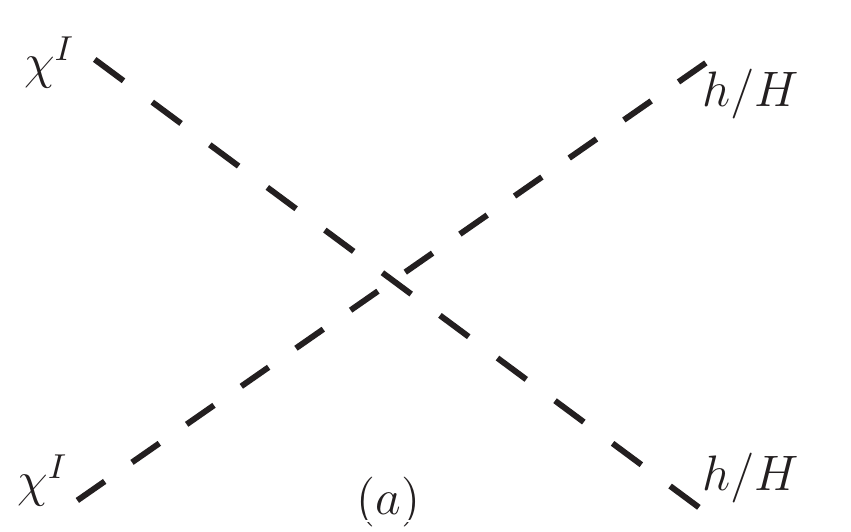}
	\includegraphics[scale=.52]{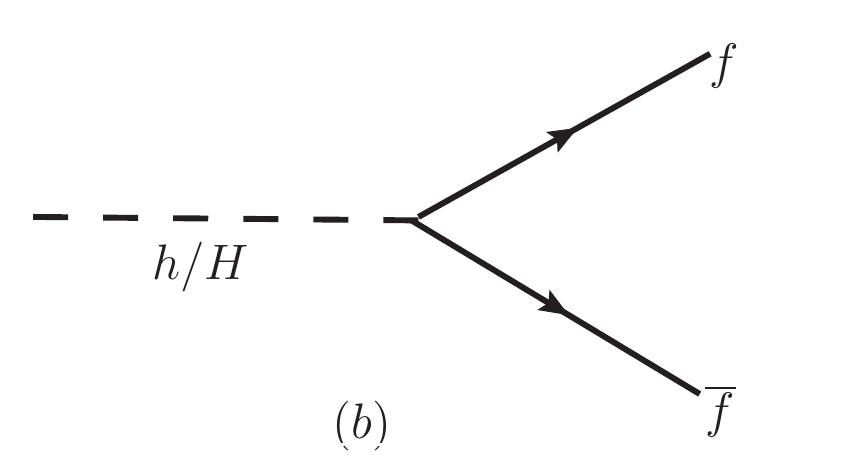}
	\includegraphics[scale=.52]{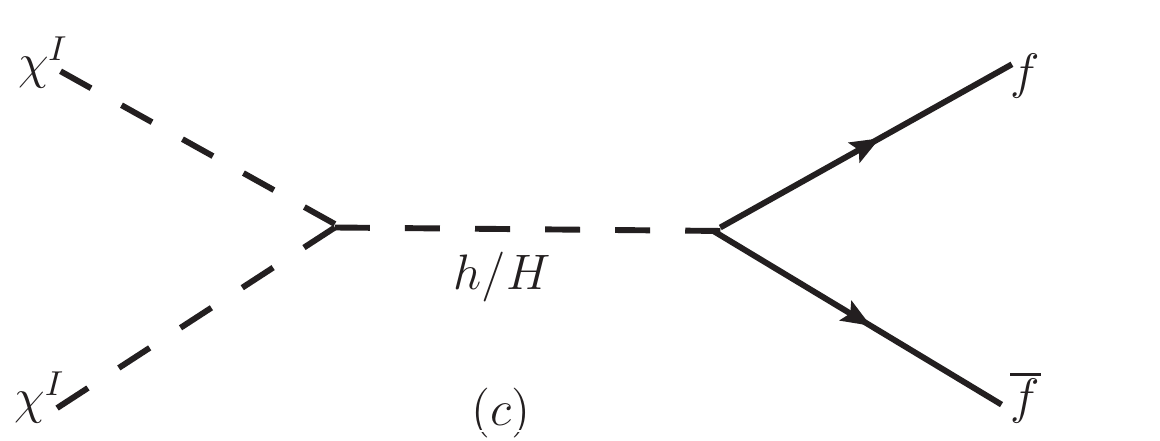}
	\caption{Annihilation processes of (a) $\chi^I\chi^I\rightarrow h/H$, $h/H\rightarrow SM~particles$ and Higgs mediated $\chi^I\chi^I\rightarrow SM~particles$. Here $f$ represents SM fermions.}\label{ann1}
\end{figure}
\begin{figure}[h]
	\includegraphics[scale=0.29]{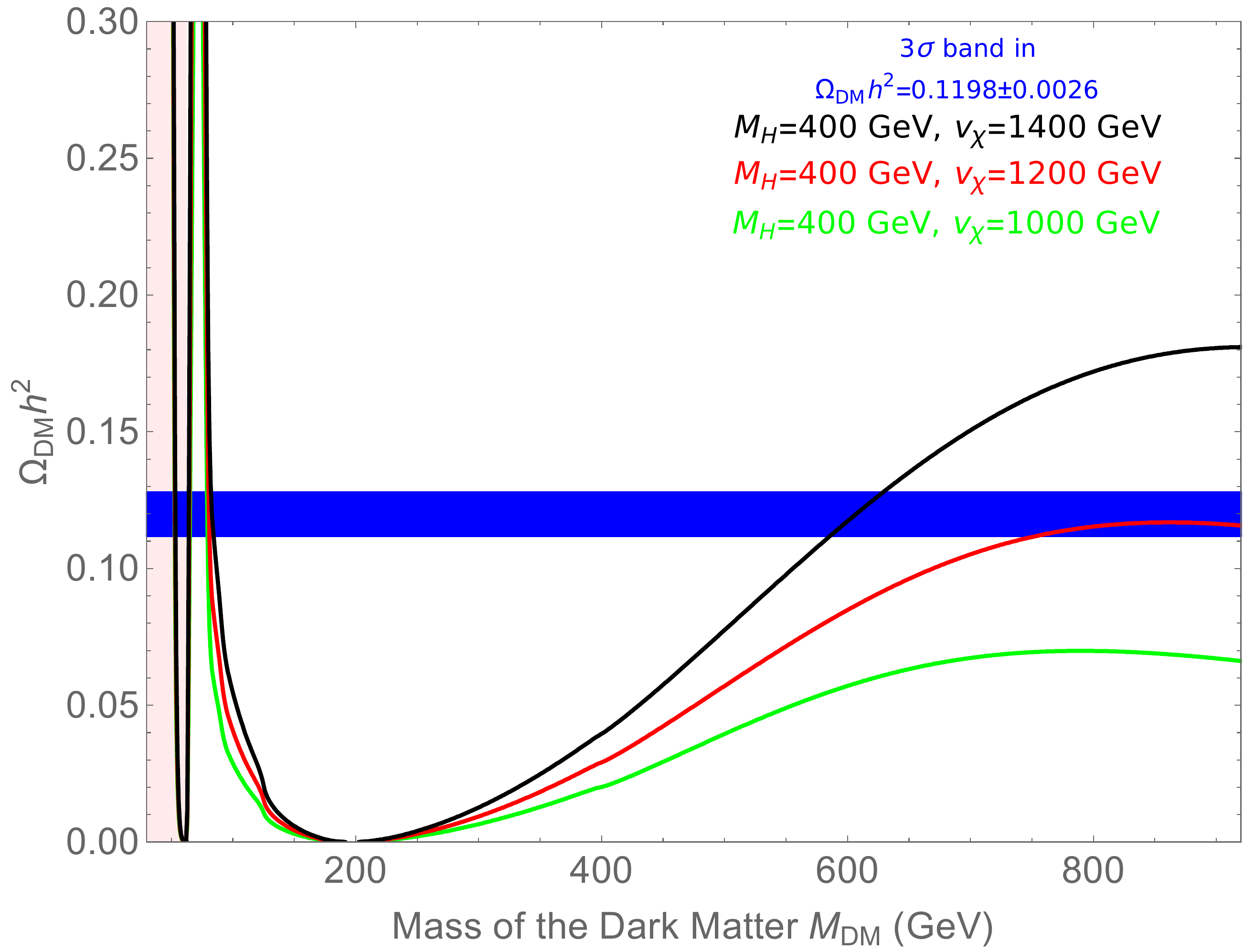}
	\includegraphics[scale=0.29]{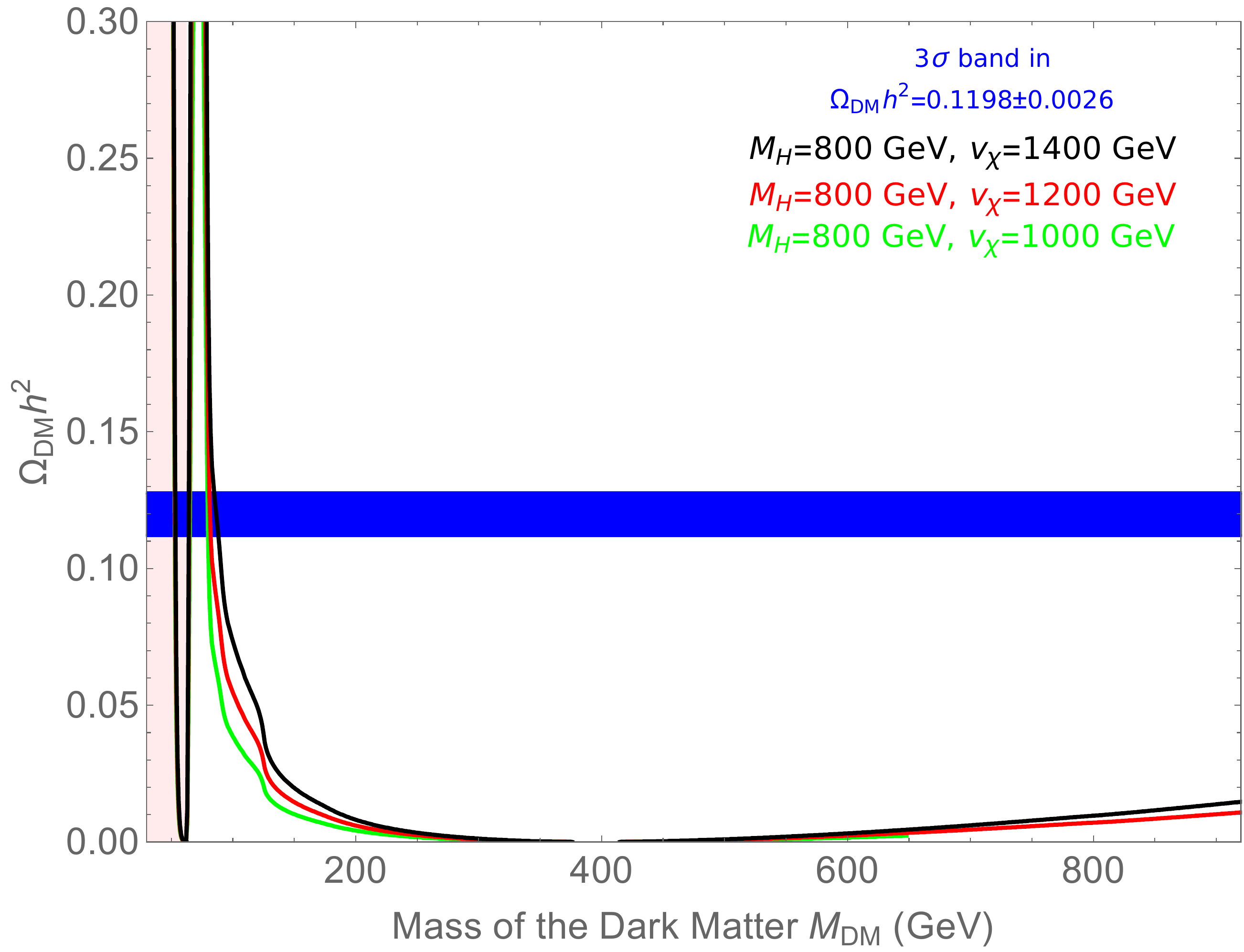}
	\caption{Dark matter density against dark matter mass for two different heavy Higgs masses ($M_H=400$ GeV and 800 GeV). We took three different VEV of the singlet scalar and the corresponding variation is shown as green (1000 GeV), red (1200 GeV) and black (1400 GeV). The blue band stand for 3$\sigma$ variation of experimental relic density data.}\label{dm1}
\end{figure}
\begin{table}[h]
	\begin{tabular}{|c|c|c|c|c|c|p{4cm}|}
		\hline
		BMP&$M_{DM}$ (GeV)&$M_H$ (GeV)&$v_{\chi}$ (GeV)&$\cos\alpha$&$\Omega_{DM}h^2$& Processes\\
		\hline
		\hline
		I&65.4&400&1000&0.95&0.119&$\chi^I\chi^I\rightarrow b\bar{b}$(67\%) 
		$\chi^I\chi^I\rightarrow W^+W^-(22\%)$\\
		\hline
		II&200&208&1000&0.95&0.118&$\chi^I\chi^I\rightarrow W^+W^-$(40\%)
		$\chi^I\chi^I\rightarrow HH~(32\%)$ $\chi^I\chi^I\rightarrow ZZ~(19\%)$\\
		\hline
		III&800&200&1000&0.95&0.121&$\chi^I\chi^I\rightarrow SS(79\%)$
		$\chi^I\chi^I\rightarrow HS(16\%)$\\
		\hline	\end{tabular}
	\caption{Benchmark points for various processes at different dark matter mass and $S$ fermion mass with relic density for $v_{\chi}=1000$ GeV.}\label{bmpt1}
\end{table}

\begin{table}[h]
	\begin{tabular}{|c|c|c|c|c|c|p{4cm}|}
		\hline
		BMP&$M_{DM}$ (GeV)&$M_H$ (GeV)&$v_{\chi}$ (GeV)&$\cos\alpha$&$\Omega_{DM}h^2$& Processes\\
		\hline
		\hline
		I&53.06&400&1200&0.95&0.121&$\chi^I\chi^I\rightarrow b\bar{b}$(77\%) 
		$\chi^I\chi^I\rightarrow W^+W^-(12\%)$\\
		\hline
		II&200&218&1200&0.95&0.121&$\chi^I\chi^I\rightarrow W^+W^-$(41\%)
		$\chi^I\chi^I\rightarrow HH~(30\%)$ $\chi^I\chi^I\rightarrow ZZ~(19\%)$\\
		\hline
		III&800&397&1200&0.95&0.118&$\chi^I\chi^I\rightarrow SS(30\%)$
		$\chi^I\chi^I\rightarrow W^+W^-(29\%)$
		$\chi^I\chi^I\rightarrow HH(21\%)$
		$\chi^I\chi^I\rightarrow ZZ(14\%)$\\
		\hline	\end{tabular}
	\caption{Benchmark points for various processes at different dark matter mass and $S$ fermion mass with relic density for $v_{\chi}=1200$ GeV.}\label{bmpt2}
\end{table}

\begin{table}[h]
	\begin{tabular}{|c|c|c|c|c|c|p{4cm}|}
		\hline
		BMP&$M_{DM}$ (GeV)&$M_H$ (GeV)&$v_{\chi}$ (GeV)&$\cos\alpha$&$\Omega_{DM}h^2$& Processes\\
		\hline
		\hline
		I&53.2&380&1400&0.95&0.122&$\chi^I\chi^I\rightarrow b\bar{b}$(77\%) 
		$\chi^I\chi^I\rightarrow W^+W^-(12\%)$\\
		\hline
		II&200&228&1400&0.95&0.118&$\chi^I\chi^I\rightarrow W^+W^-$(41\%)
		$\chi^I\chi^I\rightarrow HH~(29\%)$ $\chi^I\chi^I\rightarrow ZZ~(19\%)$\\
		\hline
		III&800&450&1400&0.95&0.118&$\chi^I\chi^I\rightarrow W^+W^-(39\%)$
		$\chi^I\chi^I\rightarrow HH(24\%)$
		$\chi^I\chi^I\rightarrow ZZ(19\%)$
		$\chi^I\chi^I\rightarrow SS(14\%)$\\
		\hline	\end{tabular}
	\caption{Benchmark points for various processes at different dark matter mass and $S$ fermion mass with relic density for $v_{\chi}=1400$ GeV.}\label{bmpt3}
\end{table}

\begin{table}[h]
	\begin{tabular}{|c|c|c|c|c|c|p{4cm}|}
		\hline
		BMP&$M_{DM}$ (GeV)&$M_H$ (GeV)&$v_{\chi}$ (GeV)&$\cos\alpha$&$\Omega_{DM}h^2$& Processes\\
		\hline
		\hline
		I&54.7&800&3000&0.95&0.119&$\chi^I\chi^I\rightarrow b\bar{b}$(76\%) 
		$\chi^I\chi^I\rightarrow W^+W^-(13\%)$\\
		\hline
		II&134&800&3000&0.95&0.120&$\chi^I\chi^I\rightarrow W^+W^-$(50\%)
		$\chi^I\chi^I\rightarrow HH~(29\%)$ $\chi^I\chi^I\rightarrow ZZ~(21\%)$\\
		\hline
		III&940&800&4000&0.95&0.119&$\chi^I\chi^I\rightarrow W^+W^-(48\%)$
		$\chi^I\chi^I\rightarrow ZZ(24\%)$
		$\chi^I\chi^I\rightarrow HH(23\%)$\\
		\hline	\end{tabular}
	\caption{Benchmark points for various processes at different dark matter mass and $S$ fermion mass with relic density for $v_{\chi}>3000$ GeV.}\label{bmpt4}
\end{table}
The coupling strength for the interaction $\chi^I\chi^Ihh(HH)$ is $\lambda_2\cos\alpha\sin\alpha$ and the Higgs portal couplings are $g_{\chi^I\chi^Ih}=\frac{\sin\alpha(M^2_{DM}+M_h^2)}{v_{\chi}}$ and $g_{\chi^I\chi^IH}=-\frac{\cos\alpha(M_{DM}^2+M_H^2)}{v_{\chi}}$. It is clear that the dark matter relic density and direct detection cross-section mainly depend on $M_{DM}$, VEVs ($v,v_{\chi}$), mixing angle $\alpha$ and other quartic couplings $\lambda$. Hence, these are depending on the masses of the scalar particles too. In this analysis we will keep the mixing angle fixed at $\cos\alpha=0.95$ and vary other parameters, mainly, $v_{\chi},M_H$ and $m_{DM}$. The $\cos\alpha=0.95$ ensures the Higgs signal strength within the experimental limits \cite{Sirunyan:2018ouh}. In the fig. \ref{dm1}, we have varied the DM mass, $M_{DM}$ along x-axis and plotted the corresponding relic density in the y-axis. We keep fixed $M_H$ at 400 GeV and 800 GeV in the respective figures. For three different values of singlet scalar VEVs $v_{\chi}=1000,1200,1400$ GeV, we have carried out the whole DM analysis. The blue band indicates the relic density band at current 3$\sigma$, $\Omega_{DM}h^2=0.1198\pm0.0026$ \cite{Aghanim:2018eyx}. One can see from these figures that the relic density in this model can be obtain near $M_{DM}\sim M_h/2$ region. This is the Higgs ($h$) resonance region, we need vary small Higgs portal coupling to get the relic density. The other coupling produces a very large $\langle\sigma v\rangle$, hence, a depletion region is occurred near $M_{DM}\sim M_h/2$. In this region $\chi^I\chi^I\rightarrow h b\bar{b}$ is the dominating annihilation process. Near $|M_{DM}-M_h|<3$ GeV region, the cross section $\langle\sigma v\rangle$ remain small due to the small $h$-portal coupling, hence we get very large relic density ($\Omega_{DM}h^2\propto\frac{1}{\langle \sigma v\rangle}$). It is to be noted that the Higgs ($H$) portal coupling, {\it i.e.}, $H$-mediated diagram has a tiny effect in this region. After $M_{DM}\simeq70$ GeV, $\chi^I\chi^I\rightarrow h\rightarrow VV^*~(V=W,Z)$ starts dominates over other annihilation channels ($V^*$ stands for virtual gauge boson). $\chi^I\chi^I\rightarrow h\rightarrow WW$ become effective at $M_{DM}>M_W$, whereas $\chi^I\chi^I\rightarrow h\rightarrow ZZ$ become effective at $M_{DM}>M_Z$. 

The effective annihilation cross-section becomes large at $M_{DM}>70$ GeV; hence the relic density becomes small. One can understand these effects from the fig. \ref{dm1}. The relic density $\Omega_{DM}h^2$ falls again near $M_{DM}\sim M_H/2$ where $\sigma(\chi^I\chi^I\rightarrow H\rightarrow XX(X=W,Z,h))$ cross-section become large in this $H$-resonance region. The relic density is again starting to increasing after $M_H/2$ with the mass of DM as $\Omega_{DM}h^2\propto M_{DM}$ too. In the left figure of \ref{dm1}, we consider small $H$-scalar mass $M_H=400$ GeV, hence, the Higgs portal coupling remain small as compared to $M_H=800$ GeV. Also the larger $v_{\chi}=1200$ and 1400 GeV gives suppression to provide exact relic density for $M_H=400$ GeV. However, we unable to achieve relic for $M_H=800$ GeV in the high mass region. We can get the exact relic density for $M_H=800$ GeV for very large $v_{\chi}>3000$ GeV. We have presented various benchmark points allowing the current relic density value and present direct detection data with corresponding contributions in tables \ref{bmpt1}-\ref{bmpt4} (DM annihilation processes with contributions below 10\% are not shown in the chart.).
% We have also presented the corresponding contribution to the same table.
%The observed relic density through annihilation is primarily depends upon the DM mass, the scalar couplings ($\lambda_1,\lambda_2,\lambda_3$) and the mixing angle $\alpha$. We have scanned DM mass from 5 GeV to 1000 GeV for three different set of VEVs for the singlet scalar. We kept fixed mass for $S$ fermion, $M_H=400$ GeV throughout our calculation. Dark matter mass below 48 GeV are ruled out by the direct detection constrains, invisible Higgs decay width including the Fermi-LAT bounds. In our study, we get a lower bound of the DM mass at $\sim49$ GeV. Few benchmark points are discussed in following tables to make the situation more understandable for readers. 

\section{Summary}\label{6sec5}
In this chapter, we have explore the possibility of five-zero texture in the active-sterile mixing matrix under the framework of the minimal extended seesaw (MES). In the $(4\times4)$ mixing matrix, elements in the fourth column are restricted to be non zero; hence, we have imposed zeros only in the active neutrino block. There are six possible structures with five zeros, and among them, only one is allowed ($T_5$) by the current oscillation data in the normal hierarchy mass ordering. There is a broken $\mu-\tau$ symmetry as $M^{\nu}_{\mu4}\ne M^{\nu}_{\tau4}$; hence, non-zero reactor mixing angle ($\theta_{13}$) can also be achieved from this structure, which we skip in our study. Here, we have constructed the desirable mass matrices with discrete flavour symmetries like $A_4,Z_3$ and $Z_4$. In addition to the SM particle content, three RH neutrinos, a single fermion and an additional low-scaled Higgs doublet are considered. This additional Higgs get VEV $via$ soft breaking mass term. Two $A_4$ triplet flavons $\psi_1$ and $\psi_2$ were associated with the diagonal charged lepton and Dirac mass matrix generation. Singlet flavons $\eta$ and $\chi$ were related to the Majorana and sterile mass generation, respectively. The flavon $\chi$ is considered to be a complex singlet, and the imaginary component of this ($\chi^I$) will behave as a dark matter candidate in this study.

The active neutrino part is skipped here, and we mainly focused on active-sterile bounds from the five zero texture. Notable bounds on the sterile mass and the active-sterile mixing angles are obtained in this study. In comparison with the global 3$\sigma$ results, we get sterile mass between $0.1\le m_4^2 (\text{eV}^2)\le3$ and other mixing angles, as shown in table \ref{com1}, which agrees with current experimental results. We expected to see verification/falsification of these model bounds from texture-zeros in future experiments.

A nearly degenerate mass pattern for the RH neutrinos have considered in such a way that they can exhibit resonant leptogenesis at the TeV scale. For successful RL leptogenesis, the mass splitting of the RH neutrinos should be of the order of the decay width of the particle, and we choose $\frac{M_{N_3}}{M_{N_1}}-1=10^{-11}$ in our study. Quantum approaches of the Boltzmann equation are used and solved to see the particles' evolution in out-of-equilibrium condition. We have included flavour effects in this study, and with our choice of mass, we get an equal contribution from each charged lepton flavours in the final lepton asymmetry. We also have checked bound from baryogenesis on the Dirac CP phase ($\delta$) as well as the lightest neutrino mass and we get $\delta$ around $180^{\circ}$ and $330^{\circ}$ while $m_{lightest}$ on 0.069 eV with current best fit value $\eta_B=6.1\times10^{-10}$.  

With the bounds obtained from texture zero, baryogenesis study and available data from global fit results for light neutrino parameters, we have studied sterile neutrino influence in the effective mass calculation. A large enhanced region from sterile contribution is obtained in comparison to the SM contribution. The sterile contribution goes beyond the current upper bound of the effective mass obtained from various experiments, which give support in favour of minimal active-sterile mixing angle in future works.

We have studied the allowed parameter space of the model, taking into account various theoretical bounds for dark matter mass $10$ GeV to 1000 GeV. The imaginary part of the complex singlet $\chi$ serves as a potential dark matter candidate in this study. Due to the choice of the mass spectrum of the particles, only the annihilation channels are contributing to the relic density through Higgs portal coupling ($h/H$). The Higgs portal mixing angle $\alpha$, $M_H$ and the VEV of $\chi$ ($v_{\chi}$) and the quartic couplings play a significant role in dark matter analysis. We have chosen three sets of $v_{\chi}$'s and two sets of $M_H$s precisely to visualize various dark matter mass regions satisfying current relic abundance. In the lower dark matter mass region ($\sim M_h/2$), both the cases satisfy relic abundance due to the Higgs resonance. However, in the high mass regions, we get satisfying relic only in the $M_H=400$ GeV case due to the small Higgs suppression. In the $M_H=800$ GeV case, a very large VEV of $\chi$ does satisfy the current relic abundance value. 

As a final word, neutrino mass generation does not directly connect to the dark matter and matter asymmetry studies under the tree-level seesaw framework. However, with flavour symmetry, the complex singlet scalar $\chi$ and the RH neutrinos are the bridges to connect these sectors under the same roof. We can see how the VEV $v_{\chi}$ controls the dark matter mass throughout various regimes as an immediate consequence of flavour symmetries.  
% Chapter Template
\begin{savequote}[1\linewidth]
\normalsize	``It is always important to know when something has reached its end. Closing circles, shutting doors, finishing chapters, it doesn't matter what we call it; what matters is to leave in the past those moments in life that are over.''
	\qauthor{\Large\it Paulo Coelho (1947-- Present)}
\end{savequote}
\chapter{Conclusion and Future Plan} % Main chapter title
\vspace{-2.1cm}

%{\it ``After five years' work I allowed myself to speculate on the subject, and drew up some short notes; these I enlarged in 1844 into a sketch of the conclusions, which then seemed to me probable: from that period to the present day I have steadily pursued the same object. I hope that I may be excused for entering on these personal details, as I give them to show that I have not been hasty in coming to a decision."---{ \Large Charles Darwin}}

\label{Chapter6} % Change X to a consecutive number; for referencing this chapter elsewhere, use \ref{ChapterX}

\lhead{Chapter 6. \emph{Conclusion and future plan}} % Change X to a consecutive number; this is for the header on each page - perhaps a shortened title

%----------------------------------------------------------------------------------------
%	SECTION 1
%----------------------------------------------------------------------------------------
\section*{}
In this chapter, we have presented the overall conclusion that can be drawn from our phenomenological works. This thesis deals with neutrino mass models and their connections with $new~physics$ such as baryogenesis, dark matter, extra generation of neutrino and neutrino-less double beta decay ($0\nu\beta\beta$). Discrete flavour symmetry $A_4$ along with $Z_n~ (n\ge2)$ are extensively used to construct desired mass structures. We have added an extra generation of sterile neutrino with mass $m_S\sim\mathcal{O}$(eV-$keV$). To execute active and sterile masses, minimal extended seesaw has been used widely throughout this thesis. Active-sterile mixing patterns are studied for sterile mass in eV as well as in the $keV$ scale. Contribution of sterile neutrino for a wider mass range ($m_S$[eV-$keV$]) in $0\nu\beta\beta$ effective mass are explored in great details. Baryogenesis $via$ thermal and resonant leptogenesis is also studied. We have proposed fermion DM ($keV$ sterile neutrino), singlet, and doublet (real/complex) scalar DM candidates in our work in dark matter study.
\section{Conclusion}
In the following, we summarize and present significant conclusions drawn from our last four chapters, respectively.

In {\bf Chapter 2}, we have investigated the extension of low scale SM type-I seesaw $, i.e.,$ the minimal extended seesaw, which restricts active neutrino masses within the sub-eV scale and generates an eV scale light sterile neutrino. In this chapter, $A_4$ based flavour model is extensively studied along with a discrete Abelian symmetry $Z_4$ and $Z_3$ to construct the desired Yukawa coupling matrices. Under this MES framework, the Dirac ($M_D$) and Majorana ($M_R$) mass matrices are $(3\times3)$ complex matrix. The Majorana mass matrix $M_R$ arises due to the coupling of right-handed neutrinos with the anti-neutrinos with non-degenerate eigenvalues. Three independent set of singlet $S_i$ (where $i=1,2,3$) are considered which couples with the right-handed neutrinos ($\nu_{Ri};i=1,2,3$) and produces a singled row  $(1\times 3)$ $ M_S$ matrix with one non-zero entry. We have addressed the non-zero reactor mixing angle with a detailed discussion on VEV alignment of the flavon fields discussed under the light of flavour symmetry within this MES framework. Three separate cases are carried out for both NH and IH for three $M_S$ structures. Within the active neutrino mass matrix, the common $\mu-\tau$ symmetry is broken along with $\theta_{13}\neq 0$ by adding a new matrix ($M_P$) to the Dirac mass matrix. This $M_P$ matrix significantly impacts the reactor mixing angle ($\theta_{13}$) and active-sterile mixing pattern. This model also predicted the favourable structure of the sterile mass matrix, giving the proper octant of atmospheric mixing angle. 

In {\bf Chapter 3}, we have studied a modified  version of neutrino dark matter model from previous chapter, where a $A_4$ based flavour model along with $Z_4$ discrete symmetry to establish tiny active neutrino mass. Along with this, the generation of non-zero reactor mixing angle ($\theta_{13}$) and simultaneously carried out multi Higgs doublet framework where one of the lightest odd particles behaves as DM candidate. Both scalar and fermion sectors are discussed in great detail. Three sets of SM like Higgs doublets are considered, where two of them acquire some VEV after EWSB and take part in the fermion sector, particularly in tiny neutrino mass generation. On the other hand, the third Higgs doublet does not acquire any VEV due to the additional $Z_4$ symmetry. 
As a result, the lightest odd particle becomes a viable candidate for dark matter in our model.\\
In this minimal extended version of the type-I seesaw, we successfully achieved the non-zero $\theta_{13}$ by adding a perturbation in the Dirac mass matrix for both the mass orderings.
The involvement of the high scaled VEVs of $A_4$ singlet flavons $\xi$ and $\xi^{\prime}$ ensure the $B-L$ breaking within our framework, which motivates us to study baryon asymmetry of the Universe within this framework. As the RH masses are considered in non-degenerate fashion, we have successfully produced desired lepton asymmetry (with anomalous violation of $B+L$ due to chiral anomaly), which eventually is converted to baryon asymmetry by the $sphaleron$ process.
The influence of the Higgs doublets ($\phi_1,\phi_2$) can be seen both in fermion and the scalar sector. In the fermion sector, the Higgs doublets' involvement is related to the model parameters $via$ the Yukawa couplings. Relations between the constrained Dirac CP-phase and satisfied baryogenesis results for two different Yukawa couplings are shown, which related via two Higgs' VEV. On the other hand, in the scalar sector, a large and new DM mass region in the parameter spaces is obtained due to the presence of other heavy particles. 
We can successfully stretch down the  DM mass at limit up to 42.2 GeV, satisfying all current bounds from various constraints, which is entirely new in inert Higgs doublet case. 

{\bf Chapter 4} is a study of $keV$ scaled sterile neutrino and related phenomenology. The MES mechanism has been analyzed, considering a single flavour sterile neutrino in $keV$ scale. This model has been used to study the connection between effective mass in neutrinoless double beta decay ($0\nu\beta\beta$) in a broader range of sterile neutrino mass, simultaneously addressing the possibility of $keV$ scale sterile neutrino as dark matter particle. We have constructed a model, based on $A_4$ flavour model with discrete $Z_4\times Z_3$ charges. Presence of an extra heavy sterile flavour has a significant impact on effective neutrino mass. One can find a broader effective mass range in the active-sterile case than the active neutrino case.  Normal hierarchy (NH) was more favourable than the inverted hierarchy (IH) mode for $0\nu\beta\beta$ in this MES framework. Consequential bound on active-sterile mixing angle is obtained for future sensitivity in effective mass, which restricts the upper bound on the mixing element up to $10^{-4}$ for $|\theta_{S}|^2$.\\
BAU is satisfied in this model framework, and NH shows more efficient in producing the observed matter-antimatter density than IH pattern. This model also successfully correlate $0\nu\beta\beta$ with BAU result. Projection of BAU with sterile mass and effective mass in sterile neutrino presence gives an unsatisfactory remark while observing IH.\\
Dark matter analysis results from decay width and relic abundance restrict sterile neutrino mass within few $keV$ to behave as dark matter. Among different bounds for the sterile neutrino's thermal relic mass, very few results are consistent with X-ray observations. Lyman-$\alpha$ forest of high resolution quasar spectra with hydrodynamical N-body simulations gives bounds ranging from $m_S \geq 1.8$ $keV$ to $m_S\geq 3.3$ $keV$ \cite{Viel:2006kd,Seljak:2006qw,Viel:2013apy}. Regardless, these bounds may vary depending upon various uncertainties affecting the constraints \cite{Schultz:2014eia}. Within MES framework, NH predicts sterile mass range from ($1-3$ $keV $) and IH results for relic abundance gives mass up to $10$ $keV$ while the decay width constraints the mass within 3 $keV$.  Hence, from these results, we conclude that sterile neutrino as a dark matter in minimal extended seesaw is still an unsettled aspect with current bounds on hand. A deeper discussion with new bounds on $keV$ sterile neutrino may resolve these issues, which is left for future studies. In conclusion, although, results on $keV$ sterile neutrino as a dark matter candidate are still on the verge of uncertainty within MES framework, we keep an optimistic hope to get better bounds from future experiments which may establish the same within MES.

In {\bf Chapter 5}, we have explored the possibility of five zero textures in the active-sterile mixing matrix under the minimal extended seesaw (MES) framework. In the $(4\times4)$ mixing matrix, the fourth column elements are restricted to be non zero; thus, we have imposed zeros only in the active neutrino block. There are six possible structures with five zeros, and among them, only one is allowed ($T_5$) by the current oscillation data in the normal hierarchy mass ordering. To study neutrino dark matter simultaneously, we have constructed the desirable mass matrices with discrete flavour symmetries like $A_4,Z_3$ and $Z_4$. In addition to the SM particle content three RH neutrinos, a single fermion and an additional low-scaled Higgs doublet are considered. This additional Higgs get VEV $via$ soft breaking mass term. Singlet flavons $\eta$ and $\chi$ were related to the Majorana and sterile mass generation, respectively. The flavon $\chi$ is considered to be a complex singlet, and the imaginary component of this ($\chi^I$) will behave as a dark matter candidate in this study.\\
Active neutrino part is skipped in that chapter, and we mainly focused on active-sterile bounds from the five zero texture. Notable bounds on the sterile mass and the active-sterile mixing angles are obtained in this study. We expected to see verification/falsification of these model bounds from texture-zeros in future experiments.\\
A nearly degenerate mass pattern for the RH neutrinos have considered in such a way that they can exhibit resonant leptogenesis at TeV scale. For successful RL leptogenesis, the mass splitting of the RH neutrinos should be of the order of the decay width of the particle, and we choose $\frac{M_{N_3}}{M_{N_1}}-1=10^{-11}$ in our study. Quantum approach of the Boltzmann equations are used and solved to see the evolution of the particles in out-of-equilibrium condition. We have included flavour effects in this study and with our choice of mass, we get an equal contribution from each charged lepton flavours in the final lepton asymmetry. We have also used bound from baryogenesis on the Dirac CP phase ($\delta$) as well as the lightest neutrino mass and we get $\delta$ around $180^{\circ}$ and $330^{\circ}$ while $m_{lightest}$ on 0.069 eV with current best fit value $\eta_B=6.1\times10^{-10}$. \\
With the bounds obtained from texture zero, baryogenesis study and available data from global fit results for light neutrino parameters, we study sterile neutrino influence in the effective mass calculation. A large enhanced region from sterile contribution has obtained in comparison to the SM contribution. The sterile contribution goes beyond the current upper bound of the effective mass obtained from various experiments, which give support in favour of minimal active-sterile mixing angle in future works.\\
We have also studied the model's allowed parameter space, taking into account various theoretical bounds for dark matter mass $10$ GeV to 1000 GeV. The imaginary part of the complex singlet $\chi$ serves as a potential dark matter candidate in this model. Due to the choice of the mass spectrum of the particles, only the annihilation channels are contributing to the relic density through Higgs portal coupling ($h/H$). The Higgs portal mixing angle $\alpha$, $M_H$ and the VEV of $\chi$ ($v_{\chi}$) and the quartic couplings play a significant role in dark matter analysis. We have chosen three sets of $v_{\chi}$'s and two sets of $M_H$s precisely to visualize various dark matter mass regions satisfying current relic abundance. In the lower dark matter mass region ($\sim M_h/2$), both the cases satisfy relic abundance due to the Higgs resonance. However, in the High mass regions, we get satisfying relic only in $M_H=400$ GeV case due to the small Higgs suppression. In the $M_H=800$ GeV case, very large VEV of $\chi$ does satisfy current relic abundance value. 
%As a final word, neutrino mass generation does not directly connect to the dark matter and matter asymmetry studies under the tree-level seesaw framework. However, with flavour symmetry, the complex singlet scalar $\chi$ and the RH neutrinos are the bridges to connect these sectors under the same roof. We can see how the VEV $v_{\chi}$ controls the dark matter mass throughout various regimes as an immediate consequence of flavour symmetries. 
\section{Future plan}
The works carried out in this thesis can be further extended into many directions. In active-sterile phenomenology, we can work on inverse and radiative seesaw frameworks with some additional symmetries. $keV$ sterile neutrino as a DM candidate is still in doubt, and if possible, we can work on some new models, where those uncertainties may resolve. Lepton number violation (LNV) and lepton flavour violation (LNV) processes can also be studied alongside those models. We can also probe our low scale models into collider studies.

We can further propose new dark matter candidates in the dark matter sector and explore new viable DM regions in various model studies. We may probe those results in future studies to explain the excess of electron recoil events by the recent XENON1T experiment and future collider experiments.
%----------------------------------------------------------------------------------------
%	THESIS CONTENT - APPENDICES
%----------------------------------------------------------------------------------------

\addtocontents{toc}{\vspace{2em}} % Add a gap in the Contents, for aesthetics

\appendix % Cue to tell LaTeX that the following 'chapters' are Appendices

% Include the appendices of the thesis as separate files from the Appendices folder
% Uncomment the lines as you write the Appendices

% Appendix Template

\chapter{Product rules and Vacuum Alignment under $A_4$} % Main appendix title

\label{appn1} % Change X to a consecutive letter; for referencing this appendix elsewhere, use \ref{AppendixX}

\lhead{}\label{a4p} % \section{Product rules and Vacuum Alignment under $A_4$}\label{a4p}

Here we will investigate the problem of achieving the VEV alignment of the two flavons. We will take the minimization potential and try to solve them simultaneously. In general, total potential will be consisting of the contribution from the field $\zeta$ and $\varphi$ and their mutual interaction. However, interaction among the fields are forbidden by the discrete charges.
The total potential will be,
\begin{equation}
	V=V(\zeta)+V(\varphi)+V_{int.},
\end{equation}
with,
$$ V(\zeta)=-m_1^2(\zeta^{\dagger}\zeta)+\lambda_{1}(\zeta^{\dagger}\zeta)^2$$ and
$$V(\varphi)=-m_2^2(\varphi^{\dagger}\varphi)+\lambda_{1}(\varphi^{\dagger}\varphi)^2$$
$V_{int.}$ term will not appear in our case.\\
The triplet fermions will have the form,
\begin{equation}
	\begin{split}
		&\langle \zeta\rangle = (\zeta_{1},\zeta_{2},\zeta_{3}),\\
		&\langle \varphi\rangle=(\varphi_{1},\varphi_{2},\varphi_3)
	\end{split}
\end{equation}
Using the $A_4$ product rules from equation \eqref{a4r} , the potential for $\varphi$ will take the form,
\begin{equation}
	\begin{split}
		V(\varphi)=&-\mu_2^2(\varphi_{1}^{\dagger}\varphi_{1}+\varphi_2^{\dagger}\varphi_3+\varphi_3^{\dagger}\varphi_2)\\
		&+\lambda_2[(\varphi_{1}^{\dagger}\varphi_{1}+\varphi_2^{\dagger}\varphi_3+\varphi_3^{\dagger}\varphi_2)^2+(\varphi_{3}^{\dagger}\varphi_{3}+\varphi_2^{\dagger}\varphi_1+\varphi_1^{\dagger}\varphi_2)\times(\varphi_{2}^{\dagger}\varphi_{2}+\varphi_1^{\dagger}\varphi_3+\varphi_3^{\dagger}\varphi_1)\\
		&+(2\varphi_{1}^{\dagger}\varphi_{1}-\varphi_2^{\dagger}\varphi_3-\varphi_3^{\dagger}\varphi_2)^2+2(2\varphi_{3}^{\dagger}\varphi_{3}-\varphi_1^{\dagger}\varphi_2-\varphi_2^{\dagger}\varphi_1)\times(2\varphi_{2}^{\dagger}\varphi_{2}-\varphi_3^{\dagger}\varphi_1-\varphi_1^{\dagger}\varphi_3)]\\
	\end{split}
\end{equation}
Taking the derivative $w.r.t.$ $\varphi_{1}, \varphi_{2}$ and $\varphi_3$ and equate it to zero gives us the minimization condition for the potential. Three equations are solved simultaneously and various solutions are found out as,
\begin{enumerate}
	\item $\varphi_1\rightarrow\frac{\mu_2}{\sqrt{10\lambda_2}},~\varphi_2\rightarrow0,~\varphi_3\rightarrow0\Rightarrow\langle\varphi\rangle=\frac{\mu_2}{\sqrt{10\lambda_2}}(1,0,0)$;
	\item $\varphi_1\rightarrow0,~\varphi_2\rightarrow\frac{\mu_2}{\sqrt{10\lambda_2}},~\varphi_3\rightarrow0\Rightarrow\langle\varphi\rangle=\frac{\mu_2}{\sqrt{10\lambda_2}}(0,1,0)$;
	\item $\varphi_1\rightarrow\frac{\mu_2}{2\sqrt{3\lambda_2}},~\varphi_2\rightarrow\frac{\mu_2}{2\sqrt{3\lambda_2}},~\varphi_3\rightarrow\frac{\mu_2}{2\sqrt{3\lambda_2}}\Rightarrow\langle\varphi\rangle=\frac{\mu_2}{2\sqrt{3\lambda_2}}(1,1,1)$;
	\item $\varphi_1\rightarrow\frac{2\mu_2}{\sqrt{51\lambda_2}},~\varphi_2\rightarrow-\frac{\mu_2}{\sqrt{51\lambda_2}},~\varphi_2\rightarrow-\frac{\mu_2}{\sqrt{51\lambda_2}}\Rightarrow\langle\varphi\rangle=\frac{\mu_2}{\sqrt{51\lambda_2}}(2,-1,-1)$.
\end{enumerate}
Similar solutions will be generated for the $\zeta$ field also.
% Appendix Template
\chapter{The Boltzmann equation for resonant leptogenesis}

\label{appen2} % Change X to a consecutive letter; for referencing this appendix elsewhere, use \ref{AppendixX}
\lhead{} 

Baryon asymmetry is generated when the RH neutrino decays out of equilibrium, and the abundance of RH neutrino along with the lepton generations are determined by Boltzmann equation. We use the quantum approach of the Boltzmann equation, which is based on the Schwinger-Keldysh Closed Time Path (CTP) formulation \cite{Iso:2013lba}. Within this formulation, one needs to derive quantum field-theoretic analogues of the Boltzmann equations, known as Kadanoff-Baym (KB) equations \cite{kadanoff2018quantum} describing the non-equilibrium time-evolution of the two-point correlation functions. These time integral of KB equations ensure non-Markovian which allows studying the history of the system as a memory effect. We followed the approaches by \cite{Dev:2014laa, DeSimone:2007edo, Pilaftsis:2003gt, Deppisch:2010fr} to worked out the flavor effect in the RL scenario.

{\bf The collision and scattering terms:}
The BEs in \eqref{be1} includes the collision processes like $N_j\rightarrow L_i\phi$ as well as $L_k\phi\leftrightarrow L_i\phi$ and $L_k\phi\leftrightarrow L_i^c\phi^{\dagger}$ scattering processes, which are defined as \cite{Pilaftsis:2003gt}
\begin{eqnarray}
	\gamma_{L\phi}^{N_j}&\equiv& \sum_{k=e,\mu,\tau}\Big[\gamma(N_j\rightarrow L_k\phi)+\gamma(N_j\rightarrow L_k^c\phi^{\dagger})\Big],\\
	\gamma_{L_i\phi}^{L_k\phi}&\equiv&\gamma(L_k\phi\rightarrow L_i\phi)+\gamma(L_k^c\phi^{\dagger}\rightarrow L^c_i\phi^{\dagger}),\\
	\gamma_{L_i^c\phi^{\dagger}}^{L_k\phi}&\equiv&\gamma(L_k\phi\rightarrow L_i^c\phi^{\dagger})+\gamma(L_k^c\phi^{\dagger}\rightarrow L_i\phi).
\end{eqnarray}
Including the contributions from narrow width approximation (NWA) as well as real intermediate states (RISs) from \cite{Pilaftsis:2003gt}, these collision terms are explained as,
\begin{eqnarray}
	&\gamma_{L\phi}^{N_j}=&\frac{m_N^3}{\pi^2z}K_1(z)\Gamma_{N_j},\\
	&\gamma_{L_i\phi}^{L_k\phi}=&\sum_{\alpha,\beta}^{3}(\gamma_{L\phi}^{N_{\alpha}}+\gamma_{L\phi}^{N_{\beta}})\frac{2\big(\bar{Y^*_{i\alpha}}\bar{Y_{k\alpha}^{c*}}\bar{Y_{i\beta}}\bar{Y^{c*}_{k\beta}}+\bar{Y^{c*}_{i\alpha}}\bar{Y_{k\alpha}^{*}}\bar{Y_{i\beta}}\bar{Y^{c}_{k\beta}}\big)}{\Big[(\bar{Y^{\dagger}}\bar{Y})_{\alpha\alpha}+(\bar{Y^{c\dagger}}\bar{Y^c})_{\alpha\alpha}+(\bar{Y^{\dagger}}\bar{Y})_{\beta\beta}+(\bar{Y^{c\dagger}}\bar{Y^c})_{\beta\beta}\Big]^2}\nn\\
	&&\times\Big(1-2i\frac{M_{N_{\alpha}}-M_{N_{\beta}}}{\Gamma_{N_{\alpha}}+\Gamma_{N_{\beta}}}\Big)^{-1},\\
	&\gamma_{L_i^{c}\phi^{\dagger}}^{L_k\phi}=&\sum_{\alpha,\beta=1}^{3}(\gamma_{L\phi}^{N_{\alpha}}+\gamma_{L\phi}^{N_{\beta}})\frac{2\big(\bar{Y^*_{i\alpha}}\bar{Y_{k\alpha}^{*}}\bar{Y_{i\beta}}\bar{Y_{k\beta}}+\bar{Y^{c*}_{i\alpha}}\bar{Y_{k\alpha}^{*}}\bar{Y_{i\beta}^c}\bar{Y^{c}_{k\beta}}\big)}{\Big[(\bar{Y^{\dagger}}\bar{Y})_{\alpha\alpha}+(\bar{Y^{c\dagger}}\bar{Y^c})_{\alpha\alpha}+(\bar{Y^{\dagger}}\bar{Y})_{\beta\beta}+(\bar{Y^{c\dagger}}\bar{Y^c})_{\beta\beta}\Big]^2}\nn\\
	&&\times\Big(1-2i\frac{M_{N_{\alpha}}-M_{N_{\beta}}}{\Gamma_{N_{\alpha}}+\Gamma_{N_{\beta}}}\Big)^{-1}.
\end{eqnarray}
Finally the BEs of \eqref{be1} are rewrite in the form,
\begin{eqnarray}
	\nn&\frac{d\eta_{L_i}}{dz}=&\frac{z}{\eta_{\gamma}H(z=1)}\Big[\sum_{j=1}^{3}\big(\frac{\eta_{N_j}}{\eta_N^{eq}}-1\big)\epsilon_i\gamma_{L\phi}^{N_j}\\
	\nn&&-\frac{2}{3}\eta_{L_i}\Big\{\sum_{j=1}^{3}\gamma_{L\phi}^{N_j}B_{ij}+\sum_{k=e,\mu,\tau}\big(\gamma_{L_i^{c}\phi^{\dagger}}^{\prime L_i\phi}+\gamma_{L_k\phi}^{\prime L_i\phi}\big)\Big\}\\
	&&-\frac{2}{3}\sum_{j=1}^{3}\Big\{\eta_{L_k}\epsilon_{ii}\epsilon_{ik}\gamma_{L\phi}^{N_j}B_{ij}+\big(\gamma_{L_i^{c}\phi^{\dagger}}^{\prime L_i\phi}-\gamma_{L_k\phi}^{\prime L_i\phi}\big)\Big\}\Big].\label{be5}
\end{eqnarray}

Here, $\gamma_Y^{\prime X} =\gamma_Y^X-(\gamma_Y^X)_{\text{RIS}}$ denote the RIS=subtraction collision terms which are motivated from past studies \cite{Pilaftsis:2003gt, Deppisch:2010fr} and $B_{ij}$ are the branching rations,
\begin{eqnarray}
	B_{ij}=\frac{|\bar{Y}_{ij}|^2+|\bar{Y}_{ij}^c|^2}{(\bar{Y^{\dagger}\bar{Y}})_{ii}+(\bar{Y}^{c\dagger}\bar{Y^c})_{ii}}.
\end{eqnarray} 

\addtocontents{toc}{\vspace{2em}} % Add a gap in the Contents, for aesthetics

\backmatter

%----------------------------------------------------------------------------------------
%	BIBLIOGRAPHY
%----------------------------------------------------------------------------------------
%\nocite{*}
\label{Bibliography}
\lhead{\emph{Bibliography}} % Change the page header to say "Bibliography"
\bibliographystyle{utphys}
\bibliography{refs}
%\bibliography{Bibliography} % The references (bibliography) information are stored in the file named "Bibliography.bib"
\clearpage

\lhead{ }

%\clearpage

% Note: This thesis is based on the first four publications from the above list.
\end{document}